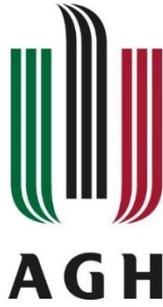

# AGH University of Science and Technology

Faculty of Computer Science, Electronics and Telecommunications

Department of Computer Science

DOCTORAL THESIS

# Methodology for development of scientific software and test frameworks in function of precision of the expected results

Tomasz Przedziński

Supervised by
dr. hab. Maciej Malawski

Co-supervised by
prof. dr hab. Zbigniew Wąs

Kraków, 18 November 2020

**Streszczenie**

Rozwój oprogramowania naukowego wiąże się z wieloma wyzwaniami wynikającymi z nieprzewidywalnej natury tego procesu oraz ze złożoności problemów, z którymi mierzą się środowiska naukowe. Celem wielu projektów naukowych jest poprawa precyzji wyników poprzedniego projektu. Niniejsza rozprawa skupia się na procesie rozwoju oprogramowania tych projektów. Prezentuje metodologię, która wyłoniła się przez lata rozwoju oprogramowania naukowego. Prezentowana metodologia bazuje na cyklu rozwoju oprogramowania, który prowadzi do nowej wersji oprogramowania z usprawnionym modelem naukowym i opisem matematycznym jak również z usprawnionym środowiskiem testów, które razem tworzą oprogramowania dające wyniki o wyższej precyzji.

Rozprawa obejmuje opis podstawowych konceptów i terminów związanych z fizyką wysokich energii potrzebnych, by zrozumieć naturę omawianych tematów. Zawiera również krótki opis różnych typów oprogramowania obecnego w fizyce eksperymentalnej oraz opisuje wyzwania związane z rozwojem oprogramowania naukowego. Pokazane jest, dlaczego najpopularniejsze metodologie rozwoju oprogramowania nie radzą sobie z tymi wyzwaniami.

Przedstawiony jest również proces rozwoju kilku projektów naukowych wraz z kluczowymi elementami tego procesu. Te przykłady pokazują, że rozwój kolejnych kamieni milowych tych projektów podąża za cyklem: usprawniania modelu fizycznego, opisu modelu fizycznego z użyciem matematycznego formalizmu, implementacji tego modelu z uwzględnieniem przybliżeń numerycznych, tworzenia struktury oprogramowania oraz dokumentowania i weryfikacji wyników.

Rozprawa opisuje historię kilku projektów, wliczając kluczowe decyzje podjęte podczas ich rozwoju oraz wskazując na których etapach rozwoju tych projektów zwiększona została precyzja ich wyników. Pokazana została relacja między podwyższoną precyzją wyników a zwiększoną złożonością testów i środowiska testowego tych projektów.

Szczególną uwagą objęty został temat testów oprogramowania naukowego. Opisane zostały testy wykonywane dla oprogramowania naukowego oraz ich taksonomia w zestawieniu z taksonomią innych testów oprogramowania. Przedstawione zostały użyteczne techniki testowania oprogramowania naukowego użyte w niektórych z prezentowanych projektów.

Rozprawa opiera się o doświadczenie zdobyte podczas rozwoju kilku narzędzi Monte Carlo stworzonych dla środowiska eksperymentów Fizyki Wysokich Energii. Autor tej rozprawy jest współautorem narzędzi opisywanych w tej rozprawie. Część z nich została z powodzeniem wdrożona w środowiska naukowe. Część stała się częścią analiz przeprowadzanych w eksperymentach wokół Wielkiego Zderzacza Hadronów. Analiza procesu rozwoju tych narzędzi może pomóc w estymacji wysiłku potrzebnego by usprawnić model i precyzję złożonych algorytmów. Jest to temat, którego popularność w środowiskach naukowych, jak i w komercyjnych zastosowaniach badawczo-rozwojowych, wciąż rośnie w związku ze stale rosnącą złożonością nowych problemów i rosnącym zapotrzebowaniem na wyniki o wyższej precyzji.

# Abstract

Scientific software development faces many challenges due to its complexity and unpredictable nature. Many scientific projects focus solely on improving the precision of the results of a previous project. This thesis focuses on the development process of these projects.

This thesis presents a methodology that has emerged over time during development of scientific software. It is based on the development cycle that leads to a new software version with refined mathematical and scientific design, as well as improved test framework, which in turn leads to improved precision of the results.

A short description of the physics background needed to understand the subjects presented in this thesis is included and the different types of software created for the physics experiments are outlined. The challenges related to the scientific software development are listed and the most common existing software development methodologies are matched against them to show in which aspects they fall short when used to manage the scientific software development process.

The development process of several scientific projects is described and its key elements analyzed. It is shown that the development of subsequent milestones of these projects follow the cycle of improving the physics model, describing the model using mathematical formalism, implementing the model with numerical approximations, creating the software framework, documenting and validating results.

The history of some of these projects is presented in relation to the increased precision of the results of these tools. Key decisions made during the development of these tools are identified. The relation between increased precision of the results and increased complexity of tests and test frameworks is also demonstrated based on these projects.

The subject of scientific software testing is addressed in great details and the taxonomy of the scientific software tests is presented in comparison to the existing taxonomy of software testing. Some useful testing techniques used in the development of the scientific software has been also presented.

The experience presented in this thesis is based on several Monte-Carlo tools created for the High Energy Physics (HEP) community co-authored by the author of this thesis. Some of these tools has been successfully introduced into the HEP community. Some of these tools gained large user base and are in active use by the community. Some of them are also part of analyses performed by experiments around Large Hadron Collider.

The analysis of the development process of these tools can help estimate the effort needed to improve the design and precision of complex algorithms. This subject becomes more and more relevant not only for scientific community but also for Research and Development teams in business environment as the complexity of their problems and demand for higher precision results grows.

*Acknowledgements*

*The process of writing this thesis, not unlike the development process of a scientific tool, spanned over many years and resulted in many intermediate versions; each an improvement over the previous one. My sincerest gratitude goes to prof. Zbigniew Wąs who endured this process with me guiding me along and helping me out with his vast knowledge and experience. Without him I would not have been able to work with such wide variety of specialists from different scientific communities and gain the experience needed to write this thesis. He also taught me the basics of particle physics without which I would not have been able to work with the scientific software. I'm grateful to prof. Wąs for all his patience towards his student.*

*I am deeply grateful to dr. Maciej Malawski for putting this thesis on the right tracks. His invaluable insight into the computing science aspects of the scientific software has greatly helped me formulate and present the core ideas behind this thesis. He has also helped me expand my knowledge on number of aspects related to this thesis. His guidance ultimately resulted in this thesis taking its current shape.*

*I would also like to extend my gratitude to prof. Elżbieta Richter-Wąs for introducing me to the scientific world and for her patronage over me through all these years. She has helped me gain experience in distributed computing on large computing clusters and has helped me in my collaboration with the community of physics experiments around Large Hadron Collider.*

*I would like to thank all of my collaborators with whom I had a chance to work and contribute to scientific community. Each of those encounters has provided me with exciting opportunities to learn something new.*

*Lastly, I would like to thank my friends for pushing me onwards, and my family, for their kind words of encouragement. I especially thank my wife, Magda, for her patience and for helping me survive many long days, evenings and nights in front of the screen for the sake of this thesis. My last thank you goes to my little girl, Danusia. Her birth was the last push I needed to complete this thesis.*

# Table of contents

# List of figures

# List of tables

# Listings

# Chapter 1.    Introduction

Development of scientific software, in disciplines such as High Energy Physics (HEP) is a lengthy process which may span over multiple decades. High Energy Physics, since at least 50 years, relies on massive numerical and in general computing algorithmic applications for simulation of physics processes. The evolution of software needs to match not only the increasing requirements coming from rapidly advancing development in mathematics, theory and experiments but it also has to follow the advances in Computer Science and Software Engineering. This translates into demand of high quality of precisely defined components. Often redefinition of the design of the project is necessary.

Use and development of Information Technology cannot be separated from physics, engineering, medical or financial projects. The demand for expertise in applied Computer Science becomes more and more apparent with the increasing complexity of problems tackled by specialties that make use of Computer Science. These are, for example, cognitive science, studies of computer vision, robotics, bioinformatics, computational chemistry or expert systems in finances, engineering and medicine. Such specialties require extensive cooperation between experts in different field of studies creating a complex system that requires proper communication and management of knowledge and competence.

> **Personal note:**
>
> As explained by Adam Kolawa, the founder of Parasoft Corporation [1], wealth of knowledge found when approaching the electrical engineering or physics problems can be used to form the strategies and principles of software management and development of complex systems minimizing the defects found throughout the whole development process. Adam Kolawa majored in Electrical Engineering at AGH in Cracow and in Physics at Jagiellonian University in Cracow. He later earned his Ph.D. in Theoretical Physics at Caltech and took part in creating the Caltech Cosmic Cube [2], arguably the most significant supercomputer system of the early 1980s. The experience taken from adapting the commodity hardware and an architecture suited for quantum chromodynamics computation into a very large-scale parallel computation unit has become the foundation of the innovative approach to software development called Automated Defect Prevention (ADP) [3].
>
> Since its creation in 1987, Parasoft has become one of the most successful corporation in Computer Science revenue introducing ADP principles to the projects of hundreds of corporations worldwide by developing software for static code analysis, data flow analysis, code coverage, requirements traceability and number of tasks related to testing and deploying the software on target platforms. The corporation holds patents for the cutting-edge memory error detection algorithms for C and C++ projects as well as service virtualization algorithms emulating the behavior of hard to access mainframes or third-party components.
>
> It is hard to overstate the influence of Parasoft's software on the modern IT industry. The practices and tools developed by this company helped to form

> the new paradigm of software development. Most of the biggest software and hardware developers, such as IBM, Intel, LG, HP, Sony, Samsung or Dell, to name a few, are all using its software to produce top-quality products. Adam Kolawa passed away in April 2011. During his life he became one of the most influential people in software development industry. His scientific carrier and following success in business industry is one of many examples proving that IT experience coming out of the collaboration with different fields of science is hard to ignore in business applications.
>
> My experience comes from similar environment. When analyzing the ADP practices after years of experience working for scientific communities, I have realized that many of its basic principles regard the same issues that I have encountered while working on these projects. Many of them would have helped me much more if I knew them earlier. This is one of the reasons why Computer Science practices based on the experience derived from the physics experiments is the key focus of my thesis.

## 1.1 Our experience

The projects mentioned in this thesis have been developed in collaboration with wide range of specialists from Switzerland, France, Germany, Canada, Russia, Italy, Australia or Japan; working in theoretical physics or experimental physics. It is this variety of different environments that allowed to create valuable tools that can benefit physics communities and can contribute to understanding of scientific software development process.

Some of the projects mentioned in this thesis are already completed and introduced to the physics community giving foundation for this thesis. Most notably, tools such as `Photos++` [4], `Tauola++` [5] and `TauSpinner` [6]. We use these tools often as an example in this thesis because the author of this thesis co-authored the first two tools since their adaptation to C++ environment and was one of the co-authors of the third tool from its very beginning. They serve as a good example because all of these tools found large user base in scientific community, as exemplified by the number of citations of the main publications related to these tools, and their development process, as well as their history, is very well known.

In this thesis we concentrate on our experience and perspective of us and our collaborators. This will unavoidably differ from the experience of other projects and other communities. However, due to the unpredictable nature of problems found in scientific projects, in some cases experience is the only available hint at what is the root cause of the problem at hand. That is why solutions presented in this thesis may offer a roadmap that could be considered when dealing with similar problems.

## 1.2 Challenges and constraints of scientific software development

Scientific software is expected to abide to a long list of restrains that come naturally due to the nature of the environment in which the project is developed and due to the nature of the project itself.

First restraints come with the inter-project dependencies as the developed product must cooperate with other products and must adjust to environments in which the product is intended to be used. The second puts emphasis on the validation and testing process. A scientific analysis cannot be conducted without a proper protocol and its results must pass an extensive validation procedure before they can be published, which forces the development team to allocate a lot of their time on the testing and validation of the tool results. These two aspects are the easiest to identify and to take into account during the development process. However, they are hardly the only ones.

One can hardly find a high-end product build without the use of knowledge and experience from the older products. A Monte Carlo generator validated and refined over years can become an invaluable asset; often a sole foundation of a new tool. The time and effort put into the design and development of such module makes its adaptation a cheaper and more reliable approach than the development of a new one. This, however, implies that the project must incorporate an already existing module. Depending on the age of the module this imposes small or severe impediments to the project design. To retain the full value of this module some of its parts must be treated as a black-box and remain intact while changes to some other parts should be minimal. This often leads to the use of unorthodox methods and legacy solutions forcing an approach that strongly values user goals over development goals. This makes the design of the product even more challenging.

The choice of appropriate team members for such project is imperative. The knowledge base and a set of skills of the team must cover all fields of expertise required to develop the final product. But, again, their cooperation requires compromises in terms of the development strategy. The bulk of the work on the project is focused not on the software or software framework made by Computer Scientists, but on the content provided by specialists from fields outside Computer Science.

When a team of software developers approaches a project they usually agree on a set of tools and a single approach. Each of them adapts their own tools and methods to this approach. It's easy for software developers to do so thanks to their experience. However, experts from other fields of science with little or no Computer Science background will have hard time adapting to any preset software environment. The ability of such experts to extend or modify the software without learning the tools or software engineering methods that he or she is not familiar with is imperative to the success of the whole project. While it normally seems counterintuitive to adapt the software and the methodology to team members, in case of scientific software this is often inevitable. This may result in an approach that may be considered improper from the perspective of the short-term activity but it is for ease of communication with experts from different fields that have to work with each other with minimal effort on their part.

> **Example:**
>
> At one stage of development of `TauSpinner` tool (presented later in Section 6.5) we started developing extensions with the cooperation of new team members. The tool was written in C but these experts were working on `MatLab` models and were most familiar with Fortran code so they tended to use `MatLab` to generate Fortran code.
>
> We discussed for a short while if the code should be switched to C for ease of importing into the project and to solve other limitations of the Fortran code, but we eventually decided to adapt the project so that it was able to use the Fortran code. This was because there was large amount of potential new models to be added to the project and it would be inconvenient for the experts to

> read and debug every one of them in C. On the other hand, it was fairly easy for computer scientists to create framework that allowed Fortran code to be added into the project, even though this required compromising on some of the aspects of the software architecture.

Learning new concepts always comes at a price of time and effort. Learning a technique outside of one's expertise should be required only if it is worthwhile. Therefore, whenever it is not necessary, the computer scientist should thrive to find a solution that does not force his or her team members to learn a new topic. Similarly, an expert from the physics department or an electronic engineer should help computer scientists tailor these parts of the system that require their cooperation. The task of dividing the knowledge between collaborators and recognizing what needs to be learned and what can be ignored is not trivial and such division is often impossible to fully define. They should, however, be performed naturally and on-the-flight to avoid wasting time and effort of the team members.

The project must also be designed with as much flexibility in mind as possible as it must accommodate for incomplete specification. In most cases the target user base cannot be fully explored at the development stage of the project. In many cases, even during development process, collaborators, beta-testers and users of the prototype versions of the product introduce new possible use cases and hint at new functionality that could expand the user base of the final product. The project must allow to easily explore these possibilities with minimal effort.

In most cases the end product will not be used standalone. It will have to be customized to meet specific end-user requests or will have to be extended by the users themselves. Sometimes even some of the more specialized modules or algorithms may need to be adapted by the end-users or business partners. Such modules should not be obscured by the project design. On the contrary, they should be well documented and accessible with the description of the details of their use to encourage such modifications.

Lastly, the deployment step of the project adds another layer of problems. Variety of target users increases number of target environments; access to most of which is either limited or unavailable. Inability to predict all possible use cases implies that the design of the project must take into account that the feedback from beta users is often the only way to flag an issue in specific application. This issue can come from improper use of the tool, an obscure use case not handled by testing framework, or from a widely different application of the tool that needs to be explored in detail. All of these possibilities must be carefully analyzed and properly handled.

To sum up, some of the challenges that a team working on scientific project must face are:

- Taking into account inter-project dependencies
- Creating extensive tests and validation frameworks
- Dealing with large amount of legacy code
- Providing software that can be easily modified by experts from fields other than computing science
- Dividing knowledge between team members
- Accommodating for incomplete specification and possible future use cases
- Encouraging modifications and documenting possible future extensions

All of these challenges put a lot of restrains on how the project can be developed and create a hard to solve problem from the software engineering standpoint. Sometimes the choice of the approach to building the software framework can become a make-or-break for the whole project as all of the aspects mentioned above impact the usability of the tool and, in consequence, the potential user base. These challenges are explored in more details further in this thesis in Section 3.1).

## 1.3 Estimating the development and testing effort

Effort, typically expressed in man-days (MD), is the key factor used for planning and budgeting a project. It is usually estimated by experts who base their judgement on historical projects and their own experience. However, each scientific project is unique and has very little common factors with the others. It is hard to apply historical knowledge to such project.

### 1.3.1 Physics precision

Physics precision can be thought of as comparison of different levels of approximation of a theoretical model. An increase in precision is like taking into account one more term in the Taylor expansion. Using this analogy, if a model takes into account only the leading order (LO) terms of the Taylor expansion of some function, the improved model could take next-to-leading order (NLO) terms. Another improvement upon this model could take next-to-next-to-leading order (NNLO) terms (see Figure 1). This nomenclature is used to describe level of approximation used in field theory calculations. For example, if perturbation expansion $\sigma^{(\alpha^n)}$ is used, then the next precision threshold would be the next expansion $\sigma^{(\alpha^{n+1})}$.

$$\sin(x) \approx x \quad -\frac{x^3}{3!} \quad +\frac{x^5}{5!} \quad - \ldots$$

Leading Order (LO)

Next-to-Leading Order (NLO)

Next-to-Next-to-Leading Order (NNLO)

**Figure 1.** Comparison of levels of approximations used in field theory calculations to Taylor expansion of a function.

An important property of such setup is that the statistical error of a difference $\sigma^{(\alpha^{n+1})} - \sigma^{(\alpha^n)}$ is $\frac{1}{\sqrt{N}}$ where $N$ is the number of samples generated during the simulation. Usually more refined techniques are used to compute it. This error should be at least three times smaller than error that comes from the physics model to be considered insignificant. For the same reason technical precision must be also three times better than statistical[1].

> **Example:**
>
> Figure 2 shows the comparison between different levels of approximation. Note that this example expresses several issues related to the complexity of the physics problems when viewed from the point of view of a computer scientist.

---

[1] Technical precision can be thought of as all practical consequences resulting from numerical calculations, such as the length of computer word, precision of numerical integration, order of mathematical operations, software issues, etc.

> First of all, it's hard to find a real-world example which could be presented to an audience that does not have deeper understanding of the physics behind the described phenomena. For this reason, we often omit physics-related descriptions in such examples so as not to lose focus and confuse the reader.
>
> Moreover, some phenomena apply to very specific use-case scenarios and are near-invisible in other cases. Compare, for example, the difference between NLO and NNLO approximations (green and red lines) on both plots presented on Figure 2. They are almost indistinguishable on the left-hand side plot and on the large portion of the right-hand side plot. The only case where the difference is clearly visible is the low photon energy of the process used for the right-hand side plot.
>
> This shows one of the fundamentals reasons why close cooperation between computer scientists and physics experts is necessary. It is near-impossible for a computer scientist to determine if the effect is properly implemented. Only the expert in the field can design appropriate projections that will render the implemented effect visible. Later, once these projections are recognized and results verified by the expert, computer scientist can devise and implement tools to automate the comparison (for example by calculating the ratio between blue and green lines and between green and red lines) which can serve as a basis for automated tests.

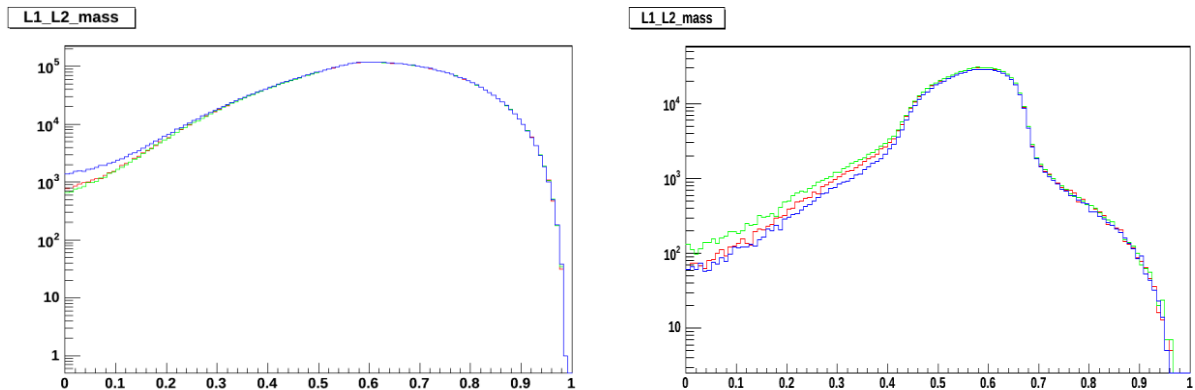

**Figure 2.** Spectrum of visible photon energy in respect to the total invariant mass of the process. Experimental cuts applied over data generated using chain of Monte Carlo tools. For simplicity, physics-related details are omitted. In short, three different levels of approximations are compared, where blue lines represent LO, green NLO and red NNLO approximation. Left-hand side graph depicts different process than right-hand side graph.

Just stating that the project should improve the precision of the tool from LO to NLO says nothing about the effort needed to construct and implement such model. The only implication is that NLO model requires algorithms and data of higher precision than LO. This also infers that complication of theoretical calculations must be higher. And, as outlined in an article co-authored by me [7], this also carries along an increased effort in building proper test framework and increased effort needed to validate the results.

As the quality and quantity of experimental data grows, the demand for higher precision theoretical models rises, compelling the scientific community to focus on improving the results of existing models. Thus, it becomes important to attempt to estimate the effort required to implement and test these requirements.

## 1.3.2 Development effort

An automated estimation, based on code metrics and other quantitative units used to describe the software cannot be applied to scientific projects. For example, function points[2] are often used for software development cost estimation. They are used, for example, in standardized cost estimation metrics such as `COCOMO` [8]. They are used to estimate the complexity of implementing single functionality. The process of their calculation is similar to code metrics and the result is strongly correlated with lines-of-code (LOC) metric. They are calculated for each functional user requirement. As such, they do not take into account internal algorithms used in the project.

Code-based metrics are ill-fitted for scientific software. They neither accurately assess the effort required to implement such algorithms nor do they provide any insight into the potential effort required to increase the precision of the results or the effort required to validate the solution. Consider, for example, the most common code-based metric used to assess software quality – Cyclomatic Complexity (CC). CC is a simple metric that measures the number of linearly independent paths that can be followed during program execution. High CC is often a sign of a badly-written code that is hard to follow when reading and which outcome is hard to predict. Common project quality management tools, such as `SonarQube` [9], trace CC in the source code of the project and mark functions with CC higher than 10 as "complex"; prime targets for code refactoring. However, when considering scientific projects, which mostly contain complex algorithms, this metric will always show high complexity (see Table 1 and Table 2). This obviously should not be treated as an indication that the code requires refactoring.

| *C++ project* | Total lines | Comment lines | Code lines | Max indent | Total complexity |
|---:|---:|---:|---:|---:|---:|
| *MC-TESTER* | 8091 | 1716 | 6130 | 7 | 1181 |
| *TauSpinner* | 3064 | 781 | 2363 | 8 | 719 |
| *Tauola++* | 4781 | 1151 | 3566 | 6 | 559 |
| *Photos++* | 8584 | 2881 | 5697 | 8 | 973 |

Table 1    Code-based metrics for several C++ projects described in this thesis and in more details in Chapter 6. Metrics has been computed using `Metrix++` tool [10] based on source code without tests. In `Tauola++` and `TauSpinner` the Fortran code has been omitted.

| *Fortran project* | Total lines | Comment lines | Code lines | Max call depth | Total complexity |
|---:|---:|---:|---:|---:|---:|
| *Photos Fortran* | 12612 | 3959 | 8237 | 7 | 566 |
| *Tauola Fortran* | 28088 | 2967 | 22456 | 22 | 2086 |

Table 2    Code-based metrics for several Fortran projects described in this thesis and in more details in Chapter 6. Metrics has been computed using `FPT`[3] tool [11]. Tests has been omitted from the metrics.

---

[2] First defined in 1979 [102], function points are systematic approach to measure effort by categorizing and estimating complexity of individual functional user requirements. They are, by their definition, ignoring internal functions and algorithms that do not directly contribute to given functionality but are needed to implement it.
[3] I would like to thank John Collins for providing the license for the purpose of this thesis. John is always happy to help any academic work performed with the help of his tools.

Similarly, function points are strongly related to functional requirements. However, most of the effort put into scientific projects is not spent on the development of the code and should be treated as time spent implementing non-functional requirements (NFR). In fact, in many cases the only goal of the team is to increase precision of the results of already existing tool or already known process.

### 1.3.3 Testing effort

In business applications typical cost of quality (COQ) estimations and metrics used to estimate the effort put on testing usually only consider functional testing. However, precision is an NFR and should follow NFR-related testing strategies. Estimating the effort required to test NFR is as hard as estimating the effort to implement it. As many researches note (see e.g. [12] and its references related to test suite reduction) NFR testing is often neglected or delayed until late stages of the development process. Often identifying NFRs themselves is omitted or reduced to bare minimum in hope that development team will identify and implement the NFR that emerge naturally during the development process. The most common requirements related to performance or usability are usually dealt with by using an automated process while other requirements, unless client states them specifically, are neglected as they are often viewed as having low impact on the overall project in comparison to the functional requirements. This uncertain, unsystematic approach to NRF means that little effort is given into their testing. There is also another factor that reduces the resources spent on the testing phase overall, and that is the overall project cost reduction.

Skimming on QA-related processes is not something that can be afforded in scientific project. Without proper validation, which proves the implementation is correct in as much details as possible, the implemented tool is worthless to scientific community. An analysis that uses such tool may very well be a waste of time. When a bug is found, if it is related to configuration options, API or other elements aside from implemented model, it can be fixed without much of an impact to the tool itself. However, if an error in implemented model is found, it may harm or completely destroy trustworthiness of the tool. Verifying the correctness of implemented solution is of utmost importance.

The same is true for many other industries. In a 2002 report on the economic impact of the inadequate infrastructure for software testing [13] highlights the amount of revenue lost due to software bugs and the liability they create. It is best expressed by a headline from Washington Technology, February 1998 quoted in this report: "If Microsoft made cars instead of computer programs, product-liability suits might now have driven them out of business." The report also shows the continuous growth of the importance of testing and the overall shift from unit testing, prominent in 1960-1970s due to limited complexity of the systems at that time, towards integration testing and variety of system-level tests required nowadays.

One thing remains apparent – the costs incurred due to inadequate testing are rising and will keep rising along with the complexity of the software products themselves. Regardless of actual COQ that has to be paid to deliver reliable software, any savings made in the QA department may result in high losses as the product hits the market; a risk that is often unacceptable.

As the aforementioned report highlights, there is a growing importance of reliability and reputation that can be lost in case of failure. Exactly the same behavior can be observed in scientific software development.

Due to the importance of the testing process, estimating the testing effort becomes as important as estimating the effort of developing the tool itself. While there are no silver-bullet solutions that allow to estimate the effort of a scientific project, the experience gathered in this thesis may be a valuable input for such estimation.

## 1.4 Scope and objectives of the thesis

The list of challenges and restrictions outlined in previous section create foundation for the design goals of scientific software. This list is very long and diverse. Estimating the effort needed to implement a solution under these conditions proves challenging at best, if not impossible. This creates an environment that is very hard to manage, unique for each project and for each set of collaborators that take part in the project.

The following thesis is presented:

> **Often scientific software development focuses on increasing the precision of the results. This process follows the cycle of improving the physics model, describing the model using mathematical formalism, implementing the model with numerical approximations, creating the software framework, documenting and validating results, testing and publishing results. Then, the effort put into testing and into building proper validation framework rises proportionally with the effort required to achieve results with increased precision.**

This thesis is an attempt to highlight the common factors of scientific projects. It aims to fulfil the following goals, which represent the main contribution of this thesis:

1. ### Present the physics background required to understand the common concepts of software development for physics experiments

   This thesis aims to analyze the development process behind that scientific software developed in physics communities. This cannot be done without presenting the physics background needed to understand the subjects presented in this thesis (Chapter 2), which includes an analysis of different types of software developed for physics experiments (Section 2.7).

2. ### Analyze the scientific software development process

   The goal of this thesis is to describe the process the project follows in order to increase the precision of the results of an analysis or a tool designed to be used in an analysis chain. For this purpose we outline the scientific software development process (Chapter 3) which requires elaboration on challenges faced during this process (Section 3.1) and a brief analysis of how existing software development methodologies relate to this process (Section 3.2).

3. ### Describe the methodology that emerged during the development of several tools

   This thesis attempts to present an approach designed to prevent most common problems encountered during development process of a scientific tool based on the cycle of improving the physics model, describing the model using mathematical formalism, implementing the model with numerical approximations, creating the software framework, documenting and validating results, testing and publishing results (Section 3.2.2).

4. Address scientific software testing process

    In this thesis we address in great detail the variety of different tests written for scientific software (Chapter 5). This is achieved by presenting the taxonomy of testing techniques (Section 5.1), as well as listing the different types of tests performed for scientific projects and show how they fit into this taxonomy (Section 5.2). Lastly, we present several testing techniques useful for scientific software (Section 5.3).

5. Present history and decisions made during development of several tools

    We present how the cycle described in this thesis relates to the development process of several High Energy Physics (HEP) projects, that I took part in, such as tools for analysis and production of heavy particles, tools for comparing different physics models and the Monte Carlo tools testing frameworks (Chapter 6)[4]. These projects represent valuable examples of different problems faced when developing scientific software.

6. Summarize our experience related to scientific software development process

    Summary (Chapter 8) lists lessons learned from the analysis of the scientific software development process. We have also enclosed a short section in which we describe similarities and differences between the approach described in this thesis and the approach used in business environments, based on my experience working in Research and Development sectors of few companies (Section 7).

---

[4] The history of the tools described in this Section served as a basis of the short description of these tools in Reference [14].

# Chapter 2. Physics background

In this section we present the most basic concepts and terminology of physics analysis from the point of view of the analyzed physics processes and data structures. This description is necessary to understand the environment of code written for physics experiments. Description presented here has been eventually, in its shorter form, published in Reference [14]. As a lot of the terms mentioned here relate to physics some of the details will be moved to footnotes leaving only the most important information in this explanation.

## 2.1 Simulations in High Energy Physics

We focus on the most common type of a High Energy Physics (HEP) simulation, which is a Monte Carlo simulation of the effects of a collision of two particles. In HEP terminology, such collision is referred to as "an event" and its data is stored in an Event Record.

An Event Record is essentially a data structure describing all details regarding the starting point of the collision and all of the products of this collision. While a variety of collision products may exist[5] for simplicity, we will refer to all of them as "particles".

The most basic properties of a particle are its four-vector[6], its mass (virtuality)[7], its identifier[8] and its status[9]. The distributions of the four-momenta and mass of the particles are, in general, calculated using complex matrix elements[10], while their identifiers are usually chosen to suit the needs of the analysis[11]. The choice of the status depends on the tool used to generate a given particle.

Event Record contains a set of such particles and describes relations between them. Namely: a particle can decay into other particles. The products of such decay are called "daughters" of the decaying particles, which is then referred to as a "mother" of the decay products[12]. It is common to record initial collision as two particles which annihilate (merge) into an intermediate boson, which then decays into other particles. The analogy to family relations is used also when referring to siblings (particles that have the same mother) and grandmothers. A particle is considered stable (or "final") when it has no decay products[13]. An example of a decay chain created from electron-positron collision is shown on Figure 3.

---

[5] Quarks, jets, partons, particles, to name a few.
[6] Four-momentum in phase-space stored as a 3-dimensional momentum (px,py,pz) and energy (e). Phase-space can be considered as compact manifold of dimension $3n-4$ depending on the number $n$ of particles/objects of final products in the decay (or hard process) tree. Obviously, description which could provide all details of one or two-dimensional projections of such multidimensional distribution may not be realistic.
[7] For some particles "mass" is referred to as "virtuality" and then it is not of a fixed value. I won't go into details regarding this property. However, it is worth noting that this value may be ambiguous or negative (due to rounding errors).
or even completely absent from an Event Record causing a lot of unique use cases that have to be considered.
[8] Denoted as PDGID (Particle Data Group Identifier). This is a standardized identifier that uniquely represents each known and theoretical particle used in a simulation. A tool uses this identifier to recognize particles of interest.
[9] A versatile and notably non-standardized property used to determine the state of the particle. In some cases it can indicate the origin of the particle, its purpose or a tool that generated it.
[10] Matrix element is a way how probability distributions over the final state phase-space are calculated from Field Theory. One can think of phase space as a manifold and of matrix elements as functions (functionals) on such space. Probability density distributions are then obtained from the squared modules of such functions.
[11] The list of decay products of a particle can also be chosen randomly based on the probability distribution of possible decay channels.
[12] Note that a particle or a set of particles can have more than one mother.

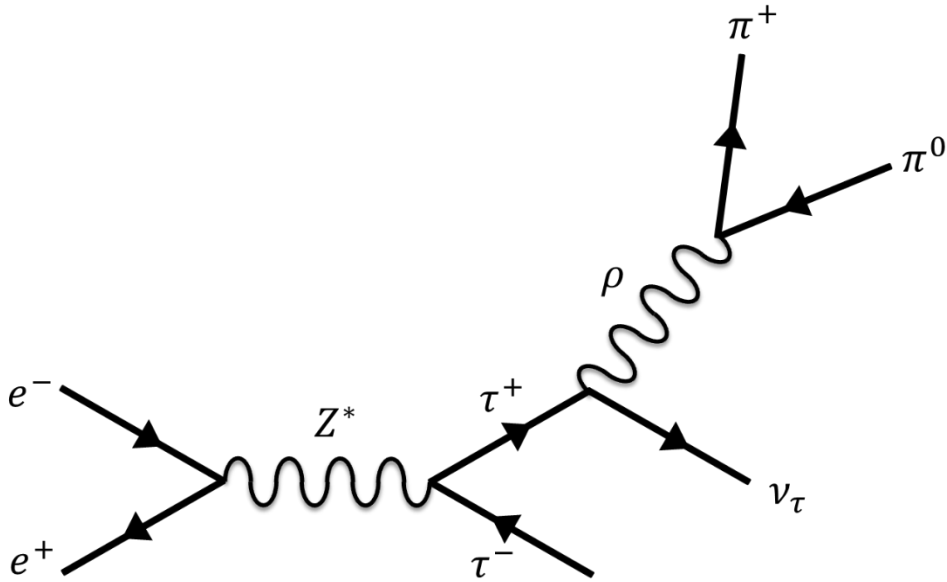

**Figure 3.** An example of $e^+e^- \to Z \to \tau^+\tau^-$ decay chain. The grandmothers of τ (electron and positron) collide forming the intermediate boson Z which decays into τ pair. In this example, the decay of $\tau^-$ has been omitted while $\tau^+$ decays to $\tau$ neutrino and resonance $\rho$ which in turn decays to a pair of pions. This is a simplified example. To put it in a scale – more complex generation process may include 2-3 thousand particles, not taking into account detector response simulation.

In an essence, an Event Record describes a "decay tree" - a cascade of particle decays from the initial collision up to the stable decay products[14]. It is worth noting that for the sake of describing some unique phenomena exceptions are introduced to this data structure that can break its usual formula[15]. Such exceptions have to be taken into account.

A simulation can be thought of as a chain of tools that produce, modify or analyze content of an Event Record. This description immediately points out the three types of tools available for physicists:

- **Generators**. Such tools are used to calculate products of initial collision or products of a decay of a particle within the decay tree. For example, a tool `Tauola` [5] is an example of a generator that produces decay products of $\tau$ leptons.
- **Post-processing tools** (sometimes called also "an afterburner"). These tools are used to modify the decay tree with respect to its predefined role. For example, a tool `Photos` [4] is used to modify the decay tree by adding additional photons.
- **Analysis tools**. These tools do not modify the Event Record but extract information encoded in it. For example, `MC-TESTER` [15] creates a set of histograms based on stable decay products of all particles with given ID found in an Event Record. Analysis tools present various projections fine-tuned to the specific purpose of the analysis, which usually requires extensive filtering of a dataset that contains huge amount of Event Records and traversing each Event Record in search for specific particles.

---

[13] Such particles have status 1. This is the only status code used uniformly by all tools.
[14] It can contain more information as well, see Section 3.1.7.
[15] They can, for example, introduce loops within the tree structure effectively transforming it into a graph.

An analysis chain can contain any number of tools of any of the above three types and in various orders[16]. Tools are usually designed to work independently. It is not uncommon for each tool in a chain to be provided by different authors. The most important factor of this cooperation is the fact that a tool has no way of knowing what other tools have been used in the preceding step. They have to carefully inspect the Event Record for information they require and do not expect this information to be always complete or valid.

This is the most common source of problems that authors of such tools must face. The tool has also no way of knowing what other tools will be used after it. If the tool modifies Event Record it must strive to produce correct Event Record at any time and to minimize potential destructive modifications. Required input and possible results of its use should be well documented as one incorrectly used tool may damage whole analysis.

Since we are talking about Monte Carlo simulation, an analysis must process a large set of events in order to obtain meaningful results. A simple example of such result is the distribution of the mass or the couplings of particles into which $\tau$ lepton decays[17]. The conclusions taken from such results must take into account statistical error of the data sample.

So, to summarize, an analysis requires a chain of tools that produces Event Records and analyzes final versions of such Event Records modified by other tools used in the analysis chain. An analysis produces results (histograms with distributions or other values) based on a sample of events. The result of the analysis and its statistical error depends on the size of this sample.

## 2.2 Scale of the physics simulation

The aspect of scale is very important to understand the complexity of the tools used in the process. The sense of the scale can be shown by looking at the end-result of an analysis. A single generated proton-proton collision can produce an Event Record that includes more than 10 000 particles.

Then, consider that to observe an NLO effect a data set containing hundreds of thousands of events is needed and few orders of magnitude more to validate that a tool correctly simulates this effect. For testing `Tauola++` generation samples of 10 million events (MEvents) are the most common ones. Final state radiation process in `Photos++` requires samples of 100 MEvents and when validating NNLO effects in `Photos++` we had to generate samples of 10 000 MEvents.

Note that such huge data samples are really not uncommon in HEP experiments, which is why the generation process is often parallelized. And yet, a single test can take hours of generation on a single 32-core machine.

Next, let us look at the size of the output. A very small event, for example for `Tauola++` tests, takes 2kb of disk space. A 10 MEvents validation sample takes 22 GB. On the other hand, a single event with 10 000 particles takes almost 3 MB. Storing a 10 000 MEvents of such samples would require roughly around 30 petabytes of storage space. The vast disk storage space required is the main reason why in tests these events are generated anew every time the test is performed and the event is discarded after processing by the analysis tool[18]. This makes the whole process excruciatingly long but there is simply not enough storage space to store all these events.

---

[16] Usually the sequence starts from a generator, or two, followed by some post-processing tools and ends with an analysis tool. However, it is easy to imagine a chain in which this order is not upheld.

[17] For example, for a $\tau^-$ decaying into three particles: $\pi^-, \pi^0, \nu_\tau$ we analyze distributions of mass of pairs $\pi^-\pi^0$, $\pi^- \nu_\tau$ and $\pi^0 \nu_\tau$.

Next problem is how to analyze such data. Not knowing what to look at may require generation of hundreds of pages of one- or two-dimensional distributions in hope of finding an anomaly. Without proper projections or comparison tools this task is almost impossible.

This is why this process is aided by analysis tools. Let's say our analysis focuses only on tau lepton decays. Given an event with 10 000 particles we are bound to find several of them in each event. We can filter out most of the particles in the event[19] leaving around 100 interesting ones. Then, observing a single property of these decays, we can reduce the output data to as little as a single one-dimensional histogram. The result of the analysis of 30 petabytes of data can be stored in a picture that takes few kilobytes.

While usually a larger set of histograms is needed, augmented by many statistics gathered from the data sample, this example shows the difference of scale between the results of the analysis output and amount of data processed by the analysis chain. It also highlights the importance of the analysis tool used at the end of the chain as improper use of such tool can mean that hours of simulations have to be redone not because of the bug in one of the Monte Carlo generators or improper generator configuration but because the resulting projections do not show the desired effect or show it improperly. The analysis tool is as important element of the analysis chain as any other.

The above example also highlights the fact that results of such analyses are tailored to the original purpose of the analysis. Usually, the results of such analysis cannot be used for any other purpose; a new filtering and projection needs to be performed and since generated data is not stored, generation has to be repeated.

## 2.3 Program, its boundaries and interfaces

As authors of Ref. [16] point out, software written by scientific community is in most cases targeted at end users who not only use software tools but expand them or build new tools on top of them. Therefore, it is important to define interfaces to the program, which means the code responsible for interacting with the program from other programs or for data exchange using common data formats.

This term should not be confused with the term Application Programming Interface (API)[20]. An example interface can be used to communicate between a simulation program (such as `Tauola`) and other programs in the event generation toolchain. Such interface can include not only the data import and export procedures for specific Event Record format but also important algorithms for processing complex structures in the Event Record or computing the necessary input parameters required to run the tool.

---

[18] In some applications we may not care about the generation process. In these cases a fake initial event containing only few particles can be manually produced without the need of a generator. This is immensely helpful but only in a very narrow applications such as technical tests that do not focus on physical content of the output.
[19] All but mothers and grandmothers of tau leptons as well as stable decay products of these leptons.
[20] Scientific programs do have an API. However, it is usually as simple as possible and is often not paid much attention to. I will not discuss or refer to such API here.

Development of interfaces is thus an important and crucial work and requires often much effort in terms of lines of code and tests, as shown in [14]. The interface plays also an important role of documentation and specification of the code, which is needed for future extensions. As a consequence, the software tool contains not only knowledge related to the problem at hand, but a description of its potential extensions as well.

The amount of work spent on the future interfaces is one of the aspects of the scientific code development missing from all related publications. It is common for scientists to document both existing analysis as well as its possible follow-ups. It is as natural as addition of a "future work" section to a paper describing the theory behind an analysis, and to add comments to the code with similar motivation in mind.

Parts of the code and comments in code are created to aid in development of the next steps. Some of these comments may not be precise, but nonetheless of a great potential value. They document intellectual effort performed for the main purpose of the development, which in the end may not be useful for the fulfilment of the original purpose but provides hints for the future extensions of this work.

The interface (or a prototype for the future interface) is thus the code written specifically for this purpose. Such an interface of a scientific application exposes its inner algorithms and structures describing how these algorithms can be used but also how to modify them. Such interfaces, in contrast to APIs, are not meant to be immutable. With each new development step, previously existing interfaces are refactored and refined as only when reaching the new step, the developer gains understanding of which parts of these interfaces are essential. They provide a starting point and part of the knowledge base which usually need to be investigated modified to be useful for the following steps.

## 2.4   Evolution of algorithms driven by precision demands

As mentioned in Section 1.3.1 physics precision can be thought of as comparison of different levels of approximation of a theoretical model. An increase in precision is like taking into account one more term in the Taylor expansion.

Results of the measurements of high energy accelerator experiments are of stochastic nature. Increasing precision of these experiments is intimately mixed with increasing statistics of collected data sample. Increasing statistics means not only decreasing corresponding errors but also leads to necessity to consider details of multi-dimensional phase-space. This brings into light details which at earlier steps were not significant thus also not available for the measurements. The early steps of a project design could have avoided going into such details, which were beyond detector sensitivity anyway. This was often the reason for algorithmic simplifications. These solutions had to be revisited later once the demand for higher precision arose.

It's important to keep in mind that such future steps may be needed so as to avoid, or at least minimize, necessary later efforts. This affects not only the design of the analysis, or the design of the tool, but their test frameworks as well.

> **Example:**
>
> Impact of higher statistics samples can be seen in our work on RChL theoretical model (see [17; 18; 19]). At first, due to small statistics of the data and nature of the experimental analysis the available distributions were at most 1-dimensional. The model was built based on these distributions.

> However, when 2-dimensional data became available they showed obvious flaws of the model. Both the model and the framework were not properly prepared for validation using 2-dimensional data which ultimately became one of the reasons why the work on the project has stalled.
>
> It's worth noting that experimental validation of systematic errors was by far more demanding for multidimensional distributions than it was for one-dimensional distributions.

As stated by the thesis formulated in Section 1.4, the majority of the focus of scientific software is stimulated by the constant need to increase the precision of the results. One of the many examples is the history of W boson mass measurement. The first discovery of this particle in 1983 resulted in a Nobel prize in Physics in 1984. At that time, the measurement showed the mass to be around 80 gigaelectronovolt (GeV). The most-recent study from 2018 [20] determined the mass of the W boson to be 80.379 ± 19 GeV. The amount of work devoted to measurement of this property throughout the years shows its importance for all physics analyses but it also shows how great amount of effort is put into increasing the precision of the results. The fact that over 30 years of physics have improved this result roughly by two orders of magnitude also indicates how hard this process can be.

## 2.5  Serializing measured data

As described before, all of the event information is stored using an Event Record. An Event Record can contain more information than the most basic properties of the particles. For example, an Event Record can (but does not have to) contain information about space and time position of each decay. Depending on the capabilities of an Event Record, it can also store other significant information about particles or initial collision.

No Event Record can uniformly describe every phenomena that physics analysis tries to simulate or verify. This results in a number of unique solutions and workarounds introduced in an attempt to represent a relation that an Event Record cannot represent using its default structures.

> **Example:**
>
> Two most common Event Records, `HEPEVT` and `HepMC`, lack the support for handling quantum mechanics entanglement; a property that can introduce relation between any two particles within the whole structure, making the data structure a directed graph instead of a tree.
>
> The solution to this issue is not trivial and many experiments introduce their own unique approaches that allow them to include such information within the limitations of the data structure. For example, one of the Monte Carlo generators introduced a mechanism called "self-decay" in which particle decayed to itself; a purely technical notation that had no basis in physics. This mechanism was applied to a group of particles which were not directly related to each other. Such relation could not have been described by a tree data structure.
>
> Our tools provided improper results when acting on such decays because they tried to treat it as a usual, physics-related decay. We had to implement several mechanisms to detect these scenarios.

Some of these solutions are easy to account for while others may completely break the algorithm. Handling such cases correctly takes a lot of time, especially that in some cases the issue may be hard to notice.

> **Example:**
>
> Some tools use negative status codes to determine different generation steps of the particle. These status codes can be safely ignored by a tool that does not need this information. In fact, one of our tools only searched for specific status codes ignoring all others.
>
> That was fine until one of the experimental groups started using the specific status code -3 to encode historical entries. This caused errors that, due to the error-handling policy of this group, became unnoticed for some time. Only when invalid results became clearly visible the issue came to surface. We had to update our tool to handle this case.

On the other hand, some changes break the algorithm almost instantly, rendering the tool unusable.

> **Example:**
>
> One of the Monte Carlo generators introduced a relation in which a daughter points to a mother as its decay product. This creates an infinite loop in an algorithm that traverses Event Record expecting a tree data structure. For our tool to work on the events produced by this generator we would have to introduce a mechanism that detects loops which would slow down the processing significantly.

Lastly, there are custom changes to the Event Record introduced by experiments to circumvent the existing mechanisms of this Event Record to include information specific only to these experiments. Since analysis within these experiments rely on this extra information, the tool must adapt to these changes as well.

> **Example:**
>
> One of the experiments used to encode generation-specific data in their event numbering scheme. Changing event number meant that this information has been discarded.
>
> Similarly, one experiment used to encode specific meaning in particle numbers within the event. Our tool used default policy to attribute particle numbers to newly generated particles which for this experiment meant the particles were attributed wrong numbers.

There are many examples of both small and large-scale changes over the years of evolution of Event Records and their use by different experiments. Algorithms for traversing Event Records have to be frequently updated to handle the new use scenarios. Otherwise, there is a risk that the tool will break for a specific analysis performed by specific experiment.

## 2.6 Parallel computations

It is worth noting that most physics analyses are embarrassingly parallel. The most common way to speed up calculations is to separate the job into few smaller jobs and run them simultaneously as separate instances of one application. In some cases, it is done on separate threads of one application. This approach requires only proper management of seeds for random generators used by each job and a proper way of merging the results. As such, parallel computation is not challenging to achieve. Most of the effort here is put into designing and managing large computing clusters, such as LCG [21].

## 2.7 Working with old codebase

As previously mentioned, scientific software deals with large amount of old codebase, written mostly in Fortran or C. Unfortunately, when working with codebase that uses very old standards at some point you are bound to deal with issues which are one of the common pitfalls of the language and coding standard in which the code was written. The information on them is scarce as these issues are native to very few environments. It's hard to find any sources or methodological approach to finding them.

### 2.7.1 Compiler issues

Scientific software is very often a mixture of Fortran and C or C++ code. Usually both Fortran and C/C++ parts of such code are written using old standards and were compiled using old versions of the compiler. The note regarding the compiler is important because newer compilers usually warn about the most common issues that can occur, not to mention they perform more rigorous checks of typical misuse of the language. However, when compiling large codebase with new compiler the mass of warnings that pop up may foster the idea that they are harmless and can be safely ignored. After all the code worked so far without any problems.

There are many reasons that force us to push the supported compiler version higher and higher. One is the deprecation of old compilers and the lack of support of therein on new platforms and operating systems. At one point you may be forced to resign from f77 and gcc 3.4, or maybe you have already forgotten about them.

The switch from f77 to gfortran and from gcc 3.4 to 4.1 was, for me, the first experience with the issues related to differences between compilers. However, the following changes to newer gcc versions were, at some points, even more dramatic. Anyone that tried to upgrade the toolchain at least once knows that new compilers can break old software in various ways. This was especially apparent in the gcc 4.X era as the newer compilers implemented more and more features of new C++ standard, back then known as c++0x. New standard was needed for a very long time so everyone was keen on adapting as many of its features as possible. This, unfortunately, only created issues for the older codebase.

Often the failure is simply compile-time error that previously was not detected or was treated as warning. Such issues are simple to solve. The other ones are missing, unsupported or deprecated compiler flags. They are also easy to fix. The trickier issues are changes to the default settings of some underlying options. The most obvious one is the default C++ standard used by gcc. Since gcc 6.1 the default standard used for C++ code is no longer c++98 but c++14. Obviously old codebase does not use c++14 features so it does not benefit from this change and the new language standard is not fully compatible with the old one, especially as it redefines some miniscule details that previously were treated as undefined behavior.

```
1  int i = 0;
2  int j = 0;
3
4  i = i++;
5  j = ++j;
6
7  std::cout << i << " " << j << std::endl;
```

**Listing 1.** Example of undefined behavior which gcc handled in the same way for decades but stopped since gcc 4.8.1.

Consider, for example the code as seen on Listing 1. Line 4 of this code creates an undefined behavior as it is invalid for the right-hand-side of the of the assignment operator to modify the variable that is on the left-hand-side of the same assignment operator. And yet, a line of code that relied on this undefined behavior remained in our codebase for years as gcc always compiled the code in the same way – the result of line 6 was "1 1". However, in gcc 4.8.1 the approach changed and the result was no longer the same. It changed to "0 1".

Obviously, the core issue here was the mistake in code but the author of this line made a hasty mistake and, since the code worked in variety of tests and on variety of different platforms, never looked back. The mistake came up only when we compiled the code with new gcc. This is actually the most common reason why old code base works only with old compilers – the use of non-standard code that was not corrected due to the limited capabilities of static code analysis at that time, or a notation that has become non-standard over time.

Note that gcc 4.8.2 with all warnings turned on warns about out-of-sequence operations that may cause undefined behavior[21]. There are many reasons why compiler warnings should never be ignored. However, when working with old codebase new compilers tend to report huge amount of warnings, especially when behavior that was previously permitted now became deprecated or is considered invalid. The amount of warnings reported inclines developers to ignore them. After all, the code compiles and, at least at first glance, works correctly.

This is especially true when compiling a tool authored by someone else. Updating the tool to new compilers is usually very far away on the priority list of the tool's authors so if user has newer toolchain, there is a chance that the tool won't work correctly for that user. This was exactly how we have discovered the issue described above – it was reported to us by someone who tried to compile our tool with new, at that moment, gcc 4.8.1 and found it not working correctly.

---

[21] As a side note, gcc 4.8.X series is infamous for other peculiarities as well. For example, gcc 4.8.2 caused infinite loops in SPEC CPU benchmark code. The reason was similar as example presented here – the benchmark code uses common patterns that coders use in their software. In this particular example it used undefined behavior in loop iteration. An interesting conversation came up from that issue: gcc team did not change the behavior of gcc in response to this issue by stating, rightfully, that since this behavior is undefined the compiler can do anything with such code. On the other hand, the benchmark authors state that since their code use commonly used patterns, compilers should strive to deal with them properly even though the code does not adhere strictly to the standard. From their standpoint, to achieve benchmark neutrality they cannot modify the code of their benchmarks. For more details regarding this issue see https://gcc.gnu.org/bugzilla/show_bug.cgi?id=53073. You can follow the responses of SPEC CPU team and gcc team from there. The dilemma of what to do with old code that not always complies to language standard is an interesting one and has been discussed in the context of many other issues, especially when dealing with C89 and C99 standards.

To summarize this section, we would like to note two most important points:

- Be aware that compiler change can break your software. Test it against older compiler version.
- Always keep all warnings turned on and never ignore a warning in your own code.
- Consider if a warning reported by an external test platform or external generator can influence your code.
- When possible, compare results with the environment compiled with older toolchain.

### 2.7.2 Pitfalls of mixing Fortran and C

Programs that breach the boundary between two languages have to deal with number of issues unique to this predicament. In the case of Fortran and C, for example, the use of "extern" directive alleviates the compiler from performing the usual checks done when linking against non-extern symbols. For example, one of the common mistakes we have encountered, especially with the Fortran code built in the era when double precision floating point variables were used with caution, is declaring a parameter of an extern function with different precision then it actually is. Consider two examples presented on Listing 2 and Listing 3.

```
test.f:
1        SUBROUTINE FOO(X, Y, Z)
2        INTEGER X, Y
3        REAL Z,Z2
```

```
test.cxx:
1  extern "C" void foo_(int *X, int *Y, double *Z);
```

**Listing 2** Example of incorrect C++ and Fortran mixing. The content of function `FOO` defined in `test.f` has incorrect declaration in `test.cxx`

```
test2.cxx:
1  extern "C" complex<double> bar_()
2  {
3    ...
4  }
```

```
test2.f:
1        COMPLEX C, C2
2        C = BAR()
```

**Listing 3** Example of incorrect C++ and Fortran mixing. The function `bar_` defined in `test2.cxx` has incorrect use in `test2.f`

In both examples the files will compile and link correctly. However, both calls to `FOO` function and `BAR` function presented in these examples will produce undefined behavior. While the reason for that is obvious in the first example, the second is less self-explanatory but also very simple – the default complex variables in Fortran have single precision, not double precision.

These types of mistakes seem easy to spot when writing an interface but consider large codebase or a case in which someone has already created the C++ interface and the software seems to work correctly. In reality, the result is often a memory overwriting of another variable. Knowing that a Monte Carlo process has large complexity and many of its paths are frequently unused during single generation of an event, this mistake may be hard to spot as it may occur only in very specific setups. The time from the moment the bug was introduced to the moment it has been detected can be very long which makes finding such mistakes really painful.

> **Example:**
>
> Consider that in test2.f from Listing 3 variables C and C2 belong to a common block and only C is used (and computed) within this function. When after long hours or days of debugging finally an invalid value C2 is found it will usually be found when used in completely different function, different stage of the generation and probably when processing different event than the one that set invalid value of this variable in the first place. There will be no way to correlate invalid value of C2 with setting value of C.
>
> This is, in fact, a simplified variant of an error that we had when writing a C++ interface to one of our Fortran tools. The mistake remains unnoticed for many months. Only after new blocks of code has been added and memory alignment changed, we could notice the issue created by memory overwriting. Tracing it back to the precise line of code that caused this issue took days of careful analysis as this issue could not have been reproduced by generating a single event. It required a specific sequence of events to occur.

But problems with floating point variables passed between Fortran and C can generate something even worse – a precision loss stemming mostly from the fact that Fortran implicit variables are single precision while in C and C++ they are of double precision. Debugging precision loss in a Monte Carlo generator may take weeks as such issue is very easily disregarded as statistical fluctuation. Numerical instabilities in floating point numbers are hard to pinpoint by themselves in pure Fortran[22] not to mention when Fortran and C is combined. For such issue to be noticed an accumulation of errors must be found and depending on in which part of the generator the issue occurs, this may only happen in very narrow scenarios. Lastly, even when the issue is finally visible, and a predictable reproduction process is available it will still require days of breaking the generator apart piece by piece to find the source.

The lesson that we have learned from numerous issues such as these is to never use implicit variables in Fortran and always double-check that the precision of function arguments and return variables is always consistent in both Fortran and C code. This is something that you cannot rely on compiler to verify as it would be the case in a code written in single language.

## 2.8 Software in physics experiments

The physics community offers a range of different tools aimed either to help others perform their own analyses or to reproduce the results of analysis presented by the authors of the tool. This means that software produced by this community varies greatly in terms of scope, size of the development team and amount of work put into the tool itself. Following Section lists most common types of software found in high energy physics divided into the categories based on how much focus is shifted from software architecture and overall software complexity of the tool towards the physics content of it.

---

[22] For example, declaring a constant 1.0f-3 declares a single precision floating point number constant which, by the nature of floating point, is not stored precisely. Casting it to double precision causes the error of imprecise representation to accumulate. In one case in our analysis such value was then multiplied by 2000.0. This resulted in value 2.000000008446551902 – a value that is close enough to 2 to escape our 8-digit precision debug printouts and yet exceeding the maximum threshold of 2.

### 2.8.1 Software frameworks

The largest software projects written for physics experiments are data analysis frameworks[23] designed as a starting point for more complex analyses. Their main purpose is to deal with the problems common to all physics analyses, such as those related to the data on which the analysis is performed. Physics analyses are embarrassingly parallel. They follow Single-Instruction-Multiple-Data (SIMD) approach by running parallel instances of the same analysis on large computing clusters. Each instance processes different data set. Frameworks used to manage parallelization focus on storage access optimization, efficient filtering or management of datasets for distributed computing and tasks related to persistency formats or gathering, merging and persisting the results of parallel computations. More advanced systems provide end-to-end solutions which include library version management, package configuration, deployment on the computing cloud and monitoring the computation process.

In order for such systems to work effectively, they can enforce code organization of the libraries that are included in the framework and may require specific version management process or even unified test and build environments. Most of the experiments surrounding CERN (such as ATLAS [22], CMS [23], LHCb [24]) and many other experiments (such as BaBar [25] experiment, Belle [26] experiment or CDF experiment [27]) use such framework to organize chains of user analyses and manage access to the data gathered from the detectors. These frameworks, however, are mostly private to these experiments.

The most prominent example of an open-access framework is `ROOT` Data Analysis Framework [28] (see also: [29]). Its history dates back to Physics Analysis Workstation (`PAW`) project started in 1986. `ROOT` suffered wide spectrum of problems in its development process due to its complexity and very long lifetime. Over the years, vast amount of legacy code had to be adapted to new coding standards and development strategies. `ROOT` has seen transition from Fortran to C++ with, now defunct, `CINT` C++ interpreter made specifically for `ROOT`, and very recently from C++ to C++11 (replacing `CINT` with `Cling`, which is a part of the LLVM project [30]). Many of its radical changes often resulted in sub-optimal, in terms of software architecture, solutions that had to be implemented to preserve the original context of the project and tests. Some of the old and obsolete solutions had to slowly fade away from the project while others can no longer be removed for project to retain its support of its original user base. This is one of the most common problems to befall all projects with such long lifetime.

Currently, `ROOT` is a foundation of a vast amount of analyses in particle physics experiments. It is focused on analysis, storage, representation and manipulation of huge amounts of data with multitude of algorithms for I/O access optimization and efficient data analysis. Over the years, it has been extended with new modules, such as `PROOF` [31] parallel computing framework, `RooFit` [32] toolkit for data modeling or `PyROOT` [33] Python interface. New functionality and new modules are still being developed and added over time. It is, by far, the most complex computing science project in particle physics.

---

[23] In broad terms, data analysis framework provides tools for data access, processing, analysis, filtering and visualization using variety of methods suited for given type of data. Most of the data analysis frameworks are optimized for parallel computation and designed to efficiently process high amount of data.

While `ROOT` serves as an all-purpose data analysis framework, most experiments require use of a specialized framework built for specific sub-process of the overall data analysis. An example is `GAUDI` [34] framework for High Energy Physics data processing applications. `GAUDI` defines basic components of an analysis and creates interfaces that allows to easily create modules that can be connected to form an analysis chain. It enforces use of its structure to create new modules in return for data access, persistence and parallel computation management. `GAUDI` is extensively used in LHCb experiment and, similarly to `ROOT`, serves as a basis for several more complex tools, such as `Gauss` [35] detector simulation tool.

`ROOT` and `GAUDI` are accessible for everyone which is why they are good examples that give an indication on how other data analysis frameworks are built for private use of the experiments. Such frameworks are the most complex, from Computer Science point of view, projects in particle physics. They do, however, contain the least amount of physics context in comparison to other projects. For this reason, they belong to the smallest category of physics software.

### 2.8.2 Detector simulators

Monte Carlo generators described in previous section base their computations on a physics theory. They aim to match the theoretical expectations as close as possible. However, a collision observed in a detector will never match these expectations as its observation is skewed by the architecture of the detector itself. The physical shape of its elements and the materials used to build them apply their physical limitations on the observed phenomena. They have to be taken into account when analyzing results of collisions registered by a detector.

One other important factor related to detector data gathering is the frequency of event generation that takes place in the Large Hadron Collider (LHC). A detector is expected to measure multitude of events in a very short period of time. For example, ATLAS detector can register a million events per second[24]. At this rate the energy stored in calorimeters inside the detector has no time to fully dissipate between the events. This changes the way subsequent events are perceived. Lastly, the detector response strongly depends on the initial collision parameters as well as the process that results from the collision. The detector will behave differently depending on the types of particles that collide within it as the resulting set of particles that hit the detector will change.

Needless to say, a sophisticated tool is needed to reproduce what scientists can observe in the detector. The tools used for that purpose are called detector simulators. In short, these tools simulate passage of particles through matter. Combined with a detailed model of a detector, such as the one used in Atlas or CMS experiments, these tools can modify the events produced by a chain of Monte Carlo generators by reenacting how this event would look like if the initial collision occurred inside the modeled detector[25]. The data processed this way closely resemble events that scientists can actually see in the data gathered by experiments around Large Hadron Collider.

The most renown detector simulator is `Geant4` [36]. It is used both by ATLAS and CMS experiments at LHC as well as by many other experiments around the world. It handles detector geometry, tracks the movement of the particles, simulates the detector response and provides an interface to visualize the results.

---

[24] Several layers of triggers is used to remove events that do not pass certain quality criteria. Only around 2000 events per second pass these criteria. Out of them, around 10% is marked as "interesting" and stored.
[25] Note that detector simulators use Monte Carlo processes as well.

`Geant4` is built in a generic way allowing its use outside of high energy physics. The tool can be used in any experiment that uses simulation of particle interaction with matter, which includes simulation of radiation in medical equipment, ionizing effects of radiation in microelectronics and nuclear physics.

### 2.8.3  Monte Carlo generators

Monte Carlo generators are the heart of the physics simulations. Each analysis uses a toolchain of such generators that produce a long series of events; a vast amount of data that has to be meticulously filtered to fit the analysis criteria and thoroughly analyzed using variety of different projections focused on one or few specific details. The overall process is described in Chapter 2 which focuses on the physics background needed to understand the development process of these tools. In this section we would like to briefly present few of such tools as an example of the variety of the scope and functionality that a single Monte Carlo generator can provide.

#### 2.8.3.1 Large, general-purpose generators

`Pythia` [37] is one of the oldest and most popular general-purpose hard-process Monte Carlo simulators[26] able to simulate particle collisions. Its main objective is to generate the decay tree based on the initial collision parameters provided by the user. The configuration options are overwhelmingly exhaustive allowing to tweak every single element of the generator. As of version 8.201 the tool lists 1274 options, excluding individual particle properties and decay options, which by themselves would easily triple this number. This impressive number shows how massively customizable this tool is, and yet despite that huge amount of options, basic event generation can be configured using just four lines of configuration script.

`Pythia` is one of the best documented HEP Monte Carlo tools and its popularity is easy to understand. Apart from its configuration, it allows to turn on or off many of its sub-processes or use a hard process generated outside of Pythia and only apply specific sub-processes on top of it. This is a crucial aspect of the tool that allows combining `Pythia` with many other Monte Carlo generators by replacing part of the Pythia functionality with functionality provided by another tool. `Pythia` has been used as a basis of many analyses, including most of my work on testing the Monte Carlo generator tools.

Playing very similar role, `Herwig` [38] tool offers an alternative to `Pythia` providing similar basic functionality, albeit with less configuration flexibility. The use of `Pythia` or `Herwig` in a simulation strongly depends on the environment. Since these tools use different models, they provide different results even if the modeled process is basically the same. That is why the LHC experiments tend to use one or the other tool almost exclusively when producing their benchmark Monte Carlo samples.

Due to the scale of these projects and number of contributors, they are not a very good representation of a typical Monte Carlo tool developed for scientific community. In fact, they are rare examples of a large-scale projects with strong Computer Science aspects created by HEP community. Most of the commonly developed and used tools are much smaller.

---

[26] It is an SHG generator, which stands for "shower+hadronization generator" to emphasize that, contrary to Matrix Element (ME) generators, they use a factorized approximations called "parton showers". SHG generators operate at Leading-Order precision (see Section 1.3).

### 2.8.3.2 Small, single-purpose generators

Most of the smaller Monte Carlo tools work on already generated event by analyzing and modifying its content. Such tools are often called post-processing tools or "afterburners". This thesis focuses on this type of tools the most as they are the most prominent, yet least documented, in HEP community. They are often written by a small group of contributors. Their focus is not to provide the whole out-of-the-box solution but a small tool that focuses on one specific property under study. Often a custom tool or a modification to existing one is developed for the purpose of validating single physics theory. This can later become a basis for developing a more complex tool.

Due to their narrow scope these tools either provide higher precision of the results than the general-purpose generators or generate a process that cannot be easily produced by such tool. This is the main reason why physics community uses a junction of several tools for specific analyses. Two examples of such tools are `Tauola++` [5] and `Photos++` [4].

`Tauola++` is focused on a very narrow part of the large event generation process – tau lepton decays. This greatly narrows down the user base of the tool, but the scientific community is vast and it has experts and experiments around the world focused only on tau lepton decays. Each team that performs an analysis that requires accurate tau lepton decays, even if only for background calculation, at one point in time should consider augmenting their toolchain with `Tauola++` as it provides more accurate decays than any general-purpose generator. It is widely used in many analyses that strongly rely on tau decays, including those related to Higgs boson.

On the other hand, `Photos++` is focused on final state radiation. This tool has much broader application as any realistic simulation requires this process. The algorithm of `Photos++` includes many corrections related to specific generation processes that produce much better results than general-purpose generators. The comparisons of `Photos++` with other tools have been well documented over the time and the tool is very popular in many HEP experiments.

The history of these tools is described in Sections 6.2 and 6.3 respectively. Here, we would only like to point out that both of these tools started off as one small Fortran file written by one author. The scope of their functionality is a fragment of the scope of a large-scale generator and yet they have become the basis of many analyses. This is not something unique in scientific community; a small tool, often not written for the purpose it is ultimately used for, can quickly gain popularity as long as it provides a better or more applicable solution to an ongoing or new problem. The impact of the small Monte Carlo tools written, sometimes spontaneously, by a single researcher cannot be stressed enough. Other similar examples will be presented in this thesis.

### 2.8.4 Custom data analysis tools

One last type of tool worth mentioning relates to data analysis. As mentioned in Section 2.8.1, large frameworks can help physicists to organize the whole process and analyze the results but when a detailed, custom analyzes are performed it is necessary to precisely manage the data analysis process or tweak the visualization in a form that may be hard to obtain when using a generic framework. For this purpose, a number of tools has been developed over the years that automate the analysis process and help create custom projections. The most well-known tools are `Rivet` [39] and `CheckMATE` [40]. They have been briefly described in Section 6.5 when describing the tool `MC-TESTER` [15]. All of these tools are designed to automate the analysis process and to create custom projections.

## 2.9 State of the art

When looking at the overall approach to scientific software development, it is known that most of scientists are not familiar with the software development methods [41]. However, the awareness of tools and good practices is growing as can be seen by publications like reference [16] which lists five software engineering techniques and approaches which are specifically useful for computational scientists. These are: code separation, organization, review, testing, and simplification. In [42] the authors emphasize the need for testing not only for the correctness of software in terms of agreement with theory but also to test the implementation once the project gains a wider user base. These ideas have been further developed into more formal models of scientific software development [43].

The authors of reference [42] note an important factor that influences the software development process in scientific communities, which is the fact that more than 25% of the code of the scientists is written only for themselves. In total, more than 50% of the code is aimed at a group of less than 10 people. This matches the descriptions I've presented in Section 2.7 whereas I've stated that most of the software produced in scientific communities are small tools developed in small teams. This has great impact on the development process as it shows that the usual rules aimed at building a code that can be uniformly understood, modified and tested by large communities does not apply here. It is enough if a handful of experts will be able to efficiently work with the code. That is the main reason why scientific model and key elements of the numerical algorithms are commented in great details, some even worthy of their own publications, while the structure of the project is mostly presented "as is" in hope that its implementation is readable enough, at least in comparison to other complexities present in the project.

### 2.9.1 Precision of scientific software

Since the majority of scientific work is focused on accuracy and precision of the result, this subject has been researched by software engineers and scientific programmers. One of the early studies of accuracy of scientific software was presented in [44]. The authors investigated several software packages used for seismic data analysis. They have discovered a disturbing non-random numerical disagreement between the results of different packages. The growth of this disagreement was estimated to be of around 1% per 4000 lines of Fortran code. It was shown that the results of this feedback can improve the quality of software, which is why the authors recommend using these results to provide feedback to software developers in order to correct the software errors.

There are also examples of solutions to problems related to detection of software numerical instability. One of them, presented in [45], combines stochastic and infinite-precision testing. Similar techniques, or at least techniques that were feasible in the earlier years, were used in projects that gave foundation for the Monte Carlo tools described in Section 6.2. Here, analytic computations with twelve digits of precision were used to validate the results. Semi-analytic results were used as well. However, one should carefully use such results when comparing them to results of a Monte Carlo generator as integration algorithm can introduce potential errors that are similar in nature to statistical fluctuations visible in Monte Carlo data samples. This may produce false-positive results.

### 2.9.2 Scientific software testing

Number of issues unique to scientific software has been listed in the literature overview on testing scientific software [46]. These descriptions are quite accurate when related to software for physics communities. The authors of this overview state that none of the primary studies reported the complexity of the software in terms of measurable unit such as coupling, cohesion, or cyclomatic complexity. This is connected to the fact that the software engineering techniques are often not efficient in scientific environments. As mentioned before, code metrics have no meaning for scientific projects. They do not reliably show parts of the code that could be simplified or refactored. Such work would likely not have been done anyway.

Although there are many approaches to testing of scientific software, see for example aforementioned reference [46], there have been no studies on the relation between the effort put into tests or test frameworks and the precision of the results or studies related to the variety of tests performed in scientific software communities.

### 2.9.3 Methodologies

During research on software developed in particle physics communities we searched for publications related to software organization and methodologies used to develop such projects. However, while there are many publications describing the project's architecture, information about methodology used to develop them is scarce or not publicly available. For example, `ROOT` framework documentation [29] provides a very descriptive short overview of the design of the tool. The GAUDI [34], the LHCb data processing application framework technical report as well as CMS computing platform technical report [47] also offer a very exhaustive description of the design of the framework. However, while some documents regarding the development process are referenced in [47] they are published internally within CMS collaboration. This is not an uncommon approach and results in a lot of hidden knowledge, some of which is inevitably lost over the years. This knowledge is either overwritten by new one or slowly gathered in a form of online documentation not always publicly available.

Note that the documents mentioned in last paragraph relate to few of the largest software frameworks available in scientific communities. Such documentation is rarely available for smaller tools as the development process or the methodologies used during development aren't usually documented. The key reason is that the software developed for physics communities is rarely the focus of the scientific project. It is the physics content inside this project that is of major concern. Software becomes a necessity and is written mostly without much planning or methodological approach used in software industries for decades. That is partially because community of particle physics does not encourage mixing scientific research with solutions developed in the business industry, even though these solutions are often inspired by scientific projects[27]. The lack of technical documentation of smaller Monte Carlo tools that

---

[27] Stephen Wolfram, for example, left particle physics to build his business career based on the experience he gathered from scientific communities. Similarly, Adam Kolawa used his experience from work on Lattice QCD to form his own company. There are many other examples, some more difficult to trace than the others, but they

focuses on the computing science aspects of the projects is the reason why the process of developing such tools is not solid nor easy to formulate. It is also one of the key motivations for writing this thesis.

Since there is very limited knowledge of the methodologies used in scientific software, we will briefly present the most popular approaches used by software engineers. While for many software engineers this may be considered a common knowledge, this short description is needed so that we can later, in Section 3.2, show how these approaches relate to the scientific software development process.

As a side note, in Section 7.1 we present how these methodologies relate to business environment.

### 2.9.3.1 The waterfall approach

The oldest and the most well-known methodology on the market is the Waterfall model. It is said that its first formal description dates back to year 1970 with Winston W. Royce's seminar "Managing the Development of Large Software Systems" [48]. This, however, comes from misinterpretation of the Royce's seminar in which he presented such approach as the one that is often used yet deeply flawed. In reality waterfall model, by itself, was never formalized. However, its concept is so recognizable that many other methodologies compare or contrast themselves against it which is why it is worth describing in short details.

The development cycle of the waterfall model starts with definition of requirements, followed by creating the design, using this design to implement the desired functionality then testing the software. The last step is the maintenance of the final product (see Figure 4).

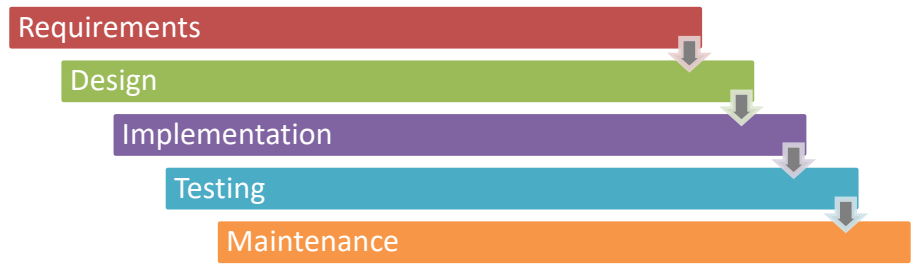

**Figure 4.** Five steps of the waterfall model. In the original model there was no way to introduce feedback to the process or to correct errors of the previous steps. Several solutions to these issues have been introduced in future variations of the original model.

The reasoning behind this methodology is that if enough time is spent on scoping the functionality and highlighting the possible risks, the documentation created in the process should be sufficient to create a proper design. Then, if design is well thought-out mitigating the risk as much as possible and ironing out all of the issues raised during the previous step, the development will be smoother and faster. While this reasoning has its good points, it hinges a lot on the proper project specification created up-front. If the specification is detailed enough, and assuming that the communication between the team and the client was as best as possible, then the finished product is expected to have full functionality declared by the customer,

---

are always related to the migration of people from scientific communities. In scientific communities these people are sometimes called "renegades".

should be made within the budget and delivered on time. This premise, however, is so optimistic that it is hard to find a project that can fulfil it.

When faced with the real-world projects the original waterfall model often fails short of fulfilling its goal. Over the years many variations of this model have been created to address the most basic issues such as lack of feedback between the phases and no possible way to address risks and other issues that come up during the later phases. Even the final version of the model presented by Winston W. Royce has feedback loops that introduce modification to the requirements and design (and thus also to the implementation) if an issue is found in testing.

Contrary to the popular notion that early software was mostly developed using this approach, the so-called "pure" waterfall model without any feedback was rarely used as obvious flaws in the projects were quickly fixed and subsequent phases were iterated over. It was unfeasible to push the project onto the next phase if significant defects were found in previous ones. After all everyone wanted to produce a marketable product not to blindly follow directives that would lead to an end product with obvious defects.

### 2.9.3.2 The agile methodologies

The agile software development methodologies, greatly popularized by the publication of agile manifesto in 2001 [49], are known best for their focus on development through small, incremental change. Each development cycle starts with the analysis of the requirements that the software is supposed to fulfil at the end of the cycle and ends with an increment that can be evaluated (see Figure 5). The cycles can be executed concurrently by several teams working on the same project. The development ends when there are no more requirements to fulfil.

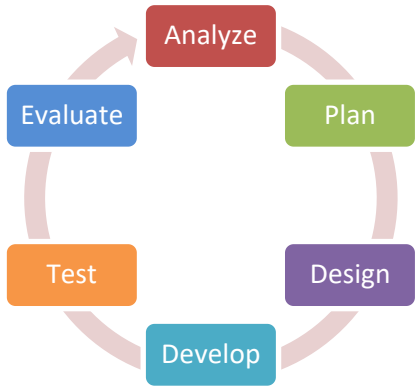

**Figure 5.** The agile software development cycle.

While there are many other aspects to agile methodologies than just this cycle, this key element is what they are most known for. They aim for the new versions of the software with new, complete functionality, to be released in matters of weeks. The team responsible for this functionality must be able to demonstrate it to the client so that the client is constantly aware of the progress being made and can easily influence the direction of the development team without much impact on the whole development process.

There are many benefits to this approach. The most prominent one is the ability to adapt to the changes proposed by the client without significant impact to the overall project. The other is the flexibility towards the budget and time. In case either runs out, the client remains with a product that may not have its full functionality but still has a set of viable use cases that he can make use of. In most of the other approaches such client might have been left with nothing as the development team has no obligation to plan its work in a way that allows for the project to be cut short at any time. The opposite is also true; given a bit more resources the team may effectively implement few additional features that were originally not planned.

While the benefits of all agile approaches are similar regardless of the actual methodology used, the drawbacks vary depending on the methodology the team strives to follow and how the team perceives it. In fact, this variety is the common drawback for all agile methodologies; there are so many of them and the agile approach itself is so imprecise that it is up to the interpretation of each team which practices to follow and how to make use of them. This results in the use of an unspecified mixture of independent practices as shown by the number of different agile methodologies on the market such as Scrum, Extreme Programming (XP), Kanban, Crystal, Feature-Driven Development (FDD) or Dynamic Systems Development Method (DSDM), just to name the most popular ones.

This is most apparent with the Scrum methodology [50]. Described in all of its details in just below 20 pages, Scrum is the most frequently used agile methodology (see Fig. 1 from ref. [51]) and yet, very few teams that claim to follow Scrum practices follow all of them[28]. The most frequently used practices revolve around Sprints – a cycle that lasts from a week to a month and is the heart of the Scrum methodology. Sprint starts with planning and ends with review and retrospective. Each day of the sprint a team meets for a 15 minutes daily meeting during which each team member briefly presents what have they done the day before, what are their plans for today and what problems they face that need to be solved. Most often than not, just the fact that the team plans their work for two weeks ahead, and has the daily meetings is enough for a team to claim that it follows Scrum methodology.

Similarly, roles of Scrum Master and Product Owner, as defined by Scrum guideline, are often neglected. Especially the former. Most teams take from Scrum methodology only the elements that suit their needs. And for a good reason. Developers should be able to adapt their approach to the needs of each individual project. After all, methodology is there to aid the development process not to pose rigid structures which harm more than help.

There are other popular agile methodologies, such as Extreme Programming (XP), Crystal or Feature-Driven Development (FDD) but this one example is enough to outline them. Agile methodologies are known to speed up the production of the software in around 30% of completed projects. However, they are also known from its lack of scalability; they are well fitted for small or medium projects, but they are hard to use for larger projects. They are also often blemished for the number of different meetings they introduce and are not always accepted by the client due to the amount of feedback and cooperation required on the client side. Some clients prefer not to spend their time guiding the development, even though it may be to their own benefit, and instead prefer involving them only to show the overall progress.

### 2.9.3.3 Rational Unified Process (RUP)

The Rational Unified Process [52], created by the division of IBM in 2003, is a framework for an iterative software development process. Similarly, to Waterfall approach, RUP presents a predefined, step-by-step project development outline with a firm line that divides every phase of project lifecycle. These are, however, the only similarities between these two methodologies.

---

[28] Note that the end note of the Scrum Guideline states explicitly that no methodology can be called "Scrum" unless all of its roles, events, artifacts and rules are taken into account.

The four phases defined by the RUP are:

1. **Inception** – the phase in which the project is discussed, initial business plan is formed, and requirements are broken down into details to find potential risks and estimate the costs. This work allows stakeholders to decide the feasibility of the project and establishes the baseline to which the project progress will be compared. If initial analysis produces unacceptable results, the project is dropped at minimal cost or redesigned to minimize the risk and costs.
2. **Elaboration** – the phase in which all major risks and concerns are addressed, all necessary research done and missing details in project specification filled out. The architecture of the project takes shape and initial feasibility studies or proof-of-concept (POC) solutions are prepared. The end of this phase marks the moment where it is still acceptable to drop the project if it turns out unfeasible or way over the initial estimations. Afterwards, any changes will have high impact on the project and its cost.
3. **Construction** – the phase focused on development of the final product. This is the longest phase of the project resulting in a product that should fulfil all requirements defined in the initial phase of the project.
4. **Transition** – the phase that transitions the project into final product that can be delivered to the client. In this phase the product is validated against the client's expectations.

RUP, being an iterative methodology, assumes that every phase of the project can be split into as many iterations as needed; each iteration providing a subset of the artifacts that a whole phase should produce. This is somewhat similar to a common modification of the Waterfall approach called iterative Waterfall model. Contrary to Waterfall, however, RUP focuses on the content that should be produced in each phase and the effort that should be put into providing such content (see Figure 6) which limits how feedback from one iteration can influence the project during given phase.

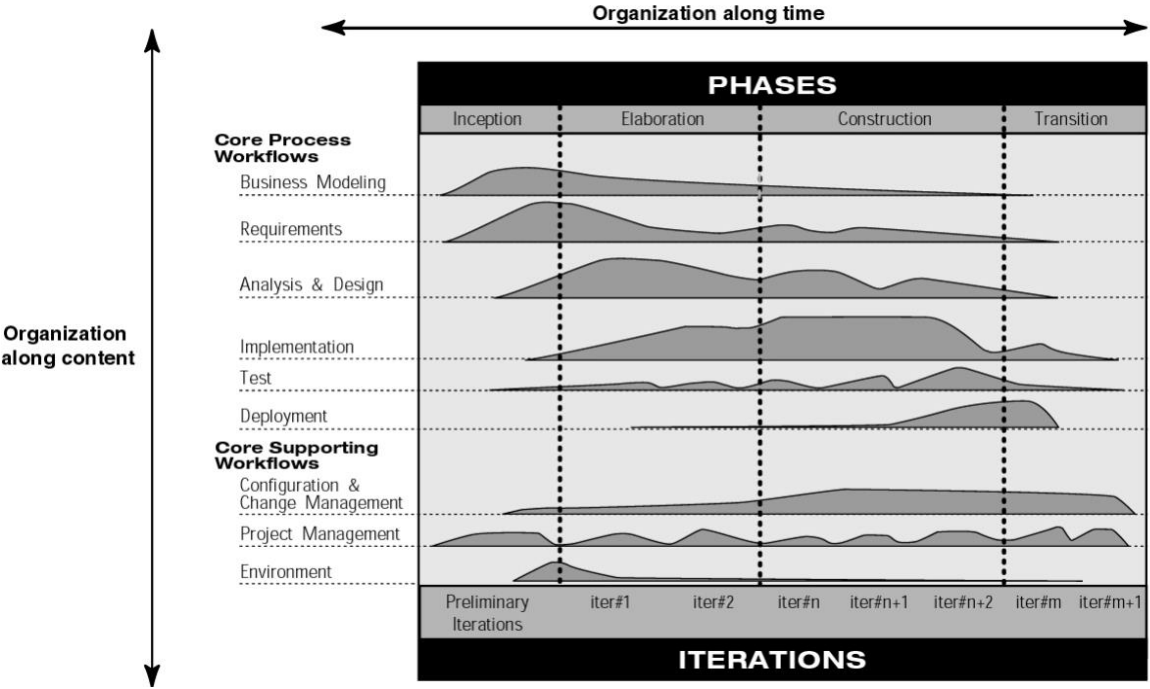

**Figure 6.** The two-dimensional process structure presented on page 2 of ref. [52] showing how much effort should go into different content in different phases of the project lifecycle.

The RUP approach is well-suited for projects that requires initial research and feasibility studies. It tends to be well-balanced in terms of traceability and amount of paperwork required to manage the project. The clear division into phases helps to build milestones needed to keep the project on track while the iterative nature of each phase creates an adaptive, agile-like environment. This is why this methodology is very popular in large-scale projects developed by companies with well-established process.

### 2.9.3.4 PRINCE2

PRINCE2 is a popular structured project management approach that uses a clearly defined framework for the project development process. The name of this methodology is an acronym for "PRojects IN Controlled Environments". The original methodology, called PROMPT2, dates back to 1989, while the PRINCE2 was originally published in 1996. Since then it has become very popular and has been made a standard for project management in many corporations, especially in UK. It has been updated twice, in 2009 and most recently in 2017.

PRINCE2 focuses on six aspects of the project: scope, timescale, risk, quality, benefits and cost. These aspects are treated as performance goals of the project. They are also used to determine the project health at any point of the project development. The initial values established for the project can vary within certain ranges, which is why these aspects are also thought of as tolerances. If a project exceeds the range in one of these aspects, a correcting action should be performed to guide the project back on track.

There are seven principles of PRINCE2 that should remain unchanged regardless of the product for which the methodology is used. The authors of the methodology state that these principles should be treated as mindset needed to achieve the complete, controllable development and they state that project must adhere to all of them. Otherwise it cannot be said that the project is managed using PRINCE2 approach. These principles are:

1. **Maintain business justification** – a principle that states, in short, that the business viability should be monitored all the time and once the project becomes unviable, it should be dropped. A special document called business case is provided that helps to enforce this rule.
2. **Learn from experience** – a straightforward rule that aims to reduce the risk and the work log by re-using previous solutions and avoiding risky scenarios. Several documents are prepared to help aid in this process.
3. **Define roles and responsibilities** – PRINCE2 structures roles into four layers: corporate management, project board, project management and team management. This principle clearly defines the responsibilities of each role from each level. It also separates roles from people that perform these roles allowing many people to share one role or one person to take on several roles as long as the responsibilities are fulfilled.
4. **Manage by stages** – this principle divides the project into stages, each of which ends with an update of the project documentation followed by the planning of the next stage based on the updated documents.
5. **Manage by exceptions** – this point stipulates that as long as the project's progress is within the previously mentioned tolerances no special action needs to be performed. However, if there is a risk that an action will exceed one of these tolerances, such risk should be escalated to the higher management for a decision on how to resolve this issue.

6. **Focus on the product** – the overall goal at all times should be to deliver a product that meets predefined quality requirements.
7. **Tailor to suit the project** – the methodology should be tailored to suit each project by taking into account all six aspects of the project mentioned before. Tailoring should be the first activity that starts the project and its outcome should be reviewed at each stage.

PRINCE2 also defines seven themes which are built on top of these principles and are used to look at them from a practical standpoint. These are: business case, organization, quality, plans, risk, change and progress. The methodology also describes in fine detail all stages of the project, documents metrics and actions performed at each stage of the project.

The amount of details provided by this methodology is one of the reasons for its popularity as it alleviates the team that develops the project from defining them by themselves. It is also the reason why the methodology is popular in large companies undergoing large projects but is not well suited for small companies and small-scale projects for which the amount of documentation and processes that need to be done is a disadvantage that outweighs benefits of a clear and controlled environment that the methodology gives. It is also distinctly not suited for constant change of requirements, which is why modified versions of the methodology exist. Most notably, PRINCE2 Agile treats agile methodology as project environment and uses special techniques to help with project management, risk management and risk response performance estimation. On the other hand, XPrince [53] mixes Extreme Programming (XP), PRINCE and RUP to create a new, agile methodology that tries to balance between the rigidness of PRINCE and flexibility of XP by fitting them into a framework outlined by RUP.

### 2.9.3.5 The automotive Spice (ASPICE) model

The automotive industry almost universally uses the automotive SPICE process (abbreviated as A-Spice or ASPICE) [54] which at the first glance looks like it is on the completely opposite side to the agile approach. It is for a good reason – the automotive industry, similarly to many hardware-related industries, work in long cycles dependent on the production cycle of the hardware unit. The ASPICE model greatly focuses on traceability and quality assurance. The importance of traceability in this industry is hard to understate and methods the industry uses are very effective. An anecdote says that Ford, if needed, can track a supplier of a screw in any given car manufactured up to twenty years back; a feat that, if true, would be immensely valuable when such screw causes a failure as it may help preventing similar failures in other cars with screws from the same supplier.

ASPICE model enforces traceability through extensive documentation of each step and each part of the system that is being built. While it does not strictly force the waterfall top-down approach, it requires certain steps to be done in correct order, such as software requirements analysis, design and test planning to be done before implementation can begin (see Figure 7). The amount of work related to planning the test at this early stage of the development is quite unique to this methodology, to the point that each requirement has to have assigned acceptance criteria and testing method before it is implemented.

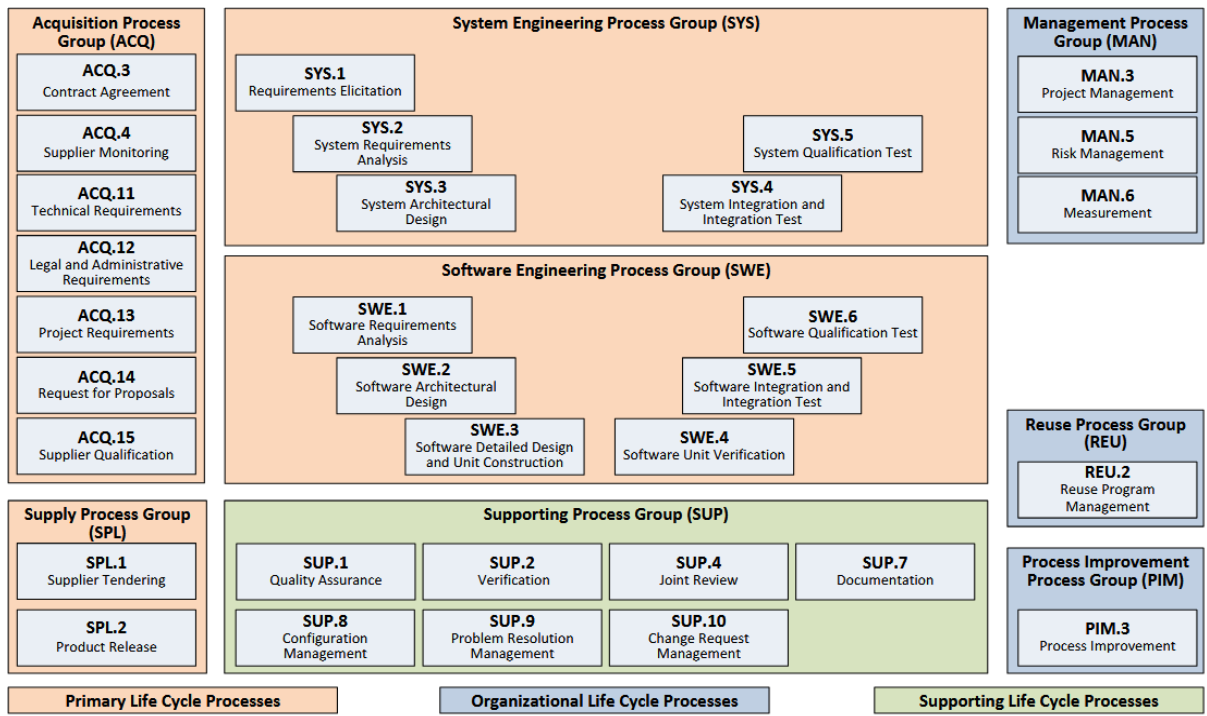

**Figure 7.** The ASPICE reference model, so-called V-model, presented on Figure 2 of ref. [54]. The right-hand side of the V-model (SWE.4, SWE.5, SWE.6, SYS.4, SYS.5) are processes designed to verify the left-hand side of the model.

The benefits of this approach to the development process is twofold; developers know exactly how to implement that feature as they know what criteria such feature has to pass and testers can work independently from the developers creating the test alongside the development team without the need of these two teams to communicate. In fact, the separation is so great, that testers can even write the unit tests themselves, based on documentation and design diagrams alone.

There are drawbacks as well, starting from the amount of paperwork that has to be done up-front. Creating such detailed specification of the whole system so early means it is hard to ensure feasibility of the designed solution or to verify that specification is correct in all use cases. There can be number of errors in the specification and fixing each of them has to follow appropriate process, impacting both the development and the testing team. It also means that introducing a change is very hard; harder even than in modern-use waterfall approaches. That is why a lot of time is spent on the design and requirement specification and only the most qualified team members are designated to these tasks. Efficiency of the overall process strongly depends on these two steps. While the ASPICE model is perfectly fitted for automotive industry it requires great amount of documentation which may be too excessive for many other projects.

# Chapter 3. Managing scientific software development

In order to attempt to describe how scientific software development can be managed, we need to first establish our goals and challenges that we will have to face. In this thesis we will focus on the smallest and most common type of tool listed in Section 2.8 – a small-scale Monte Carlo tool. Such tools are built for specific purpose driven by current and possible future analysis performed by physics communities.

The complete product consists of:

- A set of libraries – this is the main resource that can be included in other projects
- Source code – used to extend or modify the product
- Examples of the product usage – used for initial, technical tests and to present the product functionality
- Validation framework – used to verify scientific basis of the product and technical correctness of applied solutions
- Scientific paper – describes the implemented solution and validation procedure
- Technical documentation – descries the intricate details of the implemented solution and serves as a reference manual on the tool usage

The development team consists of:

- Theoretical physicists
- Experimental physicists
- Computer scientists

Additionally, the development process is aided by:

- Potential users
- Experts from different fields of physics

The difference of experience between the team members is a source of many additional challenges related to work organization and communication not present in common projects. Finding a methodology that can take into account all challenges of such software development proves to be quite a challenge.

## 3.1 Challenges

In introduction, in Section 1.2, we have listed some of the main concerns that need to be addressed when developing scientific software. To reiterate summary of this sections, some of the challenges that a team working on scientific project must face are:

- Taking into account inter-project dependencies
- Creating extensive tests and validation frameworks
- Dealing with large amount of legacy code
- Providing software that can be easily modified by experts from fields other than computing science
- Dividing knowledge between team members
- Accommodating for incomplete specification and possible future use cases
- Encouraging modifications and documenting possible future extensions

Following sections address this list and expands upon it by going in more detail into the most important challenges found during development of a small-scale Monte Carlo tool.

### 3.1.1 Multi-layered structure

All projects described in this section are based on layers built from different fields of expertise. Each of these layers comes with its own set of problems and needs throughout validation before the next layer can be built on top of the previous one (see Figure 8).

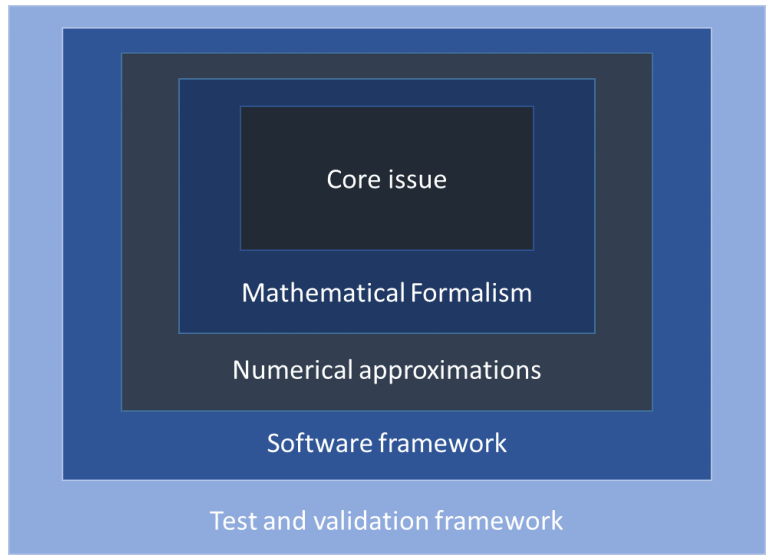

**Figure 8.** Example of layers from which the scientific software can be built.

The core of the project is its scientific context, which needs to be identified and defined. To find the solution to the problem at hand it has to be described using mathematical formalism. This is needed so that software implementation can be made, which in turns requires numerical approximation so that the formalism which operates on infinite precision can be computed using finite-precision machines. In order to test, validate and use the algorithms built for the purpose of solving this problem, a software framework must be constructed which then enables building test framework.

It is easy to imagine mathematical formalism built as closely to the physics context as possible. On the other hand, it is hard to create a well-organized structure that does not obscure this context.

> **Example:**
>
> Let's say our **core** is a complex theory that describes existence of a resonance "Y". Let's say that Leading-Order (LO) approximation of the probability density function (PDF) of this resonance is the gaussian distribution. In short, this means we can use gaussian distribution to estimate probability of generating the resonance of given energy.
>
> Our **mathematical formalism** will then be to describe these resonances using gaussian distributions.
>
> We can use one of the well-known algorithms, such as Box-Muller transform, to generate random numbers with gaussian distribution. This will be our **numerical approximation** of the mathematical formalism.

> Next, we will write an API that allows to initialize our generator and provide parameters for setting the width and mass of our resonance (they correspond to $\sigma$ and $\mu$ of the gaussian distribution). Lastly, we create function for generating the resonance. This becomes our **software framework**.
>
> Now, we can write variety of test based on this framework and embed them into our **testing framework**[29].
>
> Note that the final product that the user has access to completely hides the internal model and potentially obscures the core problem. Not to mention that this approach is closely tied to the original assumptions, therefore it is not extensible.
>
> Consider that NLO model requires a new set of parameters and the resulting PDF can no longer be approximated using gaussian distribution. This requires changes to be made to all of the layers of the project.
>
> An alternative approach would be to create a framework that allows to provide a custom PDF and custom parameterization of the PDF. In such case the software framework, and thus the initial test framework, would not change. The tests would have to be eventually extended to include the more complex model but it would not lose the effort previously put into its creation.
>
> Lastly, if the authors of this tool were to come up with mathematical formalism as close to the core problem as possible, then potentially, its implementation would take into account any future extensions of this model. It's implementation, however, will most likely be very complex, if not impossible.

There are number of things that need to be considered when constructing each of these layers. The most important ones are relations between the layers, restrictions that they are able to fulfil and the quality they are able to provide as well as the flexibility to ever-changing requirements that they are able to maintain.

The relations between different layers within the project are the main source of challenges of the complex systems. Developers have to validate each layer separately in as much independent way as possible. Such testing procedure is mostly unique for each problem and requires the development team to spend a lot of time constructing specific test frameworks.

### 3.1.2 Challenging human resources management

While project management in scientific community differs from commercial applications, in terms of organization and financing, they are driven by similar goals. They represent similar level of complexity and, by extension, they are plagued by the same problems. It is a challenge for community of hundreds of people (or in case of the biggest experiments such as `ATLAS` or `CMS,` even thousands). The work on a single project takes years to complete, with new members introduced mid-work and current members leaving the project, taking their experience and acquired knowledge with them. This forces the development teams to divide their attention between work and training of the new team members. On the other hand, a smaller team of experts is much more prone to lose the accumulated knowledge when a team member leaves the project. That is why they have to divide the knowledge between the team members, which is especially hard to achieve considering their different fields of expertise.

---

[29] Note, that this test framework is used for end-to-end tests as it operates on a software framework. Such framework is needed to present results that users can recreate themselves. Users are not required to know the details of implementation. However, during project development, each layer has to be validated and tested upon creation. Tests of each layer should be available to the developers in case additional tests are needed.

The rapidly changing environment creates a unique challenge from software engineering point of view. The project's purpose expands over regular business applications as one of the goals is to create a knowledge base for both the current team members, the part-time collaborators, and all future members of the project. Different contributors have different goals within the project and different level of engagement in the project (see Table 3).

| *Member* | **Role in the project** | **Cooperation level** | **Contribution** |
| --- | --- | --- | --- |
| *Expert* | Content provider | Involved at all stages of the project. High engagement in the project | Core content creator<br>Gathers knowledge about the project content<br>Teaches less experienced team members |
| *Computer Scientist* | Software development | Involved at all stages of the project.<br>High engagement in the project | Creates and organizes software<br>Gathers knowledge about the project content |
| *External collaborator* | Content provider | Involved for short periods of time at different stages of project development.<br>Focused on contributing to selected project functionality | Supplementary content creator<br>Provides analysis and test data |
| *External expert* | Content provider | Involved for short periods of time at different stages of project development | Core content creator<br>Provides missing competencies and expertise |
| *Initial user* | User | Involved for a short time at last development stages | Performs tests<br>Provides usability feedback<br>Reports issues<br>Identifies potential new use cases |
| *Future user* | User | Uses published software<br>Can potentially become an external collaborator | Provides usability feedback<br>Sometimes reports issues<br>Sometimes identifies potential new use cases |

**Table 3** List of different types of project contributors, their role, engagement in the project and their contribution to the project.

Managing the knowledge spread across the whole team becomes an important task of the project. Experts from other fields often join the project for a short period of time to help solve a particular problem. Such collaborators are based in different universities and are members of different communities creating problems in communication and organization of work. This is especially difficult when considering the information barrier that must be maintained between collaborators from different environments[30].

Such cooperation may require some steps of the projects to be performed by only those collaborators that have privileges to access the corresponding data. This in turn requires the project to be used and modified, sometimes extensively, by such people. All keeping in mind that they do not know the full structure of the project and have no time nor should be forced to learn it. The collaboration with such experts is usually limited to one or two projects over relatively short amount of time, so the time spent on adapting to the infrastructure of the project would only introduce delays with little to no benefit in the long term.

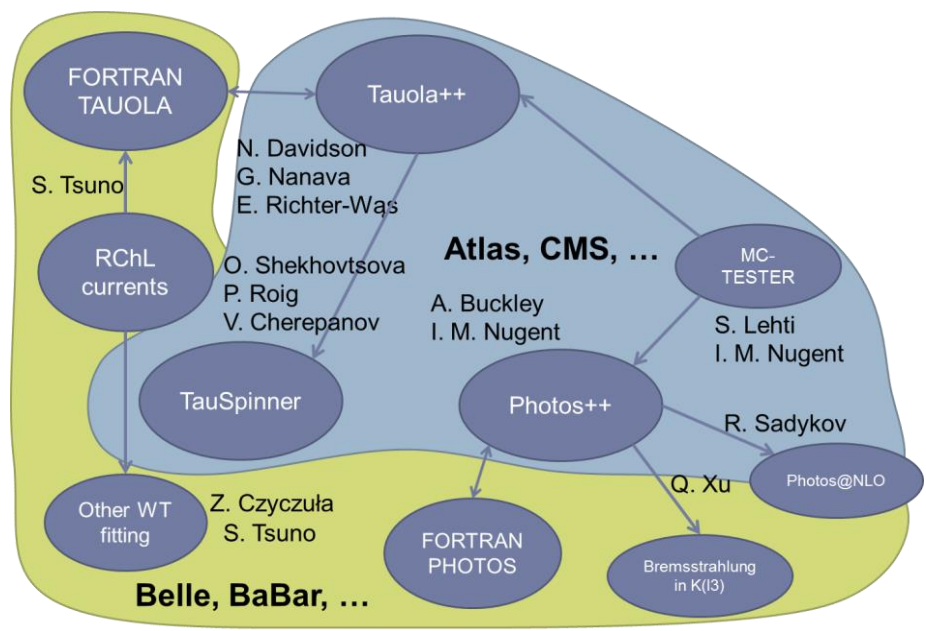

**Figure 9.** Relation between different projects, environments and collaborators of the projects mentioned in this thesis. This list is in no way exhaustive. The green background color represents low-energy experiments[31], the blue one represents high-energy experiments[32].

Figure 9 presents some of the relations different projects of the scientific team have with their collaborators and their environments. As each of those collaborators are members of larger group or experiment, for which the software is intended, they represent the stakeholders of the project. Their collaboration is of utmost value; thus, the infrastructure of these projects

---

[30] For example, collaborators from different experiments are not allowed to share their data. Their collaboration is restricted to sharing only the results obtained from these data. Similarly, in business environments, experts hired to fill the gap in experience of the business team, have restricted access to the data and the knowledge base of the whole project.

[31] Low-energy experiments focus on microscopic phenomena regarding arrangement of electrons in the atom or relation of protons and neutrons within the atomic nucleus. These phenomena concern energies around several millions electronvolts.

[32] High-energy experiments are focused on understanding the properties and behavior of elementary particles, such as quarks, gluons and bosons. This is done mostly through studies of collisions of particles with energy of hundreds of megaelectronvolts.

must provide a way for them to work without the needless knowledge of the rest of the project.

Since the team responsible for the project may change completely throughout the development process, this force changes of approach to the project itself. History of the project becomes the source of knowledge and every change must be well documented, including its reason and purpose.

The lifetime of the projects written for physics experiments is even longer and can sometimes last decades[33]. In contrast, the lifetime of a software version use becomes inappropriately short with versions lasting weeks or months. This poses a significant problem from the perspective of the software development strategy.

Lastly, it is important to note that while all modern agile methodologies [49] allow to adapt project to numerous changes through incremental updates introduced throughout the project development, they are not able to account for changes that will be introduced five or ten years from now as it is impossible to predict how the project will be used in future. This requires fundamental changes in the approach to the project design. Modularity and ability to change every aspect of the project, even the most fundamental one, becomes the crucial element. Use of the experience or intuition of experts in physics can be important.

### 3.1.3   Non-uniform, non-standard software structure

Leading edge scientific projects can be distinguished from many other applications by one key element: science requires novelty. Each project represents a new, unique solution. As stated before, the most significant content of the project is provided by specialists in different fields of science. When talking about the project software, the code produced during early stages of the development of the project can be effectively analyzed and debugged only by its original authors as such code may be developed prioritizing consistency and correctness of the results over code organization. In fact, it might be developed completely outside of the target framework. With time code may be adapted to better suit software framework and allow other specialist to analyze its content more effectively. It is crucial that software organization do not impair these first steps of the development of the proper algorithms.

Knowing that field of specialization of the author of such code follows different principles than code written by computer scientists, its organization may not adhere completely to the rules set by the rest of the project. This notion may be as simple as matching variable naming style to the equations taken from source publications, so as not to confuse the reader about the meaning of such equations, or it may relate to the style of the interface of the code written to match parts of already existing solutions[34]. Such code is far easier to understand by original authors and other specialists that have used it in the past. It is their input and their future use of this code that is the focus of such project. That is why, to some extent, such legacy solutions should be permitted inside the project, adapting to its format as much as possible without impairing the ability to analyze and modify it by its authors and initial users.

---

[33] The τ decay Monte Carlo generator TAUOLA, up to this day used by many experiments, was first introduced in 1990. Many significant changes have been introduced over the years, including the C++ interface introduced in 2005, but the core generator, written in Fortran, is still included with all distributions.

[34] A good example is `RooFit` package [32], where parts of the code evolving from `FORTRAN` was left with a C-style API instead of moving to object-oriented approach of C++ framework to which the package was adapted.

> **Example:**
>
> An expert was asked to provide the results of a set of algorithms. The output should account for future new algorithm and should be easy to serialize. The expert has prepared a generic data structure to provide the results of these algorithms as shown on Listing 4.
>
> The result of each algorithm is provided through a separate structure. A union of all result types is created and a list of such unions can then be serialized and sent between modules. The definition of different types of the results is stored in a header file. The receiving module has to properly cast this structure to retrieve the data as shown on Listing 5.
>
> Taking aside the obvious flaws of this approach, it does fulfil its purpose: the expert can relatively easily expand the structure with new types of results for new algorithms and can provide the rest of the team with appropriate header file describing data within these structures. He or she can also send any number of results in any order and, to some extent, provide backward compatibility[35]. The goals, that expert was asked to fulfil, are met. Needless to say, this solution would have been different if it were to be introduced by a computer scientist.

```c
struct result1 {
    int value1;
}

struct result2 {
    double value1;
}

union single_result {
    struct result1;
    struct result2;
}

struct all_results {
    union single_result[10];
}
```

**Listing 4.** Example structure in C used to store the results of a set of abstract algorithms.

---

[35] Older receiving modules can parse the results of a new sender module provided that the size of the union has not changed. The switch statement from Listing 5 allows the older module to ignore unsupported (new) algorithms.

```c
for (int i = 0; i < 10; ++i) {
  switch (get_result_type(i)) {
  case 1:
    use_result1((struct result1 *)all_results[i]);
    break;
  case 2:
    use_result2((struct result2 *)all_results[i]);
    break;
    (...)
  }
}
```

**Listing 5.** Example of use of the structure from Listing 4 which requires switch-case and type casting of the union to properly interpret the results.

The reason why the problem has been solved this way is that the aforementioned specialist works almost exclusively in `MatLab` [55] and knows how to efficiently combine `MatLab` results with the framework that he has built for this purpose. Any complex structure introduced by a computer scientist might not have been well understood by the specialist and he or she might not know how to properly incorporate `MatLab` results into such framework. He or she might end up wasting a lot of time learning the pre-determined structure or adapting his or her own code, possibly creating a layer of similar code to the one above just to ensure that his or her work remains undisturbed.

This issue must be carefully balanced. Such specialized code, if left with no supervision, can with time evolve into a project which may be hard or near impossible to apply into the framework of the project. Not to mention that combining the results of the work of several specialists may, with time, become troublesome or near-impossible. In the extreme cases, development of the core functionality must be halted in favor of code reorganization, which is both very problematic and damaging. Such situations should be avoided as much as possible. Proper communication between specialists and computer scientists must be maintained throughout the whole development process so that specialists can develop their parts of the software in a method as much suited to their own needs as possible while maintaining overall consistency and interoperability of the project submodules.

This is actually a very important issue that is similarly addressed by scientists from different fields of expertise. If the majority of focus is given to something other than software, any obstacle found when writing the code may be solved using fast, short-time solution that deals with the issue as quickly as possible without distracting the scientist from his original goal.

> **Example:**
>
> Recall examples of issues related to different use of the Event Record by different experiments presented in Section 2.5. Last of these examples mentions how variables with different purpose (such as Event Record number or particle number) have been given more meaning than they were originally supposed to have. Particle number, which was supposed to only describe a relation between particles (e.g. particle 7 decays into particles 8 and 9) was used to encode additional information about the particle.
>
> This additional information was needed for the analysis purposes and the Event Record could not include it in any other way. The simplest approach was to

> introduce a hack which repurposed unused bits of some variable to include missing information.
>
> Such customization must have been well documented within the collaboration that introduced it internally and was probably followed without any issues in all internal analyses. The problem occurred when external tools were used. These tools had no knowledge of such non-standard modifications.

What is important to point out is how the behavior of the users affects the developers of the tools. Usually, when user raises similar issue, the developers of the tool can answer twofold:

1. Incline that this is a non-standard modification which the tool does not support
2. Modify the tool to allow for this non-standard behavior

First approach would force the user to either introduce a workaround specifically for this tool or to switch to another tool which allowed for this non-standard behavior.

Second approach would require developers to introduce a mechanism that may require lengthy implementation and hacks to the internal data structures of the tool in return for functionality that may be useful only to a single user.

The second option may seem less reasonable from the point of view of the developers of the tool but is, in fact, most often the one that the developers will follow.

This ties to the conclusion that comes from the previous example: if the solution is useful, it is worth compromising on its clean implementation, robustness of the structure of the tool or the use of best practices in software design. For this reason, the software structure of the tool isn't always following the best practices and may contain solutions that may look like temporary workarounds. Such solutions may provide functionality which, to an outsider, may seem useless, but is in fact crucial to some narrow user base.

### 3.1.4 Complicated testing process

Typical project requires at least as much time to be spent on testing as on its development. Sometimes even drastically more time goes into the testing process. Unlike typical projects, software that consists mainly of physics content and data analysis, requires testing procedures that are almost always unique and have to be designed and written for each problem separately. And when discussing scientific projects, they must pass complex validation procedure[36] driven by the scientific approach. This requires, among many other things, that the validation procedure needs to be defined and described in a way that allows it to be reproduced in different environments. In this context, an environment means not only the technical differences of different compilers or operating systems but mainly the differences between the different analysis frameworks in which the tool will be used. This is an important requirement as it drives the development process towards building a robust testing framework.

The most standardized forms of testing, such as writing unit tests or behavioral tests, are the least useful as they only test the functionality of the technical aspects of the project. Majority of the tests concern the physics content and the algorithms themselves. In case of Monte Carlo tools this issue is even more tricky. One can easily write basic assertions and test the boundaries of the most crucial parts of the algorithms but the overall validity of the algorithm can only be verified by producing a statistically significant data sample and testing it against analytical results or benchmark data. However, not often analytical formula can be derived or benchmark data are present. And if they are, it is also possible that these kinds of tests are still not enough. Figuring out the proper method of testing and validating the algorithms becomes a troublesome task.

---

[36] One of the most problematic but also the most important steps is to confront the theoretical model with the experimental data. This process requires a lot of effort and fine-tuning of the parameters used in the model but allows to validate the theoretical assumptions used in this model. Also, fine-tuning even the trivial technical parameters, such as precision of numerical integration, may require a lot of attention.

One of the side-effects of this process is that the test environment carries extensive knowledge of the tested subject. Capabilities of the testing framework of such projects grows along with the project itself. As such, it should be properly documented and maintained as it is invaluable when approaching similar problems in different environments or when moving to different architecture or programming language, as they hint to what may be the root cause of the problem.

A large part of this thesis is devoted to the subject of testing scientific software. It has been addressed in detail in Chapter 5.

### 3.1.5  Lack of beta-testers and scarce user feedback

It is important to stress how dependencies between different projects influence the development process. When chaining several tools in a single analysis, it is hard to fully control each of the tool's setup. Most of the users spare no time to read the documentation of an external tool that they are using. Sometimes even reading a `README` file explaining short procedure becomes a hassle. User starts reading documentation only when the results seems not to correspond to what they expect. That is a very precise statement – user can ignore warnings, non-critical errors and other information as long as the result *seems* correct.

In case of physics experiments, when the setup is incorrect, the result may still be correct due to the number of overlapping coincidences or it may seem correct from the point of view of the analysis performed by the user. In both cases any type of message issued by the project may very well be ignored and underlying issue, if it exists, may lay undiscovered until much later. And of course, the later the problem is discovered, the harder and more expensive it is to fix.

That is why it is very important to let the user know about every single issue or possibility of wrong configuration. One has to design the project in a way that will properly inform the user what is wrong with the configuration on which the project is running. This also poses its own set of problems. Throwing out too many warnings and errors proves to be counter-productive as users tend to ignore repetitive output that they think they know the source. There is also the issue of what kind of information should be provided to debug the problem, which often is not apparent. Between the information flooding and the lack of the information, proper design of the debugging interface and error messaging system is a very important issue.

As mentioned also in Section 3.1.2, the social relations of the project cannot be disregarded. The goals of our users do not always coincide with the goals of the developers and they may not feel obliged to nor be willing to spend time, however short it may be, to help improve the tool. Having different goals and priorities, they can even willingly risk using the tool in an improper way for a chance that it may actually work correctly.

> **Example:**
>
> During installation of one of our tools we have been asked by one of the important users of the tool to turn off the critical stop procedure that was triggered when the project was working using configuration it was not supposed to process. This request came despite the fact that this error was the only reason we've received the feedback about the problem and was able to track its source in the first place.
>
> We have informed the user that introducing such option may cause improper behavior of the tool and this option will limit our ability to debug such issues in the future.
>
> The option has been introduced nonetheless.

While such extreme cases are very rare, the less severe ones are much more common, creating an environment in which collaborators from different experiments can sometimes be the only beta-testers that provide feedback about the performance of the project in their environment. This is far from exhaustive tests of possible use cases and yet, in some cases, this has to suffice.

Since the future users of this tool must be able to understand and adjust the tool to their own needs, their feedback during the development process is a key to the proper design of the interface of the project. This is also a very important aspect of the project as more often than not such interface must be adapted to allow experts contributing to the project development to freely add their own input without the need to adjust their own tools and methodologies.

### 3.1.6 High impact of the project dependencies

Due to the lifetime of the projects mentioned in this thesis, their user base may change or expand drastically. A tool that cannot adapt to these changes and forces the user to adapt their own environment instead, can be easily pushed away in favor of a more flexible one. This problem may require constant adapting of the project to new standards but with proper approach can be easy to manage. The more complex problems arise when looking at the software environment from a larger perspective.

The tool should be applicable to as many use cases as possible. This is mostly limited to the quality and availability of the data on which the tool is used[37]. Figure 10 presents the relations between number of projects described in this thesis and few of the external projects used by the team to perform their analysis. This list, of course, is not comprehensive. It ignores the experimental physics side of the projects and takes into account only the most basic applications. When Monte Carlo generators such as `Tauola++` [5] or `Photos++` [4] are placed in context of the environment of different experiments (see Figure 9), they are subjects to new set of constraints and dependencies coming out of these experiments. This relation diagram could expand endlessly.

---

[37] For example, the data structure used to describe the event might not be sufficient to describe all of the effects needed to perform the complete analysis or the event description might not contain all of the required data. As will be shown later, the tool might be able to produce partial results, still useful in some use cases, based on such incomplete information.

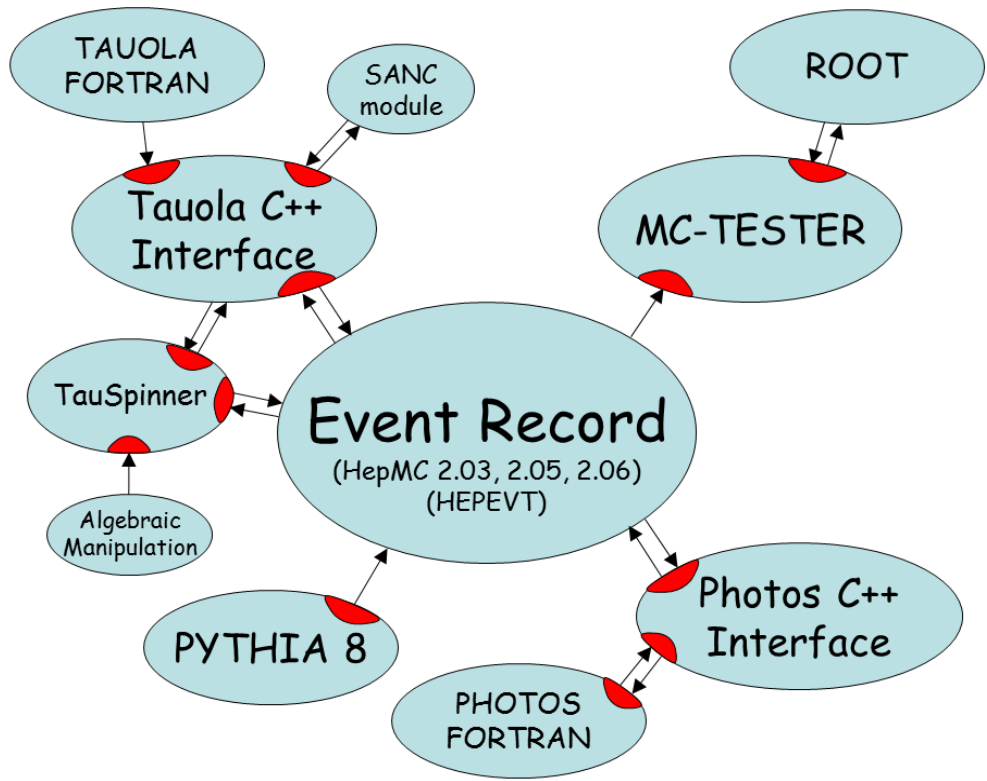

**Figure 10.** Relations between different projects that the scientific team have taken part of or used in their analysis. Red segments symbolize interfaces used for the communication. The lines indicate the direction of the flow of information. The size of the elements indicates the significance of the element in the communication process.

However, already at this level, some more complex problems can become apparent. For example, most of the communication between different modules is done through the Event Record. As described in Chapter 2, Event Record represents a single event simulated by the Monte Carlo tools or registered in the detector. Many experiments use their own Event Record within their projects. However, to communicate with other experiments or use external tools, they use one of the more standardized Event Records, such as `Les Houches Event (LHE)` [56] or `HepMC` [57]. A change of the standard in how the information is stored can force modification in number of different modules in different projects. It is crucial to protect key parts of the project from such changes[38].

This is not always trivial and sometimes even impossible to arrange. Let us consider another example of storing the mass of the particle in the Event Record. Not every Event Record stores such information as it can be derived from the particle's four-momentum[39]. In earlier times, when disc space was far more precious than today and the data was compressed as much as possible, such information was not included in the Event Record and was instead calculated whenever needed. Unfortunately, some of the applications, such as final state radiation performed by `Photos++`, requires information about the mass to be provided with high precision. Due to numerous manipulations of the kinematical configuration of the particles stored in the Event Record, the information about four-momentum may lose its precision. It becomes crucial to know the original mass of some of the particles.

---

[38] As described in Section s 6.2 and 6.3 in our projects it is done by separating event record implementations from the abstract interface containing minimal functionality needed to work with the data.
[39] Some of the Event Records store mass instead of energy.

This has been resolved in number of ways by different projects. For example, `HepMC` Event Record [57] introduces the concept of `generated mass` – mass provided by the generator that added the particle to the Event Record. This is, supposedly, the precise information about the mass of the particle, as opposed to the value calculated from its four-vector[40].

Considering the above example, the project must take into account all possible exceptions to how the mass information is stored – the Event Records that provide precise information about mass and those that don't provide this information at all. While it is easy to make an abstract method for getting the information about the mass, it becomes harder if one considers the fact that the project must be able to treat such information differently depending on its source. And in the case of `Photos++` this difference determines whether the algorithm will be able to process an event containing such information or not, influencing the overall usability of the project.

Some of the projects, however, do not require such information and this, again, causes problems for `Photos++`. Since this information is not needed, an analysis performed by the user can completely omit this information from the Event Record. This is another example of how even the simplest dependencies between the projects can cause complex problems.

### 3.1.7 I/O interface design issues

Lastly, the most important challenge is to maintain proper communication between different modules of the same project. No project in physics community is intended to work as a standalone software. Be it the Monte Carlo generator used in the generation process or a tool for the analysis, every project has its own set of dependencies and will always be used in combination with modules or results taken from other projects. Let us consider the event generation process. A simple Monte Carlo simulation can include all or some of the steps listed in Table 4.

|   | **Generation step** | **Tool used for this step** | **Output** |
|---|---|---|---|
| 1 | Hard process | General-purpose MC Generator (GMC) or specific hard-process generator or input file | 5-10 particles describing the hard process |
| 2 | Parton showers | GMC | Depending on the process and settings, from few to 1000 particles added to the event |
| 3 | Initial state radiation | GMC or separate tool (afterburner) | Photons and lepton pairs added to the event |
| 4 | Heavy particles decay | GMC with separate tools for specific processes | From few to 10000 particles added to the event |

---

[40] This is of course not always true, for example in case of broad resonance.

| 5 | Final state radiation | GMC or afterburner | Photons and lepton pairs added to the event |
| 6 | Calculating the influence of the interference in radiation | Afterburner | Modification of existing particles, additional photon emissions |
| 7 | Applying electro-weak corrections | Afterburner | Modification of existing particles in the event |
| 8 | Saving the event | Analysis software | Event data saved to file |

**Table 4** Steps performed in an example Monte Carlo simulation chain.

At each step, the information from previous step is used as an input for the next. Considering the fact that each of those points can be performed by a separate tool (or several tools), each managed by different team and each developed using different design strategy, this introduces a chain of dependencies in which changing one of these steps has influence on some or all of the next ones, at the same time imposing restrictions on some or all of the previous ones. For example, the information about particles generated in one step serve as an input for the other and can be modified as a result of its work.

Let us consider that in step 4 we add $\tau$ lepton[41] decay to the previously generated hard process by using `Tauola++`. The information used to decay $\tau$ is presented in Table 5.

| Information needed for single $\tau$ decay | Mandatory or optional data | Result |
|---|---|---|
| Charge and four-momentum of the decaying $\tau$ | Mandatory | Decay products |
| Information about process that generated $\tau$ (or $\tau$ pair) | Optional | Corrections based on the initial process |
| Information about $\tau$ neutrino $\nu_\tau$ (or second $\tau$ In case of $\tau$ pair decay) | Optional | Corrections due to correlation between $\tau$ and $\tau$ neutrino ($\nu_\tau$) (or two $\tau$'s in case of $\tau$ pair decay) |
| Polarization vector of the intermediate boson | Optional | Longitudinal polarization |
| Information about "grandmothers" (incoming beams, particles from which the simulation has started) | Optional | Electroweak corrections |

**Table 5** Information used in single $\tau$ lepton decay.

The product of $\tau$ decay is a list of particles ("children") of the decayed $\tau$. An example of adding $\tau$ decay to the process is showed on Figure 3. As listed above, `Tauola++` relies on the information about the $\tau$ decay stored in the Event Record provided as an input. This setup implies that the validity of the output of the generator depends heavily on the input. The first problem that arises from this setup is that we cannot control the validity of the input.

---

[41] $\tau$ leptons have a long lifetime compared to other particles and thus they serve as a good example of well-defined process that can be easily separated from the rest of the simulation.

This is one of the most common problems in all complex systems – teams that develop parts of the project independently have to work out an efficient way of communication between each other to form a full product. The communication must be robust and subject to multitude of changes. What aggravates this issue in the case of physics experiments is that the developers of a particular project have no way of knowing the full user base. The project is designed with several target experiments in mind and if it is successful, more others can find their use with it as well. To allow that, a great deal of focus should go into flexibility.

This means that the tool has to be able to understand wide variety of possible inputs; variety of standards in which different generators store the information. This is both technical problem (related to the Event Record structure) and physics problem (related to the standard of how the physics information is stored). The first and foremost dependency is to comply with these standards.

Now, depending on the previous steps of the analysis, some information may be missing from the Event Record and the project must take this into account. Note that only the first information listed in Table 5 is mandatory. In fact, in the most simplified example, to produce the decay it is enough to know the charge of the $\tau$ and its four-momentum. If some of the information mentioned in the table is not present, the respective effects will not be included in generation. The design of the Monte Carlo generator must create a communication interface that will take this lack of information into account and provide a simplified solution in the case some of the information is missing[42].

This may not look like significant problem at first, but keep in mind that the above description is true for every tool in the analysis chain. While the tool has to adapt to the variety of inputs, the output will also vary. Therefore, the next tool in the chain has to take this variety into account. If the output of a tool is not compatible with another tool in the chain, one of those tools cannot be used in a chain.

The goal of the team of developers is to comply with as many different standards as possible, as ability to adapt to variety of analysis chains is the basis for expanding user base. However, not knowing the exact user base means there is no way of knowing what tools (or analysis software) to adapt to; what standards the input and output has to comply with. Since multitude of standards exist, an interface allowing to adapt to as many of them as possible must exist.

### 3.1.8  Summary

In this section we have presented the most common challenges faced by the team developing scientific software. We can summarize them as follows:

- Project spans over several layers; one has to be aware of complex relations and restrictions imposed by these layers

---

[42] There are also number of reasons one needs to exclude additional effects from the generation. For example, if they are to be applied later by yet another tool in the generation chain. It is also crucial for testing and validation of the Monte Carlo generator (see Section 5.3.4).

- Project team consists of specialists from different fields of expertise. Project must allow efficient collaboration between them.
- Experts provide the most valuable input. Project should allow their input to be added with minimal effort on their side.
- Testing and benchmarking become a crucial necessity. Algorithms used in the project must be well-tested and constantly validated.
- Project must be able to adapt to as many use cases as possible. This includes potential future use cases.
- Unreliable input and lack of information in the input source should be analyzed. Impact of the varying output on other tools and equipment should be well described.
- Lack of beta-testers and user feedback. Has to be compensated with rough testing process covering as many fields of the expected use cases as possible.

We can extend this list by common challenges faced in development of the software for physics experiments:

- Extensive amount of legacy solutions used in the project.
- Work is focused on the algorithms built outside of the project framework. Ability to modify and update these algorithms comes before optimization and efficient architecture.
- Numerical issues. Any of the algorithms present in the project can become unstable under some circumstances (non-standard input, untested use cases, edges of the algorithm stability, etc.).
- Problems defining proper tests. The default testing techniques (unit tests, integration tests, behavioral tests, etc.) are useful for validating that the project fulfils the designed goals, however the vast majority of the tests regard the algorithms used in the project. These tests are in many cases hard to properly define and may produce different results when tested outside of the target platform.
- Long (sometimes very long) lifetime of a project. This challenge restricts available tools and environmental dependencies. Such dependencies may cause project to be unusable in the future, as tools may no longer be available (or usable in development/testing framework) years later.

## 3.2  Methodology

In last section the list of challenges faced by scientific software has been presented. In order to find the most suitable methodology to manage development process of such project one should first try to consider the well-established approaches to software development (see Section 2.9.3).

In following sections, we will try to relate these approaches to the challenges listed before. We will also try to compare the scientific project to a similar process found in business environment and present a short example of how scientific software development process can be partially managed.

### 3.2.1  Common methodologies in relation to scientific software

The similarity of scientific software development and R&D narrows down the choice of methodology as some of the most commonly used ones are not applicable to this type of software. For example, let us consider the waterfall approach.

### The waterfall approach

The waterfall approach is ill-suited for scientific software due to amount of planning that has to be done up-front. If you recall description presented in Section 2.9.3.1, in waterfall approach work is planned based on initial requirements. In scientific projects it is near-impossible to know from the start what will be needed to achieve the goal nor is it possible to expect the result to go exactly in the direction assumed at the start. In fact, if the methodology used during development would try to steer the development process in just one direction, instead of exploring all possible options, the project could very well end up as a failure. In high energy physics it is impossible to plan full functionality of the finished project up-front as even the basics of the whole project can change overtime. Moreover, the project itself is not made for any specific client, organization or specific experiment. The stakeholders are not known at the start of the project. There are only speculations regarding the possible use cases. There is no way to tell what functionality end users might request.

### The agile methodology

The agile methodology was created specifically to deal with the problems that regular methodologies cannot solve. However, it has many requirements that high energy physics projects cannot comply with. For starters, it requires good management and well-coordinated team to be fully successful. Writing software for physics experiments requires cooperation with specialists around the world and implies that some of the team members can join in during project development while others may leave the project due to lack of time or other issues that require their attention. The other problem is that the agile methodologies rely on constant interaction with the stakeholders, or their representatives, to provide the team with new functionality that can be planned for future iterations. In the case of high energy physics projects such constant interaction is hard to establish. For starters, it is hard to tell which experiments might actually be interested in the software as well as what potential user might expect. The requirements for new iteration may appear unexpectedly as some of the users find functionality that they would like to be implemented, while for the most of the time user feedback may be scarce at best.

Also, it's worth noting that agile methodologies use incremental improvement of software through small cycles, usually between a week and a month. Scientific software can rarely be split into such short cycles and rarely can be divided into evenly-timed phases. These are the most important factors why agile approach is almost nonexistent in scientific communities. Even projects that focus heavily on software, such as those described in Section 2.8.1, focus on few well-validated releases rather than on small updates released on a bi-weekly or monthly schedule.

### Rational Unified Process (RUP)

The Rational Unified Process shows some promise when it comes to its use in scientific software. Considering only the basics of this process its first phase, inception, could be applied only to the physics part of the project without any attempt to plan or estimate the project's effort other than in its rough outline. If the theory and the physics model shows promise and it looks like the project would be beneficial to the scientific community and could be feasibly implemented, the project could pass the life cycle objective milestone and go into elaboration phase, where actual feasibility studies are performed, risks are tackled and prototype solutions prepared. This would be the phase in which necessary research is performed and validity of the theoretical assumptions verified. This phase also allows the team to form development plan, which is then used in construction phase. The last phase of the RUP, the transition phase, would be the moment in which the software framework of the project is finalized and necessary changes are made resulting in a tool or library that can be used by wider audience.

Already at this point, when only the rough points of RUP are presented, the above description seems a bit stretched. RUP seems to be a good guideline but does not perfectly fit the scientific software development process.

One point worth highlighting in this analysis is the narrowing down of the scope of the inception phase. It may seem illogical to strip RUP from its most important safeguard process that aims to reject unfeasible projects but contrary to business applications, scientific community can still benefit from projects that failed to meet their original goals. A result in which a model has been built that does not manage to fulfil its task is still a valuable result provided that the foundations of the model are well-based in theory, the model was correctly implemented and reasons for failure were well researched and understood. Such result may simply show that the theory is wrong and yet still be very valuable[43].

A process of proving the theory wrong never starts with an assumption that the theory is, in fact, wrong. Otherwise there would be no point in working with this theory in the first place. However, if the result of an analysis shows that the data do not back the theory (fully or in specific cases), such project can still be considered successful. The process of proving the failure may results in tools and methods that can still benefit the scientific community[44]. This greatly contradicts the typical business approach and goals of an R&D team where there is little benefit from a failed project other than experience and know-how that may or may not be applicable to other projects.

### PRINCE2

When we think about the approach presented by PRINCE2 we immediately find a fundamental flaw of this approach when trying to apply it to the projects discussed in this section. The first principle of PRINEC2 is to always have one eye on business feasibility of the project. Both the R&D projects and scientific projects have no way of estimating the impact of the final product on the market. In fact, most R&D projects are unfeasible from the start and R&D branch of any company is bound to bring losses most of the time. It is near impossible to start an R&D project by accurately estimating the potential gains if the researched project hits the market as a final product because the assumptions that went into the project can change drastically based on the research itself. Similarly, the business viability of a scientific project cannot be determined as scientific community does not operate on a gain or loss basis. This means that the "cost" aspect of the project has very large margins and becomes unusable as a project performance metric.

---

[43] See Section 6.4.2 where a large project aimed at comparing the theory to data ultimately has shown gross discrepancies ultimately proving that more work on the theory needs to be done before it can be used to model the data. Several tools were created in the process, including a tool that allows to universally compare different models using re-weighting algorithm that gets rid of the sample statistical bias.

[44] An example of such project is one of the applications of `TauSpinner` tool described in Ref. [92]. What originally conceptually was an analysis attempting to challenge the Higgs discovery by analyzing the spin of the boson observed in data has failed to show discrepancies between observation and the theory. However, the analysis resulted in an extension of `TauSpinner` tool that allows to universally ascertain the spin of new resonances, which ultimately became the end-result of the project. `TauSpinner` is an exemplary project showing how unpredictable scientific analyses can be. I discuss this project in more details in Section 6.5.

Even if we strip the PRINCE2 model of business-related aspects, or at least limit them to the feasible concepts used in these two types of projects, we are left with a very rigid, tightly controlled structure that may help manage the overall process but produces a lot of excessive work and documentation as a by-product. All of these rules are very restrictive to the more natural and unpredictable nature of R&D project or physics experiments. Most of them are not applicable in such environments. At some point the fifth principle of PRINCE2, "manage by exceptions" would become the default approach as frequent changes introduced to the project can impact all of the aspects of the project.

The attempt at gaining full control of the project development may result in a large overhead of processes just to keep the project documentation on track. These costs can quickly outweigh the benefits of having well defined process for each step of the project development and knowing very well where the project stands at each point in time; values that seems unnecessary for a scientific project.

### The automotive Spice (ASPICE)

Lastly, let us consider the ASPICE model. However, knowing that the model is focused on ensuring software quality and is far more based on initial requirement analysis and planning than any waterfall approach. It involves vast amount of planning and documentation that has to be done up-front before any implementation can even start, offering in result full traceability of each element of the project and high quality of the final product. However, even the quality control would be limited only to the technical aspects of the scientific project as the physics content cannot be validated using predefined metrics.

It seems ASPICE has glaring disadvantages when applied to scientific projects and offers them no useful benefits, not to mention that its process would be hard to apply to any aspect of the scientific project other than its software. It is, by far, the least fitted method to manage such projects.

### 3.2.2  Scientific software and R&D environment

When compared to business applications, the nature of scientific software development is closest to the Research and Development (R&D) branch of a business company. The term "Research and Development" is often used to reference a branch of company focused on researching new technologies or new applications to the existing products that the company offers with a goal of jump-starting a new product. Most often such work builds on already existing experience and knowledge and does not aim for a breakthrough in any specific field of science but in finding new application for existing technology to form a new product. However, in some cases research of focused on a product that is not yet on the market is undertaken which may result in emergence of a bleeding edge technology. Teams developing such projects have little to no initial knowledge and expertise directly related to the project. They base all of their approaches and estimations on their experience or extensive research. This often means that the development process itself does not exists and has to be established ad-hoc. Such projects carry significant risks, including high chance that they may fail during feasibility studies.

R&D teams are usually treated as the company's expense or a long-term investment with high uncertainty [58]. The company invests in such teams for many reasons; be it an attempt to expand the market or simply for the potential profit of a successful, new product. The difference is that both the risk and the potential profit is greater in R&D projects than in regular ones. This observation matches my own experience.

The R&D team may also be part of a company producing custom software or product for variety of clients. Such companies put great effort into management and estimation of each R&D project as R&D becomes one of the initial steps of each of their new product. These companies have a process and common outline of an R&D project that they use to estimate time and effort as well as to plan the work.

For example, common approach in R&D projects for hardware manufacture is two-stage prototyping similar to the one presented on Figure 11. Note that the R&D process in other fields may differ from the one for hardware manufacture.

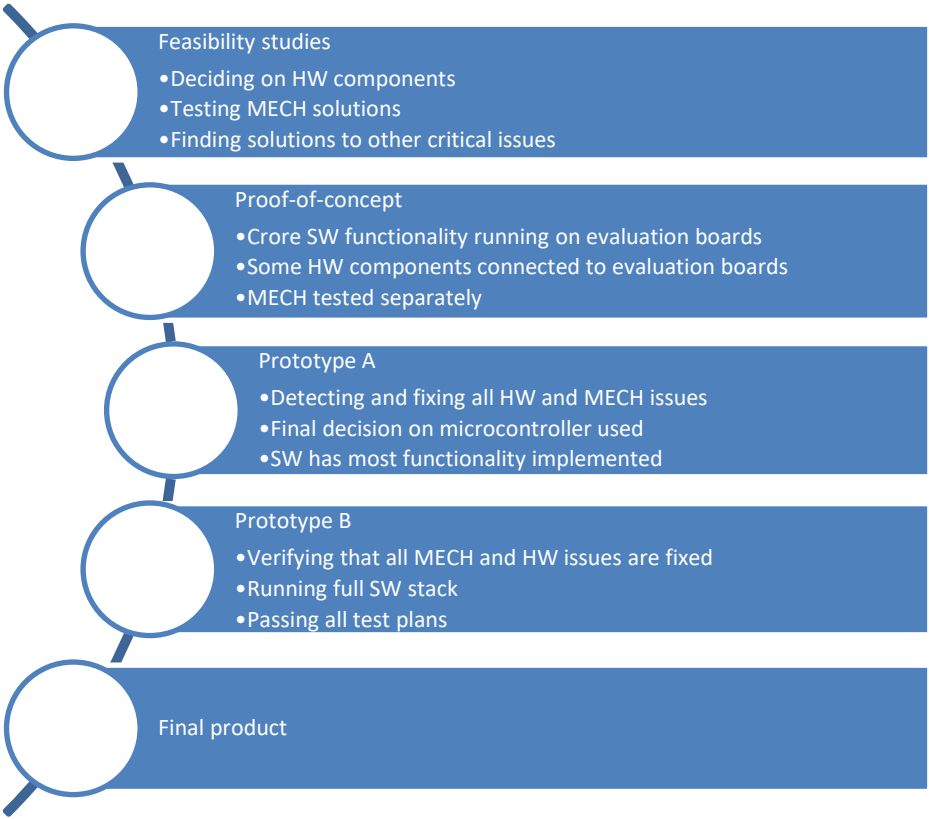

**Figure 11.** A rough outline of a typical R&D process of creating new hardware. Project is divided into three categories of components: hardware (HW), software (SW) and mechanics (MECH).

Work on hardware R&D project is often split into three categories: hardware (HW), software (SW) and mechanical components (MECH). Work on these categories often intertwines with each other but not always and certainly not during all stages of the development process. Moreover, company can be asked to produce only HW or HW and SW components with mechanical components being provided by the client.

However, when a research into new technology is involved, this ramification often cannot be used as the prototyping phase and feasibility studies take far longer and require multiple concurrent trials and multiple iterations.

### 3.2.3 Managing scientific software

While researching this subject we have found that there is no single software development methodology that can be used to effectively help managing a scientific project. It becomes obvious as soon as one realizes that all of these methodologies focus on tangible, measurable improvements of the software product. In most scientific projects software is one of the last elements that is being built and is just a small portion of the results of the project.

However, while the whole project cannot be evenly managed using one methodology, parts of it can be divided into manageable steps. For example, if the development of a new tool requires several algorithms to be created, some of them may be researched and implemented separately and some concurrently with the others. Then, the work on these algorithms can at least be partially estimated, partially scheduled and a Gantt diagram can be constructed outlining the relations between algorithms and sub-modules that will eventually form the final tool. While even this outline may change very drastically it can help focus on what is important to achieve the final goal.

> **Example:**
>
> Software described later in Section 6.4 required several components to be developed. In very simplified terms, these components were:
>
> - Physics model (including all the research that goes into building it)
> - Validation framework (used to verify that model matches the theory)
> - Mechanism for comparing results with data
> - Interface to fitting framework
> - Parallel computation mechanism
>
> Fortunately, all of the above steps could have been developed simultaneously. In our case we did not formally created the Gantt diagram but we did estimate how long each step could take and what we need for each of them. The work on the model, the validation framework and the other three tasks were done simultaneously and were managed separately until we were ready to merge them. Afterwards each team member still continued on their own part of the project but at this point we had to cooperate more closely which meant the development process became less formal.

While tasks related to scientific software usually cannot be well divided and planned into two-weeks or even monthly sprints, they can be managed using Kanban board. A four-column board (To-do, Blocked, In Progress, Done) with cards (or sticky notes, if physical board is used) where the card color signifies different types of tasks is often the simplest and most useful technique to control what needs to be done in order to progress with the development.

I have found the simplicity of Kanban board to be very useful in many R&D projects, including embedded software development. Most of the advanced issue trackers and project management tools, such as the overwhelmingly common in the IT industry tool `JIRA` [59], have the support for managing a project using Kanban boards. But even the simplest freely available tools, such as `Trello` (see Figure 12), are often enough to create a high-level project status overview.

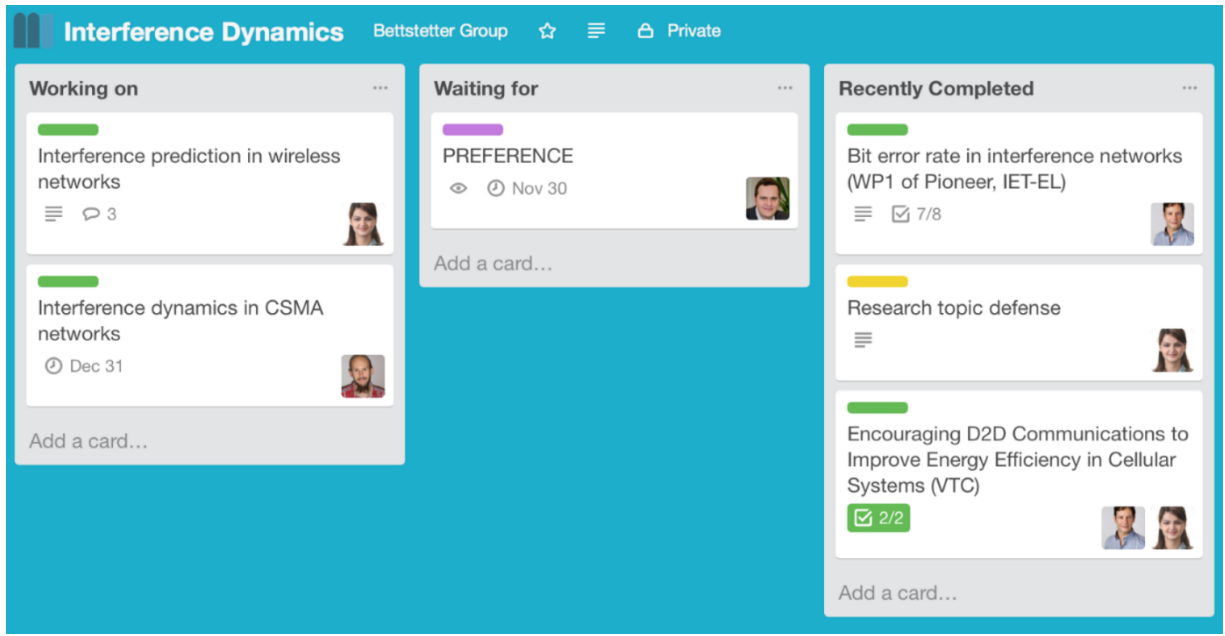

**Figure 12.** An excerpt of the board created in `Trello` tool ([www.trello.com](www.trello.com)) of the research project "Interference dynamics" from November 27, 2016. Picture taken from Ref. [60].

### 3.2.4 Summary

In this section we have presented how commonly known development methodologies fail short when it comes to managing development process of scientific software. Adding to that the lack of research done on the subject and it quickly becomes obvious that it's hard to find a systematic approach to managing the scientific software development process.

In next section we will present an approach that may guide the developers towards a successful solution.

# Chapter 4. Development cycle

Section 3.1 listed the most important challenges scientific project has to face, summarized in Section 3.1.8. This Section will describe scientific software development process that emerged during the development of the tools described later in Chapter 6. As with any practices, the ones described here evolved over time and took many forms and iterations. Many times, over the years we have realized that our current approach is not good enough for our purposes and we had to adapt to the new approach and have found new ways in which we can improve the process. The result of this evolution can provide valuable starting point for any new and existing projects.

When considering scientific software, majority of which is focused on solving a problem deeply rooted in different fields of science, one has to consider a project with a unique solution implemented by specialists outside of field of Computer Science. The bulk of the work on the project is focused not on the software or software framework but on the science behind the problem, the mathematical model needed to represent the problem, numerical implementation of algorithms needed to solve the problem and the validation process of the implemented solution. In short, it is focused on the content provided by the specialists, not the software itself. The result is a library of custom-tailored, fine-tuned algorithms with an overbearingly large testing and validation framework. Such library is rarely the end of the project's lifetime. Often, the constantly growing demand for higher precision results pushes scientists to improve their tool further.

Scientific teams often introduce new solutions or improvements on the basis of the results of their previous projects. This new, improved tool becomes a new starting point for future work. The cycle that describes this process has been presented on Figure 13.

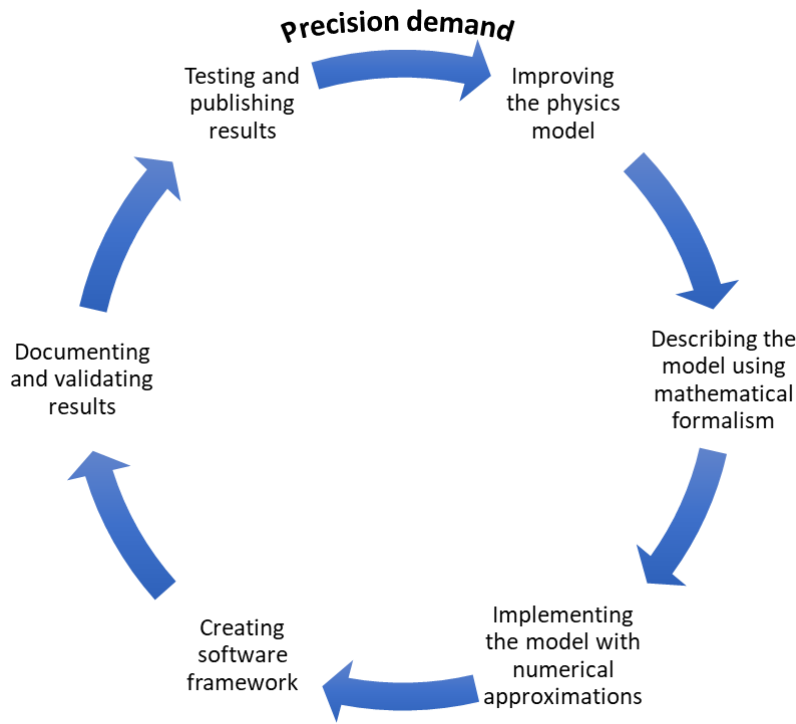

**Figure 13.** The cycle leading to the improved scientific model which ultimately results in higher precision of the results. The development of the first version of the tool can be described by the need to solve a problem instead of the precision demand.

The fundamental source of the content of the project comes from the field of physics; a theory behind the solution that project is designed to implement. However, in order to be able to represent this solution, a mathematical formalism must be introduced to describe the theory in a coherent way that can be transcribed as a step-by-step algorithm. Only then such algorithm can be processed through numerical solutions, applying changes needed to adapt the formalism to limitations of a processor. Finally, the whole project is embedded in an architectural framework managing its structure and a testing framework is built.

What is important to note is that the same cycle is needed to improve precision of the results, in which case the first step is to improve the physics model used to describe the problem instead of designing it.

> **Example:**
>
> Recall the example provided in Section 3.1.1 in which we have presented a simple project that implemented a model describing a resonance that can be approximated using gaussian distribution. This project was built from layers: core issue, mathematical formalism, numerical approximations, software framework and validation framework. Note how the cycle presented here is strongly connected to these layers.
>
> This example also indicated the possibility of increasing precision of the results generated by the project which required more sophisticated model than just a gaussian distribution. This implied changes to all of the layer of the project. Note how, again, the same issue is represented by this cycle.

This cycle represents the process that I have outlined when presenting my thesis in Section 1.4. Following sections will briefly describe each of the steps of this cycle.

## 4.1 Creating (or improving) the physics model

The first step in this cycle is creation or improvement of the physics model. In the case of a new project, the driving force for starting the cycle may be as simple as the need to find a model that will accurately describe given data. In the case of second and further loops, the driving force is the precision demand.

Consider following examples of this process.

> **Example:**
>
> One of the examples, that I'll present in more detail in Section 6.2, is the `Tauola` tool for simulating $\tau$ decays. The project started from the need of a model that would describe the $\tau$ decay process as part of other tools, `KORALB` and `KORALZ`. At that time (around year 2000 and earlier), technical model precision was at 1%. At best, 100k events could have been generated. The year 2001 witnessed the introduction of technical precision of 3‰ which drove the authors of the tool to improve the precision of its model. This happened a few times already; each iteration adding more and more details to the model.

> **Example:**
>
> The development of all of the modern extensions of the tool `Photos` described in Section 6.3. `Photos` started as a branch of `Tauola` and was considered just an addition to the `Tauola` Monte Carlo generator but once the demand for higher precision results raised, `Photos` become a widespread tool. Its development has become more important than the development of `Tauola` and the tool has been separated from `Tauola`. Further updates incrementally introduced electroweak corrections that increased precision in specific sub-processes handled by the tool.

> **Example:**
>
> The iterative improvement of the model was key in fitting the Resonance Chiral Lagrangian (RChL) currents[45] to the experimental data (see Section 6.4). The first model used to solve this problem offered results that weren't really satisfying. Only after the model has been iterated over and significant improvements introduced the results matched the available data much closer.

The important aspect of scientific project development worth taking into account is the fact that usually the documentation of the modeled process and proposed solution is drafted very early in the development cycle. It is not uncommon to start the outline of the document once the problem is identified and possible solution recognized even before its implementation starts. That is because as mentioned several times in this thesis, the physics problem itself is the core element of the project. Even if the solution does not exist or the results of implementation are unsatisfactory the description of the physics problem and attempted solution is still valuable to the scientific community. Therefore, the solution is documented through the whole development cycle even if the technical details of its implementation are usually omitted or restricted to the bare minimum.

## 4.2 Describing the model using mathematical formalism

Mathematical formalism allows construction of a numerical algorithms and the analysis of its stability. In some cases, such analysis can show that the algorithm based on this mathematical description will result in the accumulation of errors.

> **Example:**
>
> The first implementation of the model assumed in `Photos` tool for generating final state radiation assumed single photon emission. Equation for this model was quite simple and was easily translated into mathematical formalism. Later, when precision demand required the tool to take into account photons with increasingly smaller energies, an algorithm generating multiple photons had to be introduced.

---

[45] A "current" refers to leptonic or hadronic current. In simplified terms, it can be thought of as a part of the description of the underlying process used in simulation. Current should be backed up by theory. Changing the current changes the description of the process and impacts the results. Each current can have its own parameters. Part of these parameters can be shared by other currents and part can be unique to this current.

> This became a hard task as the mathematical description required increasingly higher precision of the computation. It was soon obvious that this approach could not have been used for more than several photons. Seeing as calculations with three or four photons introduced large errors, a new mathematical description was needed to solve this problem. This eventually resulted in a new algorithm used to these days.

> **Example:**
>
> An extensive analysis of the mathematical approach was needed when designing the framework for fitting the multi-dimensional distributions used in research regarding RChL currents. In this case the straightforward approach used during first iteration resulted in lengthy computation that took weeks to obtain a single result. During our work on this project we iterated several development cycles in hope to speed up computation.
>
> The analysis of the mathematical model helped us find the approximations that will not impact the fitting strategy while will allow us to produce preliminary results much faster (see ref. [61]). Without this effort a lot of time would have been wasted waiting on the results of the computations instead of improving the results.

While these examples present widely different issues, they both show how additional effort put into mathematical model helped push the project forward. In some cases, it is the only way to progress with the project.

## 4.3 Implementing the model with numerical approximations

Next step in the development of the project is the actual implementation of the solution designed in previous steps. This often requires solving the numerical restraints of the algorithms that have to be implemented.

> **Example:**
>
> Continuing on with the examples from previous sections, at each phase of Photos development the assumed model had to be implemented using numerical approximations. This was as important back, when single-precision floating-point operations were used, as it is now, when required precision is so high that even double-precision arithmetic introduces significant errors. The original algorithm worked for one or two photons but was unusable for more photons due to errors introduced to the four-momentum of modified particles. An algorithm to correct these errors was needed. The implementation required error correction algorithm.

Later, when the mathematical model changed, the iterative algorithm was introduced. Thanks to the error-correction algorithm it could have been used with any number of photons. The precision limitation of floating-point arithmetic still introduced some errors but with diminishing energies of each generated photon, the significance of these errors was smaller and smaller. Still, the approximations used in this algorithm cannot be ignored, especially when considering future extensions of the model.

> **Example:**
>
> A great deal of attention had to be put into numerical approximations used in the fitting algorithm mentioned in previous section. There were several key points that had to be analyzed for the fits of our model of RChL currents to converge. One was the fact that the fitting algorithm internally calculates the derivatives of the distributions with respect to all model parameters. In our fitting framework we used variable step size 16-point gaussian integration to integrate over these parameters. However, function that used variable step size could create artificial discontinuities in the fitting function at points where step size changes. Such artifacts were problematic for the fitting procedure when near the minimum. To obtain more accurate results in next cycle we had to change the integration function to one that used fixed step size at the cost of the precision of the integration. To offset this cost, and further improve convergence, we smoothened the integrant by changing the integration function to a function commonly used to describe resonances in particle physics[46]. This is a very good example of how much consideration has to be put into numerical approximations when dealing with complex computations which are basis of most of the scientific projects.

## 4.4 Creating the software framework

When the first steps of the development of the project starts software is usually built ad-hoc without any planning or architectural consideration. No use cases are taken into account as authors of the project still don't know if the project will even produce any meaningful results. Later, however, when the first phases of the project are completed and show promise, the documentation of the physics theory and implementation begins and the effort is put on building a software that can be used by others. This is the moment when software is refactored and put into some usable framework with more-or-less coherent division of the software into modules.

Refactoring of the original code is always limited to as little as possible to avoid damaging the physics context of the project. After all, this content is the most important part of the project. Later, configuration options are added, which often origin from debug options used to validate the code. Lastly, tests and validation frameworks are made understandable by scientists other than the original authors.

During subsequent iterations similar approach is used. While core functionality is already more-or-less organized, new software is usually not built into this existing framework but rather developed as a new analysis that uses the software from previous iteration as if it were an external project. Again, only when the results of this analysis show promise and the project begins to be formalized and documented, the process of moving the analysis as part of the internals of the project begins.

> **Example:**
>
> A good example of the iterative approach to building software framework is `TauSpinner` [6] project described in Section 6.5. It started off as a single file

---

[46] Details are presented in ref. [61].

> fully dependent on `Tauola++` project, including its framework and event processing algorithm. However, once the core algorithm began to take shape it was necessary to separate it from `Tauola++` as much as possible. After all, the application of `TauSpinner` is completely different than that of `Tauola++`. Therefore, in next iterations a separate directory structure has been created to store `TauSpinner` source code, documentation, tests and examples. The interface of `TauSpinner` has been also updated and the code that provides data to `TauSpinner` abstracted so that data can be provided from any source, including `HepMC` events and files.

## 4.5 Documenting and validating higher precision results

Once the outline of the algorithm is designed and software framework is prepared, the algorithm is ready to be tested in terms of its correctness. This is the crucial step that determines the worth of the algorithm itself and it is a complicated process unique for each tool and each iteration. It cannot be automated and requires expert knowledge of the subject matter.

> **Example:**
>
> The process of validating `Photos++` started by comparing the results of the tool to results of similar tools. Such comparison is often hard to perform as it is hard to match the exact condition the other tool was used in and create exactly the same projection as the authors of the other tool used to describe their results. The comparison was performed as close to the other tool's environment as possible and the differences were carefully examined from the physics standpoint.

> **Example:**
>
> On the other hand, when studying the applications of `TauSpinner` tool the algorithm was stripped apart and analyzed by switching off and on number of options available within the tool. The results were compared against theoretical predictions and with each other to observe if the effect with and without specific option matches the expectation. The tool was also compared against existing tools. Data with experimental cuts were used to present the results of the tool when describing the application of the tool for ascertaining the spin of new resonances.

There is really no rule deciding what type of validation process the tool must pass. The important part is to gain confidence of the future users of the tool by showing how the tool was validated. That is why this step is heavily documented.

At this point an outline of the documentation of the modeled process and implemented solution is mostly ready and is now updated and properly formed based on preliminary results and changes introduced during this step.

## 4.6 Testing and publishing the results

Lastly, once the algorithm has passed preliminary validation tests, the solution can be tested on as many use case scenarios as possible. Note that these technical tests are performed at the very end as they are far less important than the validation process described in previous step. These tests are focused on the obvious errors and technical issues. At this point the typical use case scenarios are known and it is a good moment to spend some time building an automated framework to aid with the testing process.

> **Example:**
>
> In the case of the tools mentioned here we based our tests on the tool `MC-TESTER` described in Section 6.5 to automate the process. This, of course, required the work on `MC-TESTER` tool itself. While its basic functionality was ready, we had to expand it for the purpose of our tests. This is often inevitable as each tool requires different kind of tests. Finding a tool that allows to test every aspect of a Monte Carlo generator out of the box is near impossible.
>
> We have built a setup that generates test events used to check if the tool correctly handles various use cases. Number of variants of the test were created and the process of generation of large data samples for different sub-processes and comparing them against benchmark distributions was automated. However, in order to produce this setup, results of each test had to be carefully analyzed manually in order for the generated sample to be accepted as a benchmark result.
>
> Once this set of benchmarks was ready and framework for tests was prepared, we were able to use it with in each iteration to test for regression. In each iteration it was subsequently expanded with new cases and variants of the existing cases to show the result of different options.

A tool that passes the technical tests is ready to be published. Usually at this point the documentation of the physics content in this new (or first) version of the tool is ready and requires only minor improvements or details to be filled in. Final results are generated and added to the documentation. Then, the documentation and the tool are published finalizing this iteration.

## 4.7 Summary

In this section we have presented the development cycle that the developers of the tool follow in order to create a first version of a tool or new version of the tool that increases the precision of its results.

At a first glance Figure 13 of the cycle of improving precision of the results can be compared to the development cycle in agile methodologies. However, agile methodologies use incremental improvement of software through small cycles, usually between a week and a month. The cycle shown here spans over months, up to years. This time barrier is hard to overcome; it is extremely hard to divide research tasks into incremental steps and to be able to release an incrementally better version of the software every few weeks.

Knowing that the tool follows this cycle, it is possible to partially track its progress and estimate what needs to be done and how much work is still needed to complete the cycle.

# Chapter 5. Testing scientific software

Testing is a crucial step of the project development. So much so that today industry standards estimate that 30% to 40% of the development time is spent on tests alone. The subject is so vast that the variety of different techniques, types of tests and approaches to software tests is hard to follow. Most developers, however, are familiar with some of the best and worst practices related to software testing. For example, the test pyramid is a well-known model of testing a software project (see Figure 14).

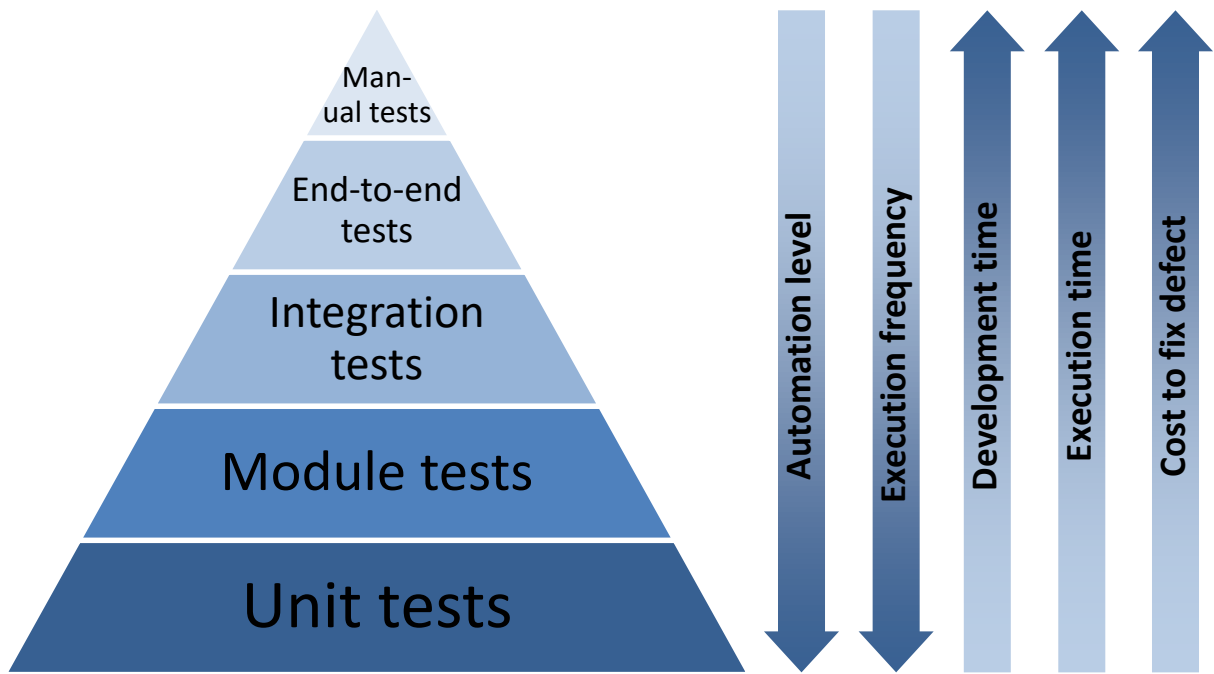

**Figure 14.** The test pyramid – common pattern recognized as best testing strategy. Many variants of test pyramid exist, see e.g. Ref [62], which contains an overview of some of them. Version presented here is more suited to software without GUI or extensive UI.

When analyzed from bottom to top, the diminishing size of the portion of the pyramid allocated to a given test type represents how much effort should be spent on such tests. Many variants of this pyramid exists, but the overall concept remains the same for all and is represented by the arrows added on the right-hand-side of Figure 14.

The base of the pyramid contains automated tests that can be written quickly, executed quickly and are able to prevent large number of common mistakes. Mistakes detected on this level are easiest to fix and have smallest impact on the overall project. The resources spent on these tests have the highest Return On Investment (ROI) as they prevent large costs from being incurred later. These tests are executed very often, usually with each commit to a repository, which depending on the size of a team, can be from few up to a hundred executions per day. The feedback from these tests is instantaneous and often the development process prevents putting into the code changes that cause even one of these tests to fail.

Going up from there, the amount of automation decreases and the development effort, as well as execution time, increases. The tests are increasingly hard to execute, requiring more time to setup and gather results. Feedback from these tests is delayed. Integration tests can take several hours and are usually done overnight. In more complex systems they require more than a day, pushing feedback even further away in time and increasing the time needed to pinpoint the sources of defects. Same goes for system tests which can take even a week. Manual tests are performed scarcely. Apart from regression testing, they can be executed as rarely as once per release.

As we go up this pyramid not only does the execution time, the effort to develop the tests and the delay between the changes introduced and feedback from the tests increases but the effort, time and cost required to fix a defect found on these levels increases as well. A critical defect found in manual tests can stall the release by weeks, whereas if the same defect could have been found in automated tests, it could have been detected as soon as it was introduced and possibly it could have been fixed much faster. But, of course, we cannot remove the possibility that the reason the defect was detected so late may be because of the defect complexity – no other test could have found it.

As well-known as the test pyramid is, the common anti-pattern is less frequently talked of, which it shouldn't, knowing that this pattern is more frequently seen in the industry than the pyramid. The test cone anti-pattern presented on Figure 15 presents result that is exactly opposite to the ideal scenario.

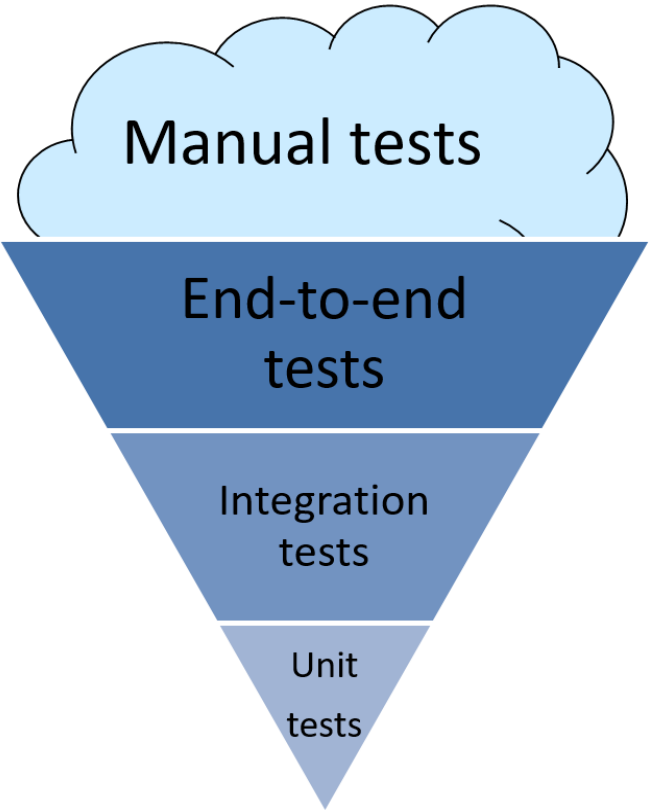

**Figure 15.** The test ice cream cone – common anti-pattern that results in severe project maintenance problems. Many variants of test cone exist, see e.g. Ref. [63]. Version presented here is more suited to software without GUI or extensive UI.

This setup is often a result of an eager development in which test strategy was not introduced early or not enough effort was put into the test process through the whole project development. In some cases, the effort needed to create proper test structure seems too high, especially at the beginning of the cutting-edge or R&D project, where obtaining results is more important than quality assurance. This can result in a defective approach that is never fixed; tests are introduced only when defects appear and usually the easiest tests to introduce are the manual ones; the simplest tests require no framework and no effort from developer's side.

They can be quickly organized by a newly formed QA team. Such tests will often be sloppy and won't cover project requirements but can be effective as an ad-hoc solution to most imminent problems. Such approach can very easily lead to the test cone.

In other cases, the project might have started from the pyramid-like test structure but without maintenance, the strategy turned into the cone or a hybrid of both.

Lastly, there are cases where a lot of legacy code is incorporated into the project inhibiting effective unit testing or integration testing. Such projects may or may not have a large set of automated tests but only on system or end-to-end level, which do not require special tools other than, for example, `Selenium` [64] or similar test automation tool. Expanding second smallest block of the pyramid while neglecting the bottom-most layer will in time result in a cone-like structure.

In scientific software tests are used not only to verify technical aspects of the code but mostly to verify the results of the computations and to validate the assumptions put into the code. Many of the tests performed for scientific software does not fit into common types of software tests. To show how scientific software is tested in relation to these common practices we would like to first present how different types of tests are categorized. In the following subsection we summarize the already existing knowledge to present the taxonomy of software tests. We then present where tests written for scientific software fit into this taxonomy.

## 5.1 Taxonomy of software tests

There exists number of software testing taxonomies, each focusing on different aspects or different approach to software testing. For example, reference [65], among other Software Engineering taxonomies, lists several references related to software testing taxonomies. Similarly, the online sources such as Reference [66], attempts to classify and categorize available software tests in a meaningful way. The variety of approaches shows how much this subject is analyzed and researched.

There are many common factors in these taxonomies and some types of tests have similar definition despite having different naming conventions. The taxonomy presented here lists the most common test types and categorizes them using generic quantities relevant both in scientific software as well as other software engineering applications. These quantities are: scope of the test, the level of automation that given test provides and weather given test is functional or non-functional.

This section focuses solely on software testing, ignoring the evaluation and other aspects of QA process. It also narrows down the testing part of QA process to software testing, ignoring hardware tests, model validation and other types of tests.

### 5.1.1 Scope of the tests

Each type of test covers part of the scope of the final product. For the purpose of this taxonomy following scopes that can be covered by the tests are defined:

Unit –  smallest amount of code that can be thought of as a block of code that solves one specific task or fulfils one role. Depending on the programming paradigm used, this usually is a single function or a class. If a function or a class fulfils more than one role, they most likely should be separated into more than one unit.

Module – a set of units that share similar roles, fulfil a set of similar functionalities or provide a set of coupled or closely related services. Usually this set of units is bundled into a single library. An API of this library can be separated into another library. Units within a module do not have to interact with each other. There can also be a number of back-end components that do interact with each other while another set of front-end components provide services to other modules.

Application – a set of modules working together to fulfil large set of goals. It has well defined means of communication with other applications or with a user. An application can be a service that runs in the background, a plug-in for other application or an actual standalone application that has its own front-end.

System – a single application or a set of applications that interact with each other. While a system can interact with other systems, and these interactions should also be tested, for simplicity we assume that our requirements are fulfilled by a single system.

### 5.1.2 Types of tests

This section lists all types of tests that are commonly found in different references. Note that depending on the source, usually only some of these tests are mentioned; no single reference that we've found lists all of them. We have included some types of test inside others as they are conceptually identical. We have made a choice to group such tests or use one name rather than the other knowing, that this choice is subjective. However, without it the list would be longer and would contain test types that are very similar, if not redundant. The following list of types of tests is ordered by the increasing scope covered by the test.

Smoke tests - used to check the most trivial conditions required for a unit, a module or a system to work. Smoke tests performed on a higher level, such as during integration or system tests, are usually called sanity tests (or sanity checks). Their role is to verify that the environment is sane (i.e. the system under test works in expected environment).

Unit tests – used to verify single unit within the complex architecture of a project. The interaction of this unit with others is not tested and if needed, a fake environment is created in order to test this specific unit in a controllable and fully predictable manner.

Functionality tests - used to verify single functionality or requirement. As functionality may span over several units, this kind of test is often closer to integration tests than to unit tests.

Integration tests - used to validate if units or modules correctly interact with each other. Integration tests usually require a large set of units or modules to be set up before each test but does not necessarily require full system setup. For example, when testing middleware library, the database and the front-end can be stubbed. If integration tests are performed at the system level, nothing is being faked and full application is running.

Regression tests - performed after each addition of a module or modification of a module to check if this change has not introduced any errors. Usually this

| | requires running unit and integration tests for all previously implemented functionality in all modules that interact with this modified module. |
|---|---|
| Interface tests - | verifies that communication between different modules works accordingly to specification. This usually means that one side of the communication channel (the user of the interface) is mocked so that the other side (the interface) can be tested. It is worth noting that unlike in compatibility tests, where communication between different modules can be tested, in interface tests no interaction between modules is actually tested – just compliance to the interface documentation. |
| Compatibility tests - | designed to verify that a module or a system can work with compatible modules or systems. These tests are usually based on documented interfaces (and may be expanded version of Interface tests) but can also be part of integration or manual testing process. Backward-compatibility tests can be thought of as compatibility tests that verify the system or a module can work with older systems or modules. |
| Acceptance tests - | designed for a specific acceptance criteria posed in a requirement. Usually tests for these criteria can be covered or already are covered by other tests but the requirement states that some sort of compliance must be visibly tested[47]. A sub-group of acceptance tests are user acceptance tests which are designed to be performed by the user (client) so that he or she can manually verify that requested functionality is implemented. |
| Compliance tests - | verify that the software complies to specific norms or directives. These types of tests are closely related to acceptance tests, however usually acceptance tests are created in response to specific needs of a single requirement while in order to fulfil one requirement (e.g. a requirement to comply with some norm) a large set of very detailed tests are needed. |
| Localization tests - | performed for software that supports different locale or different languages. Each supported locale and language must be tested to verify that translation is present for each UI element and that the translation is correct. Units (currency, metrics, dates, etc.) used in different locale are also tested, especially if conversion mechanisms are in place. Globalization (or: internationalization) tests are expanded version of this type of tests. They try to verify that the product supports every type of international input and every locale (but not every language). These types of tests are more focused on checking if the product breaks under specific locale (or is partially non-functional) rather than if every locale is actually supported. |

---

[47] For example, a requirement may state that four instances of the application may work simultaneously. Given as such feature may be tested as part of the performance or stress tests, there should be no need to do an additional test to verify that. However, the client may ask to present that this requirement is fulfiled, so a special test has to be created for this purpose.

Installation tests - check the installation process of the software for each target platform. In case of smaller products these tests are performed scarcely and often manually by one of the developers. These tests are especially needed if the software has complex, multi-stage installation process.

Configuration tests - verify that the system works under different software and hardware configurations. These tests also verify that differences between $3^{rd}$ party libraries or features native to some operating systems are correctly handled.

Usability tests - used to understand how user interacts with the application. These tests are performed on an application or system level and cover variety of topics related to application front-end, such as design, responsiveness, layout and availability of different options within a system. Feedback from these tests is used to improve user satisfaction.

Performance tests - used to verify requirements related to quality of service, such as the response time for different tasks or resources usage (memory, CPU, GPU, network load). These tests verify how software behaves under normal (anticipated) working conditions.

System tests - used to verify if an application fulfils system-level requirements. A large variety of tests is performed and no part of the system or its environment is faked. They are usually running on a number of target platforms, or in a set of simulated or emulated environments. These tests are usually scripted to follow use case scenarios, which means they will not be able to check for unusual or corner cases as well as lower-level tests.

Security tests - performed to validate if critical data on which the system operates are protected against malicious entities that try to obtain the data that they should not have access to. Nowadays this term usually is used in relation to a system connected to the Internet, but security tests cover any system which limits, in any way, what kind of data a user can access. For example, in medical devices medical personnel that sets up the device may not have access to patient data, only to the device setup. If such requirement exists, it should be tested. Penetration tests are a type of security tests, usually made by external company to verify if it is possible to gain access to a system from the internet (by "penetrating" its defenses or by bypassing its access control system). Experienced penetration testing team will use the same tools and techniques as a real hacker would use.

Stress tests - designed to see how the system will behave under heavy load. This is usually done as part of performance tests but contrary to regular performance tests, these tests are designed to check how system behaves under peak conditions. This type encompasses flood tests, load tests and volume tests as all of them have common goal. Due to the time and effort needed to perform them, they are usually performed for the whole system only.

Reliability tests -    used to verify stability of the system, its availability and failure recovery mechanisms. This type of tests can be called stability tests or availability tests. Regardless of the name, these tests are designed to run for a very long time by continuously running scripted scenarios of common tasks that system is designed to handle. They verify that there are no errors that accumulate over time causing the system to fail. They are also used to measure average Time To Failure (TTF). System recovery tests are sometimes separated as in some scenarios an error is introduced manually just to see if, and how fast, the system will recover.

Scalability tests -    used to verify if the system scales (as in: increases or decreases its resources usage in response to the load) according to assumptions. This type of tests is similar to stress tests, but during stress tests the system is not allowed to scale as this would defeat the purpose of testing the system under peak load. In these tests the system's scaling policy is tested against varying loads in numerous scenarios, including the policy for scaling down. For example, the scenario may simulate the increased load during rush hours and rapid load decrease afterwards. Another scenario may constantly increase load to see how the response time changes or to find the limit after which the system stops scaling properly.

Maintainability tests -    designed to check if the software architecture aims for the ease of software maintenance. We have placed these tests at the end of the list as this is the only type of tests that is focused on software architecture. As such, it should be thought of as covering every part of the final product but are not in fact localized in any of previously mentioned scopes. Maintainability tests are usually divided between tests performed by automated tools and tests performed by developers. Tools can check, for example, code complexity, reporting modules that are overly complex and should be split into several smaller modules; can report use of magic numbers or other hard-coded configuration options (e.g. database connections or queries that can cause a problem if database layer changes) and can check if code is sufficiently covered by tests (showing that the code is testable, which easies its maintainability) or check for code duplication. Manual methods are mostly based on code review, during which developer checks if the code follows (when possible) design patterns rather than invents a hard to maintain unique solution; verifies that solution re-uses existing code when it is applicable and check for proper use of language features or other places that could cause problems in the future.

### 5.1.3 Testing techniques

Lastly, it is worth mentioning the different testing techniques that can be used when choosing what kind of tests to implement and how to implement them. Some of these choices may significantly impact the architecture of the system and the design of the tests.

Black box testing –    an approach in which tests do not take into account the internal implementation of a functionality, just the functionality itself. Black box tests are sometimes called functionality tests as they focus on the functionality, not its implementation. This approach requires an architecture that allows to fully test functionality through black-box tests, which is often not present. This approach can be used when writing unit tests, integration tests or system tests.

White box testing – an approach in which internal implementation of a functionality is tested. White box tests are sometimes called glass box tests as the internal structures and classes are transparent to them. They can test and set any value and any state of any class. This approach allows for very detailed tests but strongly couples the tests with the implementation of the functionality. This approach can be used when writing unit tests, integration tests or system tests.

Agile testing – an approach in which tests are designed and implemented along with implementation to go along with the agile incremental delivery approach. This approach requires that some tests are planned and executed ad-hoc and may not be repeated during every test session as test sessions are strongly limited in time.

Exploratory testing – a manual testing technique in which a tester tries not to follow any specific scenario and tries to find out valid use cases which may potentially hide a defect. This often means that a tester will try to find a corner case that developers have not thought of or try options not designed to work together just to see the result. The tester may also try to run the software in an environment in which it has not been tested before but is supposed to work (or can be inferred from documentation or other sources, that it should work).

Negative testing – a manual testing technique in which a tester tries to break the system under test in any way possible. This includes following a scenario that is contrary to the documentation, purposefully providing wrong inputs or using the software in an environment it is not expected to work. The goal is not to verify if the system works correctly under these circumstances but if the software can correctly handle or report erroneous inputs or report a critical failure in a way that user can understand. It also verifies that error reporting process for developers or artifacts of such errors (such as core dumps) are generated correctly and can be used to correctly identify the problem.

End-to-end testing – an approach in which the full flow of a use case scenario is reproduced. This can be done in an automatic way but most use cases are tested manually. This test covers both functional and non-functional requirements.

Static testing – a testing technique that utilizes only static methods; methods that do not require running any code. It narrows down testing to code reviews, walkthroughs and automated tools for semantic code analysis. Due to limited scope of defects this technique can detect it is used only under special circumstances.

## 5.1.4 Unit, module, application and system-level tests

Since we have decided to order the list by the scope of the tests, the first distinction should regard this subject. While some of the tests can be performed on one level, others can spread across different levels. Usually this means the approach to these tests is different. Integration tests at module level can use fake components to isolate part of the whole system while integration tests performed at the system level cannot. These differences are not subtle, but they do not change the fact that the test type remains the same. Figure 16 presents the most common scopes in which given test type is used.

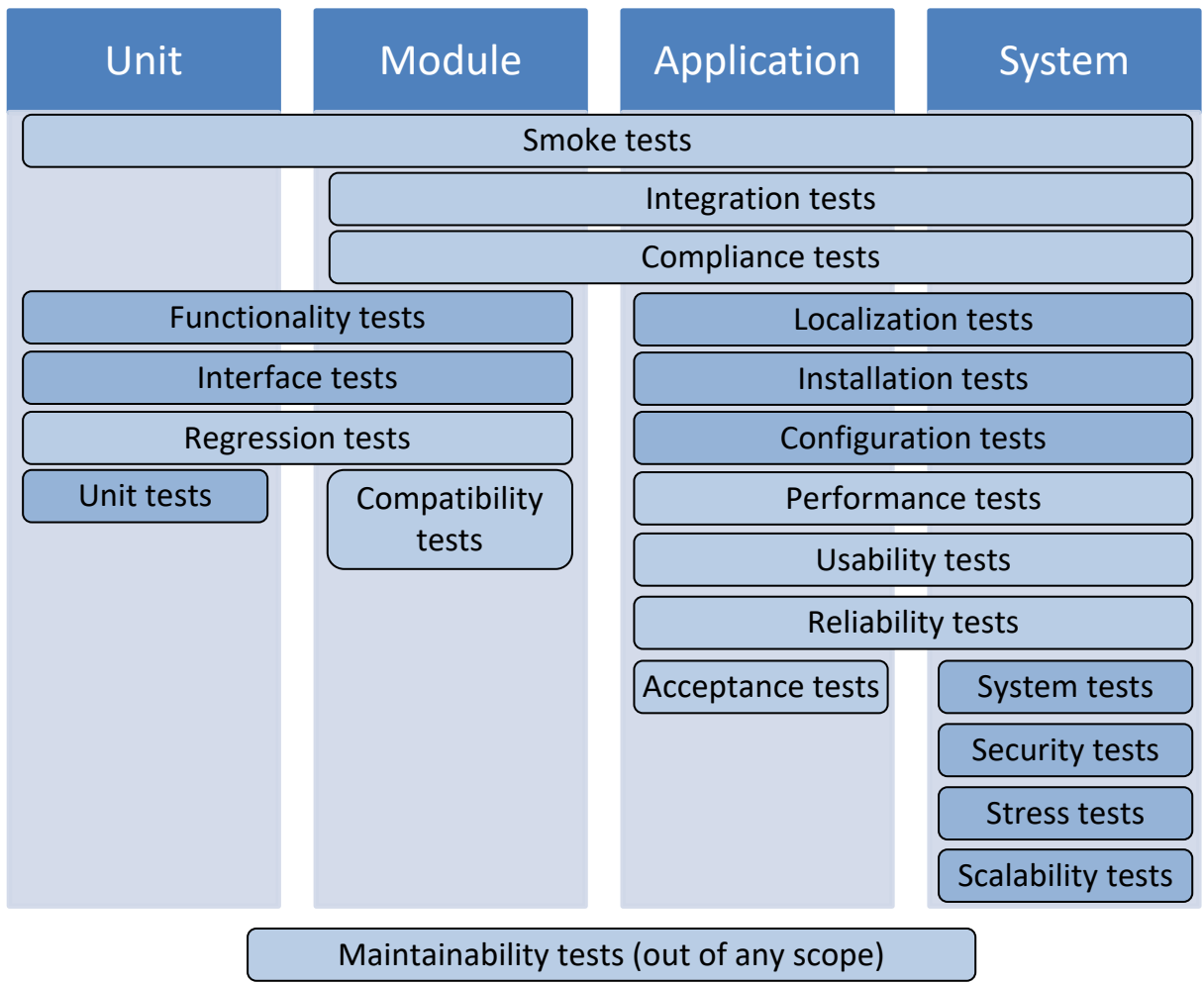

**Figure 16.** Taxonomy of tests based on the scope they can cover. The scope of the test types shown in dark should not be changed. The scope of other test types may vary.

Note that, to some extent, strict adherence to the test scope does not have to be followed. For types presented in light-blue the scope is an indication on which level given test type should usually be performed but exceptions can occur and should not be forcefully blocked. For example, time-critical algorithms can undergo performance tests on the level of a single module, if needed. Especially if we are considering database module or similar I/O or memory-intensive module. Similarly, regression tests can be executed at application and system level (for example manual regressions tests) but they lose their efficiency outside of their original scope. The types presented in dark should be performed only in the indicated scope as they have no purpose in other scopes.

### 5.1.5 Functional and non-functional tests

As I've pointed out several times through this thesis, most of the requirements of scientific software belongs to the non-functional category. As such, it is important to note what kind of tests or testing techniques can be used to validate them. Figure 17 presents the division of test types into those that are used to test functional requirements and those used to test non-functional requirements. This division is unambiguous. The requirements of these two types are vastly different from each other, so are the tests that cover them.

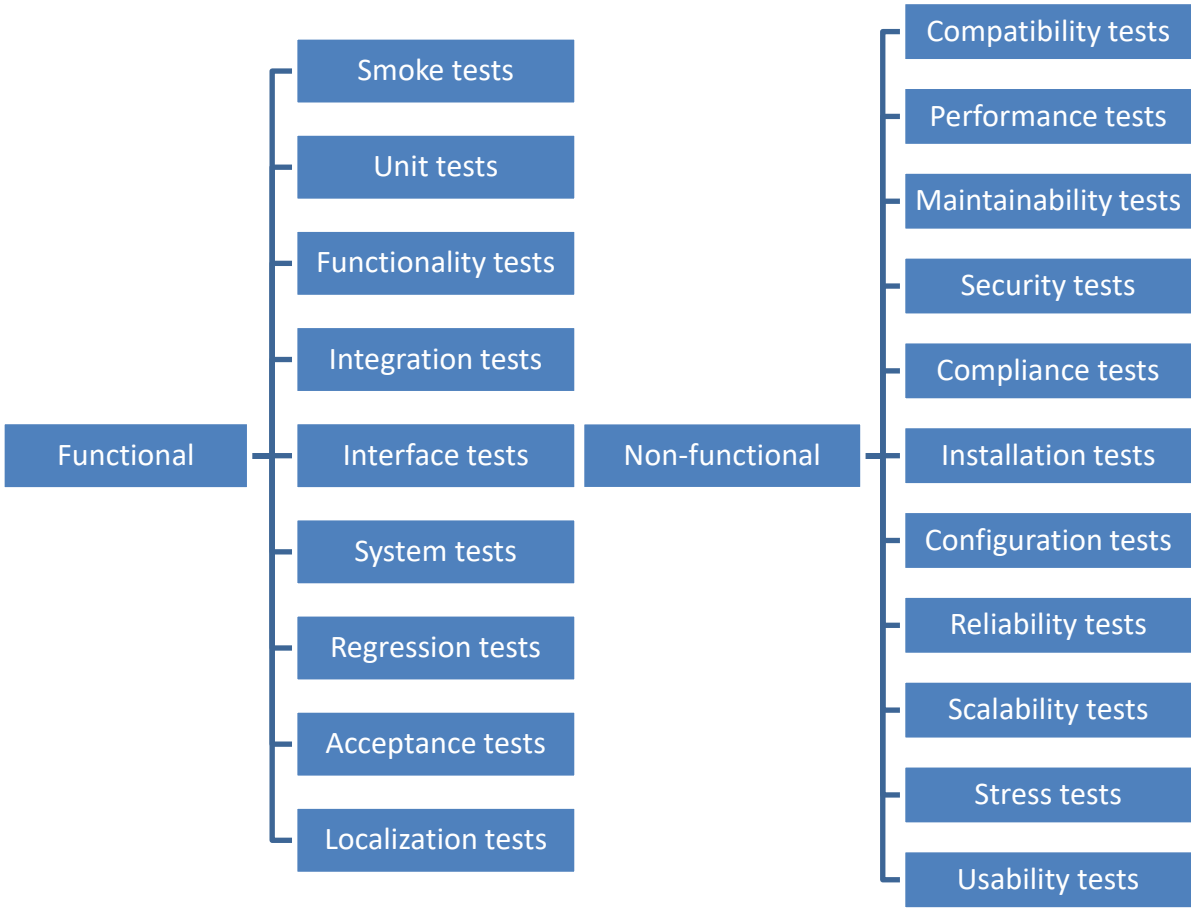

**Figure 17.** Taxonomy of tests based on the adherence to functional or non-functional tests category.

### 5.1.6 Manual and automatic tests

Tests can be split into those that can be executed automatically and those that have to be performed manually. However, this distinction is not very strict as some types of tests can be performed both in an automatic and in manual way. In some cases, they even have to be split into the automated and manual parts. There are enough of such test types to warrant a separate category – automated and manual. Figure 18 shows which tests are executed manually, which are executed automatically, and which require both manual and automatic approach.

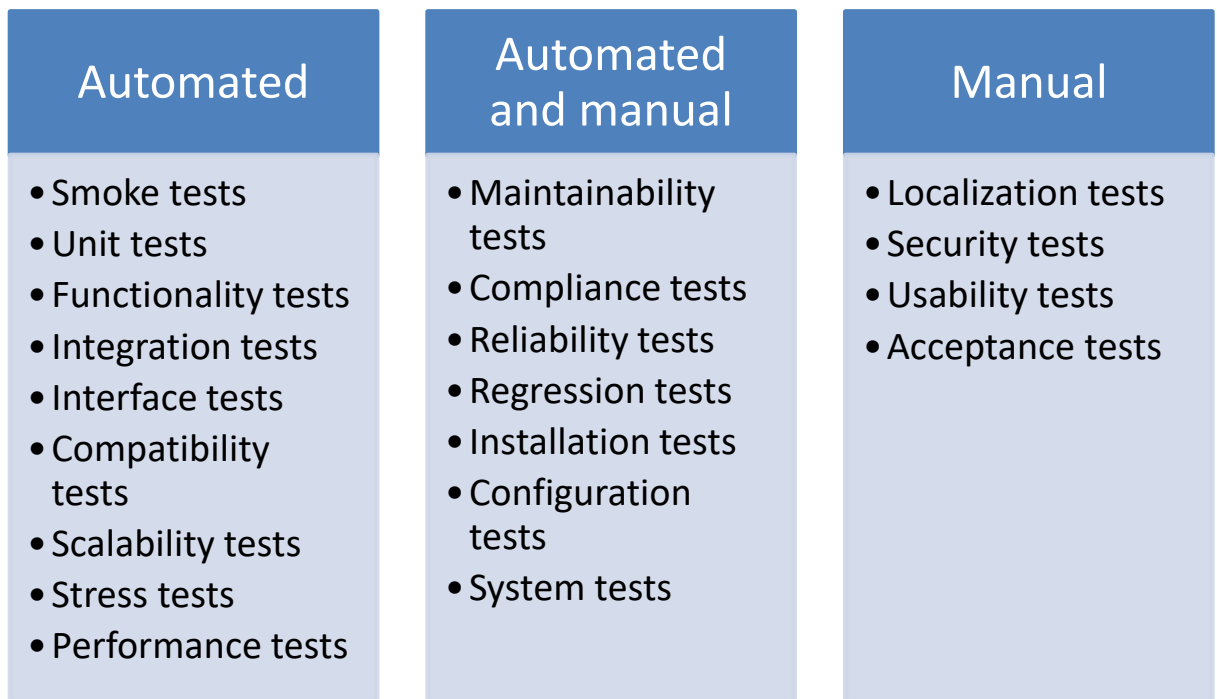

**Figure 18.** Taxonomy of tests based on the level of automation used. The middle column contains tests that are usually in part implemented as automated tests in part performed as manual tests.

The middle column contains tests that usually require both automated and manual solutions. For example, part of maintenance tests can be automated, while part cannot; automated tests are based on code metrics and manual on code review. Similarly, some regression tests may require manual tests of the final product. On the other hand, compliance tests are usually performed manually due to their nature, but the most technical aspects of followed procedures can be automated, and report of such tests used to show conformity.

There is also one other category of tests in relation to test automation, which is: semi-automatic tests. These are automated tests that require some input from the tester; at least one of the steps in the test scenario is manual. For example, the test may require user to turn on a device or to click a button in the application or website at specific time, be it at the start, in the middle or at the end of the test procedure. Otherwise the test executes without human input.

The reason why these kinds of tests are not always categorized as manual tests is that one tester can perform tests of multiple devices or execute multiple similar test plans of an application without each of them needing his or her full attention. Tests that are executed automatically, but the results have to be verified manually, also fall under this definition and are perceived as semi-automatic. The approach to such tests is different than to other manual tests. For example, a batch of tests can be executed overnight, and the results can be verified in the morning. This still means that the information about test results will not be available without tester's input, but such tests require much less of the tester's time than other manual tests.

That being said, every automatic test can be treated as semi-automatic if, for example, the testing framework is not prepared to automatically parse the test results. The other relation is also true – parts of manual tests can often be automated but for variety of reasons they are not; the team may not be experienced enough to automate the process or the testing procedure would be too complex or automation would not be worth investing time or effort for given project[48]. Keeping that in mind, we have omitted semi-automated test category as it could, technically, contain almost every type of test.

## 5.2 Tests used in scientific software

When talking about the testing procedure setting up the automated tests, such as unit tests, integration tests and deployment tests, is a crucial first steps of the project development. They are, however, the easiest to perform and least valuable from the point of view of the project itself. They verify the technical aspects of the project and are mostly used as tools that prevent regression. The semi-automated or behavioral tests are much more important as they validate that implemented functionality works as designed. Usually, when working on a product test specification, most of the time is spent building and documenting these tests, as the fully automated tests rarely need specification other than the description of procedures needed to run them. The most important are the tests that verify if in particular condition a program can be used. Such tests extend beyond computing aspects. The nature of the algorithm and the problem solved by the program become the dominant ones that guide the development of the testing framework.

In all of our projects we strived to provide the user with a set of tests, either basic or complex, that can help validate potential problems that may occur when integrating the project into target environment. In this Section we describe the different types of tests that we have used in our projects during whole development process.

### 5.2.1 Numerical technical tests

The technical tests are by far the least important but also the easiest to write. They are also easiest to automate and update. While their usefulness is very limited, they played the most crucial role back when compilers could not have been trusted to always follow standards and produce correct output. A set of numerical technical tests could very well produce different output in different user environments. These differences were either negligible, related to compiler optimization, or catastrophic to the point, where the output showed only NaNs. These inconsistencies were one of the reasons why the LHC Computing Grid (LCG) project was used as a source of software distributions of a tool. It provided users with a compiled and validated set of libraries for each platform. The automatic validation process compared results of a sample program executed on each of these platforms. The difference of these results had to be within a predefined limit for the tool to be considered stable. This automatic process was used for each new version release of a tool.

The need for such validation remains, if only to catch sudden changes in compiler behavior[49]. However, with constantly expanding list of supported platforms and compilers, this process became inefficient and ultimately the LCG project stopped the activity of releasing distributions of physics software, leaving the responsibility of validating a tool in target environment to the user and the responsibility of providing such tests to the tool's developer.

---

[48] This is often true when considering embedded software. The manual tests that require pushing buttons and checking the status of the device under given circumstances can be automated if proper test tools are created but this may not be worth investing time and effort, especially at prototyping phase or small-scale production.
[49] For example, gcc v4.8.2 introduced a feature in which cpp preprocessor adds C++ comments with copyright information to every preprocessed file. This turned out to break compilation of projects in which cpp was used to preprocess Fortran code; the solution followed by many projects for 20 years.

Numerical technical tests fall into following categories:

- Application level
- Non-functional
- Automatic

Conceptually, they are closest to configuration tests, but their preparation requires much more finesse and requires complex verification process; some level of numerical fluctuations that depend on platform and compiler used are acceptable, but the acceptance threshold depends on the tested subject[50].

### 5.2.2 Data flow tests

Before Event Records became widely used, Monte Carlo tools were tested by preparing a separate program for each sub-case handled by the tool. There was no need to test variety of possible inputs. This, however, became a necessary step when input for the Monte Carlo generator started to be produced by different tools and in various formats. As a result, a loss of information or invalid data could have been easily provided to the tool. A tool had to be able to distinguish between invalid cases to avoid problems for users who use it in such way by mistake. After all it was not easy to set up whole generation chain correctly.

Since the generation process varies greatly depending on its starting condition, separate tests had to be created, for example, for proton-proton collisions and electron-positron collisions. This mostly concerned algorithms used to traverse the Event Record. An interface to Event Records had to be able to handle all exceptional cases and divergence from the standards used by scientific community; if only to alert the user that the tool might be used incorrectly.

Data flow tests fall into the following categories:

- Module, application or system level
- Non-functional
- Automatic

Tests of this type are very similar to integration tests or installation tests, with the latter fitting more to the overall purpose of these tests as developed tool is tested in an environment which uses a set of other tools. On system level these tests are performed by building a chain of external tools that provide input and analyze output of the tool under tests. The result of the whole chain is then analyzed. This is done in variety of different configurations. On application level, data flow within the tool is tested by using preset data input and by analyzing the final result of their processing, which validates internal communication between modules.

On module level these tests are closer to compatibility tests as modules related to data flow are usually tested in conjunction with one other external tool to see if they are able to correctly communicate with each other both ways using given Event Record.

---

[50] For example, a number of generated particles is fixed and cannot ever change on any platform, but their four-momentum can vary from expected values up to some ε related to algorithm's numerical stability.

Taking into account the nature of the Monte Carlo processes, all of the above tests produce results that cannot be tested one-to-one against benchmark. To automate the process other techniques, have to be used (see Section 5.3).

### 5.2.3 Output metadata tests

A natural follow-up of the previous test category are tests of how metadata of the data output by the tool present themselves when used with different Event Records. These types of tests are distinct from the data flow tests as metadata will likely not be used by any test tools in any significant way. The types of metadata that can be produced vary depending on the Event Record. They can be, for example:

- The generated decay vertex position (e.g. `HepMC v2` [57], `HepMC v3` [67], `LHE` [56])
- Backup of unmodified particles as historical entries (e.g. `HepMC v2`, `HepMC v3`)
- The attribute describing the Monte Carlo generator used (`HepMC v3` only)

Usually an analysis does not need this information, especially that its presence is strictly related to Event Record used. However, some of the analyses rely on it. Their correctness has to be verified in every supported Event Record format.

These kinds of tests are usually hard to automate as the standards regarding metadata are volatile. While the tool may pass automated tests, the tests may not verify the up-to-date standards. For example, as mentioned before, the status code of the particle can be used to determine more of its properties than just the fact that particle is stable or decaying and it has been, over the years, overused by many experiments to encode additional information about the generated particles. This, however, was not standardized by anyone, as it was considered to be a "hack" used to surpass the limitations of the Event Record. Nonetheless, the tools were expected to use these hacks when possible[51].

Output metadata tests fall into following categories:

- Application or system level
- Functional
- Automatic or Manual

Metadata tests are very similar to interface tests as they validate that the tool provides correct information according to the standard available in given Event Record. Given the narrow uses of metadata it is possible that issues related to them will be reported scarcely and very late into project development, sometimes even months after some update to the metadata storage or an update introducing new Event Record support is published.

### 5.2.4 Core algorithms validation

This category contains most important and hardest to write tests. Results of the Monte Carlo tools are validated through comparison with other tools, comparison with other configuration options and through comparing the same tool with different level of physics assumptions or with different implementation, e.g. semi-analytical integration instead of Monte Carlo simulation. This last technique is especially interesting. If we recall description of physics precision outlined in Section 2.4, by comparing results of the same tool produced using two different levels of precision, validation of the higher precision computation can be performed regardless of initial conditions on which the tools is running. In fact, it helps validating that these calculations work correctly in all tested initial conditions.

---

[51] One of such usage was addition of the history entries in Photos++ (see Section 6.3). Not knowing how the standard will change in future we provided the user with ability to modify the status code of such particles.

One of the significant problems encountered during core functionality tests is the choice of benchmark distributions to which the tool is compared. In some cases, this problem can be bypassed by comparing the tool to analytic or semi-analytic results. Such results, when present, offer great advantage as they allow developers to control precision and numerical limits of the computation.

Core functionality tests fall into following categories:

- Module or application level
- Non-functional
- Manual, then semi-automatic

We can think of the core algorithm tests as of compliance tests. They validate that the implemented model matches the theory behind it and produces expected results. Both parts of this statements are equally important and need to be verified. It is not uncommon to see correct results driven by incorrect model; aggregation of errors can sometimes cancel each other out. In other cases, the assumed approximations introduce large enough error that the results seem correct. That is why verifying the result itself is not enough.

This is also the reason why these tests are very hard to automate. Parts of the testing process can be automated, especially by using a test framework that can help produce necessary results (see Section 6.5), but the results almost always have to be verified manually. If any discrepancies between the results and the benchmark occur, its source must be thoroughly analyzed as it may be the correct result of recently introduced change.

The impact of a change is not always apparent in complex project. Differences in results may emerge without initial realization of the person implementing the change. Their nature can be, however, verified. If the difference is understood and origins from the model change, benchmarks are updated.

### 5.2.5 Systematics studies

A test, or rather a full study, that is often repeated for scientific software is the analysis of the systematic error. Often parts of the project use a well-known approach to some problem which, assuming the method used is numerically stable, has predefined, bounded systematic error. If custom approach is used, bounds on its systematic error should be established.

For example, an 8-point Gaussian integration with variable step size has a bounded error as the step size is decreased when this difference between 8-point and 16-point integration surpasses desirable threshold. However, multi-dimensional gaussian integrations with predefined step sizes do not have a bounded error. Their error can be estimated but it strongly depends on the integrated function. To correctly calculate the error of such integration when used on the particular function that we want to integrate, systematics studies should be performed.

Similarly, when taking into account approximations of the modeled process, the impact of these approximations can usually be calculated or otherwise estimated to a certain degree. If not, error introduced by the approximations should be analyzed.

Systematics studies fall into following categories:
- Application or system level (rarely, Module level)
- Non-Functional
- Manual, then semi-automatic

These tests can also be thought of as compliance tests as their goal is to show that the systematic errors of the methods used in the project are bounded and at acceptable level. While the methodology of systematics studies is well known, there is no automated process to perform such tests for each project. However, once the framework for systematics studies is built for a given project, it is usually relatively easy to update it for next iteration.

Note that systematics studies are most often performed on the complete, near-final setup. It can be performed on a part of the overall algorithm but only if the correlation with the rest of the project is well defined. Otherwise the method of error propagation is unknown.

### 5.2.6 Multi-dimensional comparisons

In case of physics analyses the available data are often lower-dimension projections of actual data. There are number of reasons for that. The two most important is the low precision of the detector, which makes high-dimension data hard to use due to statistical fluctuations, and the ability of the scientists to visually observe the modeled process, which often requires 2D or 3D projections as higher dimensions are hard to perceive.

For that reason, the analysis often uses a set of 2D projections and the model is compared against them. However, when new data is gathered and precision of the results increases, there might be a time when 3D projections become available. The analysis should take these data into account and the test framework should be adapted to perform comparison to these data. This often reveals many issues with the model or may require change of parameterization.

Given these changes often tests against 2D projections have to be repeated, which means that test framework should be able to easily switch between 2D and 3D data.

Multi-dimensional comparisons fall into following categories:

- Application level
- Non-functional
- Automatic

Conceptually, they are closest to robustness tests as change from 2D to 3D data and vice versa shows that model is not dependent on the data. However, the impact of issues revealed through this test may be quite significant often causing significant changes to the project.

### 5.2.7 Testing with experimental cuts

Some of the phenomena described by the physics analysis is hard to see in data gathered by the experiments. This often calls for custom observables to be made so that the effect is exposed in data as much as possible.

However, it's hard to see if the implementation is correct when viewing the result after experimental cuts are applied, which is why almost all of the tests performed on Monte Carlo tools do not use experimental cuts but instead show the results comparable to the theory. Nonetheless, to verify validity of chosen observables or to present to experimental community how the observed effect can be seen in data, a test with experimental cuts have to be performed. This usually requires collaboration with data analytics of some experiment who can provide the data for comparison.

Figure 19 shows an example of a presentation of the effect which uses detector-oriented cuts on the generated data. We will not go into the details of the physics content of this comparison. Note, however, that on the left-hand-side plot values below 15 on the X axis are omitted and there are visible empty spaces in the data on the right-hand side plot. These are signs that the data that normally cannot be observed by the detector has been cut out of these figures. While in these examples the effects of applying experimental cuts are obvious, usually there are other subtle differences which makes the analysis of the data observed in the detector much harder than the analysis of data predicted by the theory.

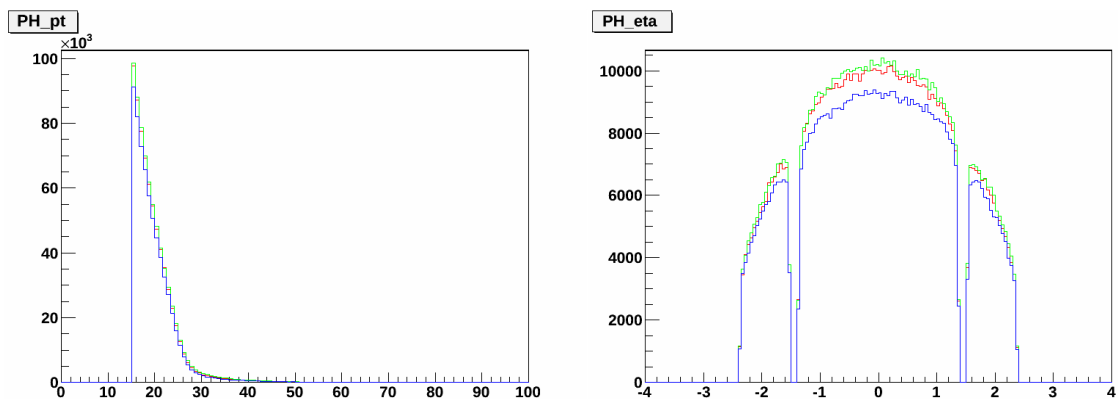

**Figure 19.** Example comparison from Ref. [68] which shows show the impact of the correction weight introduced to `Photos++` using nearly-realistic observables with detector-oriented cuts. Physics details of this comparison omitted for clarity.

Tests with applied experimental cuts fall into following categories:

- Application or System level
- Non-functional
- Automatic

Conceptually, they combine usability tests and compatibility tests, as they show weather the effect modeled by the tool is visible in the data gathered by the experiment or they present how selected observables express the modeled effect. Note that not all analyses require the effect to be visible in the data. In many cases this is not possible.

### 5.2.8 Random scan of parameter space

The random scan of parameter space is tantamount to manual exploratory tests. Their main application is to find outliers that can cause issues (e.g. a parameterization which causes division by zero or other issue that causes NaNs to appear in the results) which at the same time verifies that the whole parameter space supported by the tool is correctly handled. They can also be used to identify potential interesting cases which may require further analysis.

Random scan of parameter space falls into following categories:

- Unit or Module level
- Non-functional
- Automatic

Conceptually, they are closest to sanity checks (smoke tests). Issues exposed by these tests are usually very fundamental. They are also used as the first step of more advanced tests such as systematics studies.

### 5.2.9 Parameter impact tests

In many cases it's important to see if the parameter change is correctly propagated through the system. The impact of a parameter can be tested by comparing the differences between two data samples that have exactly the same setup with one parameter changed by a small value. If the impact of the parameter on the generation process is known, comparing the difference between these two samples can be very useful for determining if the parameter is working as expected.

Unfortunately, not many parameters can be tested this way and when discussing Monte Carlo simulation each parameter change will, most likely, shift the set of random numbers used in the process. This means the difference between these samples will include strongly correlated statistical fluctuations which is why this method is best used with parameters that introduce significant changes and with large sample size. However, with large data sample this test is very useful.

This test is especially crucial in case of models which have no known parameter impact. For example, when developing a new model and comparing it to data using known observables, there might be a case in which some of the parameters are either incorrectly used within the model, their use is not implemented (e.g. values were hard-coded in early stage of the model development) or, in worst-case scenario, the impact of these parameters is not seen in the projections used to validate the model.

Parameter impact tests fall into following categories:

- Unit or module level
- Functional
- Automatic

Conceptually, they combine interface tests and functionality tests, as they test if the software is providing the interface to change parameters which have actual impact on the observables is and they verify the functionality (the model) is correctly implemented and takes into account all effects described by the parameters of this model.

### 5.2.10 Taxonomy of scientific software tests

This section presented variety of tests unique to scientific software that has been developed to ensure the correctness of the algorithms included in the software. As stated in the thesis, which I presented in Section 1.4, and demonstrated by examples shown in Chapter 6, implementation of tests and development of proper test frameworks requires a lot of effort and time.

Following diagrams follow the same taxonomy used to present commonly used software tests to classify scientific software tests. Figure 20 classifies these tests based on their scope, Figure 21 presents the division of test based on whether they test functional or non-functional requirements and Figure 22 identifies which tests can be automated and which have to be manual.

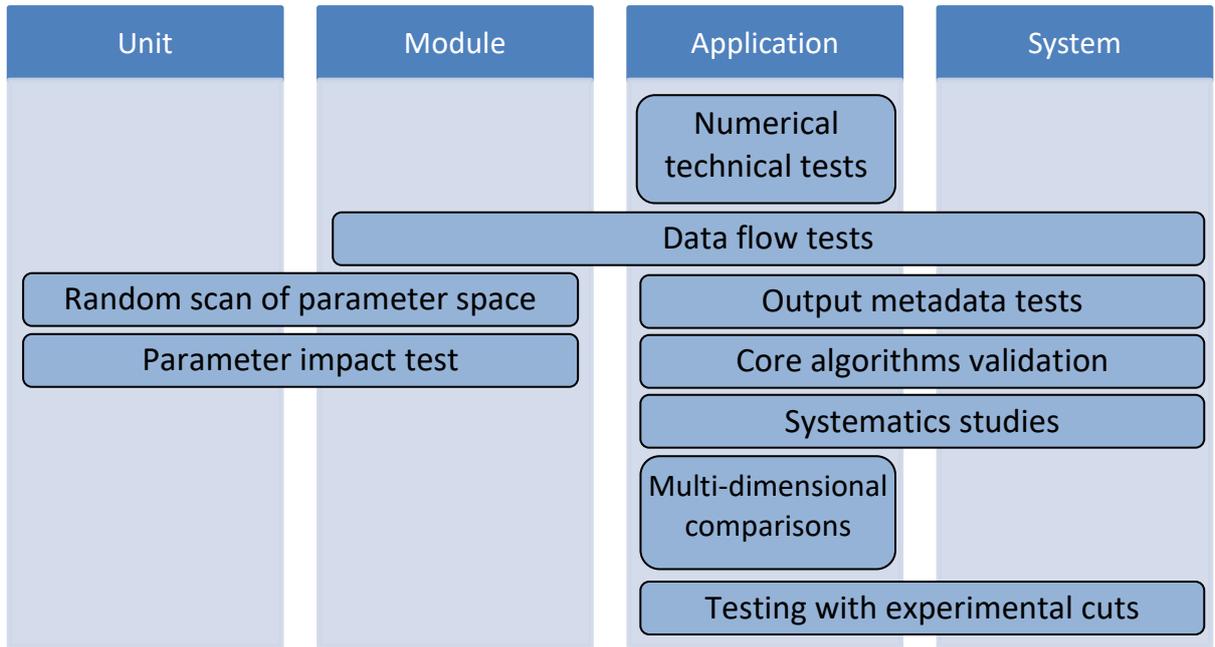

**Figure 20.** Taxonomy of scientific software tests based on the scope they can cover. The scope of these tests should not be changed.

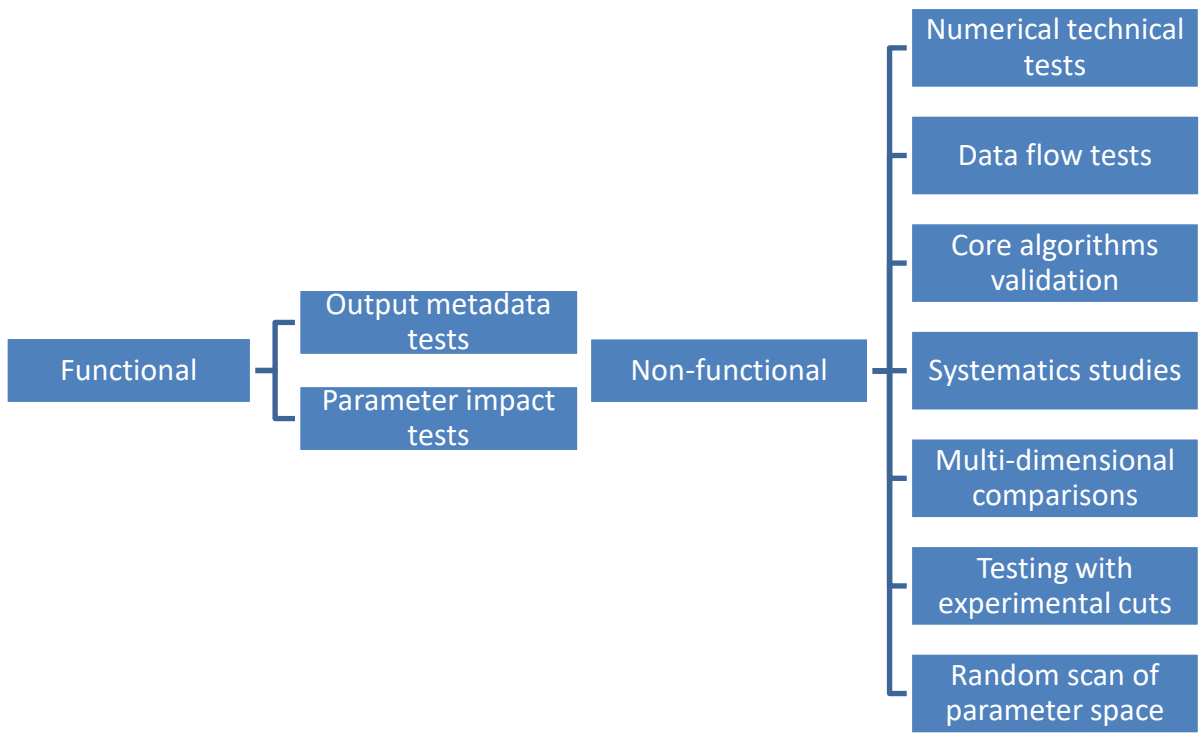

**Figure 21.** Taxonomy of scientific software tests based on whether the test is functional or non-functional.

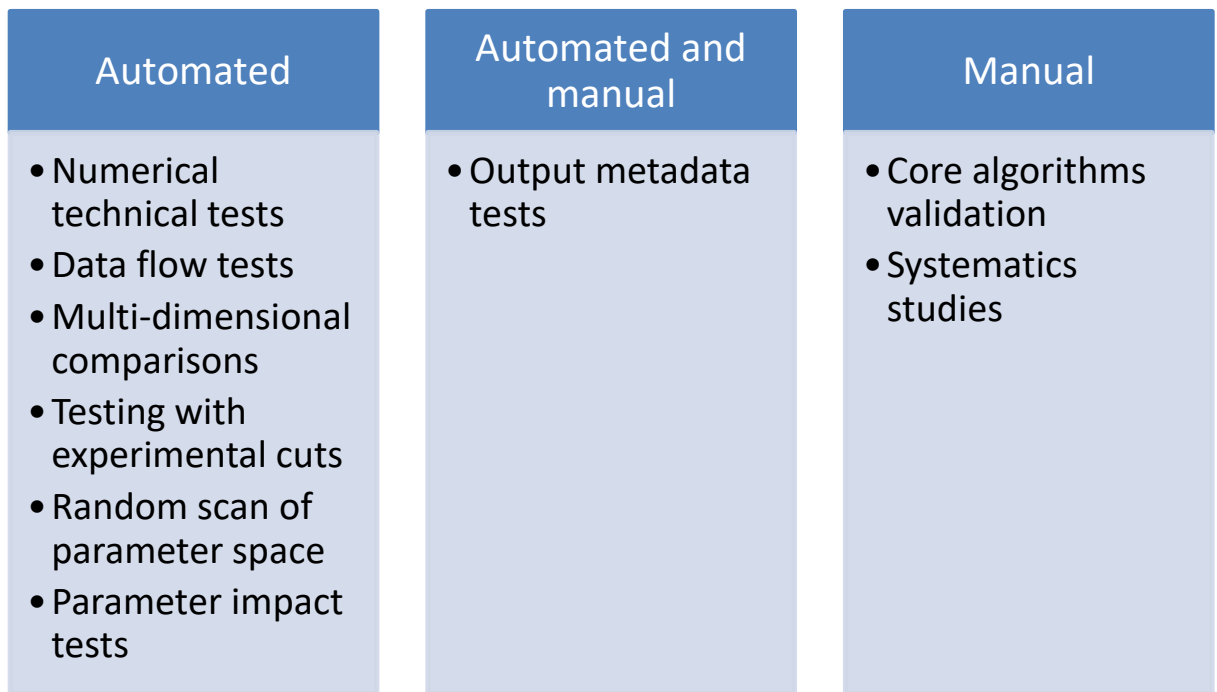

**Figure 22.** Taxonomy of scientific software tests based on the level of automation used. The middle column contains tests that are usually in part implemented as automated tests in part performed as manual tests.

Following are some conclusions and final notes that immediately come to mind when faced with this taxonomy:

1. Figure 20 shows that most of the scientific software tests are done on application or system level. This backs up the approach presented in the cycle described in Chapter 4 in which the most time invested in building the test framework is done after software framework is created. It simply makes more sense to create testing framework that helps automate the testing process based on the final results produced by the application rather than build unit tests or module tests early when the final goals of the project are not yet clarified. That being said, in some types of tests building intermediate test frameworks is often necessary, though such code is much harder to maintain and can often be neglected in future updates (see e.g. example mentioned in Section 5.3.2)
2. Figure 21 shows that most of the scientific software tests are non-functional tests. This has been already stated few times before in this thesis – most of these tests validate the output or analyses the precision of the results which is a non-functional requirement. In fact, while non-functional requirements of Monte Carlo generators are often unique for each tool, their functional requirements are usually in part common for all the tools. These are, for example, the ability of the tool to work with other tools, the ability to configure parameters of the tool or to use different input and output data formats.
3. Figure 22 shows that majority of scientific tests can be automated. This may be a surprising conclusion but one has to keep in mind that once the specific testing process has been identified, the test is often easily repeatable. This is the reason why ready-made frameworks for data analysis are often used in these tests and their generic approach is often enough to build proper projections or comparisons that can be used to validate the results of the tests. In other cases, smaller tools such as those mentioned in Section 2.8.4 can be developed and reused. However, it is worth mentioning that while the testing process can be automated, the creation process of these tests is unique for most of the analyses. They require a lot of effort to prepare every time.

## 5.3 Useful testing techniques

One of the key problems in testing scientific code is the lack of an "oracle" that could reliably verify the results of a test. That is why a "fuzzy" measurement such as code's trustworthiness is used instead of the binary value determining if the code works correctly or not. However, to measure trustworthiness of the code, its results must still be compared against some reliable benchmark; be it an output of another tool, results of a previously conducted experiment or something else entirely. There are two techniques that we have successfully used in our projects. They are similar to the techniques presented in related work [45] but they are specifically adjusted to the nature of the code for the physics experiments. Here we will briefly present these two techniques as well as several other techniques that help test and debug common issues found in scientific software.

### 5.3.1 Probing results at different steps of the computations

One of the very useful property of the Monte Carlo simulation is that the computation process can be tested at its different steps. Such partial computations still produce meaningful results and, in some cases, they can be compared to something meaningful as well.

> **Example:**
>
> First step of the multi-dimensional integration using Monte Carlo approach usually is the random scan of the integrated function. This step narrows down the range in which the integration function is non-zero in all of the dimensions. The error of such boundary estimation is strongly correlated to the number of samples used in random scan.
>
> When testing if such algorithm works correctly, knowing the exact details of the integrated functions, we can verify if the bounds fall within the errors before going to the next step.
>
> Such approach uses the knowledge of the algorithm to test the intermediate result of the algorithm. A similar approach is used in many Monte Carlo generators in which the first step is a phase space generation[52]. For this distribution a simple analytic formula exists. With high enough statistics, we can easily check if the Monte Carlo simulation can reproduce such a formula by turning off all computations that take place after phase space generation.

The existence of such "benchmark-points" is what drives developers to construct configuration options that allow to switch on and off various parts of the Monte Carlo generators. Turning enough options off may allow user to compare the results of the generator to a bare-bone model of their own or results of an older analysis that did not take into account such new options.

There are also more benefits of introducing such options. Knowing well what kind of effect a particular option introduces, a physicist can validate if this option works correctly by comparing results with and without this option. This is as close as one can get to unit-testing a Monte Carlo generator[53].

---

[52] Phase space is a subset of space in which all possible events are included. Generating this subset is analogous to finding a domain of a function.

### 5.3.2 Comparing results to semi-analytic calculations

An extension to the method described above is a comparison of a model to semi-analytic calculations. At higher levels of complexity of the model it is not viable, and at some point -- not possible, to produce analytic formula for comparison. However, other means may exist.

One of the bases of our work on the RChL project mentioned before was a set of semi-analytic distributions to which we could compare our results. As described in [61] an effort spent on exploiting these semi-analytic distributions allowed to speed up the calculations by several orders and greatly simplified distributed computations. Sometimes it is possible to obtain results for the predictions in a form of fully analytic results without any numerical integration. Such results are of a particular value for benchmarking project as systematic error of numerical integration is very different from Monte Carlo simulation. Such formulae can provide results with many significant digits. Such tests must be repeated, whenever technical precision is supposed to be improved substantially.

> **Example:**
>
> `Tauola` source code, especially its extension called `Tauola-bbb`, lists a number of unused decay channels. These are exactly these benchmarks mentioned here. Some of them present the result of analytical function while others are simplified models used for test purposes. They not always provide physical results but are suited to test parts of the generator.

The ability to cross-check parts of the project with results computed by different means, such as numerical integration of an analytic formula, greatly helps in building up the confidence in the code.

### 5.3.3 Testing against fake data

Often, the straight comparison to the data is not trivial. There can be several reasons for that:

- The data is of too low quality
- The exact interpretation of the data is not known
- The data present observables that are hard to prepare at the earlier stages of project development
- The data contain effects that are not yet modeled by the developed tool
- The data contain experimental cuts that make the modeled effect hard to observe

For these, or other reasons, the team may decide to compare the results of the tool to fake data to validate some steps of the development before the tool is ready for comparison with actual data. For this reason, a Monte Carlo sample generated using another tool, or the same tool but with different settings, can be used as fake data. Since the generation process is well-know, this eliminates the issue of not knowing exactly how to interpret the data. This method also gives control over observables and contains one important factor that semi-analytical approach presented in previous section does not – it contains statistical fluctuations which simulates how the real data will look like. Comparing against such sample can help validating that the methodology works even if the sample has large statistical fluctuations.

---

[53] A good example of such tests are analytic tests of TAUOLA. Benchmarks used for these tests are of infinite precision. Semi-analytic distributions are produced to tests results with nearly complete implementation. Then precision of numerical integration routines needs to be considered too.

> **Example:**
>
> When testing the stability of our fitting algorithm used for comparing RChL currents to the data (see Section 6.4.4) we generated MC samples from the parameterization that we have obtained after the fit. We then repeated the fits to this sample, instead of to the data, from different starting points to make sure the result will give us the same parameterization.
>
> Similarly, as a part of systematics studies, we have generated MC samples using different parameterization, then started the fits from a single starting point to these different MC samples to see if the fit will converge to the parameters for which the sample was generated. This verifies that the fitting algorithm is not biased towards a starting point and not biased towards the data.

### 5.3.4 Function switches

In the case of the tools used in physics there is one other type of tests that is far more important than all of the above test groups. That is, the tests of the algorithms which form the most crucial content of the project.

When designing the simulation, one of the key problems, apart from how to describe the effect that the simulation is going to model, is how to observe the modeled effect and test that it is implemented correctly. A lot of the time of the development process is focused on this fundamental problem. The overall model is usually very complex and previously existing functionality can shadow the new effects. Some of the effects introduce changes at a 0.1% level or lower. When implemented, such effects increase the precision of the results but in many cases their influence can be only seen when considering very large statistics or only the corner cases. And even then, they may be hard to observe, hard to consider being well modeled or invalid.

Very often a project has to be able to reproduce older result that does not include the new effects. The ability to test these new effects and show that its core functionality remains correct is one of the reasons why most of the Monte Carlo generators used in physics are built in a way that allows switching off parts of its functionality. Comparing the samples with and without these effects gives clues to whether the generator works correctly and that physics assumptions are met.

The ability to compare lower-order version of the Monte Carlo generator to the old and well validated benchmarks is even more important. Compliance to these benchmarks should be verified at every step of the evolution of the project especially when changes had been introduced to the core of the project or to the legacy code as well as for new applications. Therefore, while a lot of new effects cannot be fully separated from the whole generation process, these that can should always be added with a switch that allows turning them off. This allows for a multitude of different tests, applied with different sets of new effects, to be performed and redone whenever new functionality or new use case for the project application is added.

Building functionality switches is one of the older techniques that is still applicable today, especially in projects with continuous delivery approach[54]. Such switches serve two purposes. First of all, they allow to reproduce the results of the well-documented, described parts of the algorithms for which a number of tests and benchmarks are already present. This is especially important if the algorithms are used in more than one tool and only one of the tools is updated or the update of the other follows with a longer delay. And the second purpose is the backward-compatibility that allows the algorithm to be used in an environment that does not use or benefit from the improvements. In fact, in many cases improvements to the algorithms pose an increased restrains on the algorithm input which in some case may be detrimental to the tool's application.

As usual in such cases, breaking crucial backward-compatibility may result in a client (or in this case: tool or algorithm user) not willing to update to the newer version of the software or switching to a more stable solution, decreasing the tool's user base.

### 5.3.5 Plain text output

When combined with one of the key problems mentioned in this thesis, that most of the algorithms are imported into the project as a legacy code that should not be modified unless absolutely necessary, the role of function switches becomes even more important. Legacy code of an algorithm usually comes with its own set of benchmark tests that, similarly to the legacy code of the algorithms, may or may not be easy to adapt to the new environment.

Old tests are the fundamental aspect of any test environment designed for complex algorithms. They contain both knowledge about how to test the project and the crucial benchmarks that always need to be reproduced. That is why the test environment should not be extensively dependent on external tools used for testing/development as such dependencies can become troublesome in the future. Because of the long life of the projects, tools used to create the environment may no longer be available or be incompatible with the new technologies.

This is a well-known problem in scientific environments, that is why a large set of tools for physics experiments were built with a test framework that included all of the libraries necessary to reproduce the results of a paper describing the tool's usage. The output of these tests was usually presented in text format that can be parsed by other tools or interpreted simply by reading the text file with the results. This in turn allowed to easily design new tools for handling these results. This approach, while very old and in many cases cumbersome to work with, played a very important role in the scientific environment and allowed more than twenty years old tools to be re-evaluated after adapting these tools to a modern environment and the most up-to-date compilers.

There is a reason why every implementation of an Event Record used in physics produces a human-readable output. That is, with variety of target environments in which they are being used it is very hard to assert that the Event Record will be created or parsed correctly in every possible use case. The manual validation, at least the first time the tool is being used, is very important as it may reveal the most obvious errors or missing components in the data structure. `Pythia` [37] Monte Carlo generator always prints out the first event that it generates, starting by explicitly stating the hard process so that the user may validate that his or her settings are correct. This is doubly useful if one takes into account that the text output of the long run of the Monte Carlo simulation is usually kept alongside the generated events or results of the analysis. The information stored in these files is not enough to debug the tool in case of a problem but is enough to eliminate one possible cause: that that the setup of the tool was incorrect.

---

[54] An interesting note on using functionality switches can be found in Reference [101].

I would like to emphasize on the most important point of this section: the console output, usually treated as a log of the program run, in the case of the Monte Carlo tools used in physics experiments has far greater meaning then a simple log. It always, with very few exceptions, contains crucial physics information that can be used to verify the result. As such, it should always be treated as a part of the output of the program run, not as a log that should be deleted after the program finishes its run without errors.

A human-readable text output designed to be easy to parse has several benefits. One is that one can validate updated or new version of the test environment by manually verifying the results, which allows users to easily pinpoint any possible bugs in the test environment. Second is that given proper documentation, that may already exist or can be created once and may never need to be updated, these results can be interpreted regardless of any test framework overlaying the project. Nothing prevents more than one reporting tool to be used on the same project. In the case of `Tauola++` project, the old results of the core Monte Carlo generator called `Tauola Fortran` have been presented using in the text format, that included histograms printed in text and in as ASCII graphics (see Listing 6 and corresponding graphical representation shown on Figure 23), but has been also adapted to use `MC-TESTER` [15], a tool that can be used for extensive tests and comparison with variety of benchmark data.

At various stages of the work on the `MC-TESTER` tool we have used the text output to verify that `MC-TESTER` produces correct results. Moreover, in environments where `MC-TESTER` could not have been used due to its dependency on ROOT [28], we relied on the text output and produced parsers that allowed us to efficiently validate `Tauola Fortran`.

```
       50002           Ene total ALL
         nent            sum           bmin           bmax
        10000       .23197E+04     .36968E+01     .13203E+04
         undf            ovef           avex
    .00000E+00      .13917E+03     .23197E+00
 .0000     .762588D+03                                 I
 .0200     .698660D+02 XXXX                                                                I
 .0400     .404208D+02 XX                                                                  I
 .0600     .265671D+02 XX                                                                  I
 .0800     .150141D+02 X                                                                   I
 .1000     .185329D+02 X                                                                   I
 .1200     .184496D+02 X                                                                   I
 .1400     .188818D+02 X                                                                   I
 .1600     .923089D+01 X                                                                   I
 .1800     .964242D+01 X                                                                   I
 .2000     .789082D+01 X                                                                   I
 .2200     .887629D+01 X                                                                   I
 .2400     .157628D+02 X                                                                   I
 .2600     .820623D+01 X                                                                   I
 .2800     .136436D+02 X                                                                   I
 .3000     .112596D+02 X                                                                   I
 .3200     .903831D+01 X                                                                   I
 .3400     .102579D+02 X                                                                   I
 .3600     .799273D+01 X                                                                   I
 .3800     .102686D+02 X                                                                   I
 .4000     .802897D+01 X                                                                   I
 .4200     .761555D+01 X                                                                   I
 .4400     .111744D+02 X                                                                   I
 .4600     .139865D+02 X                                                                   I
 .4800     .699277D+01 X                                                                   I
 .5000     .955897D+01 X                                                                   I
 .5200     .369679D+01 X                                                                   I
 .5400     .100589D+02 X                                                                   I
 .5600     .538731D+01 X                                                                   I
 .5800     .152659D+02 X                                                                   I
 .6000     .167880D+02 X                                                                   I
 .6200     .101528D+02 X                                                                   I
 .6400     .111203D+02 X                                                                   I
 .6600     .200503D+02 X                                                                   I
 .6800     .195234D+02 X                                                                   I
 .7000     .304833D+02 XX                                                                  I
 .7200     .291010D+02 XX                                                                  I
 .7400     .659979D+02 XXXX                                                                I
 .7600     .179060D+03 XXXXXXXXX                                                           I
 .7800     .132027D+04 XX
 .8000     .394172D+03 XX                                                I
 .8200     .125546D+03 XXXXXXX                                                             I
 .8400     .816726D+02 XXXX                                                                I
 .8600     .541034D+02 XXX                                                                 I
 .8800     .389092D+02 XX                                                                  I
 .9000     .361515D+02 XX                                                                  I
 .9200     .396443D+02 XX                                                                  I
 .9400     .295317D+02 XX                                                                  I
 .9600     .339152D+02 XX                                                                  I
 .9800     .421026D+02 XX                                                                  I
```

**Listing 6.** Example plot generated using `glibka` – an updated version of `HBOOK` [69] Fortran tool for statistical analysis and histogramming. The histogram is scaled to min and max values printed at the beginning.

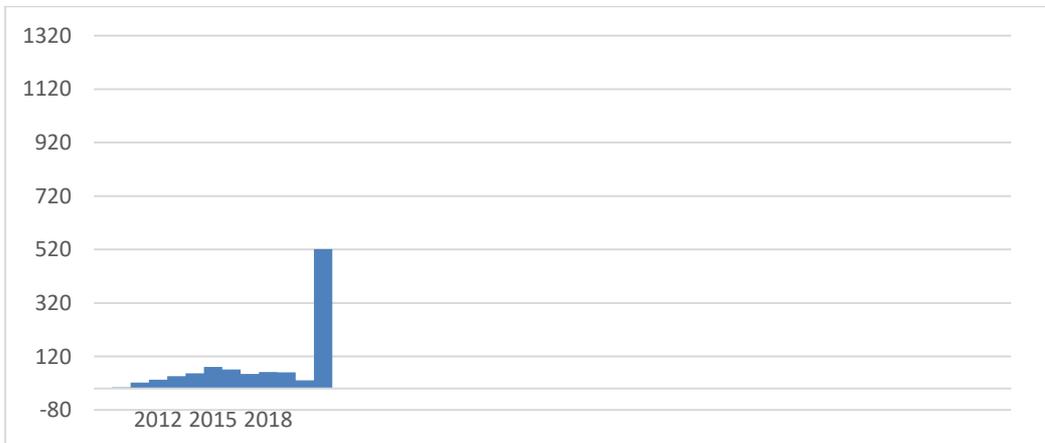

**Figure 23.** Example histogram generated from the first two columns of text from Listing 6.

### 5.3.6 Using static analysis tools

As mentioned before Section 2.7.1, old tools can break when compiled using newer versions of the compilers. The reasons for that are twofold. One is the fact that compilers are not always fully backward-compatible with old standards and can introduce bugs that can break the old code and the other is the fact that old code is not always standard compliant and thus can be handled differently when the new compiler is used. These reasons alone are enough to push back against the idea of compiling old codebase with new tools. Such decision should not be treated lightly. A proper set of technical tests needs to be prepared before such transition could take place and, in some cases, it may not be worth the effort, as long as the old Fortran or C compiler is still available on new platforms.

However, the insight from the new tools and compilers can be used to find potential issues in the old code. The new compilers are known to report increasingly more warnings and errors, especially when all reporting flags are turned on. The LLVM project [30] introduces several static code analysis tools. Using properly configured clang-tidy tool on a codebase can reveal a lot of issues that should always be reviewed. This method can be used on a project built with a very old toolchains in which compilers will easily miss a lot of instances that cause undefined behavior and do not fully validate the code against the language standard. While this may produce huge amount of warnings and errors, they should be carefully examined at least once.

The market of static analysis tools is vast, especially for such old and popular language as C++. There is number of commercial static analysis tools that do far greater work than any open-source software but if nothing else, clang-tidy is a well-established tool that does a great job as a bare-minimum static code analysis tool.

### 5.3.7 Saving random number generator state

It is often tempting to ignore a one-time issue reported by user as most of the user issues are due to improper use of the tool or improper setup or generation process. Such issues are often near-impossible to reproduce and the information regarding the user environment or the generation process may be scarce or invalid. Only if we are lucky, the bug report will contain an event or few that caused the issue. All of this discourages developers from seeking the root cause of the issue.

When developing Monte Carlo tools, we have to remember that we are dealing with stochastic algorithms. In order to reproduce the issue, we have to maximize the chance of its occurrence. Running the most up-to-date version of the algorithms on such events may yield no results plainly due to the margin of error the tool includes or the fact that the reproduction requires exactly the same seed and order of random numbers generation as the user had. The lack of reproduction can be wrongly interpreted as "the bug has been fixed in new version." The issue looks even more hopeless if the bug cannot be reproduced in development environment even on the older version of the tool. This is often the case in many products on market as user environment is always unique while the development environment offers a controllable and clean setup for each test.

A setup that allows to constantly process similar types of events increases the odds of reproduction. As much information on the generation process should be used as is available. Especially the Monte Carlo generator that produced the events and the same production setup should be used. Lastly, the same types of events should be processed over and over in hope of hitting an event that will trigger the issue. If no generation information is given but a single event that caused the issue is given, the same event may be used as source for the tool in hope that after it has been processed many times the issue is reproduced.

In such case, a very useful tool is a random number generator that allows you to save and restore its state. If your random number generator does not have such option and you cannot easily add it, set the seed of the random number generator each time before processing the event. This way, after lengthy generation of the aforementioned two million events, you should be able to reproduce the issue by setting the number generator state to the state before the event was generated.

> **Example:**
>
> Most issues found in `Photos++` require lengthy processing of the same events over and over. One of the issues we had with `Photos++` tool caused numerical instability in a corner-case scenario. We were able to reproduce this issue only after processing the same event two million times in a row.
>
> `Photos++` which uses a standard `RANMAR` [70] random number generator. This generator is easy to extend so that its state can be saved and loaded as needed. This way after reproduction occurs once, however long it takes, next reproduction is instantaneous.

Thanks to this and other techniques described in this section we were able to find root causes of all of the user issues reported for `Photos++` so far. That being said, not many actual issues ever get reported which is why each report should be carefully examined as some issues can only be found thanks to the unique environment of the users.

# Chapter 6. Designing tools for scientific experiments

In this section we would like to outline the history of several Monte Carlo simulation tools developed for scientific experiments with the focus on software development aspects. In all of these cases the physics precision and applicability domain were essential driving force. For each project we try to find key decisions made during their development in an attempt to recognize which factors contributed to their success or failure. These descriptions show the practical application of the process described in Chapter 4 as well as the use of the testing techniques presented in Chapter 5. The content of this section has been compiled into much shorted description of these tools published in ref. [7].

Figure 24 outlines the history of the tools described in this section in terms of the growing technical precision of the experimental data. It also summarizes the amount of work needed to adapt the tool to higher technical precision.

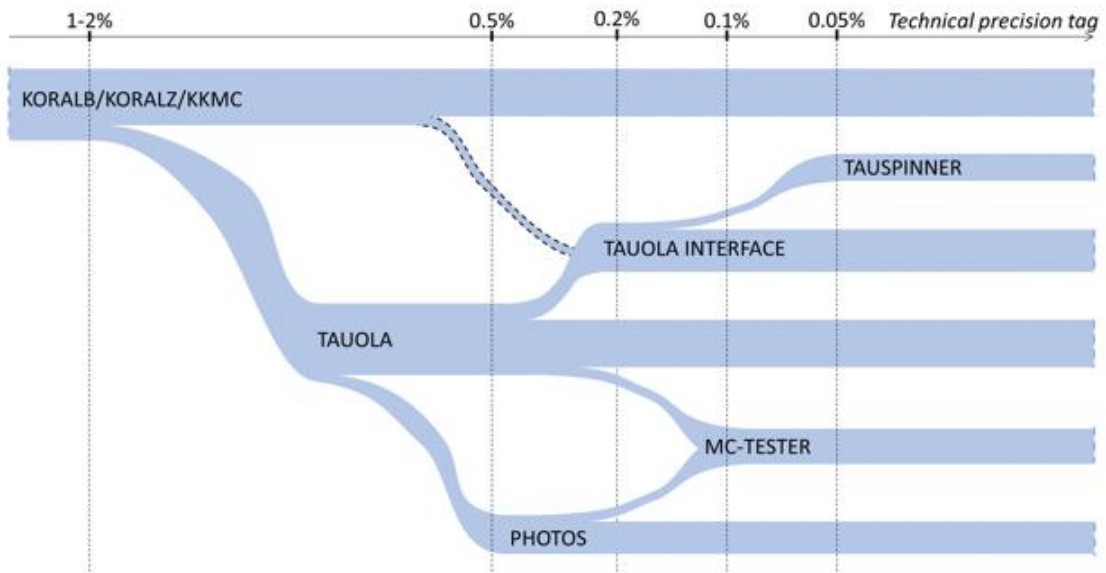

a)

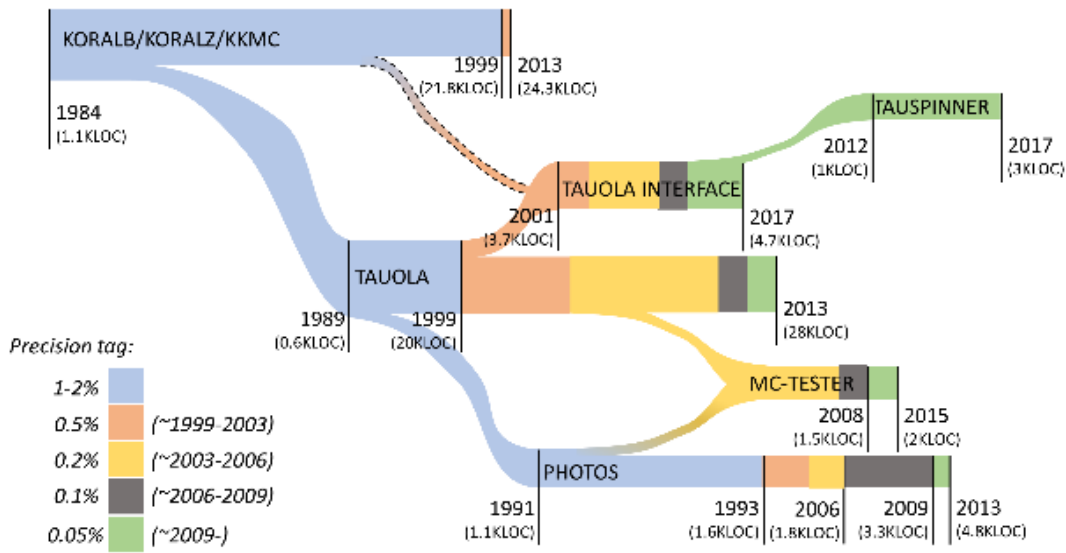

b)

**Figure 24.** Relations between the projects as function of time (precision). Thin band represents preliminary step in project development, in particular its establishment. Broad band is for evolution of mature project where applicability region is defined and dominant activity is for precision improvement. Graph 19a shows map of relation of the project in relation to precision of the results, graph 19b shows project code-base size (in Kilo Lines Of Code, KLOC) overlaid on the map of relations between the projects.

## 6.1 Physics origins

Physics experiments strongly rely on the knowledge build by the others. While the usual principle of using already existing solutions instead of writing it from the beginning is already deeply rooted in the minds of computer scientists, in terms of physics it gains additional meaning. Most of the problems solved by physicists are on a higher level of complexity than the algorithms used by computer scientists as the fundaments of their work. The ability to use solutions to such problems in more complex projects is the fundamental functionality of these projects. New projects can build upon the code that has been written twenty years ago and then, this code can be used again twenty years from now. Most of the authors of the code have already forgotten the principles on which the code has been build, not to mention the details on how the problems have been solved. The only way to get more details is to search through all available documentation, comments in the code and any notes left inside the project. The original authors can help by reading the code and trying to remember its origin. That is, of course, if these authors are still present.

There are many reasons why rewriting such code would be disadvantageous. Recall, for example, the challenge described in Section 3.1.4. A tool that already has an established user-base in the community most definitely has a large set of benchmarks and results that validates its use. By rewriting its code one risks to lose these very important benchmarks and will most likely be forced to validate the tool anew. This is a very time-consuming process that poses many problems. It also adds the risk of introducing problems or numerical instabilities in use cases not covered by current benchmarks. It is especially true if one considers that a lot of legacy code is not well-documented nor tested well enough to ensure safe refactoring[55].

This is one of the reasons why it is so important to keep the proper structure of the code leaving the older code intact. Code build from blocks written in different era create a new level of complexity forcing to mix old code with the new one and solve problems introduced by this process. While these problems are sometimes severe and take days to solve it is still more beneficial than rewriting the code or writing the solution anew. That is because rewriting the code would destroy all the knowledge gathered by the people who wrote it and, would introducing far greater number of bugs and issues. Rewriting such code takes huge amount of time wasted on something that in turn gives no benefit from physics point of view; it is merely solving the problem that has already been solved once. And to top it off, in some cases it might even be impossible due to the complexity of the project. Mixing the new and legacy code becomes a necessity and variety of problems arising from this process have to be addressed.

All of this makes the task of rewriting or extensively modifying legacy code more disadvantageous than beneficial. That is why, if possible, changes to legacy tools should be minimal. This is the approach we took when building all of our Monte Carlo tools. We decided to treat legacy code as a standalone module and write an interface through which new modules can communicate with the old ones. New functionality can then be added, depending on its purpose, either to the legacy module or to the new one.

---

[55] Reference [103] extensively describes the matter of dealing with legacy code and how to approach different problems within this task. The scope of this book is also a strong indicator that the need to work with legacy code is a source of a variety of problems. That is why in order to deal with the most common ones one should prepare the project framework and its testing environment accordingly.

## 6.2 Tauola and Tauola++

In my MSc thesis [71] I've focused on the design and implementation of C++ interface to `Tauola Fortran`, the Monte Carlo generator for τ lepton decay. I have detailed the goals and purpose of the architecture created for the interface as well as the problems that we faced during installation of the project in the environments of the experiments. That is why in this thesis I will only briefly introduce the history of this project, instead focusing a bit more on the next project that followed the same principles as `Tauola++`.

### 6.2.1 `Tauola`

The roots of `Tauola` go back to `KORALB` Monte Carlo tool first published in 1984 [72]. This program was used for simulating $e^+e^-$ collisions and for simulating production of lepton pairs. Center of mass energies of generated events was below 45GeV. For higher energies `KORALZ` [73] and later `KKMC` [74] was developed. A lot of prototypes for solution used later in other projects were present in this tool. Micro-algorithms for decays into leptonic and $\pi^\pm\mu_\tau, \rho^\pm\mu_\tau$ channels were developed within this project. It also presented first distributions of benchmarks in tau decays. It is also worth noting that complete theoretical basis for algorithm for calculating spin correlations, later used in `Tauola` and `TauSpinner`, was analyzed during work on this project.

`Tauola` was an offspring of `KORALB` and `KORALZ` projects. It started as a small module for the KORALB Monte Carlo generator to produce τ decays. Unexpectedly, it has become a widely used tool. At that time (around year 2000 and earlier), technical model precision was at 1%. At best, 100k events could have been generated. `HEPEVT` was the most common Event Record[56] and `Tauola` was built to work on this Event Record. At the beginning, the only tests of the standalone generator which were performed were using small tools `HBOOK`[57] and `glibk`[58] to plot 1D distributions. Semi-analytic and analytic reference results were also exploited. The program was used by all LEP experiments and CLEO.

The year 2001 witnessed the introduction of technical precision of 3 permilles. `Tauola` was used and modified by new experiments – BaBar, Belle. To handle building of different variants of `Tauola` suited for these experiments, a software solution imported from CLEO collaboration was used. A new directory 'Tauola-F' was added. It was a "factory" for generating the `Tauola` code from different parts of the code pre-processed using cpp[59]. At this point, the tests had to substitute single precision floating point arithmetic by double precision. Simply a big number of significant digits was lost in the big sums[60]. At this step `Tauola Interface` as independent project started to appear, first in Fortran [75; 76].

---

[56] In fact it was the first event record format used widely in HEP community. Its introduction was result of broader evolution of High Energy Physics approaches.
[57] `HBOOK` [69] is a Fortran tool for statistical analysis and histogramming with a history dating back to year 1970. This tool is able to create ASCII output of 1D and 2D histograms with variety of options.
[58] `glibk` is a compact version of HBOOK for 1D histograms only but with double precision floating-point variables needed to handle $10^7$ events.
[59] Note that `Tauola` is fully written in Fortran. Cpp was used only as a preprocessor.
[60] This may seem like a trivial observation but such mistakes are painful to find if attention is concentrated elsewhere.

### 6.2.2 Tauola C++ interface

More recently, when a C++ Event Record HepMC has gained larger user base, it quickly became obvious that it would be beneficial if Tauola could use this Event Record. In 2010 the interface for C++ was written, introducing the Tauola++ project into the environment of Large Hadron Collider Computing Grid (LCG) [5]. This interface kept the legacy code as a standalone black-box module called Tauola Fortran.

The legacy module had to be treated as a black-box for the C++ interface to allow low-energy physics communities (e.g. CLEO, BaBar) to work on it independently from this project and to benefit from their work. This allowed the original module to be extended and modified by a large community that was using it, while allowing access to the newer community that used C++ Event Records. Different Tauola Fortran variants could have been easily used without modifying C++ interface. This interface has created an abstract layer to any C++ Event Record.

The evolution of HepMC created many new use cases and also new problems[61]. A lot of development effort went into adapting the interface itself and a lot of technical tests had to be added. Complete separation of the interface and the main code made it easy to test new changes to the interface. New interface also allowed for easier addition of new modules, such as SANC electroweak corrections [77].

To this day, Tauola++ is being expanded with new modules, both in Fortran and C++. As a consequence, the code became a mix-up of different programing styles and different techniques used to solve both old and modern problems. Currently, none of its authors knows the full content of Tauola++. Introducing new changes takes a lot of time and effort remembering all of the details and even the slightest change can cause hard to find errors.

When planning new extensions to the project one has to keep in mind that some of its parts are still used by the low-energy experiments and thus cannot be modified. This principle was true when designing the C++ interface and remains true now, when Tauola Fortran has been extended to use Resonance Chiral Lagrangian currents (see Section 6.4.2) and to allow user-defined currents to be easily used during generation (see Section 6.4.3).

While the project is hard to modify due to its complexity and mixture of different approaches introduced over the ages, its design allows for a lot of flexibility in terms of adding new functionality. Thought the years, Tauola++ has been updated with changes suggested by the pilot users. Several technical updates have been made and new configuration scripts have been introduced, taking into account the possible future updates considering the Resonance Chiral Lagrangian currents[62].

Each version of Tauola++ comes with an extensive testing and validation platform that can be used to verify proper installation of the library within user environment. Tauola++ v1.1.1 has installed and validated by GENSER project [78]. Finally, the Tauola++ distribution has been extended by the TauSpinner tool for the analysis of different applications of spin effects in τ decays.

---

[61] See Section 2.5.
[62] See: Section 6.4.

It is worth noting that `Tauola++` was the first of our projects that introduced the Fortran Monte Carlo generator to the LHC experiments through C++ interface and the first publication regarding the interface [5] has to this day over 100 citations.

## 6.3 Photos and Photos++

`Photos` [79], a tool that started off as a part of `Tauola` but later evolved to much significant, standalone tool with many applications in scientific communities, followed similar path as `Tauola` in its adaptation to C++ environments. Its development benefited greatly from the experience gathered during `Tauola++` development. As such, `Photos++` shares a lot of its solutions with `Tauola++`, expanding upon them when needed.

### 6.3.1 Photos

Before `Photos` publication time (1990), precision of data was at a percent level. Precise theoretical simulations were not needed. In particular photon emission in decays of resonances was not needed, nor could have been included with an overall normalization correction. Nature of dynamics for physics processes lead nonetheless to final states where additional photons (bremsstrahlung photons) are present. In approximation such additional photons are described by simple and universal dynamics, so called eikonal factorized terms. In general, every event is accompanied by infinite number of small-energy photons. Those photons, up to certain precision level, can be ignored. Once precision improves more of those photons need to be taken into account.

When precision required by experiments improved to 3 permilles, it drove the demand for higher precision solution. `Photos` [79] started off as a part of `Tauola`; a Fortran routine named RADCOR. Only single photon emission was available. RADCOR worked on internal variables of `Tauola`. Later on, it was extended to work on HEPEVT Event Record used for interfaces. The motivation was first of all not to proliferate complexity of the data structure but also to save memory space and computer time. Tests were performed using only a tool for plotting histograms, glibk and a tautest program originally written for testing `Tauola`.

An algorithm for 2-photon emission has been added soon after [80]. At this step yet another change of data structures on which `Photos` operates had to be performed. `Photos` had been updated so that it creates a copy of Event Record, extracts the particles that it needs, modifies them and copies them back to HEPEVT if needed. This was because of the need to manage flaws present in the event content. As it turns out, this became useful throughout the whole development of `Photos`, including its switch to `HepMC`.

Resulting algorithm was quite successful and program became widely used despite severe problems due to rounding errors. Algorithm was iterative and its consecutive application lead to aggregation of rounding errors. At this time the first attempt to translate the algorithm to C++ was started. It was bound to have limited applicability because no C++ Event Record has yet been established in the community. Nonetheless, as a consequence formal analysis of the code was performed [81]. As it turns out, this attempt provided essential tests for future development of the project. At that time development of Fortran compilers slowed down and the risk that they may disappear was considered.

When precision improved to 1 permille, `Photos` become a widespread tool. Due to precision demands its development has become more important than the development of `Tauola`. It has been extracted from `Tauola` extensions as a separate generator. Soon after an algorithm for 4-momentum rounding error corrections have been added. Only after these corrections were applied an algorithm could have been extended to produce more photons. Thanks to this algorithm error aggregation would no longer break the 4-momentum conservation.

Later, when precision improved to less than 1 permille, the focus of `Photos` development shifted almost entirely on testing and validation of the generator. `MC-TESTER`, originating from tests of `Tauola` was used to automate tests at next-to-leading order (NLO) precision[63]. While `Photos` provided NLO precision, tests were performed with next-to-next-to-leading order (NNLO) reference distributions obtained from KKMC.

### 6.3.2 Constantly expanding functionality

Around year 2009 data structures of `Photos` have been modified to include information about particle's grandmothers. This allowed to introduce Matrix Elements, which was not developed in Fortran because such precision was not required. Later, it was extended with an algorithm for pairs emission[64]. This algorithm itself has a long history as its code had existed long before it was integrated into `Photos`. It was not used without a reliable multi-photon algorithm as it then did not provide any gain in precision. Only after all other steps required to get to 0.1% precision this algorithm have was worth to be introduced. As such, it has been implemented 15 years after its original concept was first envisioned, coded and tested [82].

Thanks to the modifications of the data structures of `Photos`, necessary for the identification of particle's grandmothers, a framework for prototyping new applications became available. At the time of writing this thesis another extension of `Photos` is being considered. It would expand the project's functionality and its interface to allow introduction of user modification of QED matrix element. There are two classes of applications for this functionality; each may lead to the creation of new projects:

- Implementation of genuine weak loop corrections into `Photos` Matrix Elements.
- Implementation of anomalous couplings.

Both use cases lay outside of the `Photos` application domain which is QED bremsstrahlung. In both cases new expertise can be brought in and results of rather large physics projects can be introduced into the code. Once applicability region is defined, this may lead to pressure on change of the `Photos` algorithm and data structure organization. If community of potential users will be established, then the beginning of the new project may start. Similar processes could have been identified at the earliest steps of evolution for nowadays mature projects such as `Photos`, `MC-TESTER` or `TauSpinner`.

### 6.3.3 Photos C++ interface

Following the successful design of the C++ interface for `Tauola Fortran`, a similar project has been developed for `Photos`, the Monte Carlo generator for final state radiation. In 2009, parallel to the `Tauola++` project, `Photos++` project has started with the first stable version ready in 2010 [4]. During recent years a number of important tests have been performed validating the physics content of `Photos++` [83; 84]. These are important results validating proper installation of the tool within the environments of different experiments. They have also been published online (see Figure 25).

---

[63] This nomenclature is used to describe level of approximation used in field theory calculations. Putting it simply, next-to-leading order (NLO) requires algorithms and data of higher precision that leading order (LO). This also infers that complication of theoretical calculations must be higher. We exclude this aspect, even though essential, from our consideration as it is out of scope of the present paper.

[64] Bremsstrahlung up to certain precision level concerns emission of photons only. However, at higher precision level emission of nearly-real photons needs to be taken into account. Such nearly-real photons manifest themselves as light lepton pairs, thus pair emission has to be taken into account.

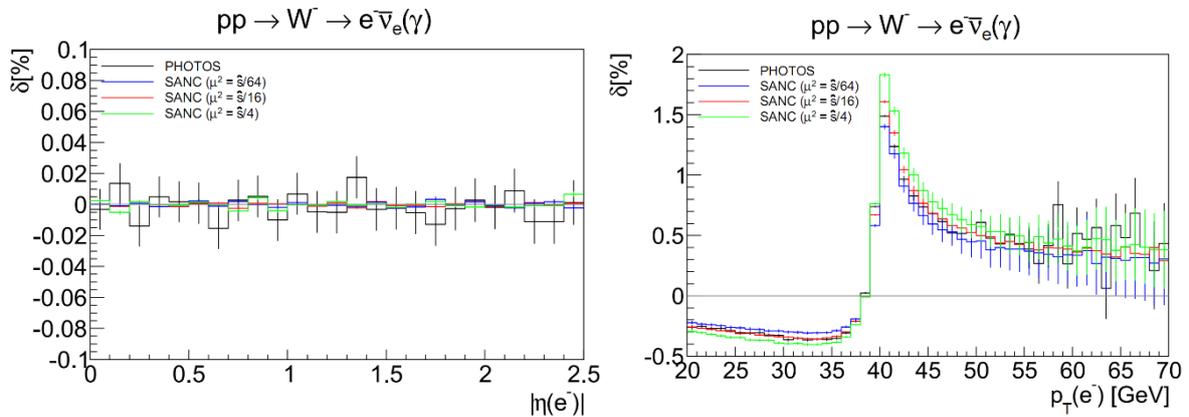

**Figure 25.** Example comparison of SANC to `Photos`. An exhaustive comparison has been published online. See Ref. [85].

With every new version of `Photos++` came an extensive number of tests validating both physics and technical side of the project[65]. However, the most significant technical problems can only be found when `Photos++` acts on events provided by users.

`Photos++` was designed to deal with such special cases with minimum technical difficulties. The hardest part is to determine the physical nature of such cases, how they should be processed and what influence do they have on the rest of the Event Record, as specific problems introduce subsequent issue of how output of `Photos++` should look like.

Below are some of the problems we were able to find thanks to user feedback. They serve as an example of problems that appear when adapting software to different conventions and different usage by variety of users working in different environments.

- **numerical precision and mass of the electrons and muons** – as `Photos++` expects that particles have proper mass and four-momenta. In cases where the precision is lost, numerical instability occurs. An option compensating numerical instabilities or faulty information (such as mass presented in TeV instead of GeV) has been introduced.
- **status codes and history entries** – Event Records often contain particles that do not describe the actual event, such as history entries or particles before/after processing by external tool. Such particles are indicated by different status code, and different experiments use their own set of status codes for such purposes. The option of ignoring specific status codes had to be introduced.

    An option to store particles before `Photos++` processing has been introduced as well.

---

[65] This was also true for the `Photos Frotran` version, see e.g. [68] or [104].

- **non-canonical event structures** - Fortran `Photos` was designed to work on tree structure. However, with time, Event Records allowed much wider flexibility and such structure was no longer sufficient. This poses a set of new problems; each of them must be carefully examined.
  In one of such cases, decay vertices with three mothers had been found, which cannot be correctly processed by `Photos++`. In other case events where, subsequent particles form a loop, were present. In the most basic example, two daughters of one particle are produced and subsequently annihilate, which in Event Record is marked as both particles having the same decay vertex. In some cases, this caused `Photos++` to incorrectly boost particles inside such decay vertex twice, creating NaNs.
- $pp \to t\,\bar{t}$ **and self-decay vertices** – another problem was found when working with Pythia8. In the case of $pp \to t\,\bar{t}$ process, for each decay product of $t\,\bar{t}$ decay Pythia8 introduces a self-decay vertex boosted to different frame than the original particles. `Photos++` had to deal with it by finding out the correct frame to which photons had to be boosted. To retain the same behavior as this of Pythia8, an additional self-decay vertex for photons added by `Photos++` had to be introduced.
- **pomerons and their diffractive states** – while this was not the problem in Fortran, as such events did not exist at that time (or `Photos Fortran` was not used on events containing such particle), it is physically incorrect for `Photos++` to work on pomerons and they had to be added to the list of particles ignored by `Photos++`.

All of the above problems had been analyzed and solutions have been introduced. It's important to stretch that these problems could not have been found and fixed if not for the user feedback. Most of the above problems come from the different conventions in use of the software in analysis chain. As such analysis differ greatly from one environment to the other, it is impossible to take these issues into account when testing the project.

Physical content of `Photos++` remain stable since June 2011, when the last electroweak corrections algorithm has been introduced. Since then, only technical changes were introduced. `Photos++` v3.51 has been thoroughly tested and were installed and validated by `GENSER` project. Thanks to user feedback, new options have been added, installation scripts has been adjusted and `Photos++` has been prepared to work with new types of Event Record structures as well as compensate for some of the inconsistencies within Event Record. Finally, all code of `Photos` was rewritten to C++ while retaining all of its naming conventions[66].

## 6.4 Building framework for comparing physics models

An important problem in stochastic model comparison is the statistical error of the samples produced by these models. MC samples are generated using random numbers, so the accuracy of the predictions is limited by the statistical fluctuations. Two MC distributions generated with the same parameters will differ if different seed for pseudo-random number generators is used. They will also differ if one of the parameters of the generation process will differ slightly, though the difference may be subtle for small data sample.

---

[66] Fortran coding conventions remained intact. As such, most of the code is actually written in C and all necessary data is kept in static structures as if they were Fortran common blocks. The main goal of this step was to allow users to link both to old `Photos Fortran` and the new `Photos++`. If the need arises, the new versions of the project may use more of the C or C++ futures.

These differences are an obstacle for the fitting procedure, which is sensitive to small fluctuations in the predictions. To overcome this issue an approach of applying weights to already generated events has been devised which enabled comparison of multiple models to each other eliminating the problem of statistical fluctuations of such comparison.

### 6.4.1 Re-weighting algorithm

First version of our re-weighting algorithm was created for the purpose of comparing multiple models to the data taken from Belle experiment (KEK, Nagoya, Japan, [86]) or BaBar (SLAC, Stanford, USA [87]) experiments. The algorithm works as follows:

1. Events generated with model X are used as the input data
2. The tool is set up to calculate weight for model X
3. Algorithm calculates weight
4. The tool is set up to calculate weight for model Y
5. Algorithm calculates weight
6. The ratio of weights for model Y and model X is the weight that represents the probability of this event in model Y with respect to model X

An example of this algorithm, which computes ratio of weight of the RChL model to CLEO model, which is default for `Tauola`, has been shown on Listing 7.

```cpp
int main() {
    // Switch to  RChL currents and initialize Tauola
    inipcht_(1);
    Tauola::Initialize();

    (...)

    readParticlesFromFile(&tau, &tau_daughters);
    prepareKinematic(tau, tau_daughters);

    inipcht_(0); // Switch to Tauola CLEO currents

    // Determine decay channel and calculate weight
    double WT1 = calculateWeight(tau_pdgid, tau_daughters);

    inipcht_(1); // Switch to RChL currents

    // Determine decay channel and calculate weight
    double WT2 = calculateWeight(tau_pdgid, tau_daughters);

    // Calculate weight RChL / CLEO
    double WEIGHT = WT2/WT1;

    (...)
}
```

**Listing 7** An example of the reweighting algorithm used to reweight a sample generated with CLEO currents to RChL currents. In case of $\tau \to 2\pi$ or $\tau \to 3\pi$ events without detector effects simulation reweighting events generated with CLEO currents to RChL currents is 6 to 20 times faster than generating new data sample with RChL currents.

This approach has a number of benefits. First of all, it eliminates the long process of generating new Monte Carlo sample for each model used for comparison. This is especially useful when taking into account full detector simulation which often takes much longer than the generation of the τ decay itself. This algorithm can be applied after detector simulation eliminating the need of redoing this lengthy process. It can also be applied in cases where other approaches fail, e.g. when experimental cuts are taken into account.

Moreover, since we are using only one data sample for comparison of different model, we have only the statistical error for the difference between the two assumptions used in comparison. Statistical error of the sample itself is eliminated. Finally, this algorithm can be used to compare any number of models simultaneously by using one input data sample and producing several sets of weight. The only thing that remains is how to systematically perform such comparison on a multi-dimensional distributions.

This algorithm has served as a basis for the tool `TauSpinner` (described in Section 6.5). It has been also used for fitting the Resonance Chiral Lagrangian currents to the data[67]. Figure 26 shows conceptually how this algorithm has been implemented in variant of `Tauola Fortran` for BaBar collaboration.

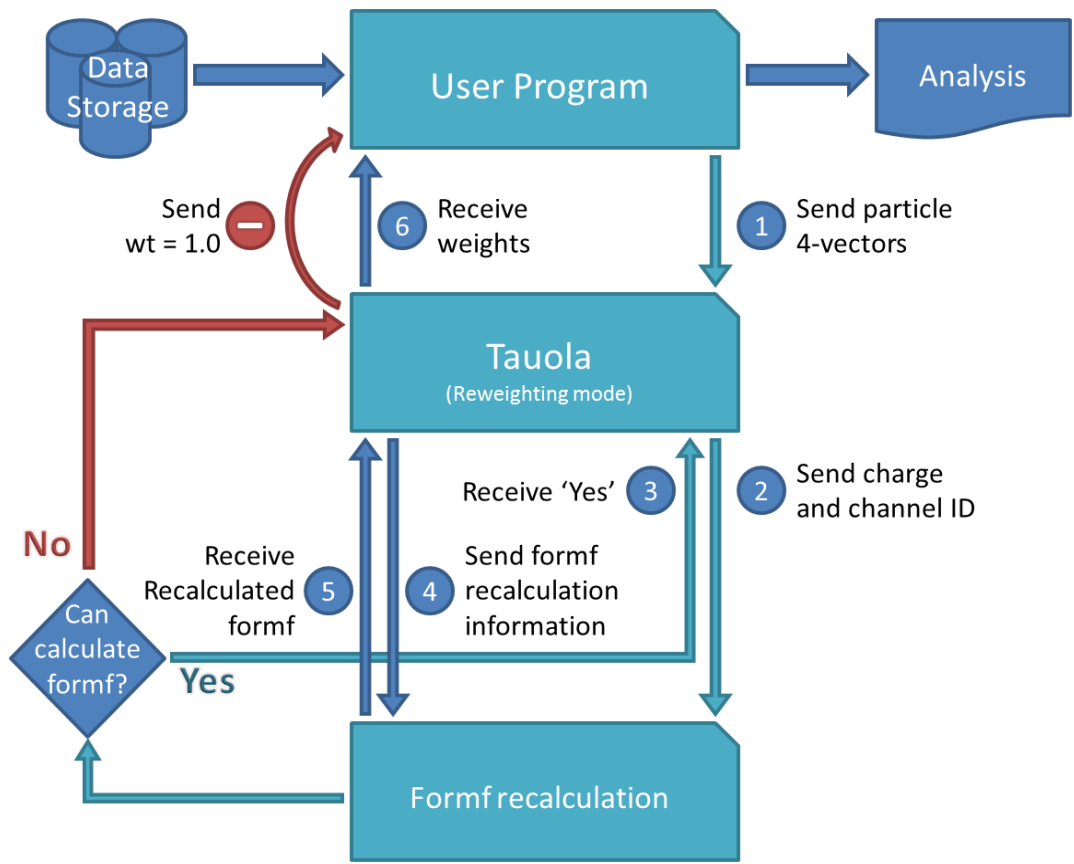

**Figure 26.** Implementation of re-weighting algorithm used in variant of `Tauola Fortran` for BaBar collaboration. Each event follows steps 1 to 6 shown on the figure. This results in a calculation of a weight for that event using different model. Note that `Tauola` will not calculate weight for an unsupported decay channel, as highlighted by the presence of the red exception path.

---

[67] Technical description of both of our approaches to fitting RChL parameters to the data can be found in [61].

The algorithm was built without fully formalized frame for future applications. That is why we focused on flexibility when designing the algorithm allowing for input to be of any format and returning the weight that can be used for few different purposes. However, the tool proved to be useful in number of applications outside of its original goal. It is often the case that a simple idea can lead to construction of a very interesting tool that helps in number of different tasks. One of the most notable examples is the use of `TauSpinner` in the first measurement of the $\tau$ polarization in W boson decays in ATLAS detector [88].

The Appendix A2.3 includes excerpts from the publications describing some of the technical solutions used in our projects. They are concentrated on the Computer Science side of the project. Due to this, the understanding of the naming conventions or some other aspects mentioned in these excerpts may be crippled. If needed, these details can be found in the appropriate references.

### 6.4.2 Resonance Chiral Lagrangian Currents

The first of our projects that fully benefited from the re-weighting algorithm was related to the Resonance Chiral Lagrangian (RChL) theory. The goal of this project was to confront RChL currents with the data and to allow the experiments to use the new currents in their project. If the confrontation with data taken from low-energy experiments brings any significant results, the currents should be available to high-energy experiments as well. Thus, the project needs to accommodate for the restrictions given by both types of environments. This is important restriction as it implies the changes to the original `Tauola Fortran` must be minimal. Different experiments use their own version of the `Tauola Fortran` library which often differs significantly from the distribution from year 2005. Such users must be able to use their own version of the core library as well as be able to modify or expand it further. For that reason, rewriting or imposing heavy modifications upon this library would be detrimental to achieving the goals of the project.

The project should take into account as many other options as possible. In the case the comparison with the data proves to be unsatisfactory, there must be an option of introducing external modification as well as replacing the RChL currents for specific decay channels with the new ones. Taking all of those restrictions into account, we have grouped the old currents, the new currents and tests into corresponding modules:

- `Tauola-Fortran` – core of `Tauola` CLEO as of year 2005.
- `RChL-currents` – separate files with currents for different modes
    - $2\pi$
    - $K\pi$
    - $KK_0$
    - $3\pi$
    - $KK\pi_0$
    - $KK_0\pi$
- `other-currents` – placeholder for external currents other than `RChL`. These currents can be used in future to replace `RChL` currents for some or all of the decay channels.
- `value_parameter.f` – separate file for fit parameters for easier modification
- `wid_a1_fit.f` – linear interpolation of pretabulated function for $a_1$ width. Used to speed up calculation of time-consuming integration

- `tabler` – program for automatic generation of routine `wid_a1_fit.f`
- `cross-check` – a suite of several numerical and analytical tests
- reweighting algorithm
- fitting algorithm

To allow new currents to be applied with as few modifications to the already existing `Tauola Fortran` distribution, the new currents were completely separated from the previous ones with only minimal changes introduced through patches that can either be applied automatically through patch tool or in the case of more complicated modifications made to the `Tauola Fortran`, they can be applied manually by following the detailed instruction and comparing the changes with the original files provided with the installation.

As outlined on Figure 27, the patch introduces minimal changes to the `Tauola Fortran` distribution. This was done to accommodate for the fact that every low-energy experiments have their own version of `Tauola Fortran` library which often differs significantly from the distribution from year 2005.

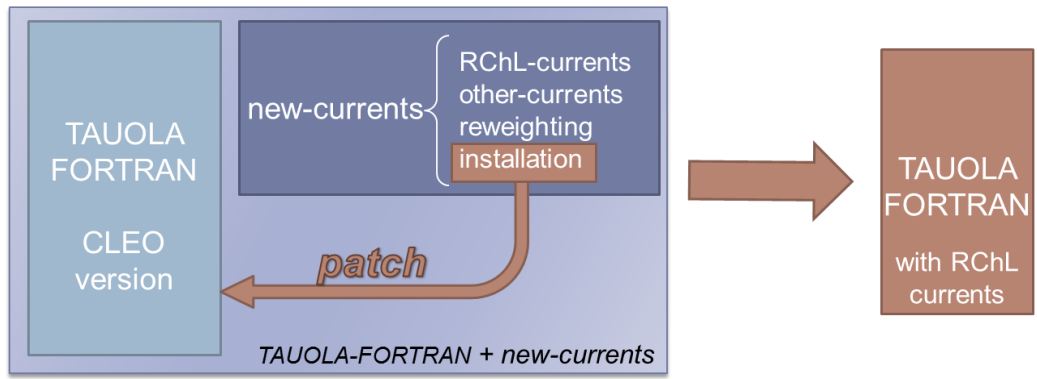

**Figure 27.** The RChL currents update for `Tauola Fortran`

For technical tests, an option of switching between old and new currents has been introduced. By using the old currents after the patch is performed, the user can verify that the installation has not damaged anything in the original `Tauola Fortran` distribution which is an important technical test and the first one that should be performed after the patch. As shown on Figure 28, the installation procedure for high-energy experiments is even simpler.

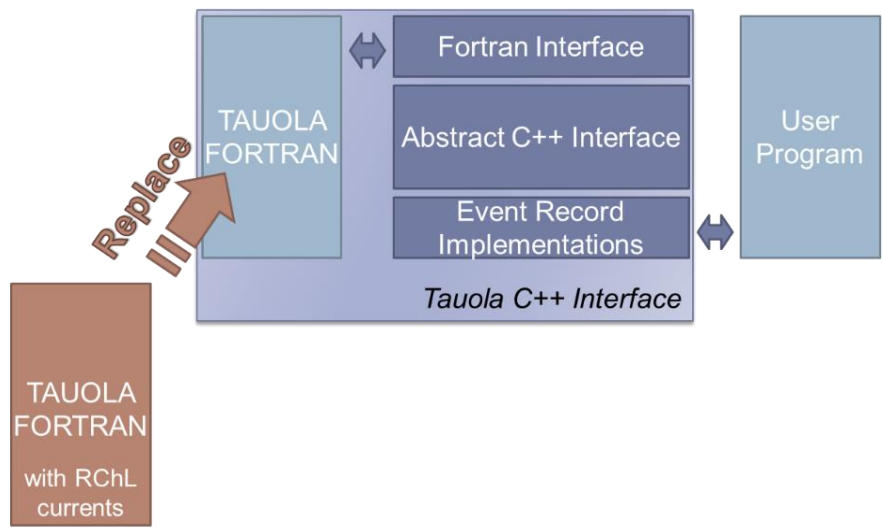

**Figure 28.** The RChL update for `Tauola++`

Thanks to the design of `Tauola++`, where `Tauola Fortran` is completely separated from the rest of the project, it is enough to replace existing `Tauola Fortran` with the patched version. The automated build system introduced in `Tauola++` version 1.0.7 takes into account the newly added extension and new currents is available in C++ projects. It is enough to add one line of initialization to turn the new currents on. Of course, after this procedure it is necessary to perform all technical tests to ensure that the installation has been completed without any errors.

### 6.4.3 Changing existing hadronic currents and matrix elements

Many times, throughout the years we were required to extend `Tauola Fortran` with new hadronic currents. In fact, this subject came up on many other occasions when working on different topics in the past. For that reason, a branch of `Tauola Fortran` called `Tauola-bbb` has been created for BaBar collaboration [89]. This was an attempt to include a long list of placeholders intended to be replaced by user currents. However, this approach was still lacking in its functionality and flexibility. Addition of new currents required changes to the library itself, which in some cases becomes a large inconvenience, nonetheless it served well BaBar collaboration for years.

For many experimental environments `Tauola Fortran` is just one of many libraries treated as a separate module imported to the user analysis through a software management tools, either as a set of precompiled libraries or source code intended to be compiled on target platform. These tools rarely allow to modify such modules as modifications included by one user may disturb the work of the others. In that manner, new currents could not have been included in `Tauola Fortran` without breaking this encapsulation, which meant that researching and testing new currents by individual experts was much harder than it is supposed to be.

There were other issues. For example, adding a current written in C or C++ required a workaround approach and a set of wrappers to be made. All in all, a new more user-friendly interface was needed. Basing off the `Tauola-bbb` approach we have started a project that would allow both Fortran and C/C++ currents to be included in `Tauola` with as little effort as possible. The new interface would have to allow users to write hadronic currents and matrix elements in an object-oriented environment and use own currents in `Tauola Fortran` without the need to modify the original library.

### 6.4.4 Fitting new currents to the data

Once new version of `Tauola Fortran` that included the new hadronic current based on RChL theoretical model was ready, we attempted to compare the results produced by the model to experimental data. The model's prediction is a system of complex functions of three variables and many parameters. One of our goals was to fit these parameters to experimental data. The technical details of the fitting strategies that we used in this project are described in [44].

The hadronic current was implemented in our Monte Carlo generator and distributions generated by this hadronic current were validated against analytic calculations with precision up to 1 permille for 1-dimensional distributions. This was the first step in model validation and it has shown excellent results, so we began fitting model parameters to the data. At this point we have not yet analyzed correlations between model parameters.

We prepared a fitting strategy based on linearization of the matrix element and event weighting technique and we published the results of these fits (see [26]).

These results were far from successful. We experienced numerical instabilities and the minimum found by our fitting framework was ill-defined. This has showed us that our approach is not good enough for our application and showed us that our model needs improvement. Furthermore, our approach required significant CPU power and results of a single fit could take up to a week. We began validating the results and searching for the ways we can improve our methodology. This work consisted mostly of tests and validation on different levels:

1. Cross-checks of results obtained from semi-analytic distributions against analytic computations (see Section 5.3.2).
2. Tests of Monte Carlo fits to fake data (see Section 5.3.3). Used to verify convergence of our initial method.
3. Tests of the fits of semi-analytic distributions to data (variation of the test described in Section 5.3.2). Used to verify that linearization does not introduce significant errors.
4. Random-scan of parameter space (see Section 5.2.8). We used this method to search for a better starting point for the fits.
5. Tests of stability of the fit results (see Section 5.2.9). These tests showed us that with slight variation of starting point we could have ended in different minima, showing that our fitting method produced unstable results.

The results of these tests showed us the direction in which we should follow to improve our results. Based on them we decided to introduce some changes focused on:

1. Finding new approach in physics of the model. This resulted in partial but still insufficient improvements.
2. Improving numeric stability so that integrals of our derivatives are stable. This involved changing the integration method from a Gaussian integration with a variable step size to a constant step size.
3. Use of approximation instead of the CPU-demanding computations. These approximations introduced error to the computed functions. However, since we were using these approximations to fit the parameters to data, we only had to make sure that the errors of the derivatives of these functions in respect to the model parameters are acceptable.
4. Change of fitting strategy from fitting Monte Carlo results to analytic approach. This greatly improved computation speed and overall stability of our fitting strategy.

After applying these changes, we started our fits anew, testing, validating and expanding the scope of the tests along the way. This also included creating a framework for systematic studies. To show the scale of testing involved, the fitting framework had less than 4 KLOC (thousands (kilo) of lines of code) and the hadronic current that we fitted had around 6 KLOC. Compared to that all of our test code had more than 11 KLOC while our current archive of raw results from different stages of the fits takes more than a gigabyte.

The results of these fits were much more promising (see [45]). The model required more work but fits were finished, including statistic and systematic error computation.

### 6.4.5 Consequence of uneven development

RChL project is an example of a project which after initial progress and interesting initial results slowed down to a halt. The reason for that was unequal development in different areas of the project. The code was maintained and reorganized but this process was not accompanied by the work in other aspects of this project, in particular flexibility for fit arrangements software and parameter correlation. The work of the experts focused on phenomenology model building was not consolidated with the work of other team members and further steps became awkwardly difficult.

At some point training of the younger members of the project has stalled and experts only partly associated with the project lost interest in its completion. This demonstrates importance of interactions between all fundamental activities: physics, mathematics, algorithms, program design and implementation. Lastly, it is worth noting that had we been using better model from the very start, we would have avoided strong correlations between model parameters. Also, if we had access to experimental multi-dimensional data the faults of the model would have been exposed earlier[68].

## 6.5 TauSpinner

What started off as a simple algorithm for attributing weights to events calculated by using different models (see Section 6.4) has unexpectedly grew into a powerful tool for numerous spin-related analyses in τ decays.

First possible application of the reweighting algorithm was brought to light when a tool was needed for validation of tau lepton polarization used in `Tauola` tool. We realized that for this comparison we could use samples without polarization and could apply polarization to them to compare results.

For this purpose, a module called `tau_reweight` has been created. It allowed to apply weights to an event calculated as a ratio of matrix element of one model to the other. Such weight can be thought of as an element of Monte Carlo integration for one model with respect to event distribution generated for the other model. Weight is calculated for each event. Then Monte Carlo integral of weighted event represents the sample of the new model. Same integral of unweighted events represents the sample of the old model.

This approach opens doors to many powerful techniques especially if weights are close to 1 as such comparison reduced statistical error for the model difference only as both models used the same event sample. This also allowed to compare two different models without the need to generate new data sample. This algorithm was the basis of the `TauSpinner` algorithm which uses weights to add or remove spin effects to tau decays [6].

### 6.5.1 Modular approach

The modular approach to `TauSpinner` created a flexible tool. It used `Tauola` as a separate library. As such, it could have been used with any version of `Tauola`. It applied weights to an event taken from any Event Record. `TauSpinner` has been extended to allow generation of arbitrary weights, including those generated by models outside the Standard Model (non-SM).

---

[68] To this day multi-dimensional data are not publicly available due to discussions regarding their systematic errors.

Part of the `TauSpinner`'s success was in its design. By using `Tauola++` (and by extension, `Tauola`-Fortran) as its base libraries, it benefits from all the experience and technical work made to validate `Tauola Fortran` and `Tauola++` ensuring that results produced using `TauSpinner` are the same as the results of the use of these base libraries. This is a very important point as it tackles the problem of complicated testing process described in Section 3.1.4. By using well-validated libraries, instead of introducing new ones, the validation procedure for `TauSpinner` is largely simplified. The tool is only required to show that it can reproduce results of its base libraries.

The other benefit of this approach is that `TauSpinner` benefits from many extensions that `Tauola++` provides. This includes ability to change the model or modify currents (see Sections 0 and 6.4.3) as well as other modules, such as `SANC` module for applying electroweak corrections (mentioned in Section 6.2).

Thanks to the variety of possible applications and the design that allows `TauSpinner` to be easily incorporated into analysis, the tool quickly gained popularity and users have shown interest in expanding its functionality.

### 6.5.2 Rapidly expanding applications

Unexpectedly, `TauSpinner` became very popular (see Figure 29). It allowed to apply spin effect on already generated samples, saving the time needed to generate new ones. It also allowed to quickly test results for different variants of particular model or different models and compare them against each other. Due to this, wide range of use cases have been prepared and several papers [90; 91; 92; 93] that present new functionality of the tool with example of its application, as well as introduced variety of tests and benchmarks providing a starting point for new analyses.

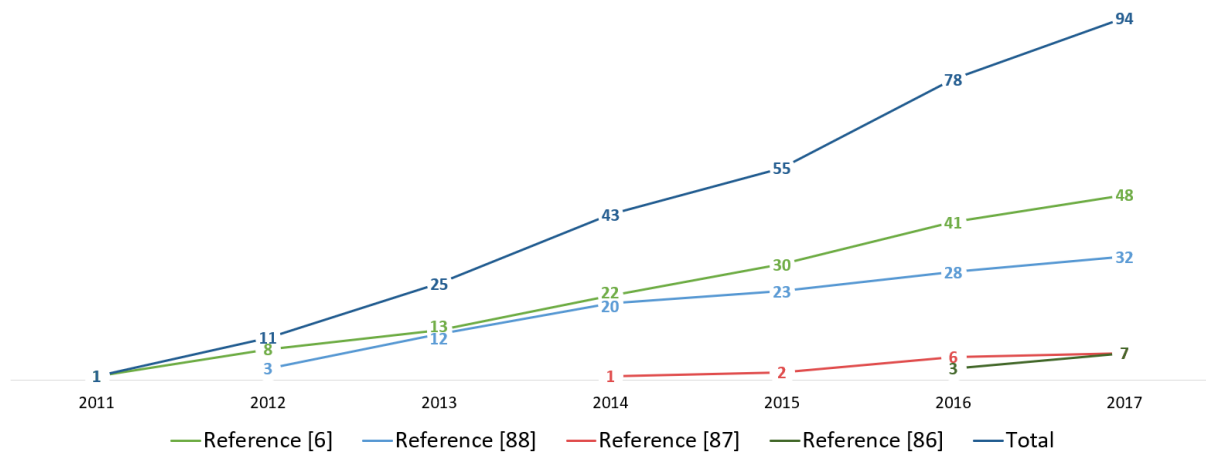

**Figure 29.** The expression of the `TauSpinner` tool popularity through the increase of the number of citations of our publications related to the tool in time.

`TauSpinner` has become our fastest-growing tool in terms of possible application. Considered that it started off as a simple re-weighting algorithm for `Tauola` this outcome was hard to anticipate.

## 6.6 `MC-TETSTER` - tool for comparison of Monte Carlo generators

The origins of `MC-TESTER` can be traced back to the beginnings of the `Tauola` tool, where a small program `tautest` was written to perform the most basic technical tests of the tool. As mentioned in Section 6.2, this function used a `glibk` tool for plotting histograms. On the advent of C++ Event Record era tautest, working only on Fortran `HEPEVT` Event Record, was quickly becoming obsolete. To aid this problem, tautest was rewritten to make use of `ROOT` histograms and became starting point for `MC-TESTER`. This effort started as a satellite project of an attempt of translation of `Photos` to C++. `MC-TESTER` was written using C++ only, which at that time was a paradigm shift in HEP environments. It was built to handle any type of Event Record and used an abstract layer of its own classes to represent event structure. Originally it handled only `HEPEVT` event but was quickly extended to work on `HepMC` as well. This project survived for quite long time because it was able to perform some crucial analyses despite inability of both `HEPEVT` and `HepMC` Event Records to handle quantum mechanics entanglement[69].

First version of `MC-TESTER` was using `ROOT` as its basis for creating histograms and storing data. This approach had its risk, as it was too early to tell if `ROOT` would remain to be used in HEP. At this stage `ROOT` was changing rapidly, including its most basic API, which meant that `MC-TESTER` had to be updated to take these changes into account. Furthermore, external dependency on `ROOT` caused many problems as one could not use `MC-TESTER` in an environment that had different `ROOT` version than the one with which `MC-TESTER` was compiled. This severely limited use of `MC-TESTER` as a set of precompiled libraries, which was the default approach of distributing software for LHC. To deal with this dependency `MC-TESTER` has been updated so that it can be built without `ROOT`. However, the usability of this option was very weak. With time, it has not been supported at all. Too much of its functionality depended on `ROOT`. After some more time one could not even compile `MC-TESTER` without `ROOT`. This is a good example of an idea that should have been pursued, but was handled incorrectly, predominantly because of manpower limitations.

`MC-TESTER` produces easy to navigate booklet with comparison of two generators or two versions of the same generator. It gathers information about the invariant masses of each of the combinations of the decay products for each of the decay channels of the analyzed particle. This automates the most basic technical tests of Monte Carlo generators and some physics tests as well. Figure 30 presents the second page of the generated pdf. First page contains basic analysis information, including sample generation dates, their sizes and information provided by the user for both generators.

---

[69] `MC-TESTER` was analyzing all of a branch starting from a given point, thus its data structures could take into account all correlations between elements of the branch, including those that cannot be expressed by a tree structure.

| Decay channel | Branching Ratio ± Rough Errors | | Max. shape dif. param. |
|---|---|---|---|
| | Generator #1 | Generator #2 | |
| $\tau^+ \to \pi^+ \pi^0 \tilde{\nu}_\tau$ | 25.3029 ± 0.0159% | 25.0750 ± 0.1584% | 0.00000 |
| $\tau^+ \to \tilde{\nu}_\tau \nu_e e^+$ | 18.1201 ± 0.0135% | 18.1030 ± 0.1345% | 0.00000 |
| $\tau^+ \to \tilde{\nu}_\tau \nu_\mu \mu^+$ | 17.6046 ± 0.0133% | 17.7180 ± 0.1331% | 0.00000 |
| $\tau^+ \to \pi^+ \tilde{\nu}_\tau$ | 11.1618 ± 0.0106% | 11.2340 ± 0.1060% | 0.00000 |
| $\tau^+ \to \pi^+ \pi^0 \pi^0 \tilde{\nu}_\tau$ | 9.2787 ± 0.0096% | 9.3770 ± 0.0968% | 0.00029 |
| $\tau^+ \to \pi^- \pi^+ \pi^+ \tilde{\nu}_\tau$ | 8.7599 ± 0.0094% | 8.6890 ± 0.0932% | 0.00000 |
| $\tau^+ \to \pi^- \pi^+ \pi^+ \pi^0 \tilde{\nu}_\tau$ | 4.5342 ± 0.0067% | 4.5480 ± 0.0674% | 0.00014 |

**Figure 30.** The top of the summary from the booklet generated by `MC-TESTER`. The decay channels of τ+ are ordered by the probability. The new sample (green) of 10M events was generated using new version of `Tauola++` and compared against the benchmark distributions (red). The comparison shows agreement on the level of the statistical error.

Thanks to the flexibility of designing new comparisons and new analysis `MC-TESTER` can be used to validate the modifications introduced by new versions and to observe new effects introduced by these changes. Using `ROOT` scripts compiled on-the-flight user can easily add new plots that will be stored in the booklet along with the rest of the results. Figure 31 presents example plots generated using `MC-TESTER`.

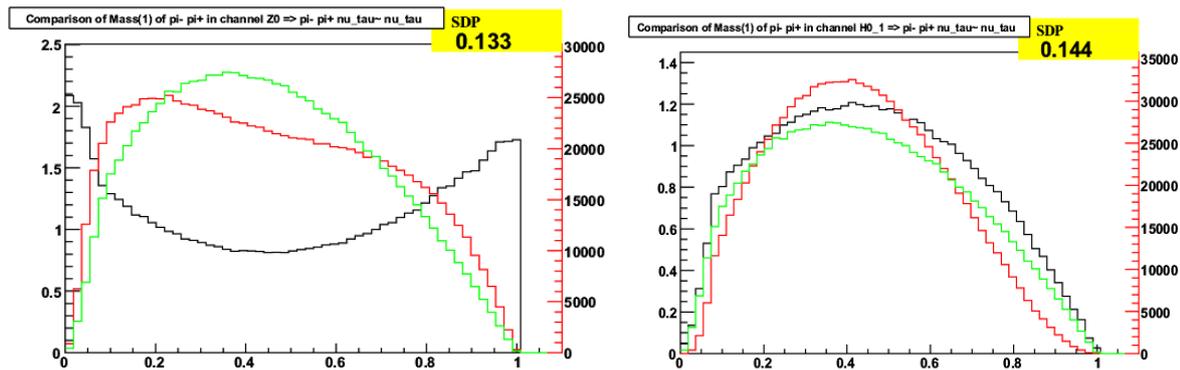

**Figure 31.** The longitudinal correlation tests. Example for Z and H bosons. Histograms present the mass distribution $\pi^+ \pi^-$ pair. Red line shows the distribution with longitudinal correlation on, green with off. Black line is the ratio of the two distributions.

One of my first tasks when I was being introduced to the Monte Carlo simulation software was to expand functionality of `MC-TESTER` and adapt its build process to match the LHC Computing Grid requirements[70] in order for the software to be easily available to other users.

Expanding functionality of `MC-TESTER` was limited due to the use of `ROOT` C++ interpreter, `CINT` and its limited capabilities at that time. `CINT` was introducing random errors in scripts[71] and made it near-impossible to debug such scripts. We had to wait for `CINT` to become stable. It is important to note that `CINT` failures were one of the easiest to spot. Technical and precision tests were unaffected. Program was running correctly or was giving faulty results in a very obvious manner. That was a very lucky situation as more often incorrect behavior of the algorithm looks more like a statistical fluctuation than an actual error in the code.

---

[70] At this time, the now defunct LCG project was a standard for distributing libraries for analyses that were running on the computing cluster and was used as a source of well validated, stable releases of libraries for local analyses as well.
[71] To the point when comments influenced code execution.

Later, when `ROOT` and `CINT` became more stable, `MC-TESTER` could be extended with new functionality. A module for writing user histograms had been added. User could add them in a form of C++ scripts and operated on the abstract layer of `MC-TESTER` to prepare data for the histogram. When running multiple jobs became more widespread a module for merging `MC-TESTER` output files was added. There was no need to focus extensively on multi-threaded analysis as it was not used by HEP community. The common approach was to generate samples on multiple devices (or separate processes within one device) then merge the results so a script that automatically merges results was useful and good enough for this purpose.

### 6.6.1 Missed opportunity

`MC-TESTER` is an example of the project that sprouted in answer to the need of a validation tool for Monte Carlo generators and despite its potential never really outgrew its initial goals. Over the years its functionality has greatly expanded and its interface has been enhanced to allow more versatile use. While the aim was to provide the tool to the larger audience, this goal was never reached.

The reason is that `MC-TESTER` was always treated as a side project; a tool used to validate the core development projects. Once its development reached the state that satisfied all requirements of its use in `Photos`, `Tauola` and `TauSpinner` projects, its development was pushed to maintenance stage. No active work on promoting `MC-TESTER` or adapting it to user needs was performed. This made the tool less widely used when compared to other testing environments used within different physics experiments, such as `Rivet` [39] or `Check-MATE` [40]. `MC-TESTER` was built upon similar concepts as Rivet, aiming to provide an analysis tool that requires minimal effort to use from the user's perspective. It required only three to four lines of code to be added to the user's program in order to gather event data. It was also easier to install and set up, as the `MC-TESTER`'s only dependency was `ROOT`, widely used in LHC experiments, while Rivet had to be distributed with a bootstrap script that installed and compiled several packages before finally building Rivet libraries. Rivet was also tied to `HepMC` Event Record, while `MC-TESTER` could be used with any. These design decisions limited the environments in which Rivet could operate.

However, the most notable difference was that while `MC-TESTER` was only dependent on `ROOT`, it required the analysis to be recompiled with `MC-TESTER` and `ROOT` libraries present. Rivet excelled by being a completely external tool that did not require any modification to user code. It operated solely on the result files produced by these analyses. Similar solutions were envisioned for `MC-TESTER` as well[72] but they were never pursued because of manpower issues. In a very complex environment prepared for distributed computing on the LHC Computing Grid adding a single dependency to an analysis may very well disqualify the tool from being used at all. That is because the whole chain of the analysis and all architectures of PCs on which the analysis will be running must be taken into account. Using an external tool, such as Rivet, allowed to process data after an analysis has ran or to add Rivet as the last step of the analysis chain limited to machines on which Rivet could operate. Rivet did not require modifications to the user code and did not impact any other steps of the analysis, therefore was far easier to include in the analysis chain than `MC-TESTER`.

---

[72] At one point a solution based on fifo pipes was prepared for Belle experiment. It was needed exactly for these purposes: no user code were to be modified and no linking dependencies were to be introduced into the analysis chain. However, it was a special setup that was far from easy to adapt for broader use.

The `CheckMATE`, on the other hand, offers similar approach to `Rivet` but fulfills a completely different role. It is aimed at automating comparison of a new models with previously published results of many analyses. It includes detector simulation and data processing needed to perform comparison with benchmark results. Its main goal is to relieve the user from checking the setup and the comparison method itself, allowing him or her to focus on providing proper model and properly generated data. While this tool requires quite an effort to install and set up, it offers and extensive documentation and tutorials that can help with these steps. Something that `MC-TESTER`, admittedly, lacks. `MC-TESTER`'s universal way of creating comparisons requires both analysis and benchmark data to have a preset format which often cannot be used to directly compare such results with results of already published analyses. If users want to add their own `MC-TESTER` analyses and plots, this often tends to be more cumbersome than simply adding a plot to the analysis code. The predefined set of histograms provided by `MC-TESTER` are simply unfit for such comparisons, diminishing any benefits the tool can provide.

Modern analysis frameworks, such as these, are focused on distributed computing, versatility of configuration and ease of use. They define a new standard which `MC-TESTER` cannot compete with[73]. This situation is similar to the one which project reached during the transition from Fortran to C++ appeared. The solution would be to adapt the project, once more, to the new requirements of the scientific community. On the other hand, the advantage of `MC-TESTER` lies in its ability to work both with Fortran and C++ Event Records. This is of lesser importance for CERN experiments but may be again of use for other communities.

### 6.6.2 Common experience of Rivet and MC-TESTER

It is worth noting that the developers of Rivet considered `ROOT` for histogram creation as well. Section 4.1 of `Rivet` user manual [39] mentions the reason why `ROOT` was rejected. The main reasons were not to introduce the dependency on such monolithic system and the fact that `ROOT` has well known problems with memory management and object ownership. The team behind Rivet agonized over the decision whether to use `ROOT` or not, especially when the tool that `Rivet` used so far, `AIDA`, stopped being useful and became awkward to use. Eventually they decided to write a separate tool, `YODA`, that handled all functionality related to histograms that Rivet needed.

Despite the fact that the development of `YODA` took several years (mostly due to very limited manpower) this proved to be an excellent decision and one of the reasons why Rivet succeeded. The problems related to the fact that `MC-TESTER` relies on `ROOT` is especially prominent now when the old version of ROOT has been superseded by ROOT6 and MC-TESTER has, to this day, not have been adapted to the new version[74].

---

[73] We tried to adapt to new build systems, expanded plotting and comparison options as well as added scripts for merging results from distributed analyses. These changes, however, were scarce and unfocused because of manpower issues.

## 6.7 Quantitative evaluation of software complexity and precision

As presented in Section 1.4, one of the key objectives of this thesis is to present how the relation between the effort required to produce new version of the project and the precision of the results. Such relation can give insight into future evolution of the project. Having access to the code repository of these projects we are able to present the history of changes introduced over time and calculate metrics related to the project complexity at key moments of these projects' development. Tables 3-6 present these metrics for the key projects described in this section. Depending on the project, following estimates of complexity are used:

- **code size** – number of lines of the project's source code excluding source code listed in other columns of the Tables 3-6.
- **Event Record interface size** – number of lines of the project's interface to Event Records and algorithms used to traverse Event Record structure
- **test code size** – number of lines of the executable programs used to test the project and validate the project's results
- **results files size** – total file size of the test results stored in form of histograms, plots, websites or papers. Due to lack of systematic approach to quantifying the amount of effort put into testing scientific software, we present this metric as an indicative measurement of this effort

Following tables present project evolution as function of the precision tag. Precision tag should be considered as an approximate technical precision of the computations. All estimates provided in these tables are indicative. For details regarding software versions see Ref. [7].

| *Software name* | Precision tag | Code size | Event record interface size | Test code size | Results file size | Typical experimental sample |
|---|---|---|---|---|---|---|
| *RADCOR* | 1-2% | 200 | 2 | - | 100kb | 10k events |
| *v1.0* | 1-2% | 1162 | $\leq$ 100 | $\leq$ 100 | 550kb | 30k events |
| *v2.0* | 0.5-1% | 1600 | 800 | $\leq$ 100 | 2.5MB | 100k events |
| *v2.15* | 0.2% | 1800 | 800 | 1271[75] | 2.5MB | 300k events |
| *v2.15 (new)* | 0.1-0.2% | 3315 | 1390 | 1271 | 19.5MB | 1M events |
| *v3.00* | 0.1% | 4747 | 3623 | 1573[76] | 20MB | 5M events |
| *v3.60* | 0.05% | 4747 | 3623 | 1573 | 21MB | 5M events |

**Table 6** Evolution of Photos Monte Carlo as function of precision tag. A typical experimental sample size has been provided to help visualize the amount of data on which the software was operating.

---

[74] The reason is mostly the fact that these two versions are highly incompatible and the decision weather MC-TESTER should be backward-compatible or not is hard to make, especially that both choices have severe impact on the tool itself.
[75] This code size corresponds to MC-TESTER version v1.0
[76] This code size corresponds to MC-TESTER version v1.23

| Software name | Precision tag | Code size | Test code size | Results file size |
|---|---|---|---|---|
| (unnamed) | 1-2% | 200 | 50 | ≤ 0.5MB |
| v1.5 | 1-2% | 1193 | 2215 | ≤ 1MB |
| v2.4 | 0.5-1% | 4016 | 8383[77] | ≤ 1MB |
| Tauola-Photos-F | 0.5% | 19695 | 8383 | 25.2MB |
| Tauola RChL 2012 | 0.05% | 28084 | 8383 | 35.6MB |
| Tauola RChL | 0.05% | 20194 | 9135 | 43.5MB |

**Table 7** Evolution of `Tauola` Monte Carlo as function of precision tag.

| Software name | Code size | Test code size | Results file size |
|---|---|---|---|
| v1.0.0 | 1010 | 147 | 840kb |
| v1.2.0 | 1181 | 214 | 1.3MB |
| v1.4.0 | 1694 | 2944 | 1.6MB |
| v2.0.0 | 2836 | 9661 | 8.0MB |
| v2.0.3 | 3064 | 10188 | 8.4MB |

**Table 8** Evolution of `TauSpinner` tool. In contrast to other projects evolution of `TauSpinner` was driven by the need to expand its functionality instead of the need for higher precision.

| Software name | Precision tag | Code size | Event record interface size |
|---|---|---|---|
| Tests of Tauola | 1-2% | 200 | 50 |
| Further tests of Tauola | 0.5-1% | 4000 | 100 |
| Tests of Photos v2 | 0.5-1% | 100 | 100 |
| Photos+ tests | 0.5-1% | 526 | 4370 |
| MC-TESTER v1.0 | 0.2-0.5% | 1271 | 4336 |
| MC-TESTER v1.23 | 0.05% | 1573 | 5541 |
| MC-TESTER v1.25 | 0.05% | 1963 | 5337 |

**Table 9** Evolution of `MC-TESTER` as function of precision tag. Program was an offspring of tests for `Tauola` and `Photos`.

---

[77] Includes `glibk` an new variant of `HBOOK` mentioned in Section 6.2

Lastly, Figure 32 presents the summary of the Tables 3-6 by showing the relation of code base size and test framework size to the precision of the results of some of the key projects described in this Section .

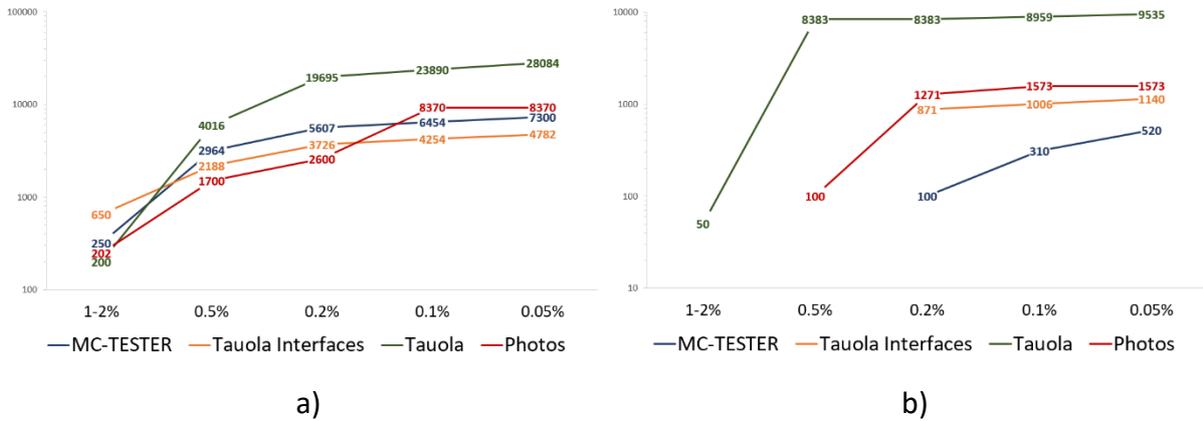

**Figure 32.** Code-base size (in lines of code) of the key projects described in Chapter 6 in relation to the precision of the results. Graph 26a displays the size of the code and interfaces to Event Records. Graph 26b displays the size of the tests and test frameworks.

The results gathered in this section back up the key statement presented in this thesis as they indicate that increasing the precision of the results requires increased effort put both into development of new algorithm as well as into development of new tests, which often comes with the need to expand functionality of existing test frameworks. This relation, while not easy to present, can be indirectly observed through increase of the lines of code and increase in the amount of code of test frameworks as well as the size of the results produced to validate the new algorithms.

# Chapter 7. Common factors related to business environments

The scientific software development process is quite unique in its nature and the experience gathered in this thesis is worth learning by anyone entering this environment if only to avoid the common pitfalls and mistakes that we have described here. This experience, however, is not fully unique to scientific community. In this section I would like to compare scientific software development with my own experience working for business industry.

When it comes to business applications one can hardly find an environment closer to the scientific one then a research and development department of a hardware company. In fact, during two years of work for R&D team developing medical equipment I have found that a characteristic of such team is quite close to the one of a high energy physics experiment; the experience gained while working for the latter could be used to benefit the former and vice versa. Both environments require an efficient collaboration of people from different fields of expertise and a proper set of tools, including an effective methodology, that drives the development of the product. The development process becomes a problem of managing inter-relations of different aspects of the product. Due to their varying understanding of the product specification, team members are focused on different, quite often contradictory, goals. Validation of the product specification must be performed on each step of the development process. To obtain high quality product an effective communication strategy must be established, bypassing the knowledge barriers between team members from different fields of expertise.

I have worked for several R&D teams manufacturing different type of hardware, from automotive industry through simple router devices and lightning equipment up to medical analytic devices and other medical equipment such as Holter devices[78]. During that time, I have noticed many differences but also many similarities between the scientific software development and development of the embedded devices.

Development process in the business environment is far more strictly controlled than the scientific one. Companies introduce a number of procedures that aim to create an environment in which each step of the process is predictable and measurable which lies ground to future optimizations. The level on which the company achieves the control over the process is a mark of the maturity of the process within the company and, by extension, the company itself.

An easy to understand and versatile metric called Capability Maturity Model (CMM) [94] is often used to asses if given company can provide sufficient quality of the products. It determines if the process within the company is predictable, manageable or chaotic. An automotive company can ask its potential subsidiaries to provide proof (via audit) that they have incorporated Automotive SPICE (ASPICE) methodology into their development process at CMM level 3 or higher. This implies that the use of ASPICE methodology in the development process within the company is well defined and has been used, or has been shown that it can be used, to develop at least a single product.

Creating a predictable and controllable development process is one of the key goals of all of the IT industries. This, however, conflicts with the unpredictable nature of the creative process of the R&D teams which is why, to some extent, it is not being applied to such teams. R&D projects are planned with larger margins of errors and are allowed not to be fully constrained by the procedures. This is why the development process in such team is distinct from other business environments; much closer to the one of the scientific communities. The nature of the work of such team is also why it shares most of the challenges described in this thesis. Here I would like to briefly present these similarities by comparing challenges faced in business environment to the ones described in Section 3.1.

---

[78] Device designed for long-time monitoring (typically for at least 24 hours) of the heart activity.

## 7.1 Comparing R&D medical hardware manufacture with scientific projects

The goal of an R&D team is to introduce a new product fitted to the business opportunity found by the team responsible for the market research. This is in stark contrast to the approach described in this thesis, as the initial product requirements are to a large extent known before the project development starts. The details, however, need to be refined through the whole development process and in many cases, they can drastically alter the original requirements.

The product of the hardware R&D project consists of:

- A piece of equipment – used to assert a specific aspect of patient's health
- User manual – describing all details related to the product use
- Certificates – describing how product conforms to the abiding norms

The development team consists of:

- Electronic engineers
- Electronic designers
- Embedded software developers
- Quality Assurance team[79]

The development process is aided by:

- Doctors
- Experts in acquisition and processing of medical data

The majority of input comes from the doctors and experts that describe what tools need to be used to properly gather the required data. This input is then transcribed by the designers into the electronic circuit and mechanical assembly of the prototype of the product. The design can be augmented with components related to other business application the tool can fulfil. The overall requirements are driven by business opportunities and limitations of the electronic design[80].

The electronic circuit will always contain at least one microcontroller for which the software has to be written. This is performed throughout the whole design, prototyping and validation procedure[81]. Similarly, a long process of assembling the documentation and performing validations necessary to certify and sell the final product can be conducted throughout the whole development process. These different fields of expertise form interconnected layers that impose constraints on one another in a same manner as the multi-layered structure of scientific software described in Section 3.1.1.

---

[79] We will use the term "Quality Assurance team" or "QA team" to describe people responsible for: managing product documentation, managing and presenting requirements for product certification, performing risk analysis and controlling the procedures used during development process, testing and validating the product's functionality based on the product specification.
[80] For example, when developing the hand-held device one needs to consider the weight of the product, which limits the battery capacity, which limits the battery usage limiting the overall functionality of the product. The other example would be the addition of wireless communication which can cause interferences in the data gathered through the analogue circuit.

The development of the device is conducted by a number of different teams, including third-party institutions performing the validations or overseeing the certification process. This also includes potential future clients or other institutions interested in the final product as they represent the stakeholders of the product. Their input greatly influences the product's functionality and its design. Relations such as these form a web of interconnections not dissimilar to the one described in Section 3.1.2 where the example of dependencies between different teams in scientific software has been presented.

As mentioned before, the goal of an R&D team is to introduce a new product on the market. While some of the R&D work goes into building the new versions of existing products or finding new applications for such products, it is always related to introducing something new to the market. As such, new solutions have to be applied which quite often relate to a very narrow field of expertise. These solutions can be provided by experts with limited knowledge of the IT industry. From my experience, such experts often work with tools that do not produce code or produce code generated by an automated tool. Such code cannot be easily incorporated into a device and requires a framework that allows the device to interface with a code provided by the experts. This situation is very similar to the one presented in Section 3.1.3, where we stipulate that scientific software requires custom, non-uniform and non-standard solutions.

The validation procedures leading to the product certification cannot be performed until the product's development is close to completion and its core functionality, as well as its electronic circuit and mechanical design, are nearly finished. If such validation procedure uncovers a major defect in the prototype it can sometimes greatly impact the development process. A new prototype has to be created and validated anew, which introduces both unplanned expenses and a delay to the product's release date. This is why a great deal of effort is being put into the testing procedure and methodologies that ensure such defects are fixed before the prototype is scheduled for validation.

The validation procedure varies significantly based on whether the device is supposed to perform real-time analysis and to alarm the medical personnel in case of an emergency, or if this is a Holter device that does not have to emit alarms in real time. Both the testing procedure and test framework changes dramatically between just these two use cases, not counting the variety of others. Moreover, large amount of the tests has to be focused on the algorithms and data gathering and processing. The business environment shares a problem of complicated testing process in scientific software described in Section 3.1.4. In both cases a robust testing environment carried over from the previous projects and expanded to fit the new one is the most valuable assets in this aspect as they provide a large amount of information on how such tests should be performed.

---

[81] Before the first prototype is available, the software is developed and tested on a prototyping board that contains all or most of the elements that will be available in the final product. Note that, when possible, the core functionality should be prepared and tested before the end of the design procedure for the first prototype.

It is worth noting that the tests performed by QA team are based on the documentation and information provided by the product developers. As such, the QA team is focused on assuring the device works as designed and their feedback rarely results in new functionality being introduced or old one being improved. This situation is also similar to the one described in Section 3.1.5 in which the lack of beta testers and scares user feedback poses a problem of limited capabilities to direct the product development or expand its functionality before the product is used by the wider audience.

The fact that the devices that beta-testers use are prototypes created on a small scale means they use the device that is usually produced using different techniques than the mass-produced units[82] and can be significantly faultier than the final products. Many errors reported in the testing period can be related to the hardware failures of single units, not the product design[83]. This significantly impairs or prolongs the beta-testing phase due to limited number of prototype devices available for testing. Faulty devices must be returned to be serviced and given back to the beta-tester to continue the testing phase.

In terms of high impact of scientific software dependencies discussed in Section 3.1.6 I have noted that while there exists a dependency in a form of the communication protocols used between several libraries within a single device, between two devices, between a device and a PC or a device and a server, these protocols usually do not pose a challenge to maintain. It is because they are not shared outside of the company and, as such, their change is easy to control. If there are protocols of communications with external tools or servers, they are usually handled by a high-level API and are not restraining the project development. Note also that the many different types of communication that device can have is mostly standardized and a widely-used protocol for communication can be used.

Lastly, the communication issues present in scientific software described in Section 3.1.7 are also shared by both environments. A typical data gathering, analysis and alarming system can include the following steps:

1. Gather data from the patient
2. Pass the data through the analogue circuit
3. Apply filters and digitize analogue data
4. Send data to the unit containing the algorithms
5. Interpret the result and generate appropriate alarms
6. Send alarms and/or data to the user interface
7. Display the data in a form that medical personnel can easily interpret

The next steps heavily depend on the previous one and each step can be performed by a separate tool. For example, the fact that a patient is connected to the device through wet electrodes, dry electrodes or through screened cables or unscreened cables impacts the set of filters used in the analogue circuit. The circuit, on the other hand, influences the filtering and digitalization process, which is performed by the software on one of the microcontrollers on the device and the alarms generated by the algorithms can be displayed either on the screen of the device or parsed by the remote server, which requires a completely different team of experts to create and manage. Miscommunication at each step can result lack of proper data, or worse, misleading the medical personnel and putting the patient's life at risk.

---

[82] For example, parts of the device that should be assembled on the production line can be assembled or even soldered manually, introducing many potential errors or weak spots prone to mechanical failure of the device.
[83] Such errors should be analyzed in terms of whether they have a chance to occur in the future production process or are limited to the procedure used on a small scale.

Based on this outline one can see how the experience gathered by designing software for the scientific community can be used to aid the development process of the R&D team. There is a strong correlation between the problems faced by the former and the problems faced by the latter. The experience gained in both environments is, to some extent, compatible and the skills gained are of common value. I hope that through reading this thesis one can gain better understanding of the nature of these problems, how to face them and which solutions can help to overcome them.

## 7.2 Methodologies used in business environments

In this Section I will briefly present how methodologies described in Section 2.9.3 are used in business environments based on my own experience. This short comparison may be useful when considering how the use of different methodologies is reflected in real-world projects.

### 7.2.1 The waterfall approach

I have not seen the use of this approach in practice. In fact, I have not saw it being even mentioned during any discussion related to the methodology for new projects or any discussions related to the changes that are being, or are planned to be, introduced within a team or a company. In my opinion, due to the variety of other practices, better suited to current generation of software projects, the waterfall model slowly fades away from software engineering even as a reference material.

### 7.2.2 The agile methodologies

Agile methodologies are undeniably the most popular ones used by software companies. Among the agile methodologies Scrum is the most wildly declared as being used. However, companies usually tend to reinterpret parts of the Scrum guideline or the role of Scrum Master and Product Owner differently.

For example, the responsibilities of Scrum Master are often divided between team leader and project manager. I have also worked for a team that had a habit of picking one of the team members as Scrum Master. Each sprint had different Scrum Master. However, the role of such person was mostly limited to keeping the schedule and leading the meetings.

Similarly, from my experience the role of a Product Owner is most frequently played by a team leader with far less importance than Scrum declares it should have. In one company the role of a Product Owner was equivalent with the role of a Project Manager.

### 7.2.3 Rational Unified Process (RUP)

The RUP approach is very common in hardware-related R&D projects as it fits perfectly into the projects that require extensive initial research and feasibility studies. It also provides necessary documentation without the feeling of significant waste of effort in case of projects that are cut off at the feasibility studies. It also accommodates the hard-lined iterative nature of the hardware prototype production.

I have seen RUP in use in two companies that I have worked for. In one of them the R&D team followed an adapted RUP approach which divided the project into following phases:

1. **Quotation** – a substitute for the RUP Inception phase in which possible risks are analyzed, effort is estimated and requirements are formalized. If needed, some preliminary feasibility studies are performed.
2. **Feasibility Studies** – the phase which role is very close to the Elaboration phase as it lights up a green or red light for the whole project. This phase may require one or more Proof-of-Concept (POC) projects done iteratively or in parallel to each other and is the major part in which research regarding the technologies used in the project is performed. Depending on the nature of the project and previous experience of the company, this phase can be skipped. For example, lengthy feasibility studies are not needed for a cost-down project[84] or a rework project[85], which are the two most common types of mass-produced hardware R&D projects. Usually, such projects do not require lengthy feasibility studies as the quality of the final result can be graded and does not have to be precisely defined at the beginning of the project.
3. **Run A** – first iteration of the Construction phase in which the first prototype is built ironing-out all missing details and rough edges. The resulting prototype should be fully functional but is allowed not to pass all quality standards of the final product. Manual rework may be needed in case of some defects done to the hardware design. The end of Run A signifies that all initial problems are solved and all functional requirements are covered.
4. **Run B** – second iteration of the Construction phase that ends with a prototype passing all quality standards. These prototypes fulfil all requirements and are ready for any certification that is needed to push the finalized product into market.
5. **Productization** – a Transition phase in which project documentation is finalized and projects becomes a product that can be mass-produced. Usually a series of prototypes is already produced before this phase or during this phase to flush out any issues on the production line or post-production test environment.

### 7.2.4 PRINCE2

While I have not seen this methodology being used to develop a single project, I have seen the approach being used by a company to manage the operation of the whole branch, including the long-lasting relations with each of its clients and subsidiaries. The business justification became the justification of cooperation with given client or subsidiary and the product, that was being developed, was the profitable relation by itself represented by the number of projects offered to given client and accepted by him or number of projects delivered and billed. This of course included maintenance contracts and other long-lasting profits for the company. Management by exception was reflected in the way company handled client or subsidiary escalations[86] and amount of such escalations was one of the metrics of the quality of the relationship with the client or subsidiary.

---

[84] A cost-down project aims at lowering the production costs of already existing product. A common saying goes: "one dollar off the production costs is one Porsche for the CEO". This is quite a descriptive saying, on more than one level, but one that clearly expresses how small savings on production costs of mass-produced electronics quickly amasses to millions in revenue. The most basic cost-down project may require that Bill-Of-Materials (BOM) of the produced device is reduced by at least 10%. This threshold is usually the bottom-line for the project to be profitable, given the R&D costs, but the overall goal is not to meet this threshold but to surpass it as much as possible.

[85] A rework project introduces a change to already existing product that fixes or mitigates some issues or optimizes some value (such as power consumption). In some cases, e.g. when producing LED drivers or power tools, a rework's aim can be to create a new line of products that offer more or less power than already existing one.

From my point of view, this was a very good example of how this methodology can be applied in practice but it only reflected how this methodology can be used for top-level management. I could not verify how it applies in smaller scale projects.

### 7.2.5 The automotive Spice (ASPICE) model

During my time working for automotive industry I was able to learn and use ASPICE model in practice. Its impact on the development process is immense; our team spent around 20-30% of the time given for implementing this functionality on quality-related tasks and around 50% on documentation. This may seem like exaggeration, after all only around quarter of the time was spent on design and development, but it is worth noting that the functionality delivered at the end of this process was, by my estimation, around a tenth of what I was used to delivering in the same amount of time in other commercial projects. There was also a significant amount of downtime when waiting for review or test results.

It is worth noting that the amount of people involved in the project, including software architects and function owners that guided the process, highlights that this approach is not well suited for small companies or small projects. In return for these drawbacks I could explicitly state who, when and why implemented each part of the delivered functionality, what kinds of tests covered this functionality. I could review the results of the execution of each test plan and see exactly what kinds of requirements they fulfilled together with the list of changes introduced along the way. It is worth noting that this documentation was readily available to me up front, meaning that when I have entered the project, I could examine every aspect of it without the need to ask anyone about it; a feat that I have not seen in any other project.

To me, the most important aspect of this model was the confidence one feels in the final product. Every step of the project was reviewed and well documented which provided full traceability of changes and made us feel that we are constantly on track. I have worked with this methodology too short to judge if the time and effort investment in the process was worth the end result but the fact that it has become an automotive industry standard required by many automotive-related companies throughout the world indicates that the benefits of this approach outweigh its flaws.

### 7.2.6 Methodologies used in embedded R&D development

My research in practices and methodologies used in development of embedded systems have found only references to the few well-known existing development models related mostly to the certification process that given industry has to follow. As such, the development process is driven by common ISO directives, SPICE (ISO 15504) or Automotive SPICE [54] model with some refinements (such as addition of MISRA software development standard [95] in case of software written in C). Not much thought is given on how well these practices fit the development process for given product.

---

[86] This is a short-hand term for an action in which client contacts the company requesting an additional measures, apart from those already presented by the company, to be taken in relation to some issue with the project. For example, if project delivery is delayed or budget has increased due to some issues found during development the client may escalate such issue requesting a compensation or pressuring the company into fulfiling the original arrangement. These discussions can often be very heated and sometimes result in legal actions.

My experience confirms this state of the art. Usually, the developers or project leaders of embedded R&D teams have little or no knowledge of the benefits or alternatives of the process they follow. As software development is not the key focus of these companies, the subject of incorporating or refining the development methodology is often neglected. This leads to long-term problems as the demand for high quality software grows. This may lead to company struggling or failing to develop a new, cutting-edge product that puts more emphasis on software than hardware, effectively stunning the company growth.

It's worth noting that MISRA has published a C++ guideline in 2008 [96], which was then adapted by other entities (such as AUTOSAR partnership) to take into account modern C++ standards (C++11 and C++14) but they are not as widespread and mostly regarded with caution, as is the overall use of C++ and modern C++ in embedded environments.

## 7.3   Learning best practices from both environments

The most valuable lesson that business environment could learn from the scientific community is to not use the methodology for software development as a set of rules that one should oblige to without consideration but as a tool that it is supposed to be. The methodology should be modified to the needs of the project, not the other way around. If the process is too rigid it can fail to properly grasp all of the aspects of the project. A forced methodology can harm the project by wasting team effort and, in consequence, reducing the motivation that the team has to fulfil its goals. After all there is nothing as demoralizing for a researcher eager to search for a solution to a new, unique problem than the thought of how much bureaucratic work is needed throughout this process.

The other key aspect of the scientific approach is often neglected by the business environments. It is the amount of effort put into validation of the results. While the code-related metrics, maintainability and unit test coverage are useful for managing the easiest types of issues, they cannot be used to validate project functionality. Validation process often requires that a custom test has to be performed and documented, which requires additional resources which, for the testing phase, are often cut short. Especially when it comes to non-functional requirements. These types of requirements are often the source of issues that show up very late into the project development. They should be addressed as early as possible in order to properly manage the risk of the project.

On the other side, when looking at what scientists could benefit from business environment the first that comes to mind is the tools and methodologies. Large part of the software development process can be easily managed and automated by the use of few simple tools (see Appendix A1) which could help decrease the number of technical errors. A systematic approach to building project architecture and technical tests can help guide the development of the software framework.

Similarly, proper definition of tasks needed to complete each step can help define project milestones and focus the team effort. Even if all of the work needed to complete a project is not possible to determine the overall process consists of a number of predefined steps that can be used as project outline. Out of it the first few smaller tasks can be defined and maybe other key elements that should be checked later can be drafted as well. All of this can be used to track the progress of the project and work left to be done at its current stage. This, along with reporting of every defect found along the way can help when reviewing the work already done and when searching for possibly missed steps. Detailed history of the tasks performed within the project also helps track the tests performed for particular part of the project and document intermediate test results. Such information can be very useful for future new projects or as an archive to which one can return to even years after the project has finished.

As usual, the mixture that takes the best elements of both environments, can result in an approach that helps overcome the issues faced by each of them separately. I hope that, if nothing else, this thesis can provide a starting point to identifying these elements and provide guideline which practices have a chance of succeeding in addressing these issues.

# Chapter 8. Summary

In this thesis we have described the methodology used in scientific community to develop Monte Carlo tools for High Energy Physics. In Section 1.4, which describes the scope and goals of the thesis, we have listed six goals that this thesis aims to fulfil.

The first goal was to present the physics background of software for High Energy Physics needed to understand the context of these projects. We have presented several different types of software developed by scientific community, which range from the large-scale data analysis framework through multi-purpose generic Monte Carlo generators to single-purpose small Monte Carlo tools. We have highlighted the common factors of scientific projects and how their development is driven by constant demand for higher precision.

To fulfil the second goal, which was to analyses the scientific software development process, we have presented the common challenges faced by scientific software and how they influence the decisions made during software development. We focused on the most common type of software written by physics communities, which are small Monte Carlo tools and analyses. We have outlined issues in managing the scientific software and showed how common software development methodologies relate to such projects.

Next, we described the methodology that emerged during the development of several Monte Carlo tools aimed at preventing the most common problems encountered during the development process of these tools. This methodology is aimed at projects focused on improving the precision of the results. We have attempted to define the key elements of the development processes of these projects that focus on obtaining higher quality design and model of the given process and, in turn, higher precision results. These elements form a cycle that consists of six steps:

1. Improving (or creating) the physics model
2. Describing the model using mathematical formalism
3. Implementing the model with numerical approximations
4. Creating the software framework
5. Documenting and validating result
6. Testing and publishing results

We have presented how these steps relate to development of several projects, such as tools for analysis and production of heavy particles, tools for comparing different physics models and the Monte Carlo tools testing framework. We have exemplified some quantitative measurements of effort needed to improve precision of the results of some of these tools.

We have also presented the taxonomy of tests and testing techniques which we then extend by the list of unique tests performed for scientific software. We have shown the purpose of these tests and how these tests fit into the taxonomy. With higher precision demand more effort is put into writing and maintaining the testing and validation environment than into development of the project itself. That is why we have also presented several testing techniques that had proven useful during development of scientific software and several important factors worth considering when working with old codebase in hope that my experience will be useful when building test frameworks for other scientific projects.

Following that, we have presented the history of several Monte Carlo tools with various scopes and purposes. Several of these projects, as well as other projects mentioned in this thesis, are in use by scientific community. For some of them we have published one or more papers describing their applications and use cases. We have also presented a quantitative relation of the complexity of these projects and the precision of their results, which was one of the goals of this thesis.

Projects mentioned in this thesis are (or were, for some time) developed with:

- A. Arbuzov
- M. Backman
- D. Bardini
- V. Cherepanov
- Z. Czyczuła
- N. Davidson
- R. Józefowicz
- J. Kalinowski
- W. Kotlarski
- G. Nanava
- E. Richter-Wąs
- P. Roig
- R. Sadykov
- O. Shekhovtsova
- S. Tsuno
- Z. Wąs
- Q. Xu
- J. Zaremba

They were also performed in cooperation with experiments:

- Atlas [22]
- BaBar [25]
- Belle [26]
- CMS [23]
- CDF [27]

We have also shown few examples of our projects that did not perform as expected as a consequence of our decisions or improper approach. They serve as valuable lessons which helped us to avoid similar results in other projects.

As a final outtake on the subjects presented in this thesis, we have shown some similarities between scientific software development process and business process of hardware manufacture R&D branches of few companies based on my own experience.

In this thesis we have focused on the Computer Science and Software Engineering side of all of the described projects while keeping in mind that it is not the only aspect of these projects nor is it their main focus. We have not referred to progress in physics, which was the central point of these projects, beyond essential minimum. These projects, despite large programming effort, are mostly focused on the domain in which the research is performed. For this reason, the development strategies and evolution of the software of these projects is often marginalized. Our goal was to address Computer Science community, in particular people who can be involved in similar projects from different domains.

## 8.1 Lessons learned

In this section we would like to highlight the most important lessons one can learn when it comes to the development of the tools for scientific community. These aspects should be considered when planning new scientific project or when facing similar challenges.

### 8.1.1 Openness to change

If a project goes over decades, it should not block potential for changes to be introduced in the future. In fact, its developers should take into account that any part of the project may require modification. This is one of the reasons why most of scientific code does not introduce any complex structure or abstract internal layers. Modification of such layers would trigger a cascade of changes. The best structure is the simplest one.

Development of such software requires compromising between attaining clear software design and allowing the scientist to develop algorithms in as much flexible way as they need. The initial design of the project may need to be redrawn and some modules may not survive the changes introduced with time. The goal of the scientists is to try and anticipate such changes ahead of time. Sometimes, as the example of `TauSpinner` project shows, such changes can also lead to the development of new projects.

This is not a trivial task as it is not always obvious what kind of components will be left or removed. Even well-established components may need to be revised when working on future extensions of the project. A good example of that is the integration algorithm that had to be changed during work on RChL project. This is something that normally we would not expect will change during project development, yet this change proved crucial to the stability of the fitting framework. One should be prepared to face such changes when needed.

### 8.1.2 Robustness to external changes

In this thesis we have noted on several occasions how changes introduced to Event Record severely impacted development of our projects. The non-tree-like decay structures caused by duplicated entries where second entry was treated as a "history" entry could break an algorithm that "did not know" that these entries should not be processed. These algorithms were entering loops as they were expecting a tree-like structure. This was trivial to fix. Other, such as introduction of multiple mothers (more than two) were not as easy. With time our project became more and more robust to such external factors. This required constant monitoring of the quality and variation of the input data.

The other prominent example is the `MC-TESTER` and its dependency on `ROOT` which in the early days of the project led to significant problems in its development. The unstable interface of `ROOT` blocked us from extending functionality of `MC-TESTER`.

It is important to consider the restraints that other tools impose over the developed project and how external dependencies can impact the project in the future. And if such dependencies are necessary, proper interfaces should be built to allow switching out the dependent library if needed.

### 8.1.3  Modular development process

While code of the scientific project cannot always be divided into well-established modules, the development process can. The team can often approach to mathematics, physics and Computer Science aspects of the project separately. However, while doing so, the team has to always maintain the flow of information between these domains. Otherwise the work produced by some team members may be unusable by others. The work of the team may become ineffective.

A good example of such outcome is the RChL project. At one point the effort put into fitting has greatly surpassed the effort put into building the model itself which consequently stalled the whole project.

### 8.1.4  Constant evolution of tests

I have stressed many times in this thesis how much effort of the team developing scientific software puts into building and maintaining the testing environment. The development of tests and testing environment is one of the elements of the development cycle. This naturally shows that the testing environment evolves and expands over time. In some cases, the tests for one project evolve as by-products of previous ones.

It is important to focus on the approach to the project validation from the very beginning. Testability is always an issue in any major project. Even more so in the case of scientific software. It is important to automate the technical parts of this process so as to leave as much time and effort into building the custom tests required to validate the project. The framework used, or built, for this purpose should allow both.

### 8.1.5  Focus on the project domain

Lastly, it is crucial to remember that scientific projects are usually focused on the domain different than Computer Science. This means that the code written for the purpose of fulfilling different goals within the project must be adapted to the task at hand and to the focus of the project, not the other way around. This is a very important element that computer scientists should remember.

In the case of physics domain in common projects theory behind the implemented model binds every aspect of the domain with each other through variety of interactions. This means it is hard to separate elements from each other and divide into smaller, more controllable modules. What's more, a model that includes all of those interactions would be impossibly complex to implement, which is why appropriate decisions have to be made to ignore some of these interactions in order to simplify the model. These decisions directly impact the precision of the results. The higher the precision demand, the more complex interactions have to be included.

Finding proper mathematical formalism, proper constructs and possible approach of decoupling part of the theory from the rest so that it can be implemented as a simplified model or the approach of including additional, more complex interactions to this model, is one of the crucial aspects of the project. These tasks befall physics experts that take part in the project. The role of computer scientists is to aid them in the process and provide feedback on what can be feasibly implemented and what can pose more problems. They also aid in building test framework for the model prototypes.

Here, again, the testing and validation process is the most important one as the model itself is often subject to change based on the results of the tests performed on its previous iteration. During project evolution all of the aspects of the model, its algorithms, its data structures and test frameworks, can be partially rebuilt or completely rewritten. However, as with other disposable proof-of-concept prototypes, this is to be expected. While the code may not survive this process, the knowledge about the modeled process and the domain of the project grows with each iteration, ultimately leading to the final product.

## 8.2 Future work

The proportional relation between the effort put into testing and effort put into achieving results with increased precision is the natural element of perturbative calculations. Because of that, tools based on these calculation are a good example of this relation. However, we could pose a hypothesis that other types of software also demonstrate this relation.

The next step could be to perform study of the variety of software from different fields of science. Such study could encompass complex analyses related to climate modeling, weather forecasting or hydrocarbon exploration. Seeing how this kind of software is also focused on increasing precision of the results or reducing the errors of the approach, it is possible that the development process of the software from these fields of science follow similar cycle as the one described in this thesis.

These are just examples of the fields of sciences that could be covered by further research. By analyzing the development history of variety of tools with cooperation of specialists from different fields of science a thesis encompassing wider type of software could be posed. The analysis of the history of such projects could help predict the effort required for further development of these tools.

Furthermore, it would be valuable to analyze the development history of larger projects, such as software frameworks described in Section 2.8.1 or detector simulators described in Section 2.8.2. Such analysis could reveal the development patterns of large scientific projects and would allow to compare and contrast these patterns with the development cycle presented in this thesis.

One of the necessary step of further research of the scientific software development process would be formalization of a systematic approach to such analysis. This could help other researches in their own analysis of different types of software.

Lastly, it is worth mentioning that any study of the scientific software development process could help fill-in the gap in this subject and could encourage scientists to document the development process of their tools. This, in turn, could lead to new research opportunities and, in future, could reveal new topics related to the analysis of the scientific software development process.

## 8.3 Author's contribution to scientific community

During my work for scientific community I contributed to number of publications. My contribution to these publication regards the Computer Science aspects of these projects. I took part in development of the appropriate software as well as contributed to the decision-making process regarding the software architecture, building the testing framework, software organization and development of automated testing processes and support tools.

Following list contains all of the publications co-authored by me ordered from the most recent to the oldest ones. The key aspects of some of these publications, that are representative to my contribution to them, are presented in Appendix A.2.

1) [The HepMC3 Event Record Library for Monte Carlo Event Generators.](#)
   By Andy Buckley, Philip Ilten, Dmitri Konstantinov, Leif Lönnblad, James Monk, Witold Porkorski, Tomasz Przedzinski, Andrii Verbytskyi.
   [arXiv:1912.08005 [hep-ph]].

2) [Software Development Strategies for High-Energy Physics Simulations Based on Quantum Field Theory.](#)
   By Tomasz Przedzinski, Maciej Malawski, Zbigniew Wąs.
   [10.1109/MCSE.2019.2947017](#).,
   Comput.Sci.Eng. 22 (2019) no.4, 86-98.

3) [Documentation of TauSpinner algorithms: program for simulating spin effects in $\tau$ - lepton production at LHC.](#)
   By T. Przedzinski, E. Richter-Was, Z. Was.
   [10.1140/epjc/s10052-018-6527-0](#).
   Eur.Phys.J. C79 (2019) no.2, 91.

4) [Study of the lowest tensor and scalar resonances in the tau to pi pi pi nu_tau decay.](#)
   By Olga Shekhovtsova, J.J. Sanz-Cillero, T. Przedzinski.
   [10.1051/epjconf/201613007022](#).
   EPJ Web Conf. 130 (2016) 07022.

5) [Tauola of tau lepton decays—framework for hadronic currents, matrix elements and anomalous decays.](#)
   By M. Chrzaszcz, T. Przedzinski, Z. Was, J. Zaremba.
   [10.1016/j.cpc.2018.05.017](#).
   Comput.Phys.Commun. 232 (2018) 220-236.

6) [Confronting theoretical predictions with experimental data; a fitting strategy for multi-dimensional distributions](#)
   By T. Przedziński, P. Roig, O. Shekhovtsova, Z. Wąs, J. Zaremba
   [csci.2015.16.1.17](#)
   Computer Science 16 (1) 2015: 17-38

7) [TauSpinner: a tool for simulating CP effects in H to tau tau decays at LHC.](#)
   By T. Przedzinski, E. Richter-Was, Z. Was.
   [10.1140/epjc/s10052-014-3177-8](#).
   Eur.Phys.J. C74 (2014) no.11, 3177.

18) [Universal Interface of Tauola Technical and Physics Documentation.](#)
    By N. Davidson, G. Nanava, T. Przedzinski, E. Richter-Was, Z. Was.
    [10.1016/j.cpc.2011.12.009](#).
    Comput.Phys.Commun. 183 (2012) 821-843.

19) [Quest for precision in hadronic cross section s at low energy: Monte Carlo tools vs. experimental data.](#)
    By Working Group on Radiative Corrections and Monte Carlo Generators for Low Energies (S. Actis et al.).
    [10.1140/epjc/s10052-010-1251-4](#).
    Eur.Phys.J. C66 (2010) 585-686.

20) [MC-TESTER v. 1.23: A Universal tool for comparisons of Monte Carlo predictions for particle decays in high energy physics.](#)
    By N. Davidson, P. Golonka, T. Przedzinski, Z. Was.
    [10.1016/j.cpc.2010.11.023](#).
    Comput.Phys.Commun. 182 (2011) 779-789.

Table 10 contains the summary of number of citations to the above publications provided by INSPIRE community; a database that focuses on publications related to High Energy Physics. Due to the focus of this database, publication 6 is not in this database.

| **Cite summary** | Citable papers | Published only |
| --- | --- | --- |
| **Total number of papers analyzed:** | 20 | 13 |
| **Total citing count:** | 1,060 | 999 |
| **Average citing per paper:** | 53.0 | 76.8 |
| **Articles based on cite number:** | | |
| Renowned papers (500+) | 0 | 0 |
| Famous papers (250-499) | 1 | 1 |
| Very well-known papers (100-249) | 2 | 2 |
| Well-known papers (50-99) | 3 | 3 |
| Known papers (10-49) | 8 | 5 |
| Less known papers (1-9) | 4 | 1 |
| Unknown papers (0) | 2 | 1 |
| $h_{HEP}$ index [?] | 13 | 11 |

**Table 10** Cite summary of papers co-authored by me provided by INSPIRE community (https://inspirehep.net/authors/1066144). This summary includes my MSc Thesis listed as unknown, unpublished paper.

Table 11 and Figure 33 contain the summary of number of citations to above publications taken from Web of Science. Publications 1, 2 and 10 are not in this database.

| | |
|---|---|
| **Results found:** | 17 |
| **Sum of times cited** | 521 |
| **Average citations per item** | 30.65 |
| **Average citations per year** | 47.36 |
| **h-index** | 10 |

**Table 11** Cite summary of papers co-authored by me. Results generated by Web Of Knowledge (https://webofknowledge.com).

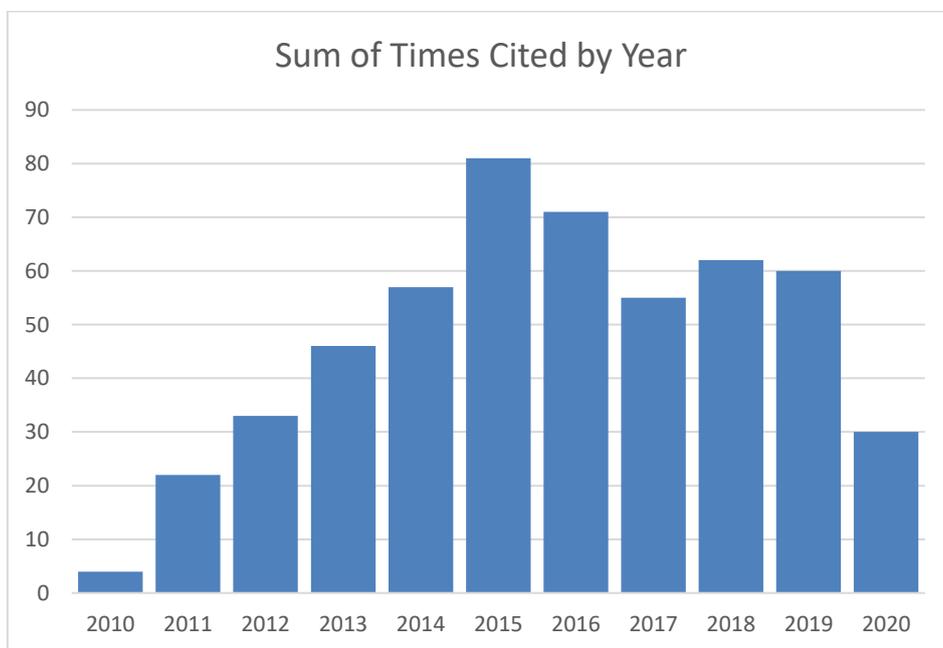

**Figure 33.** Sum of times the papers co-authored by me have been cited by year. Results generated by Web Of Knowledge (https://webofknowledge.com).

# Appendix A1 – Tools useful for scientific communities

This Appendix is dedicated to the physicists working on scientific software. Its role is to present a set of tools that can help with the software development process.

Over the years, the IT industry has worked out a set of standard development tools used within the project to ensure high software quality. A subset of these tools is used in every project involving codebase. It is worth noting, however, that the set of tools used by IT industry is much larger than this of teams working within physics environment. That is because, as I've expressed many times throughout this thesis, when working on a project in which the main content comes from the field of expertise other than computing science, tools and techniques used in computing science are only used as long as they serve to benefit the overall goal. Use of some of them may sometimes even impair productivity of the team. For that reasons, such tools are rarely used or used only in larger projects.

This Appendix focuses on the tools most commonly used in all development teams regardless of the project they are working on. But before that we would like to briefly list tools used by almost all software development teams but much less frequently used by scientific community.

## Integrated Development Environment (IDE)

This is probably the most common tool used in all IT-related projects. An IDE helps automate the most basic tasks related to code development and can include number of static analysis sub-modules that eliminate most common problems even before the project is compiled and tested. The larger the project the more necessary use of an IDE becomes. However, using an IDE requires some effort to be spent learning how it works and it benefits the team the most if all of its team members use the same IDE. For these reasons most of the specialists do not use an IDE for the code development relying only on the most basic features of an IDE that is supported by their favorite text editor.

## Issue tracker

Main software management tool used to track requirements, features, bugs, software version releases and many other aspects of software development. While use of such tool is a standard for all IT industries, in physics communities it is used only in larger projects, such as these outlined in Section 2.8.1 or 2.8.2, which make up a very small portion of all software in physics.

## Code review manager

Tool used to automate code review process. Similarly, to an issue tracker, in most applications the team does not use code review process or use it very rarely, following the manual approach of reviewing the changes by hand.

## Request tracker

Request tracker is a tool used to manage user requests, bug reports and other communication between the development team and its user base. In most physics communities there is no well-established process of handling user requests or bug reports. Most communication is performed through direct meetings, conferences, and most commonly – through e-mails.

## A1.1 Version control system

This is probably the most useful tool when dealing with code written and shared by a team working on the same project. Revision control system allows you to manage the evolution of the project providing the user with number of options for storing and viewing history of the project and modifications introduced in time. All of the changes are stored in metadata of the repository in a way that allows to retrieve any version of the code, comment the changes and tag different versions.

There are number of different approaches to how a version control system should work. Some work on a centralized repository that helps blocking concurrent changes to the same files while others base on a decentralized design that allows everyone to work simultaneously and solve any potential conflicts later, when there is a need to do so. They both have their advantages and disadvantages. In scientific community the, most popular version control system is SVN which superseded CVS many years ago. In most of the business environments SVN is no longer used, other than for storing legacy code, as it has been replaced by git. However, for the purpose of physics software, SVN is good enough and has an advantage of being simpler than its successor. Figure 34 presents kdesvn, the GUI for SVN for KDE-based Linux distributions.

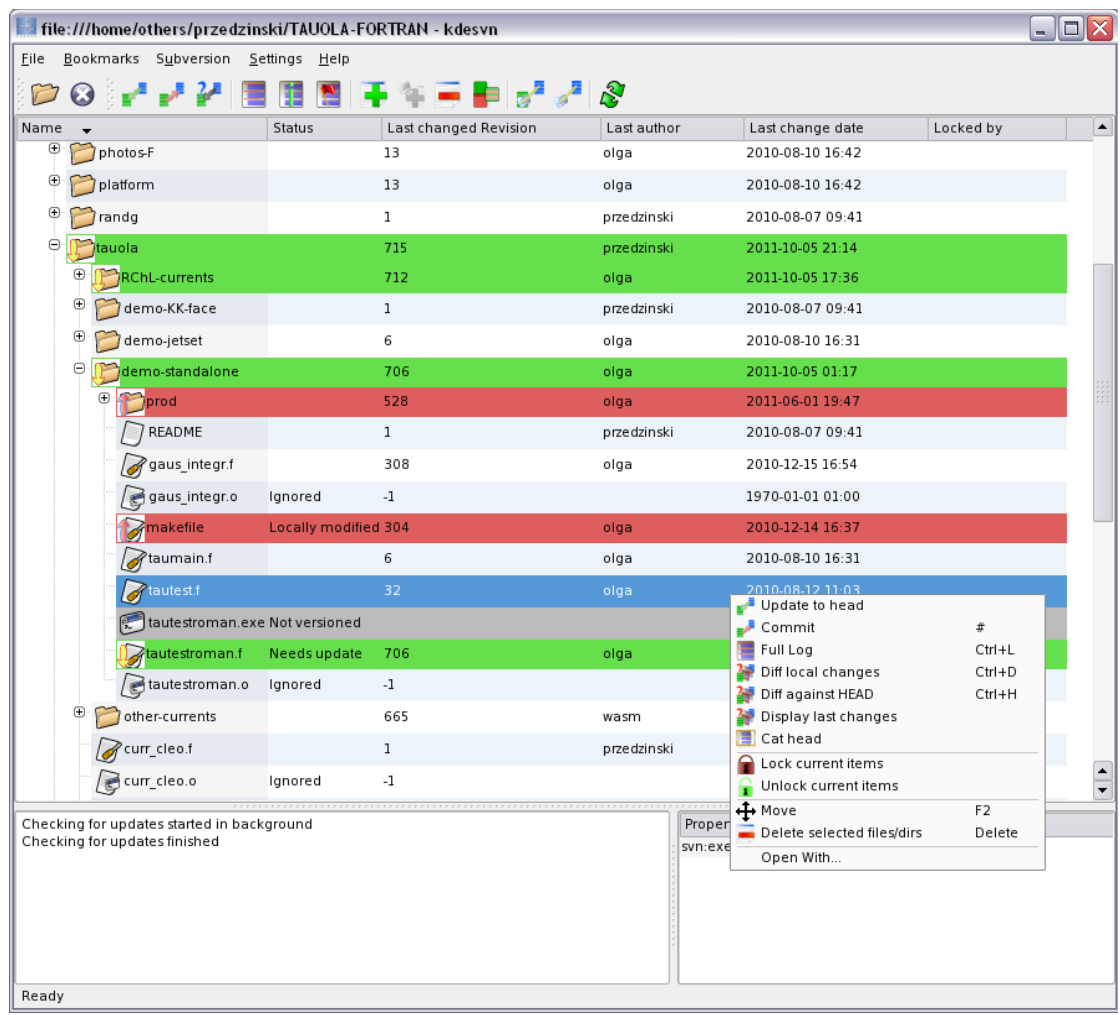

**Figure 34.** kdesvn, the GUI for SVN displaying the code for RChL currents. Red color highlights the files that are changed by the user and are not in the repository. Green color – the files that have been updated in the repository and have not yet been updated by the user. "Head" visible in the pop-up menu denotes the most up-to-date version of the code in the repository.

The most important feature of a version control system is the assertion that all team members work on the most up-to-date version of the code. New changes are immediately available to everyone with access to the repository without the need of sending them by e-mail or posting online (or other type of shared disk space).

Aside from storing the code evolution, SVN uses number of mechanisms to coordinate the work within the team. When few people work on the same part of the project simultaneously, SVN allows to merge their changes or resolve the conflict between these changes. It also allows to lock part of the code to allowing for changes to be introduced by one person at a time. When the problem arises, SVN can help to determine what change caused the problem.

This tool is so useful that most of the IT specialists use it even for the smaller projects. Even for the private ones, that they write by themselves. However, keep in mind that to fully grasp the power of the revision control system one needs to make a habit of using it all the time by submitting and commenting every step of the finished work. Nevertheless, this is something that is really worth investing your time to learn as the additional effort of using SVN quickly comes down to learning appropriate habits. Getting to know the basics takes less than an hour; even less with the graphic user interface. Then, when the need arises, the tools provided by SVN can be used to work through the number of issues related to writing the code shared by the group of people; sometimes dispersed all over the world.

Below I will briefly introduce the SVN and its main features.

**Updating, committing and reverting changes** – SVN informs the user when there are new changes available in the repository. It also traces changes made by the user that are not yet in the repository. Using this information user can view what parts of the code need updating and the details about these updates as well as keep track of his own changes. He can then commit these changes to the repository, releasing them to rest of the team, or drop them by reverting modified files to their most up-to-date state stored in the repository.

**Displaying changes** – SVN allows displaying the full log of changes for each file as well as differences between two different revisions. It also stores files and directories that have been moved, deleted or renamed allowing to trace the history of the file even when the project is undergoing directory structure changes.

**Merging changes and resolving conflicts** – when one or more people work on the same file, two things may occur – they are working on the independent parts of the file, therefore their changes can be merged without a problem, or they are working on the same code (same function, algorithm) in which case their changes are conflicting each other and cannot be merged automatically. SVN offers few methods of resolving such issues, however before any of those resolution is used it is important to view the modifications and make sure how the changes should be merged. Such problems occur rarely and should be avoided by locking the code that we plan to modify. The lock will indicate that others should wait with their changes until the person who locked the code finishes theirs.

**Tagging milestones and stable revisions** –separating key points in project development is important for future work especially when the new changes can damage part of the project. SVN can store information about key revisions and well tested, stable releases allowing to fall back to them in case of critical situation.

**Branching different approaches** – when the need arises to create several different versions of the project, SVN can create "branches" for each of those versions. Changes introduced in those branches can later be partially or fully merged into the main project if the need arises.

**Tracing the origins of the error** – SVN is invaluable in debugging, when changes introduced by one person can cause error in the code of the other. Using tool called blame one can verify who and when wrote each line of the code. This in turn can help find the origins of the problem.

**Not only for coding** – the revision control system traces changes in any type of files. This means, it can be used to store binary files or data files generated by the program. Of course, this makes little sense in terms of tracing the differences between the revisions. It may, however, be very useful allowing to fully reproduce the testing environment by storing the external libraries used at some point of the project evolution or the other. This is, however, a very limited application. More importantly – SVN can be used to store html or LaTeX files keeping track of changes in the webpages or papers documenting the project. In fact, all of the papers for the projects mentioned in this thesis were written using SVN to control their progress, including this thesis.

**Robustness and safety of use** – the algorithms behind SVN evolved over years to the point where as long as the SVN is set up correctly in the system and the privileges are correctly managed, it is very hard to damage the repository with accidental mistake. The transaction-based system rejects changes that has not been fully transferred or interrupted and disallows more than one person to commit their own changes at the same time. While both things rarely happen, they could introduce inconsistency so they are not allowed. All changes introduced to the SVN are permanently stored as a separate revision. This means the most recent changes supersede previous ones, but the information about the previous ones persists. No file can be removed from repository without a trace so even when removing files by accident, they can always be extracted from the earlier revisions stored in the repository.

The usefulness of the repository strongly relates on weather all of the team members use it or not. In cases when some of the team members are not adjusted to its use cooperation with them can become harder as rest of the team is used to different practices. For example, in one of our team single team member sent his corrections over the e-mail. These corrections were often not based on the most-recent version of the code and first few times they have been incorrectly merged until rest of the team realized that they have to merge them very carefully.

Note also that without the proper process (e.g. when all team members have the rights to modify every part of the code without review or verification) some team members may introduce unwanted changes or even break the code by introducing modifications local to its own setup. It is important to track all of the changes and review them before they can be accepted so as to pinpoint the issue sooner rather than later. Fortunately, even if such issue occurs, history of changes can be used to track and revert offending code far easier than if repository would not have been used at all.

Note that in both of the above cases the issues arise from how the repository is used and how the team cooperates when using the repository. They should not be thought of as the fault of the repository system itself. Similar issues may occur if no repository is used at all.

# A1.2 Automated code documentation system

The automated code documentation system is another useful tool worth spending few minutes to learn about. It requires very little effort to use, simply learning appropriate habits when documenting the code, while in return it greatly benefits the project's documentation.

Automated code documentation systems gather information about the classes and functions within the project and stores them in convenient to read format. In case of object-oriented language, it uses information stored in class declaration to derive relations between the different classes, automatically recognizing interfaces, virtual classes and their implementations.

It uses specifically formatted comments to gather as many information about each class and each function, as the person writing the code is willing to provide with the keywords describing the purpose, dependencies, meaning of the parameters or such basic information as the author, date of creation and last modification. Moreover, some of these systems take into account the special keywords inside the code, such as `TODO` `FIXME` or `TRICKY` to make list of possible problems and tasks that require attention.

All of this information can be used to create a detailed documentation about the project. Most of the automated code documentation systems use this information to build the website of the project with quick reference of classes and functions within the project. In our projects, we make use of `Doxygen` for this purpose. Figure 35 presents part of the main page of `Tauola++` website generated using `Doxygen`.

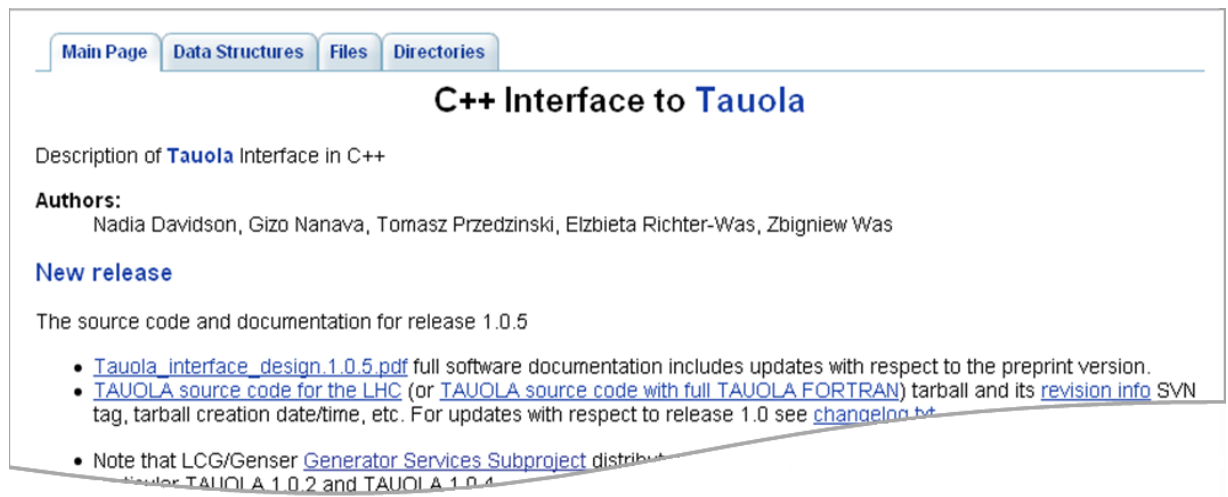

**Figure 35.** Main page of `Tauola++` website. The layout, style, as well as the rest of the content of the website available through the tabs visible at the top of the page, is automatically generated by `Doxygen`.

`Doxygen` works both for Fortran and C++ code. It provides an easy configuration file and a file to store the website's main page as well as method of adding more sub-pages. Other than that, it automatically gathers the information from the whole project creating ready-to-deploy website.

I think everyone who wrote at least two or three projects knows how important is well-documented code. This is especially true for computer scientists who know that properly defining the task solved by a class or function starts by writing this function's description. Knowing that by changing the habits of creating the comments one can additionally gain well organized, automatically generated website is both benefit and inclination to learn the habit of writing verbose descriptions.

Lastly, a good documentation of the API generated using Doxygen serves as a reference for the user who in other cases often uses header files to learn about tool's functionality.

## A1.3 Automated build system

If you're working in a larger project, it is almost sure you'll end up using an automated build system to compile and deploy your project. If you're working for ATLAS, CMS or LHCb experiment you have probably already heard of Athena control framework, `CMSSW` or `GAUDI` system for managing software builds and dependencies between project modules. These are highly advanced systems that serve more than only for building new modules. They require a lot of learning to get used to. Fortunately, most of their functionality is well documented, including tutorials for different purposes, available online to the experiment's members.

In such complex environments, equally complex management systems are necessity and there is no other option as to spend the time to learn how to use it. But let us focus on a simpler automated build system that can be used in smaller projects. One of such systems that we have been using to build distributions for `GENSER` is `Autotools` [97]. It uses a set of automatically generated `Makefiles` and `configure` scripts to control the configuration, compilation and installation procedure of the project. It helps tracking dependencies between the modules and different options inside the project as well as to configure the build system for proper architecture on which the project is being deployed.

Generally, there is no need for the physicist to learn how to manage such system. It should be set up and controlled by the person responsible for managing the code of the project. It is, however, worth spending few minutes to ask person responsible for this system how and where to add or remove new files or add dependencies to external libraries. This will speed up the process of adding new files to the project and allows physicist to easily make their own, local tests. I have learned that without knowing how the building process works, for their own tests physicists prefer to make a backup of the existing, working test (or whole directory of tests, or whole project) and modify already working program knowing that it will be compiled and linked properly.

It's a fast approach that requires no investment of time to learn anything new but leads to large amount of redundant code. While it may be hard to learn, in case of complex build systems such as the ones mentioned before, in case of `Autotools` (or simply properly written `Makefile`) it is often a matter of adding the name of the test to the list of the programs to build. It may be worthwhile to ask how the system used in the project works just to know how hard it is to learn how to modify it. On the other hand, the project manager should consider writing a `README` file or add short few lines to the internal project documentation webpage describing where these most basic changes should be located.

# Appendix A2 – Technical details of several tools

Following Appendix contains technical details of most important tools described in this thesis quoted from their original publications. These parts of the papers have been co-authored by me and they represent most of my contribution to these projects.

## A2.1 Tauola++

Following is the class diagram and excerpt from the Appendix, both taken from Reference [5] (Figure 1 from page 9 and pages 31-44), outlining the idea behind the design of the interface to `Tauola Fortran` implemented in `Tauola++` project. This part of the paper describes some of the technical aspects of its implementation. This tool has been described in Section 6.2 of this thesis.

Fig. 1 from Ref. [5] **N. Davidson, T. Przedziński, Z. Wąs.** Universal Interface of `Tauola` Technical and Physics Documentation.

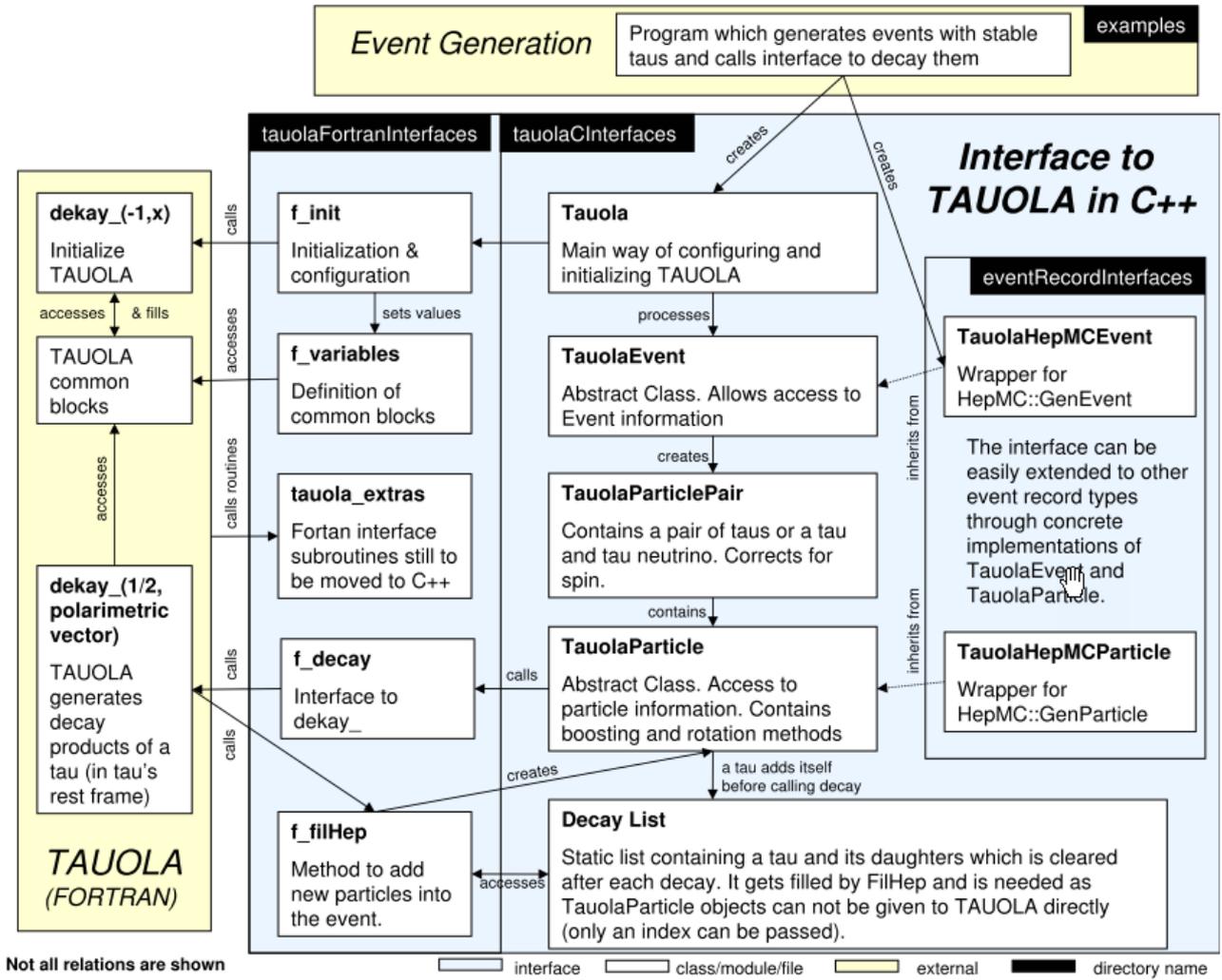

Figure 1: TAUOLA C++ Interface class relation diagram

Page 31 of Ref. [5] **N. Davidson, T. Przedziński, Z. Wąs.** Universal Interface of `Tauola` Technical and Physics Documentation.

# B  Appendix: User Guide

## B.1  Installation

The main interface library requires that `HepMC` [10] (version 2.04 or later) has been installed and its location has been provided during the configuration step. This is sufficient to compile the interface and to run the simple, standalone example.

However, in order to run more advanced examples located in the `/examples` directory, it is required to install also:

- `ROOT` [34] version 5.18 or later

- `PYTHIA 8.1` [16] or later. `PYTHIA 8.1` must be compiled with `HepMC` 2.xx so that the `PYTHIA` library `hepmcinterface` exists.

- `MC-TESTER` [14, 15] version 1.24 or later. Do not forget to type `make libHepMCEvent` after compilation of `MC-TESTER` is done.

In order to compile the `TAUOLA C++ Interface`:

- Execute `./configure` with the additional command line options:

    `--with-HepMC=<path>` provides the path to the `HepMC` installation directory. One can also set the `HEPMCLOCATION` variable instead of using this directive. This path is required for the interface to compile.

    `--prefix=<path>` provides the installation path. The `include` and `lib` directories will be copied there if `make install` is executed later. If none has been provided, the default directory for installation is `/usr/local`.

- Execute `make`

- Optionally, execute `make install` to copy files to the directory provided during configuration.

After compiling the `TAUOLA/tauola-fortran` part, the `TAUOLA C++ interface` will be compiled and the `/lib` and `/include` directories will contain the appropriate libraries and include files.

In order to compile the examples, enter the `/examples` directory and:

- Compile `TAUOLA C++ interface`

- Execute `./configure` to determine which examples can be compiled. Additional paths can be provided as command line options:

    `--with-Pythia8=<path>` provides the path to the `Pythia8` installation directory. One can set the `PYTHIALOCATION` variable instead of using this directive. This path is required for all additional examples and tests.

    `--with-MC-Tester=<path>` provides the path to the `MC-TESTER` installation directory (the `libHepMCEvent` must be compiled as well, see [15] for more details). One can set the `MCTESTERLOCATION` variable instead of using this directive. This path is required for all additional examples and tests. This option implies that `ROOT` has already been installed (since it is required by `MC-TESTER`). The `ROOT` directory `bin` should be listed in variable `PATH` and `ROOT` libraries in `LD_LIBRARY_PATH`.

- execute `make`

If neither `Pythia8` nor `MC-TESTER` are present, only the simple example will be provided. The `/examples` directory will contain the compiled example files.

## B.2 Elementary Tests

The most basic test which should be performed is verification that the interface is installed correctly, that all $\tau$ leptons are indeed decayed by the program and that energy momentum conservation is preserved. `TAUOLA` has its own database of parameters and as a consequence the $\tau$ lepton mass may differ between the program performing a $\tau$'s production and `TAUOLA` performing its decay. This leads to the sum of $\tau$ decay product momenta not exactly matching the $\tau$'s momentum. Although this effect may seem negligible, it may break numerical stability of programs like `PHOTOS` if they are applied later.

Once correct execution of the basic program steps have been confirmed, ie. $\tau$ leptons are decayed, energy momentum is conserved and there are no double decay occurrences in the event tree, step one of the program installation tests is completed[15].

In principle, these tests have to be performed for any new hard process and after any new installation. This is to ensure that information is passed from the event record to the interface correctly and that physics information is filled into `HepMC` in expected manner. Misinterpretation of the event record content may result in faulty generation by `TAUOLA`. For example spin correlations may be missing or badly represented, or some $\tau$ leptons may remain undecayed.

---

[15] We have performed such tests for all choices of the `HepMC` event record obtained from `PYTHIA 8.1` processes and listed later in the paper. Further options for initializations (parton shower hadronization or QED bremsstrahlung on/off etc.) were studied. This installation step was a necessary one of program development as well.

## B.3 Executing Examples

Once elementary tests are completed one can turn to the more advanced ones. The purpose is not only to validate the installation but to demonstrate how the interface can be used and how spin affects some distributions.

The examples can be run by executing the appropriate `.exe` file in the `/examples` directory. In order to run some more specific tests for spin effects and decays of the following intermediate states: $Z$, $W$, $H$, $H^{\pm}$, the main programs residing in subdirectories of the same name placed in the `/examples/testing` directory should be executed. For tests of all $\tau$ decay modes directory `/examples/testing/tau` is prepared. In all cases the following actions have to be performed:

- Compile `TAUOLA C++ Interface` as well as the examples.
- Check that the appropriate system variables are set: normally set by the script `/configure.paths.sh` (the configuration step mentions this script).
- Enter the `/examples/testing` directory. Modify test.inc if needed.
- enter the chosen directory and execute `make`.

The appropriate .root files as well as .pdf files generated by `MC-TESTER` will be created inside the chosen directory. One can execute 'make clobber' to clean the directory. One can also execute 'make' inside the `/examples/testing` directory to run all available tests one after another. New source code changes can easily be validated in this way. Tests are run using `examples/taumain_pythia_example.exe` and booklets will be produced with comparisons to the benchmark files.

A set of benchmark `MC-TESTER` root files are packed with the interface distribution in the subdirectories of `examples/testing/`. They can be used as examples to start new work or simply to construct comparison plots to validate new versions or new installations of `TAUOLA Interface`.

In Appendix C possible modifications to the examples settings are discussed. This may be interesting as an initial step for users physics studies. Numerical results of some of these tests are collected in Section 6 and can be thus reproduced by the user.

### B.3.1 Monitoring $\tau$ Decay Channels

It is important to check, if the $\tau$ decays themselves, are generated correctly on the user platform. For that purpose, our last demo (directory `/examples/testing/tau`) is prepared. If the test is activated, the user performs a standard `MC-TESTER` comparison of his program execution with the pre-generated one (of 10 million events). In this case all $\tau$ decay modes are activated and `MC-TESTER` is simply analyzing $\tau$ decays themselves.

## B.4 Library Linking

In order to link the libraries to the user's project, both the static libraries and shared objects are constructed. To use `TAUOLA FORTRAN` and `TAUOLA Interface` in an external project additional compilation directives are required. For the static libraries:

- add `-I<TauolaLocation>/include` at the compilation step,
- add `<TauolaLocation>/lib/libTauolaCxxInterface.a` and `<TauolaLocation>/lib/libTauolaFortran.a` at the linking step.

For the shared objects:

- add `-I<TauolaLocation>/include` at the compilation step,
- add `-L<TauolaLocation>/lib` along with `-lTauolaCxxInterface -lTauolaFortran` at the linking step.
- `TAUOLA` libraries must be provided for the executable; eg. with the help of `LD_LIBRARY_PATH` .

The `<TauolaLocation>` denotes the path to `TAUOLA` installation directory.

## B.5 Known Issues

We list here difficulties we've encountered during the testing phase and during installation for particular configurations.

The first problem occurs if a user incorrectly configures the units of `PYTHIA` to be different from the units in `HepMC`. (for example: if `PYTHIA` produces output in MeV, while `HepMC` interprets input as being in GeV). In this case, the built-in routines of `TAUOLA Interface` will treat the input as being in GeV and will adapt its output to those units as well. If this kind of situation occurs (the user will be notified by many warnings of four-vector momentum not being conserved), one can force `HepMC` to use MeV units just before filling it with the Pythia event record data. The `HepMC` event will be automatically converted to GeV when `TAUOLA Interface` is called.

Another example of a known compatibility issue arose because of a difference between the assumed default `HepMC` version 2.05 and version 2.03 (currently used, for example, in Athena[16]). In this case, the script `platform/to-HepMC-2.03.sh` will be automatically invoked during the configuration step. However, in version 2.03 methods for unit conversion are absent, therefore GeV and mm will be expected for input and used for output. The method `Tauola::setUnits(...)`, described in Appendix C.11, becomes dummy.

At present, modification to our `C++ Interface` has to be introduced eg. for use in the `Athena` system of the ATLAS collaboration software. This is to allow for backwards compatibility with older versions of `HepMC` and to prevent name clashes with the old `TAUOLA FORTRAN Interface` in environments where both interfaces are loaded concurrently. On the other hand, there is no problem with the library `/lib/libTauolaFortran.a` of the main part of the `TAUOLA FORTRAN` code itself. The version used by `Athena` can be loaded instead of ours. In `Athena`, the `binp` variant of the `cleo` initialization is used for `TAUOLA`; in this variant, for the $4\pi$ decay modes of $\tau$'s, parametrization based on Novosibirsk data is used [35].

---

[16]Software framework of the ATLAS collaboration.

All necessary changes for our `C++ TAUOLA Interface` can be introduced with use of the script `platform/to-Athena.sh`. It can be invoked by executing the `make athena` command in the main directory. Recompilation of the interface must then be performed.

# C  Appendix: User Configuration

In this section we give a description of how the user can configure `TAUOLA FORTRAN` and the `TAUOLA Interface`. All configuration is done via the static class `Tauola`. Below is the complete list of user configurable parameters and basic information on their meaning.

## C.1  Spin Correlation

By default, all spin correlations are turned on. However one may be interested to partially or completely switch off their effects for the sake of numerical experiments which validate whether a measurement will be sensitive to certain spin correlation components. This technique may be useful to evaluate the significance of spin correlations for signal/background separation as well.

Several partial treatments of spin correlations are possible. In general, the most complete intervention is to simply rewrite the matrix $R_{ij}$ for the particular channel. The following methods are nonetheless provided:

- `Tauola::spin_correlation.setAll(bool flag)`
  Turns all spin correlation computations on or off depending on the flag, which can be either **true** or **false**. Note: this should be called after Tauola::initialise().

- `Tauola::spin_correlation.HIGGS=flag`
  Turns particular spin correlation computation on or off for a given $\tau$ parent depending on the flag which can be either **true** or **false**. Implementation of this switch is provided for: **GAMMA, Z0, HIGGS, HIGGS_H, HIGGS_A, HIGGS_PLUS, HIGGS_MINUS, W_PLUS, W_MINUS**. The keywords denotes the $\tau$ parent.

**Example:**

```
Tauola::spin_correlation.setAll(false);
Tauola::spin_correlation.HIGGS=true;
```
Turns all spin correlations off, except HIGGS.

Finally one can replace density matrix following description given in Appendix D.3 also in this case one does not need to recompile of the code.

Page 36 of Ref. [5] **N. Davidson, T. Przedziński, Z. Wąs.** Universal Interface of `Tauola` Technical and Physics Documentation.

## C.2  Decay Mode Selection

By default, all $\tau$ decay modes will be generated according to predefined branching fractions. Methods to modify the default values are provided:

- `Tauola::setSameParticleDecayMode(int mode)`
  Set the decay mode of the $\tau$ with the same PGD code as set in Tauola::setDecayingParticle() (by default this sets the decay mode of $\tau^-$).

- `Tauola::setOppositeParticleDecayMode(int mode)`
  Set decay mode of the $\tau$ with the opposite PGD code as set in Tauola::setDecayingParticle() (by default this sets the decay mode of $\tau^+$).

  **Example:**
  `Tauola::setSameParticleDecayMode(Tauola::PionMode);`
  `Tauola::setOppositeParticleDecayMode(4);`
  *Forces only the modes $\tau^- \to \pi^- \nu_\tau$ and $\tau^+ \to \rho^+ \nu_\tau (\rho^+ \to \pi^+ \pi^0)$ to be generated*

- `Tauola::setTauBr(int mode, double br)`
  Change the $\tau$ branching ratio for channel *mode* from default to *br*. Note: this should be called after Tauola::initialise().
  **Example:**
  `Tauola::setTauBr(3, 2.5);`
  *Sets rate for channel $\tau^\pm \to \pi^\pm \nu_\tau$ to 2.5. All channel rates may not sum to unity, normalization will be perfrmed anyway.*

- The `int mode` enumerators which are arguments of `setOppositeParticleDecayMode`, `setSameParticleDecayMode`, `setTauBr` have the following meaning:

  - 0 - `Tauola::All` - All modes switched on
  - 1 - `Tauola::ElectronMode` - $\tau^\pm \to e^\pm \nu_\tau \nu_e$
  - 2 - `Tauola::MuonMode` - $\tau^\pm \to \mu^\pm \nu_\tau \nu_\mu$
  - 3 - `Tauola::PionMode` - $\tau^\pm \to \pi^\pm \nu$
  - 4 - `Tauola::RhoMode` - $\tau^\pm \to \rho^\pm \nu$
  - 5 - `Tauola::A1Mode` - $\tau^\pm \to A_1^\pm \nu$
  - 6 - `Tauola::KMode` - $\tau^\pm \to K^\pm \nu$
  - 7 - `Tauola::KStarMode` - $\tau^\pm \to K^{*\pm} \nu$
  - 8 - $\tau^\pm \to 2\pi^\pm \pi^\mp \pi^0 \nu$
  - 9 - $\tau^\pm \to 3\pi^0 \pi^\pm \nu$
  - 10 - $\tau^\pm \to 2\pi^\pm \pi^\mp 2\pi^0 \nu$
  - 11 - $\tau^\pm \to 3\pi^\pm 2\pi^\mp \nu$
  - 12 - $\tau^\pm \to 3\pi^\pm 2\pi^\mp \pi^0 \nu$
  - 13 - $\tau^\pm \to 2\pi^\pm \pi^\mp 3\pi^0 \nu$
  - 14 - $\tau^\pm \to K^\pm K^\mp \pi^\pm \nu$
  - 15 - $\tau^\pm \to K^0 \bar{K}^0 \pi^\pm \nu$
  - 16 - $\tau^\pm \to K^\pm K^0 \pi^0 \nu$

Page 37 of Ref. [5] **N. Davidson, T. Przedziński, Z. Wąs.** Universal Interface of `Tauola` Technical and Physics Documentation.

- 17 - $\tau^\pm \to 2\pi^0 K^\pm \nu$
- 18 - $\tau^\pm \to \pi^\pm \pi^\mp K^\pm \nu$
- 19 - $\tau^\pm \to \pi^\pm \pi^0 \bar{K}^0 \nu$
- 20 - $\tau^\pm \to \eta \pi^\pm \pi^0 \nu$
- 21 - $\tau^\pm \to \pi^\pm \pi^0 \gamma \nu$
- 22 - $\tau^\pm \to K^\pm K^0 \nu$

- `Tauola::setTaukle(double bra1, double brk0, double brk0b,double brks)`
  Change the $\tau$ sub channels branching ratio between (i) $a_0 \to \pi^+\pi^+\pi^-$ and $a_0 \to \pi^0\pi^0\pi^+$ (ii) subchannels of $K_0$ (iii) subchannels of $\bar{K}_0$ and (iv) subchannels of $K^*$. Note: this should be called after Tauola::initialise().

  **Example:**
  `Tauola::setTaukle(0.5, 0.5, 0.5, 0.6667);`
  *Set the parameters to their default values*

## C.3 Decaying Particle

The following method is prepared to impose the sign for the 'first $\tau$', that is to reverse signs of `SameParticle` and `OppositeParticle` $\tau$:

- `Tauola::setDecayingParticle(int pdg_id)`
  Set the PDG id of the particle which `TAUOLA` should decay as 'first $\tau$'. Both particles with pdg_id and -1*pdg_id will be decayed. Default is 15, one may want to use -15 for special applications.

**Example:**
`Tauola::setDecayingParticle(-15);`
*Set `SameParticle` $\tau$ to be $\tau^+$*

## C.4 Radiative Corrections

The user may want to configure parameters used in the generation of QED corrections in the leptonic decay channels of $\tau$s. For that purpose the following methods are provided:

- `Tauola::setRadiation(bool switch)`
  Radiative corrections for leptonic $\tau$ decays may be switched on or off by setting the switch to **true** or **false** respectively. By default this is **true**[17].

---

[17] Only in the case of leptonic $\tau$ decays can radiative corrections be generated in `TAUOLA` [7]. The algorithm relies on the first order complete matrix element and no exponentiation is available. If the multiple photon option is requested or if radiative corrections for other decay channels are needed `PHOTOS` Monte Carlo can be

- `Tauola::setRadiationCutOff(double cut_off)`
  Set the cut-off for radiative corrections of $\tau$ decays. The default of 0.01 means that only photon of energy (in its rest frame) up to 0.01 of half of the decaying particle mass will be explicitly generated.

**Example:**
`Tauola::setRadiation(false);`
*Switch radiative corrections off in $\tau$ decays*

## C.5  Decay of Final State Scalars

In some cases a user may want TAUOLA to decay short living scalar particles produced in $\tau^\pm$ decays, rather than invoking a host generator for the post processing step. For that purpose a special algorithm is prepared, even though high precision is then not assured. This might not be a problem if the algorithm is used for $\tau$ decays only where events with such decays are rather rare:

- `Tauola::setEtaK0sPi(int a, int b, int c)`
  The three parameters $a$, $b$ and $c$ switch on or off the decay of $\eta$, $K^0_s$ and $\pi^0$ respectively. A value of 1 is on and 0 is off.

**Example:**
`Tauola::setEtaK0sPi(1,0,1);`
*In event branch starting from $\tau$, $\eta$ and $\pi^0$ decay, but $K^0_s$ remains undecayed.*

## C.6  Scalar-Pseudoscalar Higgs

Users may wish to study spin correlations in processes involving scalar, pseudoscalar or mixed scalar-pseduoscalar decays into $\tau$'s. All options are supported by this interface. The spin density matrix will be calculated correctly for scalar Higgs (assumed PDG id of 25) and for pseudoscalar Higgs (assumed PDG id of 36) without any additional user configuration. For other cases, such as a mixed scalar-pseduoscalar Higgs or the decay of non-Higgs scalar particles, the following methods are provided:

- `Tauola::setHiggsScalarPseudoscalarMixingPDG(int pdg_code)`
  The PDG Monte-Carlo code of the Higgs which should be treated by the interface as a scalar-pseudosclar mix. The default value is PDG id 35. Please note that if $pdg\_code$ is set to the value of an existing spin case (eg. 25, the regular scalar Higgs) the scalar-pseudoscalar case will be assumed.

- `Tauola::setHiggsScalarPseudoscalarMixingAngle(double angle)`
  The scalar-pseudoscalar mixing angle. ie. $\phi$ in the coupling: $\bar{\tau}(cos(\phi) + isin(\phi)\gamma_5)\tau$. By default $\phi = \frac{\pi}{4}$.

---

used instead [13]. In [36] it was shown, that the numerical effects due to the parts not included in PHOTOS of the first order matrix element is numerically more significant than multiple photon effects. This conclusion is based on our standard numerical tests and will not necessarily be the case for other applications.

**Example:**
`Tauola::setHiggsScalarPseudoscalarMixingPDG(24);`
`Tauola::setHiggsScalarPseudoscalarMixingAngle(3.1415/3.0);`
Spin correlations will be calculated for the Higgs boson as though it is a scalar-pseduoscalar with mixing angle of $\frac{\pi}{3}$

## C.7 Helicity States and Electroweak Correcting Weight

Independent of the generation process, the information on helicities of $\tau^+$ and $\tau^-$ can be returned[18] with the help of accessors:

- `int Tauola::getHelPlus()`
- `int Tauola::getHelMinus()`

Note that these helicities are not used in the interface and carry approximate information only.

The electroweak weight can be returned with the help of accessors:

- `double Tauola::getEWwt()` - for cross section with electroweak corrections included
- `double Tauola::getEWwt0()` - for cross section at born level

These methods provide information once processing of a given event is completed.

## C.8 Use of `TAUOLA decayOne` method

In Section 3.3 an algorithm to decay all $\tau$ leptons present in the event record is explained. For that purpose `decayTaus()` method is provided. To decay a single $\tau$ lepton in a way independent of the event record content another, simple method is provided. Obvious examples when it can be useful, are processes where the hard matrix element originates from models of new physics, and different flavours of such models are to be tested. In such cases, universal methods of finding spin states of the $\tau$ to be decayed may not exist. Depending of the precision required one may need to: decay a $\tau$ without taking into account its spin state, impose its individual spin state as input information or provide a method which can be used for full density matrix generation. In the last case control over Lorentz transformations between the $\tau$ rest-frame and laboratory frame have to be available for the user.

Fortunately for all these applications a rather simple method is sufficient. It can be used to generate a decay of an individual $\tau$, without information on its parents.

- `Tauola::decayOne(`
  `TauolaParticle *tau, bool undecay, double polx, double poly, double polz)`

---

[18] One has to be careful because the actual sign may depend on the process and boosting routine.

Page 40 of Ref. [5] **N. Davidson, T. Przedziński, Z. Wąs.** Universal Interface of Tauola Technical and Physics Documentation.

> The main routine for decaying the tau. Only the first parameter is mandatory. The first parameter is a pointer to the $\tau$ that needs to be decayed.
>
> The undecay flag determines the reaction that should be taken if the $\tau$ already has daughters. By default the flag is set to **false**, which means that already decayed $\tau$ will be left unchanged. Setting this flag to **true** allows the interface to first undecay the $\tau$ and replace it with a new decay.
>
> The last three parameters are the components of the $\tau$ polarization 3-vector. In the case of `TAUOLA decayOne`, the decayed $\tau$ is treated as a standalone particle, without considering its mothers, daughters or siblings. In case the user wants to input the polarization vector (at default $\tau$ is treated as not polarized), the last three parameters have to be used.

- `Tauola::setBoostRoutine(`
  `void (*boost)(TauolaParticle *tau, TauolaParticle *target) )`
  Once executed, `Tauola:decayOne` will use the user function instead of the default one, to boost $\tau$ decay products from their rest frame to the lab frame. Such feature may be essential, in future, for use of `Tauola::decayOne` as part of a user algorithm for generation of exact spin correlations in multi $\tau$ final states.

The `single_tau_gun_example.c` is provided in the directory /examples. If polarization `polx=0, poly=0, polz=1` is chosen, then the helicity state is taken: left handed $\tau^-$ or right handed $\tau^+$. If, again as given in the example `Tauola::setBoostRoutine` is used with the proposed method, then `polx=0, poly=0, polz=1` will not mean helicity state, but rather the $\tau$ spin polarization vector oriented along the z axis of the lab frame (in fact along its space component in the $\tau$ rest-frame). Obviously spin effect chosen this way, will depend on the direction of the $\tau$ momentum.

## C.9 Logging and Debugging

This section describes the basic functionality of the logging and debugging tool. For details on its content we adress the reader to comments in the /src/utilities/Log.h header file.

Let us present however some general scheme of the tool functioning. TAUOLA Interface allows filtering out some amount of data displayed during the program run and provides a basic tool for memory leak tracking. The following functions can be used from within the user program after including the Log.h file:

- `Log::Summary()` - Displays a summary of all messages.

- `Log::SummaryAtExit()` - Displays the summary at the end of a program run.

- `Log::LogInfo(bool flag)`
  `Log::LogWarning(bool flag)`
  `Log::LogError(bool flag)`
  `Log::LogDebug(int s, int e)`
  `Log::LogAll(bool flag)`
  Turns logging of info, warning, error and debug messages on and off depending on the flag being true or false. In the case of debug messages - the range of codes to be displayed must be provided. By default, only debug messages (from 0 to 65535) are turned off. If the range is negative ($s > e$) the debug messages won't be displayed. The last option turns displaying all of the above messages on and off.

The memory leak tracking function allows checking of whether all memory allocated within `TAUOLA Interface` is properly released. However, using the debug option significantly increases the amount of time needed for each run. Its use is therefore recommended for debugging purposes only. In order to use this option modify `make.inc` in the main directory by adding the line:

`DEBUG = -D"_LOG_DEBUG_MODE_"`

Recompile the interface. Now, whenever the program is executed a table will be printed at the end of the run, listing all pointers that were not freed, along with the memory they consumed. If the interface works correctly without any memory leaks, one should get an empty table.

It is possible to use this tool within the user program, however there are a few limitations. The debugging macro from "Log.h" can create compilation errors if one compiles it along with software which has its own memory management system (e.g. `ROOT`). To make the macro work within a user's program, ensure that `Log.h` is the last header file included in the main program. It is enough to compile the program with the `-D"_LOG_DEBUG_MODE_"` directive added, or `#define _LOG_DEBUG_MODE_` placed within the program before include of the `Log.h` file[19].

## C.10 Plots for Debugging and Monitoring

This section describes the basic functionality of the plotting tool. Detailed explanations are given in the `/src/utilities/Plot.h` and `/src/utilities/Plot.cxx` files.

The `Plot` class allows generation of data for several plots we use to monitor the interface. At present, $\tau$ polarization, as taken from the `SANC` library and used by the interface (including its interpolation algorithm) is monitored in this way. The program generates data files during execution to be used later for graphic output. This code is not expected to be of a large interest to users. It is mainly for testing and debugging purposes, but may be of interest for installation tests as well.

In order to generate input data for plotting, a few methods have been provided which can be accessed after adding `#include "Plots.h"` in the main program file:

- `Plots::setSancVariables(int flavor,double cosTheta)` - sets the variables used by the first two tests. The first one is the flavor of incoming particle (by default it's set to 1), and the second is value of $cos(\theta)$ (by default - 0).

- `Plots::addSancPlot(int i)` - adds the $i^{th}$ test plot of the four tests prepared for `SANC` tables. All or only some of these tests can be run simultaneously. Figures 3, 4, and 5(a), 5(b) can be in particular reproduced with this method.

- `Plots::exitAfterSancPlots(bool exit);` - if set to **true**, force the program to exit after the plots have been created. Otherwise, the program will continue normally to its usual end.

The files generated with this tool can then be used to make plots with external scripts. For this purpose an `/examples/draw.C` ROOT script has been provided. If `root draw.C` is executed, it checks, by name, which input data files are present. For existing data files plots are drawn. Since generated files contain the test data only, without much of explanation of their meaning, to interpret them one need to look for a description inside the Plot class source files.

---

[19]Note that `Log.h` does not need to be included within the user program for the memory leak tracking tool to be used only for `TAUOLA Interface`.

## C.11 Other User Configuration Methods

The following auxiliary methods are prepared. They are useful for initialization or are introduced for backward compatibility.

- `Tauola::setUnits(Tauola::MomentumUnits,Tauola::LengthUnits)`
  Set output units for momentum (Tauola::GEV / Tauola::MEV) and decay vertex position (Tauola::MM / Tauola::CM). Methods are only available for `HepMC 2.04` or higher.

- `Tauola::setTauLifetime(0.08711)`
  Set the mean $\tau$ lepton lifetime in mm. If the user wants a vertex position to be generated by his own method, then the numerical value of the $\tau$ lifetime should be set to 0.0 here.

- `Tauola::setInitialisePhy(double iniphy_param)`
  Initializes some constants related to QED corrections. The variable `iniphy_param` is at present a dummy. It is prepared to be transmitted to some old style production code and is kept for backward compatibility.

- `Tauola::setRandomGenerator(double (*gen)())`
  In `tauola-fortran` the random number generator `RANMAR` is used. It is also used in our auxiliary methods which temporarily remain in `FORTRAN`. `RANMAR` may need to be replaced or a particular seed has to be chosen. It can be easily done and is explained in [9]. In the C++ part of the code a user can simply set the pointer to the replacement for an internal random number generator `Tauola::RandomDouble`. The generator must return a `double` from 0 to 1. `Tauola::setRandomGenerator(NULL)` will reset the program back to the default generator.

# D Appendix: Modifying Electroweak Corrections Module

## D.1 SANC Unit Initialization and Input Parameters

In this section we describe details of the `SANC` library initialization. Input parameters and constants are collected in several files located in the directory /SANC. The file `s2n_declare.h` contains a declaration of all `FORTRAN` variables used by the SANC NLO Library. The particle masses and coupling constants are initialized by the `sanc_input.h` header file. It is called by the `s2n_init` subroutine (see file `s2n_init.f`), which initializes other parameters like particle mass ratios, particle charges and weak isospin projection, as well as the value of fictitious photon mass (`thmu2`) used by IR singularity regularization and soft/hard radiation separator (`omega`). Several user controlled flags are defined:

- **iqed** = 1/0 – NLO QED correction is switched on/off. Default **iqed** = 0

- **iew** = 1/0 – NLO EW correction is switched on/off. Default **iew** = 1

- **iborn** = 0/1 – NLO correction are switched on/off; Default **iborn** = 0

- **gfscheme** = 1/0 – calculation schemes: 1 - Fermi Scheme, 0 - Alpha Scheme (default).

These flags are used to configure table preparation by the program `SANC/SANCinterface.cxx`. For the `SANC` module structure and its project details, see refs. [27, 28].

## D.2  Structure of Files with Pretabulated $R_{ij}$

In order to generate all pretabulated files the `SANC/SANCinterface.cxx` program is used. When compiled along with the `SANC` library, `SANC` FORTRAN interface and modules located inside the `SANC/modules` directory, the program generates two files - `table1-1.txt` and `table2-2.txt` for the quarks down and up respectively. The program is invoked with the command `make tables` from the directory `SANC`. The third file `table11-11.txt` representing tabulated results for electron beams will not be generated automatically.

The structure of each generated file consists of several blocks:

- Initialization block

  **Dimensions** - NS1, NS2, NS3 and NCOS values used as dimensions of the generated tables

  **Ranges** - minimum and maximum values for all three pretabulation ranges of s

- Information block:

  Date and time of generation

  Full path of the generating program

  `SANC` library information block

  `SANC` initialization parameters list

- Data block:

  **BeginRange1** statement

  tables of NS1 * NCOS lines for first range

  **BeginRange2** statement

  tables of NS2 * NCOS lines for second range

  **BeginRange3** statement

  tables of NS3 * NCOS lines for third range

  **End** statement

Lines within **Data block** consist of 4*4 numbers for $R_{ij}$, two extra numbers for the electroweak weight and an endline character. Statements used within **Data block** (**BeginRange1, BeginRange2, BeginRange3, End**) are mandatory and must be present in exact form. They mark the beginning and end of the appropriate data set. The program also checks whether the data block has been read completely and verifies if variables read at initialization were consistent.

## D.3  Importing SANC Tables into TAUOLA Interface

The three files generated by the SANC module are loaded into TAUOLA Interface during the initialization step, if they are located in the directory of the main program. When `Tauola::initialize()` is called, the interface searches for the appropriate files and if they are found - attempts to import the data.

For a file to be loaded correctly, the dimensions of the input file must match the interface settings. The content and size of the information block is arbitrary. The information of this block will be printed only, but not used otherwise. After reading all the tables from one file, TAUOLA Interface checks if the end of the data block has been reached and eventually proceeds to the next file.

If the dimensions do not match, the file is inconsistent with the structure (the end of the data block has not been reached or the file has insufficient data) - it will not be used by TAUOLA Interface even if the file was found. In that case the default density matrix will be used. TAUOLA Interface will attempt to read all of the three files, but if either one is incorrect or missing, only the data from those files that have been loaded correctly will be used.

If the need arises to modify the default tables distributed with TAUOLA, the SANC folder includes all routines needed to generate new tables along with a `Makefile` with a few options, including:

- `make` - used to recompile the library, modules and LoopTools needed for generation.
- `make clobber` - might be needed to remove previous compilation.
- `make tables` - used to compile the interface code and generate the tables.

The C++ interface to the SANC library is located in the SANC/ folder. `SANCinterface.cxx` represents the main program with setting of options for table making. Options for table layout are explained in comments within the file. The interface will be recompiled every time `make tables` is used. If needed, further program options, such as initialization of electroweak coupling constants or particle masses for calculation of electroweak corrections can be modified. The `SANC/SANCtable.cxx` file consists of routines for actual calculation of table entries from spin amplitudes calculated from SANC. Changes which may be of interest to advanced users should be done in this file.

## A2.2 Photos++

Similarly to the description of `Tauola++` project, this section contains the class diagram and excerpt from the Appendix, both taken from Reference [4] (Figure 1 from page 8 and pages 23-35) to highlight how this project benefits from the experience of the previous one and to show how the interface to Fortran `Photos` has been implemented. This tool has been described in Section 6.3 of this thesis.

Fig. 1 of Ref. [4] **N. Davidson, T. Przedziński, Z. Wąs.** Photos Interface in C++; Technical and Physics Documentation.

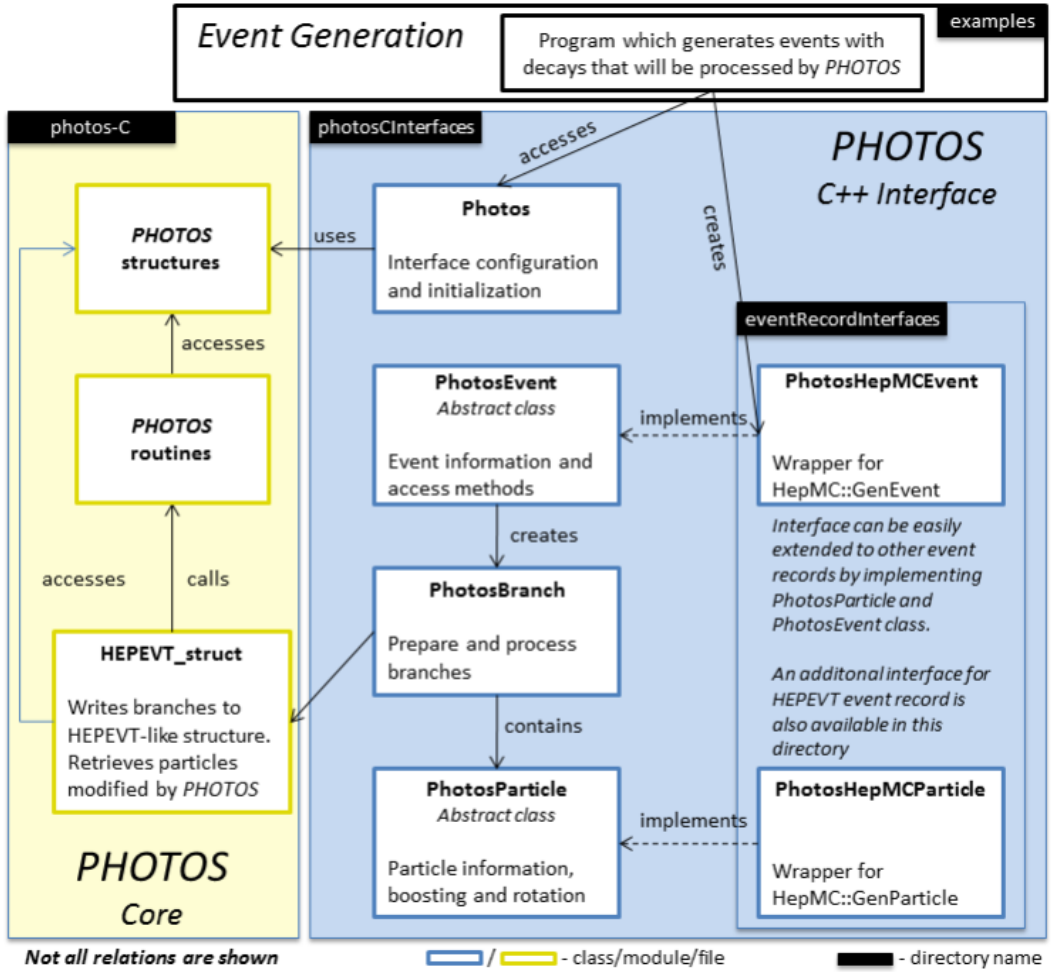

Figure 1: PHOTOS C++ interface class relation diagram

Page 23 of Ref. [4] **N. Davidson, T. Przedziński, Z. Wąs.** Photos Interface in C++; Technical and Physics Documentation.

**PHPICO** $\pi$ value definition.

    **PI** *double $\pi$.*
    **TWOPI** *double $2*\pi$.*

## A.2 Routines

In the following let us list routines which are called from the interface.

**PHODMP** prints out the content of struct `hep`.
    Return type: *void*
    Parameters: none

**PHOTOS_MAKE_C** like `PHOTOS_MAKE` from the `FORTRAN` part of the interface, but now in `C`.
    Return type: *void*
    Parameters:

    1. *int id* ID of the particle from which `PHOTOS` starts processing. In the `C++` case the importance of this parameter is limited as only one branch, reduced to the decay (process) under consideration, is in the `hep` struct at a time.

**PHCORK** initializes kinematic corrections.
    Return type: *void*
    Parameters:

    1. *int modcor* type of correction. See Ref. [4] for details.

**IPHQRK** enables/blocks (2/1) emission from quarks.
    Return type: *int*
    Parameters: *int*

**IPHEKL** enables/blocks (2/1) emission in: $\pi^0 \to \gamma e^+ e^-$.
    Return type: *int*
    Parameters: *int*

# B Appendix: User Guide

## B.1 Installation

`Photos C++ Interface` is distributed in the form of an archive containing source files and examples. Currently only the Linux and Mac OS[19] operating systems are supported: other systems may be supported in the future if sufficient interest is found.

---

[19]For this case LCG configuration scripts explained in Appendix B.2 have to be used.

**Page 24 of Ref. [4] N. Davidson, T. Przedziński, Z. Wąs.** Photos Interface in C++; Technical and Physics Documentation.

The main interface library uses HepMC [12] (version 2.03 or later) and requires that either its location has been provided or compilation without HepMC has been chosen as an option during the configuration step. The later is only sufficient to compile the interface and to run the HEPEVT example.

In order to run further examples located in the /examples directory, HepMC is required. To run all of the available examples, it is required to install:

- ROOT [40] version 5.18 or later

- PYTHIA 8.2 [16] or later[20].

- MC-TESTER [15, 7] version 1.24 or later. Do not forget to type make libHepMCEvent after compilation of MC-TESTER is done.

- TAUOLA [13] version 1.0.5 or later. TAUOLA must be compiled with HepMC.

In order to compile the PHOTOS C++ Interface:

- Execute ./configure with the additional command line options:

    --with-hepmc=<path> provides the path to the HepMC installation directory. One can also set the HEPMCLOCATION variable instead of using this directive. To compile the interface without HepMC use --without-hepmc

    --prefix=<path> specifies the installation path. The include and lib directories will be copied there if make install is executed later. If none has been provided, the default directory for installation is /usr/local.

- Execute make

- Optionally, execute make install to copy files to the directory provided during configuration.

The PHOTOS C++ interface will be compiled and the /lib and /include directories will contain the appropriate libraries and include files.

In order to compile the examples, compile the PHOTOS C++ interface, enter the /examples directory and:

- Execute ./configure to determine which examples can be compiled. Optional paths, required to compile additional examples and tests, can be provided as command line options (note that all of them are required for tests located in examples/testing directory):

    --with-pythia8=<path> provides the path to the Pythia8 installation directory. One can set the PYTHIALOCATION variable instead of using this directive.

    --with-mc-tester=<path> provides the path to the MC-TESTER installation directory (the libHepMCEvent must be compiled as well, see Ref. [7] for more details). One can set the MCTESTERLOCATION variable instead of using this directive. This option implies that ROOT has already been installed (since it is required by MC-TESTER).

---

[20] Examples can be adapted to use pythia8.1. The necessary changes are explained in examples/README-PYTHIA-VERSIONS. In fact, many of our numerical results stored in a code tar ball were obtained with older versions of PYTHIA.

Page 25 of Ref. [4] **N. Davidson, T. Przedziński, Z. Wąs.** Photos Interface in C++; Technical and Physics Documentation.

> The `ROOT` directory `bin` should be listed in the variable `PATH` and the `ROOT` libraries in `LD_LIBRARY_PATH`.
>
> `--with-tauola=<path>` provides the path to the `TAUOLA` installation directory. One can set the `TAUOLALOCATION` variable instead of using this directive.

- Execute `make`

The `/examples` directory will contain the executable files for all examples that can be compiled and linked. Their availability depends on the optional paths listed above. If neither `HepMC`, `Pythia8`, `Tauola` nor `MC-TESTER` are accessible, only the `HEPEVT` example will be provided.

## B.2 LCG configuration scripts; available from version 3.1

For our project still another configuration/automake system was prepared for use in LCG/Genser projects[21] [41, 42].

To activate this set of autotool based [43] installation scripts, enter the `platform` directory and execute the `use-LCG-config.sh` script. Then, the installation procedure and the names of the configuration script parameters will differ from the ones described in our paper. Instruction given in the './INSTALL' readme file, created by the `use-LCG-config.sh` script, should be followed. One can also execute `./configure --help`, which will list all options available for the configuration script.

Breif information on these scripts can be found in `README` in the main directory as well.

## B.3 Elementary Tests

The most basic test which should be performed, for our custom examples but also for a user's own generation chain, is verification that the interface is installed correctly, photons are indeed added by the program and that energy momentum conservation is preserved[22].

In principle, these tests have to be performed for any new hard process and after any new installation. This is to ensure that information is passed from the event record to the interface correctly and that physics information is filled into the `HepMC` event in the expected manner. Misinterpretation of the event record content may result in faulty `PHOTOS` operation.

---

[21] We have used the expertise and advice of Dmitri Konstantinov and Oleg Zenin in organization of configuration scripts for our whole distribution tar-ball as well. Thanks to this choice, we hope, our solution will be compatible with ones in general use.

[22] We have performed such tests for `HepMC` events obtained from `PYTHIA 8.1`, `PYTHIA 8.135`, `PYTHIA 8.165`, `PYTHIA 8.185` and `PYTHIA 8.201` using all configurations mentioned in this paper, all config files in `examples` directory and subdirectories of `examples/testing`. Further options for initializations (parton shower hadronization or QED bremsstrahlung on/off etc.) were also studied for different `PYTHIA 8.1` versions. This was a necessary step in our program development.

However, we do not document studies of `Pythia` physics initialization for all of its versions. That is why, distributions monitoring production processes obtained from distributed initialization for `PYTHIA 8.201`, may differ from the reference ones. See e.g. `User Histograms`, plots `mother-PT mother-eta`, in `examples/testing/ScalNLO` or `examples/testing/WmunuNLO`.

## B.4 Executing Examples

Once elementary tests are completed one can turn to the more advanced ones. The purpose is not only to validate the installation but to demonstrate the interface use.

The examples can be run by executing the appropriate .exe file in the /examples directory. In order to run some more specific tests for the following processes: $H \to \tau^+\tau^-$, $e^+e^- \to t\bar{t}$, $W \to e\nu_e$, $W \to \mu\nu_\mu$, $Z \to e^+e^-$, $Z \to \mu\mu$ or $Z \to \tau^+\tau^-$, $K_0^S \to \pi\pi$, the main programs residing in the subdirectories of /examples/testing should be executed. Note that all paths listed as optional in Appendix B.1 are required for these tests to work. In all cases the following actions have to be performed:

- Compile the PHOTOS C++ Interface.

- Check that the appropriate system variables are set. Execution of the script /configure.paths.sh can usually perform this task; the configuration step announces this script.

- Enter the /examples/testing directory. Execute make. Modify test.inc if needed.

- Enter the sub-directory for the particular process of interest and execute make.

The appropriate .root files as well as .pdf files generated by MC-TESTER will be created inside the chosen directory. One can execute 'make clobber' to clean the directory. One can also execute 'make run' inside the /examples/testing directory to run all available tests one after another. Changes in source code can be partly validated in this way. Most of the tests are run using the executable examples/testing/photos_test.exe. The $K_0^S \to \pi\pi$, $H \to \tau^+\tau^-$ and $Z \to \tau^+\tau^-$ examples require examples/testing/photos_tauola_test.exe to be run. After generation, MC-TESTER booklets will be produced, comparisons to the benchmark files will be shown. A set of benchmark MC-TESTER root files have been included with the interface distribution. They are located in the subdirectories of examples/testing/. Note that for the $W \to e\nu_e$, $W \to \mu\nu_\mu$ and $Z \to \mu\mu$ examples, differences higher than statistical error will show. This is because photon symmetrization was used in the benchmark files generated with KKMC, and not in the ones generated with PHOTOS. In the case of KKMC the generated photons are strictly ordered in energy. In the case of PHOTOS they are not. Nonetheless, on average, the second photon has a smaller energy than the one written as the first in the event record.

The comparison booklets can be useful to start new work or simply to validate new versions or new installations of the PHOTOS interface.

In Appendix C, possible modifications to the example's settings are discussed. This may be interesting as an initial step for user's physics studies. The numerical results of some of these tests are collected in Section 5.2 and can be thus reproduced by the user.

## B.5 How to Run PHOTOS with Other Generators

If a user is building a large simulation system she or he may want to avoid integration with our full configuration infrastructure and only load the libraries. For that purpose our stand-alone example **examples/photos_standalone_example.exe** is a good starting point.

In order to link the libraries to the user's project, both the static libraries and shared objects are constructed. To use the PHOTOS interface in an external project, additional compilation directives are required. For the static libraries:

- add `-I<PhotosLocation>/include` at the compilation step,

- add `<PhotosLocation>/lib/libPhotospp.a` as well as one or both of the libraries: `<PhotosLocation>/lib/libPhotosppHepMC.a` and `<PhotosLocation>/lib/libPhotosppHEPEVT.a` to the linking step of your project.

For the shared objects:

- add `-I<PhotosLocation>/include` at the compilation step,

- add `-L<PhotosLocation>/lib` along with `-lPhotospp` as well as one or both of the libraries: `-lPhotosppHepMC` and `-lPhotosppHEPEVT` to the linking step.

- `PHOTOS` libraries must be provided for the executable; e.g. with the help of `LD_LIBRARY_PATH`.

`<PhotosLocation>` denotes the path to the `PHOTOS` installation directory. In most cases it should be enough to include within a users's program `Photos.h` and `PhotosHepMCEvent.h` (or any other header file for the class implementing abstract class `PhotosEvent`) With that, the `Photos` class can be used for configuration and `PhotosHepMCEvent` for event processing.

### B.5.1 Running `PHOTOS C++ Interface` in a `FORTRAN` environment

For backward-compatibility with `HEPEVT` event records, an interface has been prepared allowing the `PHOTOS C++ Interface` to be invoked from the `FORTRAN` project. An example, `photos_hepevt_example.f`, has been prepared to demonstrate how `PHOTOS` can be initialized and executed from `FORTRAN` code. Since `PHOTOS` works in a `C++` environment, `photos_hepevt_example_interface.cxx` must be introduced to invoke `PHOTOS`.

Since version 3.54, `PHOTOS` is fully in `C++` and initialization can no longer be performed from `FORTRAN` code through the use of common blocks. In particular, information from the field `QEDRAD` localized in `FORTRAN` times common block `PHOQED` – the extension of `HEPEVT` is ignored. The `Photospp` initialization methods can be used easily instead.

Note that in the case of `HEPEVT`, the `PHOTOS` algorithm has to modify the pointers (stored as integer variables) between mothers and daughters for all particles stored downstream of the added photons or lepton pairs. This part of the code was not replicated in full detail. Also, most of our tests were performed only on cases where the modified decay was the last one in the event record, thus shifting consecutive entries was not necessary. We do not foresee the use of the program and its development for circumstances distinct from these.

# C Appendix: User Configuration

## C.1 Suppress Bremsstrahlung

In general, `PHOTOS` will attempt to generate bremsstrahlung for every branching point in the event record. This is of course not always appropriate. Already inside the internal C part of `PHOTOS`, bremsstrahlung is normally prevented for vertices involving gluons or quarks (with the exception of top quarks).

**Page 28 of Ref. [4] N. Davidson, T. Przedziński, Z. Wąs.** Photos Interface in C++; Technical and Physics Documentation.

This alone is insufficient. By default we suppress bremsstrahlung generation for vertices like $l^\pm \to l^\pm \gamma$ because a "self-decay" is unphysical. We cannot request that all incoming and/or outgoing lines are on mass shell, because it is not the case in cascade decays featuring intermediate states of sizeable width. If a parton shower features a vertex with $l^\pm \to l^\pm \gamma$ with the virtuality of the incoming $l^\pm$ matching the invariant mass of the outgoing pair then the action of PHOTOS at this vertex will introduce an error. This is prevented by forbidding bremsstrahlung generation at vertices where one of the decay products has a flavor which matches the flavor of an incoming particle.

Some exceptions to the default behavior may be necessary. For example in cascade decays, the vertex $\rho \to \rho \pi$ may require the PHOTOS algorithm to be activated.

Methods to re-enable these previously prevented cases or to prevent generation in special branches have been introduced and are presented below.

- `Photos::suppressBremForDecay(daughterCount, motherID, d1ID, d2ID, ...)`
  The basic method of channel suppression. The number of daughters, PDGID of the mother and the list of PDGIDs of daughters must be provided. There is no upper limit to the number of daughters. If a decay with the matching pattern is found, PHOTOS will skip the decay. The decay will be skipped if it contains additional photons or other particles, as long as all of the particles from the pattern are present in the list of daughters.

- `Photos::suppressBremForDecay(0, motherID)`
  When only the PDGID of the mother is provided, (the *daughterCount* is 0) PHOTOS will skip all decay channels of this particle.

- `Photos::suppressBremForBranch(daughterCount, motherID, d1ID, d2ID, ...)`
  `Photos::suppressBremForBranch(0, motherID)`
  The usage of this function is similar to the two cases of the previous function. The difference is that PHOTOS will skip not only the corresponding channel, but also all consecutive decays of its daughters, making PHOTOS skip the entire branch of decays instead of just one.

- `Photos::suppressAll()` All branchings will be suppressed except those that are forced using the methods described in the next section.

- **Example:**
  `Photos::suppressBremForDecay(3, 15, 16, 11, -12);`
  `Photos::suppressBremForDecay(2, -15, -16, 211);`
  `Photos::suppressBremForDecay(0, 111);`
  *If the decays $\tau^- \to \nu_\tau e^- \bar{\nu}_e$ or $\tau^+ \to \bar{\nu}_\tau \pi^+$ are found, they will be skipped by PHOTOS*[23]. *In addition, all decays of $\pi^0$ will also be skipped. Note, that the minimum number of parameters that must be provided is two - the number of daughters (which should be zero if suppression for all decay channels of the particle is chosen) and the mother PDGID.*

  `Photos::suppressBremForBranch(2, 15, 16, -213);`
  *When the decay $\tau^- \to \nu_\tau \rho^-$ is found, it will be skipped by PHOTOS along with the decays*

---
[23]Note that the first line of this example states that any decays of $\tau^-$ that contain $\nu_\tau$, $e^-$ and $\bar{\nu}_e$ will be skipped regardless of how many other particles are in the decay. So, for example, $\tau^- \to \nu_\tau e^- \bar{\nu}_e \gamma$ will be skipped as well. It is important to realize that excluding $\tau^- \to \nu_\tau \pi^-$ leads to exclusion of $\tau^- \to \nu_\tau \pi^- \pi^0$ as well. If it is not required it must be allowed separately.

of $\rho^-$ (in principle also $\nu_\tau$) and all their daughters. In the end, the whole decay tree starting with $\tau^- \to \nu_\tau \rho^-$ will be skipped.

In future, an option to suppress a combination of consecutive branches may be introduced. For example if bremsstrahlung in leptonic $\tau$ decays is generated by programs prior to PHOTOS, and the decay is stored in HepMC as the cascade $\tau^\pm \to W^\pm \nu$, $W^\pm \to l^\pm \nu$, PHOTOS must be prevented from acting on both vertices, but only in cases when they are present one after another. One can also think of another PHOTOS extension. If a vertex $q\bar{q} \to l^\pm l^\mp$ is found, then it should not be ignored that intermediate state can be then attributed, $q\bar{q} \to Z/\gamma^* \to l^\pm l^\mp$, and used for matrix element calculation.

## C.2  Force `PHOTOS` Processing

Forcing PHOTOS to process a branch can be used in combination with the suppression of all branches i.e. to allow selection of only a particular processes for bremsstrahlung generation.

Forced processing using the methods below has higher priority than the suppression described in the previous section, therefore even if both forcing and suppressing of the same branch or decay is done (regardless of order), the processing will not be suppressed.

- `Photos::forceBremForDecay(daughterCount, motherID, d1ID, d2ID, ...)`
  `Photos::forceBremForDecay(0, motherID)`
  The usage of this function is similar to `Photos::suppressBremForDecay(...)` described in the previous section. If a decay with the matching pattern is found, PHOTOS will be forced to process the corresponding decay, even if it was suppressed by any of the methods mentioned in the previous section.

- `Photos::forceBremForBranch(daughterCount, motherID, d1ID, d2ID, ...)`
  `Photos::forceBremForBranch(0, motherID)`
  The usage is similar to the above functions. The difference is that PHOTOS will force not only the corresponding channel, but also all consecutive decays of its daughters, making PHOTOS process the entire branch of decays instead of just one. This method can activate part of the later branch previously prevented.

- **Example:**
  `Photos::suppressAll();`
  `Photos::forceBremForDecay(4, 15, 16, -211, -211, 211);`
  `Photos::forceBremForDecay(2, -15, -16, 211);`
  `Photos::forceBremForBranch(0, 111);`
  Since suppression of all processes is used, only the listed decays will be processed, these are $\tau^- \to \nu_\tau \pi^- \pi^- \pi^+$, $\tau^+ \to \bar{\nu}_\tau \pi^+$ and all instances of the decay of $\pi^0$ and its descendants.

## C.3  Use of the `processParticle` and `processBranch` Methods

In Section 3.3 the algorithm for processing a whole event record is explained and is provided through the `process()` method. To process a single branch in the event record, in a way independent of the entire event, a separate method is provided.

Page 30 of Ref. [4] **N. Davidson, T. Przedziński, Z. Wąs.** Photos Interface in C++; Technical and Physics Documentation.

- `Photos::processParticle(PhotosParticle *p)`
  The main method for processing a single particle decay vertex. A pointer to a particle must be provided. Pointers to mothers and daughters of this particle should be accessible through this particle or its event record. From this particle a branch containing its mothers and daughters will be created and processed by PHOTOS.

- `Photos::processBranch(PhotosParticle *p)`
  Usage is similar to the above function. When a pointer to a particle is provided, PHOTOS will process the whole decay branch starting from the particle provided.

An example, `single_photos_gun_example.c`, is provided in the directory /examples showing how this functionality can be used to process the decay of selected particles. $Z^0 \to \tau^+\tau^-$ decays are generated and the event record is traversed searching for the first $\tau^-$ particle in the event record. Instead of processing the whole event, only the decay of a $\tau^-$ is processed by PHOTOS.

## C.4 Lepton pair emission and event record momentum unit

For the purpose of pair emission, introduced for the first time with Photos version 3.57, the information about the momentum unit (either GEV or MEV) used by the event has to be provided. For other Photos applications this information is not needed.

In case of the HepMC event record, this information is automatically obtained from the event. For HEPEVT events, GEV is assumed. For all other interfaces, the unit is undefined and has to be set in the event record interface by calling
`Photos::setMomentumUnit(MomentumUnits unit);`
(e.g. `Photos::setMomentumUnit(Photos::MEV);` ).
See constructors for the `PhotosHepMCEvent` or `PhotosHEPEVEvent` class for further examples.

To select available emission from PHOTOS use:

- `Photos::setPairEmission(bool flag);`
  Turn on or off emission of pairs (electrons or muons). Default is off.

- `Photos::setPhotonEmission(bool flag);`
  Turn on or off emission of photons. Default is on.

Tests and implementation of final state radiation pair emission, presently follow formulae 1 and 11 from Ref. [44]. The agreement, when pair emission phase space was restricted to the soft region, was at the 2-5 % level of the pair effect (which itself is at the 0.1 % level of the cross section). The phase space is parametrized without any mass or other approximations. This feature was checked separately with special runs (matrix element removed) of 100 Mevt samples.

The matrix element used for pair emission in decays is easy to improve. Dependence of four-momenta of final state particles is coded explicitly. Further work [45] will be devoted to this task.

Page 31 of Ref. [4] **N. Davidson, T. Przedziński, Z. Wąs.** Photos Interface in C++; Technical and Physics Documentation.

## C.5   Logging and Debugging

This section describes the basic functionality of the logging and debugging tool. For details on its content we address the reader to comments in the `/src/utilities/Log.h` header file.

Let us present however a general scheme of the tool's functionality. The `PHOTOS` interface allows control over the amount of message data displayed during program execution and provides a basic tool for memory leak tracking. The following initialization functions can be used in a user's main program. The `Log.h` header has to be then incuded.

- `Log::LogPhlupa(int from, int to)`
  Turns logging of debug messages from the `C` part of the program on and off. Parameters of this routine specify the range of debug codes for the `phlupa` routine.

- `Log::Summary()`  - Displays a summary of all messages from `C++` part of the code.

- `Log::SummaryAtExit()`  - Displays the summary at the end of a program run.

- `Log::LogInfo(bool flag)`
  `Log::LogWarning(bool flag)`
  `Log::LogError(bool flag)`
  `Log::LogDebug(int s, int e)`
  `Log::LogAll(bool flag)`
  Turns the logging of *info*, *warning*, *error* and *debug* messages on or off depending on the flag value being true or false respectively. In the case of *debug* messages - the range of message codes to be displayed must be provided. By default, only *debug* messages (from 0 to 65535) are turned off. If the range is negative ($s > e$) *debug* messages won't be displayed. The last method turns displaying all of the above messages on and off.

With `Log::LogDebug(s,e)` messages of $s$ to $e$ range, will be printed at execution time, in particular:

- Debug(0) - seed used by the random number generator

- Debug(1) - which type of branching was found in `HepMC` (regular or a case without an intermediate particle, for details see `PhotosBranch.cxx`)

- Debug(700) - execution of the branching filter has started

- Debug(701) - branching is forced

- Debug(702) - branching is suppressed

- Debug(703) - branching is processed (i.e. passed to the filter)

- Debug(900) - started check of Matrix Element (ME) calculation for the channel

- Debug(901) - ME channel value obtained

- Debug(902) - final ME channel value after checking all flags

- Debug(2) - execution of the branching filter was completed

- Debug(1000) - the number of particles sent to and retrieved from internal `PHOTOS` event record.

Page 32 of Ref. [4] **N. Davidson, T. Przedziński, Z. Wąs.** Photos Interface in C++; Technical and Physics Documentation.

The option `Log::SetWarningLimit(int limit)` results in only the first 'limit' warnings being displayed. The default for `limit` is 100. If `limit=0` is set, then there are no limits on the number of warnings to be displayed.

The memory leak tracking function allows checking of whether all memory allocated within `PHOTOS Interface` is properly released. However, using the debug option significantly increases the amount of time needed for each run. Its use is therefore recommended for debugging purposes only. In order to use this option modify `make.inc` in the main directory by adding the line:
`DEBUG = -D"_LOG_DEBUG_MODE_"`
Recompile the interface. Now, whenever the program is executed a table will be printed at the end of the run, listing all the pointers that were not freed, along with the memory they consumed. If the interface works correctly without any memory leaks, one should get an empty table.

It is possible to utilize this tool within a user's program; however there are a few limitations. The debugging macro from "Log.h" can create compilation errors if one compiles it along with software which has its own memory management system (e.g. `ROOT`). To make the macro work within a user's program, ensure that `Log.h` is the last header file included in the main program. It is enough to compile the program with the `-D"_LOG_DEBUG_MODE_"` directive added, or `#define _LOG_DEBUG_MODE_` placed within the program before inclusion of the `Log.h` file[24].

## C.6   Other User Configuration Methods

The following auxiliary methods are prepared. They are useful for initialization or are introduced for backward compatibility.

- `Photos::setRandomGenerator(double (*gen)())`  *installed in* `PHOTOS 3.52`
  Replace random number generator used by `Photos`. The user provided generator must return a `double` between 0 and 1. `Photos::setRandomGenerator(NULL)` will reset the program back to the default generator, which is a copy of `RANMAR` [19, 20].

- `Photos::setSeed(int iseed1, int iseed2)`
  Set the seed values for our copy of the random number generator `RANMAR` [19, 20].

- `Photos::maxWtInterference(double interference)`
  Set the maximum interference weight. The default, 2, is adopted to decays where at most two charged decay products are present[25] and no matrix element based kernel is used[26].

- `Photos::setInfraredCutOff(double cut_off)`
  Set the minimal energy (in units of decaying particle mass) for photons to be explicitly generated.

- `Photos::setAlphaQED(double alpha)`
  Set the coupling constant, alpha QED.

---

[24]Note that `Log.h` does not need to be included within the user's program for the memory leak tracking tool to be used only for the `PHOTOS` interface.

[25]For the decays like $J/\psi \to 5\pi^+5\pi^-$ a higher value, at least equal to the number of charged decay products, should be set. The algorithm performance will slow down linearly with the maximum interference weight but all simulation results will remain unchanged.

[26]Also in this case a higher than default 2 should be used.

Page 33 of Ref. [4] **N. Davidson, T. Przedziński, Z. Wąs.** Photos Interface in C++; Technical and Physics Documentation.

- `Photos::setInterference(bool interference)`
  A switch for interference, matrix element weight.

- `Photos::setDoubleBrem(bool doub)`
  Set double bremsstrahlung generation.

- `Photos::setQuatroBrem(bool quatroBrem)`
  Set bremsstrahlung generation up to a multiplicity of 4.

- `Photos::setExponentiation(bool expo)`
  Set the exponentiation mode.

- `Photos::setCorrectionWtForW(bool corr)`
  A switch for leading effects of the matrix element (in leptonic $W$ decays)

- `Photos::setMeCorrectionWtForScalar(bool corr)`
  A switch for complete effects of the matrix element (in scalar to two scalar decays) *installed in* `PHOTOS 3.3`.

- `Photos::setMeCorrectionWtForW(bool corr)`
  A switch for complete effects of the matrix element (in leptonic decays of $W$'s produced from anihilation of light fermions) *installed in* `PHOTOS 3.2`

- `Photos::setMeCorrectionWtForZ(bool corr)`
  A switch for complete effects of the matrix element (in leptonic $Z$ decays) *installed in* `PHOTOS 3.1`

- `Photos::setTopProcessRadiation(bool top)`
  Set photon emission in top pair production in quark (gluon) pair annihilation and in top decay.

- `Photos::initializeKinematicCorrections(int flag)`
  Initialize kinematic corrections necessary to avoid consequences of rounding errors.

- `Photos::forceMassFrom4Vector(bool flag)`
  By default, for all particles used by `PHOTOS`, mass is re-calculated and $\sqrt{E^2 - p^2}$ is used. If `flag=false`, the particle mass stored in the event record is used. The choice may be important for the control of numerical stability in the case of very light stable particles, but may be incorrect for decay products themselves of non-negligible width.

- `Photos::forceMass(int pdgid, double mass)` *installed in* `PHOTOS 3.4`
  For particles of `PDGID` (or `-PDGID`) to be processed by `PHOTOS`, the mass value attributed by the user will be used instead of the one calculated from the 4-vector. Note that if both `forceMass` and `forceMassFromEventRecord` is used for the same `PDGID`, the last executed function will take effect. Up to version 3.51, the option is active if `forceMassFrom4Vector = true` (default). From version 3.52, the option works regardless of the setting of `forceMassFrom4Vector`.

- `Photos::forceMassFromEventRecord(int pdgid)` *installed in* `PHOTOS 3.4`
  For particles of `PDGID` (or `-PDGID`) to be processed by `PHOTOS`, the mass value taken from the event record will be used instead of the one calculated from the 4-vector. Note that if both `forceMass` and `forceMassFromEventRecord` is used for the same `PDGID`, the last executed function will take effect. Up to version 3.51, the option is active if `forceMassFrom4Vector = true` (default). From version 3.52, the option works regardless of the setting of `forceMassFrom4Vector`.

Page 34 of Ref. [4] **N. Davidson, T. Przedziński, Z. Wąs.** Photos Interface in C++; Technical and Physics Documentation.

- `Photos::createHistoryEntries(bool flag, int status)` *installed in* PHOTOS 3.4
  If set to `true`, and if the event record format allows, Photos will store history entries consisting of particles before processing[27]. History entries will have status codes equal to `status`. The value of `status` will also be added to the list of status codes ignored by Photos (see `Photos::ignoreParticlesOfStatus`). An example is provided in `photos_pythia_example.cxx`.

- `Photos::ignoreParticlesOfStatus(int status)` Decay products with the status code `status` will be ignored when checking momentum conservation and will not be passed to the algorithm for generating bremsstrahlung.

- `Photos::deIgnoreParticlesOfStatus(int status)` Removes `status` from the list of status codes created with `Photos::ignoreParticlesOfStatus`.

- `bool Photos::isStatusCodeIgnored(int status)` Returns `true` if `status` is on the list of ignored status codes.

- `Photos::setMomentumConservationThreshold(double momentum_conservation_threshold)`
  Threshold relative to the difference of the sum of the 4-momenta of incoming and outgoing particles. The default value is 0.1. If larger energy-momentum non-conservation is found then photon generation is skipped in the vertex[28].

- `Photos::iniInfo()`
  The printout performed with `Photos::initialize()` will exhibit outdated information once the methods listed above are applied. The reinitialized data can be printed using the `Photos::iniInfo()` method. The same format as `Photos::initialize()` will be used.

## C.7 Creating Advanced Plots and Custom Analysis

In Section 5.2, we have presented results of a non-standard analysis performed by MC-TESTER. Figure 4 has been obtained using a custom `UserTreeAnalysis` located in the `ZeeAnalysis.C` file residing in the `examples/testing/Zee` directory. This file serves as an example of how custom analysis can be performed and how new plots can be added to the project with the help of MC-TESTER.

The basic MC-TESTER analysis contains methods used by pre-set examples in the subdirectories of `examples/testing` to focus on at most one or two sufficiently hard photons from all the photons generated by PHOTOS. Its description and usage have already been documented in [7]. The content of `ZeeAnalysis.C` is identical to the default `UserTreeAnalysis` of MC-TESTER with the only addition being a method to create the previously mentioned plot.

In order to create the $t\bar{t}$ example, an additional routine had to be added to `photos_test.c`. Since MC-TESTER is not designed to analyze processes involving multiple incoming particles, we have used a method similar to that previously used in the FORTRAN examples, `LC_Analysis`, mentioned in [15], Section 6.1. This routine, `fixForMctester`, transforms the $XY \to t\bar{t}$ process to the $100 \to t\bar{t}$ process, where the momentum of the special particle 100 is $X + Y$.

---

[27]In case of HepMC, it creates copies of all particles on the list of outgoing particles in vertices where the photon was added and will be added at the end of the list.

[28] In the past, momentum conservation was checked using the standard method of HepMC. That effectively meant that *only* momentum, but not energy was checked. This turned out to be insufficient in some rare cases.

With this modification, MC-TESTER can be set up to analyze the particle 100 in order to get a desirable result.

For more details regarding the plots created for this documentation, see README-plots located in the examples/testing/ directory.

## A2.3 Resonance Chirlal Lagrangian Currents

Following is the Appendix B from Reference [17] containing the technical details of the setup that allowed to use reweighting algorithm to fit RChL currents to data. It's important to show how the decisions made few years earlier when designing the architecture of `Tauola++` made this installation a very easy task. It's even more important that the modularity of `Tauola Fortran`, written more than 25 years ago, allowed for separation of the currents in the first place. This only serves to stress how proper architecture can mean a difference between weeks or months of adapting the code to new environment or almost no time at all.

The work described by this reference has been followed up several times during the next few years after the publication of the original work. An analysis of theoretical input and the subject of errors has been discussed [18], and results of comparison of the model to experimental data has been published [19; 98; 99]. However, while these publications discuss improvements to the model or the fitting strategy, they do not introduce a significant change to the framework itself and have been listed only for the sake of completeness.

The work that contribute to performing the comparison of RChL currents to data have been described in Section 6.4 of this thesis.

# B  Installation

Our project tar-ball, even though resulting from a rather large effort, is not designed for independent installation. This would be of course straightforward and we will return to that solution in the future, once the currents are optimized to improve agreement with the data. The parametrization will become integrated part of the `TAUOLA` distribution; for fortran or for C++ use, like in references [14, 62]. At present we concentrate on a solution which is most convenient for the experimental user e.g. from Belle or BaBar collaboration aiming at combining the code with the version of `TAUOLA` which is already being used as part of the simulation set-up. We aim at preparing an add-up[43] for already existing set-up. The tar-ball can be downloaded from the Web page [28] of our project.

Once tar-ball is unpacked inside `TAUOLA-FORTRAN/tauola` subdirectory (of user environment), the directory `tauola/new-currents` will be created, all necessary fortran files will be found there. For convenience, later on, we will use the following aliases:

- `${RCHLCURRENTS}` instead of `tauola/new-currents/RChL-currents` .

- `${OTHERCURRENTS}` instead of `tauola/new-currents/other-currents` .

- `${INSTALLATION}` instead of `tauola/new-currents/Installation` .

---

[42] Alternatively, for $\mu$ we may take $M_\rho$ for $PQ = \pi\pi$, $KK$ and $M_{K^*}$ for $PQ = K\pi$ [29].

[43] As the project is developed under svn, the tar-ball is accompanied with svn label and it should be kept for reference.

In `tauola/new-currents` further sub-directories for more advanced use or for documentation will be found:

- `${RCHLCURRENTS}/tabler/a1` - programs for pretabulations in particular of $q^2$–dependent $a_1$ width[44].

- `${RCHLCURRENTS}/cross-check` - code for technical and numerical tests.

- `new-currents/paper` - present paper.

- `${INSTALLATION}` - instructions for modifications to be introduced in `FORTRAN` files and `makefile` residing in directory `tauola` .

- `new-currents/Installation-Reweight` - instruction and example of using reweighting algorithm.

None of the directories listed above contains code which is to be loaded together with `TAUOLA` library. Code loaded with the library is located only in the main folder of `${RCHLCURRENTS}` and `tauola/new-currents/other-currents` . Programs in `${RCHLCURRENTS}/tabler/a1` can update the fortran code located in file `${RCHLCURRENTS}/initA1Tab.f` .

Once installation is completed, to invoke the calculation of our new currents the `CALL INIRChL(1)` has to be invoked[45] by user main program prior to call on `TAUOLA` initialization. If instead `CALL INIRChL(0)` is executed prior[46] initialization, old currents - as in Ref. [14] - will be used in generation.

The `CALL INIRChL(1)` may activate also new currents[47] e.g. for $\eta\pi\pi$ or $4\pi$ decay channels. At present only wrappers of currents of Ref. [14] and [33] are prepared in the directory `${OTHERCURRENTS}` for convenience of users and our future work. These currents lead to substantially different distributions, that is why, one may require adjustment of phase space presampler used to optimize speed of generation. Anyway, as a default, they are turned off. For `INIRChL(1)` the same `TAUOLA cleo` currents as for `INIRChL(1)` are used. To turn other options, `ISWITCH` located in the file `${OTHERCURRENTS}/ffourpi.f` has to be changed from its default value 0 to 1, 2, 3 or 4.

## B.1 Changes for host `TAUOLA` version

In order to use new currents, changes have to be made to the host `TAUOLA` installation. Let us document here in great detail changes to be introduced in `TAUOLA cleo` version. If some modifications were introduced and user's host `TAUOLA` installation differs from `TAUOLA cleo` of Ref. [14], then modifications prepared in `${INSTALLATION}` directory can not be used directly and some adaptation may be necessary. In either case we advice to check if at least some of the numerical results from Ref. [28] are correctly reproduced after installation.

---

[44]In the directory `${RCHLCURRENTS}/tabler` the place to calculate other pretabulated functions (as possibly the scalar form factor for $K\pi$ channel) is reserved.

[45]For a C++ user, examples of use of `inirchl_(1)` are given in `new-currents/Installation-Reweight/` directory.

[46] This can be done also after initialization as no initialization of tables is needed. Then one can revert the change again with `CALL INIRChL(0)`.

[47]Although the $\tau \to \eta^{(\prime)}\pi^-\pi^0\nu_\tau$ decays have been worked out within Resonance Chiral Theory [111], the corresponding expressions for the currents have not been incorporated yet to the program.

Let us list now changes which have to be introduced to files residing in `TAUOLA/tauola` directory of the user installation .

- `TAUOLA/tauola/makefile`
  The list of `LIB_OBJECTS` must be extended and additional objects added:
  `${RCHLCURRENTS}/f3pi_rcht.o`, `${RCHLCURRENTS}/fkkpi.o`,
  `${RCHLCURRENTS}/fkk0pi0.o`, `${RCHLCURRENTS}/wid_a1_fit.o`,
  `${RCHLCURRENTS}/frho_pi.o`, `${RCHLCURRENTS}/funct_rpt.o`,
  `${RCHLCURRENTS}/value_parameter.o`, `${RCHLCURRENTS}/initA1Tab.o`,
  `${RCHLCURRENTS}/fkpipl.o`, `${RCHLCURRENTS}/fk0k.o`,
  `${OTHERCURRENTS}/fetapipi.o`, `${OTHERCURRENTS}/ffourpi.o`,
  `${OTHERCURRENTS}/binp.o`, `${OTHERCURRENTS}/curr_karls.o`,
  `${OTHERCURRENTS}/curr_karls_extracted.o`;
  if there are no additional dependencies the `${INSTALLATION}/makefile-tauola` file can be simply copied into `tauola/makefile` .

- `TAUOLA/tauola/tauola.f`
  If the file in user's version coincides with the one of `TAUOLA cleo` distribution, the `${INSTALLATION}/tauola.f-new` file can be simply copied into `tauola/tauola.f` . To verify this, the diff file `${INSTALLATION}/tauola.f-oldDIFFupdated` may be inspected.

- `TAUOLA/tauola/formf.f`
  If the file in user's version coincides with the one of `TAUOLA cleo` distribution, the `${INSTALLATION}/formf.f-new` file can be simply copied into `tauola/formf.f` . To verify this, the diff file `${INSTALLATION}/formf.f-oldDIFFupdated` may be inspected.

Once changes are introduced the new currents will be activated and the old ones will be overruled once call to routine `INIRChL(1)` is invoked. Otherwise, or if `CALL INIRChL(0)` is invoked (at any time), old currents will be then switched back on. The routine `INIRChL(1)` has to be invoked by the user program at the initialization step. For the C++ user, a definition of `extern "C" void inirchl_(int i);` has to be included and execution of `inirchl_(&i);` performed.

An example has been provided in `${INSTALLATION}/demo-standalone` . It is based on default `TAUOLA cleo` example with the only modification being the call to `INIRChL(1)` before default `TAUOLA` initialization.

## B.2 Calculating numerical tables used by form factors

The directory ${RCHLCURRENTS}/tabler/a1 contains the program da1wid_tot_rho1_gauss.f; it creates a table of $\Gamma_{a_1}(q^2)$ according to Eq. (35). The system does not use any information from TAUOLA initialization except the pion and kaon masses.

We have also prepared a place to add tables for other functions, in the near future it will be done for the scalar form factor of the $K\pi$ mode.

### B.2.1 Executing the code

The program da1wid_tot_rho1_gauss.f produces the $q^2$ distribution of the $a_1$ off-shell width.

To compile, type make in ${RCHLCURRENTS}/tabler/a1 directory. To run, type make run. Each line of the produced output includes the value of $q^2[\text{GeV}^2]$, and the value of $d\Gamma/dq^2[\text{GeV}^{-1}]$. This table is written into the file initA1Tab.f which is the FORTRAN code ready to use. One can shift it to ${RCHLCURRENTS} directory by make move command. Text format table is written into file wida1_qq_tot_2e5.out .

### B.2.2 Setup

Input parameters and common blocks are located in ${RCHLCURRENTS}/parameter.inc . Other parameters are defined in ${RCHLCURRENTS}/value_parameter.f . These parameters may be changed by the user. If the parameters affect the $q^2$–dependent $a_1$ width (or other pretabulated functions), the tables need to be generated anew with the help of programs residing in the directory ${RCHLCURRENTS}/tabler . A list of the parameters that affect generated tables (and thus require tables to be generated again) is in ${RCHLCURRENTS}/value_parameter.f . Some of the variables used in functions from ${RCHLCURRENTS}/funct_rpt.f are declared in ${RCHLCURRENTS}/funct_declar.inc .

## B.3 Tests

Directory ${RCHLCURRENTS}/cross-check contains three subdirectories:

1. check_analyticity_and_numer_integr, it includes:

    - test of numerical stability in calculations of $\Gamma_{a_1}(q^2)$ and for the whole $\tau$ hadronic decays as described in Section 2. For that purpose it is checked if continuity of results as a function of the invariant mass holds.
    - the result for the integrated width of the $\tau \to 2\pi\nu_\tau$, $\tau \to K\pi\nu_\tau$, $\tau \to K^-K^0\nu_\tau$, $\tau \to 3\pi\nu_\tau$, $\tau \to K\pi^-K\nu_\tau$ and $\tau \to K^-\pi^0K^0\nu_\tau$. These results can be confronted with the result of Monte Carlo simulation collected in sub-directory tauola_result_modes .

2. `results_numer_integr_3pion` presents the results for the width of $\tau \to 3\pi\nu_\tau$ as a function of the 3 pion invariant mass. It is calculated by numerical integration of the analytical formula for different choices of hadronic form factors as it is described in Section 4.

3. `tauola_result_modes` contains Monte Carlo results for both differential and total width for the processes $\tau \to 2\pi\nu_\tau$, $\tau \to K\pi\nu_\tau$, $\tau \to K^-K^0\nu_\tau$, $\tau \to 3\pi\nu_\tau$, $\tau \to K\pi^-K\nu_\tau$ and $\tau \to K^-\pi^0K^0\nu_\tau$.

### B.3.1 Numerical stability tests

The directory `${RCHLCURRENTS}/cross-check/check_analyticity_and_numer_integr` contains six subdirectories with tests of numerical stability for hadronic $\tau$ decay modes and a subdirectory with the test for the $a_1$ width. Each decay channel is located in a separate directory. Details regarding each of these tests are described in README files of the directory and every subdirectory as well. That is why only the basic information is provided in our paper. We have checked using interpolation from neighbouring values that the value of $d\Gamma/dq^2$ is continuous and is not contaminated by numerical instability of multidimensional Gaussian integration. Also we present the analytical results for the partial width of every channel to be compared with the Monte Carlo ones.

Content of the directory:

- `check_analyt_3piwidth`: test of numerical stability of the distribution $d\Gamma(\tau \to \nu_\tau\pi\pi\pi)/dq^2$. Results are presented for separate modes: $d\Gamma(\tau \to \nu_\tau\pi^0\pi^0\pi^-)/dq^2$ and $d\Gamma(\tau \to \nu_\tau\pi^-\pi^-\pi^+)/dq^2$. Also the value of the partial widths for the channels is provided for the comparison with the TAUOLA results.

- `check_analyt_kkpi` - test of numerical stability for $d\Gamma(\tau \to \nu_\tau KK\pi)/dq^2$. Results are presented for separate modes: $d\Gamma(\tau \to \nu_\tau K^-\pi^-K^+)/dq^2$ and $d\Gamma(\tau \to \nu_\tau K^0\pi^-\bar{K}^0)/dq^2$. The value of the partial widths for both channels are provided.

- `check_analyt_kk0pi0` - tests of numerical stability for $\tau \to \nu_\tau K^-\pi^0 K^0$: both the spectrum $d\Gamma(\tau \to \nu_\tau K^-\pi^0 K^0)/dq^2$ and the partial width are provided.

- `check_analyt_2pi` - tests of numerical stability for $\tau \to \nu_\tau\pi^-\pi^0$: both the spectrum $d\Gamma(\tau \to \nu_\tau\pi^-\pi^0)/dq^2$ and the partial width are provided.

- `check_analyt_kpi` - tests of numerical stability for $\tau \to \nu_\tau K\pi$: both the spectrum for the total width $d\Gamma(\tau \to \nu_\tau K\pi)/dq^2$ and the partial width for channels $\pi^-\bar{K}^0$ and $\pi^0 K^-$ are provided. The partial widths for the individual decays are checked to be 2/3 and 1/3 of the total $K\pi$ width, mass effects are negligible in this case.

- `check_analyt_k0k` - tests of numerical stability for $\tau \to \nu_\tau K^-K^0$ both the differential distribution $d\Gamma(\tau \to \nu_\tau K^-K^0)/dq^2$ and the partial width are provided.

- `check_analyt_a1table` - tests of numerical stability of $\Gamma_{a_1}(q^2)$ produced by program described in Appendix B.2.

### B.3.2 Analytic integration test

The results of the analytical integration test in the three-pion case are presented in the directory `${RCHLCURRENTS}/cross-check/results_numer_integr_3pion`. They are produced by the program `totwid3pi_qq_table.f` in the directory
`${RCHLCURRENTS}/cross-check/check_analyticity_and_numer_integr/check_analyt_3pi`.
The program can be compiled by command `make` and run with `make totwid3pi_run > output.txt`.

The setup file `input_f1f2f4.dat`, in
`${RCHLCURRENTS}/cross-check/check_analyticity_and_numer_integr/check_analyt_3pi`
contains:

- `eps` - defines (relative) precision of the Gaussian integration.

- `kf1` - flag for form factor $F_1$. For `kf1=` 0, 1 or 2 $F_1$ will be set respectively to 0, 1 or to its functional form.

- `kf2` - flag for form factor $F_2$. For `kf2=` 0, 1 or 2 $F_2$ will be set respectively to 0, 1 or to its functional form.

- `kf4` - flag for form factor $F_4$. For `kf4=` 0, 1 or 2 $F_4$ will be set respectively to 0, 1 or to its functional form.

- `chan` - flag to choose the 3 pion mode. `chan=` 1 for $\pi^0\pi^0\pi^-$ and `chan=` 2 for $\pi^-\pi^-\pi^+$.

If the functional form of the form factors is used, it will be taken from the file
`${RCHLCURRENTS}/cross-check/check_analyticity/check_analyt_3piwidth/funct_3pi.f`.
If `kf1` or `kf2` is set to 2, pretabulated file `${RCHLCURRENTS}/initA1Tab.f` will be used for $\Gamma_{a_1}$ in the propagator of the $a_1$-meson[48].

Output file contains four columns:

- `qmin` (in [GeV$^2$]) - lower boundary for the integration over 3-pion invariant mass.

- `qmax` (in [GeV$^2$]) - upper boundary for the integration over 3-pion invariant mass.

- `eps` - estimate of the integration precision in the result.

- total width (in [GeV]).

Results for the different configurations of the form factor are presented in
`tauola/RChL-currents/cross-check/results_numer_integr_3pion`.

---

[48]Note that the tabulated file is generated by the program described in Appendix B.2.

## B.4 TAUOLA weight recalculation mode

Let us present now the installation necessary for the method of weighted events, which was envisaged in Section 6.1. An example of such installation code is included in our distribution tar-ball in directory new-currents/Installation-Reweight .

Before reweighting method can be used, TAUOLA needs to be adapted to new currents as explained in the Appendix B.1. Afterwards, our example program tau-reweight-test-ASCII.c, residing in the directory new-currents/Installation-Reweight, can be run with the help of the simple make command[49]. For more details regarding the reweighting examples, refer to README located in new-currents/Installation-Reweight .

The following subsection describes reweighting algorithm as well as initialization used in the example. Note, contrary to the rest of the project, reweighting algorithm, including examples of its usage, is written in C++.

### B.4.1 Weight recalculation algorithm

In order to use recalculation mode, several steps have to be performed from user program:

1. Before TAUOLA initialization, RChL currents have to be switched on. This can be done with the help of the wrapper for FORTRAN function INIRChL(IVER), by calling inirchl_(&i); with i = 1; . Two versions of currents will be used, but initialization must be done for IVER=1, for initialization of RChL-specific variables and tables.

2. Initialization of TAUOLA must be called. We are using initialization taken from the default TAUOLA example, stored in wrapper function f_interface_tauolaInitialize.

3. For each event, the information about $\tau$ and its decay products must be filled and stored in instances of SimpleParticle class[50].

4. Once the kinematical configuration for $\tau$ decay is read from the datafile (or fifo pipe), function: double calculateWeight(SimpleParticle &tau, vector<SimpleParticle> &tau_daughters) can be used to retrieve the weight.

The algorithm of the function calculateWeight is sketched in the following:

1. Particles are prepared and boosted to the appropriate frame.
2. TAUOLA decay channel is identified.

---

[49]Other example, tau-reweight-test-HepMC.c, requires installation of HepMC and optional installation of MC-TESTER, and their paths provided in the Makefile .
[50]Class SimpleParticle is used only to contain four-vector and flavour of the particle.

3. `TAUOLA cleo` currents are switched on with `inirchl_(&i); i = 0`.

4. Call to appropriate internal `TAUOLA FORTRAN` subroutine, returning weight `WT1`.

5. `RChL` currents are switched on with `inirchl_(&i); i = 1`.

6. Call to the same routine as in step 4 is performed, returning weight `WT2`.

7. Ratio of weights calculated at steps 4 and 6 gives required model replacing weight.

8. `WT = WT2/WT1` is returned to the main user program.

It is rather straightforward to extend this method to the case when more than one new version of physics initialization is to be used. Note that the examples are set up so that the weight is calculated both for $\tau^-$ and $\tau^+$ and stored in variables `WT_M` and `WT_P` respectively. In cases where only a single $\tau$ is present in the event, the weight corresponding to the second $\tau$ equals 1.0.

Alternatively, in cases when this approach cannot be used or is inconvenient, variants of the method, based on fifo pipes can be useful as well. Prototypes for such solutions can be obtained from Ref. [112].

Hadronic currents for $\tau^+$ and $\tau^-$ differ due to CP parity. The resulting effects are taken into account in the reweight algorithm.

### B.5 TAUOLA++ installation

Thanks to the modular construction of `TAUOLA C++ Interface` [62], new currents can be used in C++ projects in a straightforward way. It is enough to replace the previous `TAUOLA-FORTRAN` installation with the new one, adjusting `Makefile` with a list of the newly added object files.

For step-by-step instructions, we refer to `${INSTALLATION}/README-TAUOLA++` . Our package has already been tested to work with `TAUOLA C++ Interface v1.0.5`, but the installation procedure is similar for all previous versions and should remain unchanged for future versions as well.

## A2.4 `Tauola`-Fortran C++ framework for hadronic currents

Following are Section 3 and Appendices A and B from Reference [100]. Section 3 shows how the Fortran interface to `Tauola` Fortran has been prepared to allow switching and modifying `Tauola` Fortran currents. Appendices show how this new interface could be applied in C or C++ environment to perform the same procedure through C wrappers.

The work that contributed to development of this framework has been described in Section 6.4 of this thesis.

## 3  Data structures and wrappers

Until now, many things were hard-coded in `FORTRAN` routines. Our intention was to make explicit separation into modules: fixed multiplicity phase space generators, hadronic currents, ME calculators, initialization and interface to event records. At the same time we aimed to make changes in the code as small as possible. Our present work is a follow-up on the work started in [2] and continued in [17], which became a technical start for the hadronic current parametrizations used by BaBar collaboration. Our present arrangements profit from the opportunities available in mixed `F77/C++` environment.

Our aim was to structurize program in the object-oriented manner but at the same time retain the use of static data structures instead of switching to dynamic memory allocations to preserve performance and minimize memory usage. Such arrangement is a step towards introduction of parallelization methods while still facilitating backward-compatibility checks.

The first step of our work was to prepare `FORTRAN common blocks` designated to keep all information regarding the generation process. The changes introduced to the code are in `FORTRAN`, but as a result, each segment can be translated into `C++` independently.

We use specialized routines to generate phase-space for final states of fixed multiplicity. This is the reason why the decay channels are grouped accordingly to multiplicity. The constants `NMi`, where `i = 1..5`, denote number of memory slots reserved for channels of final states with multiplicity `i+1`. For `NM6`, this convention is overruled. This group collects decay channels with final states of any multiplicity, but for these decay channels an iterative phase space generator is used and as a consequence approximations on matrix element must be imposed. Two additional variables should be mentioned before data structures are explained. The `NLT=2` denotes number of leptonic $\tau$ decay channels. The `NMODE` denotes the total number of memory slots prepared for non-SM or non-leptonic decay channels. It's value is equal to `NM1+NM2+NM3+NM4+NM5+NM6`. In the source code provided with this paper `NMODE=196`. If needed, this value can be increased up to `500-NLT`.

## 3.1 Data structures

First, let us present these data structures that, once initialized, usually remain constant during the remainder of the program execution[3].

```
extern "C" stuct taubra_ {
    float GAMPRT[500];
    int JLIST[500];
    int NCHAN;
};
```

GAMPRT[]
: The branching ratios used to define probabilities with which choice of particular decay channel is made[4]

JLIST[]
: The channel number assigned within FORTRAN part of the code. Usually JLIST(i) == i, where i=1,...,500.

NCHAN = NMODE + NLT
: Number of all available decay channels. Parameters NLT, NMODE, NM1, NM2, NM3, NM4, NM5, NM6 are initialized in two places TAUDCDsize.inc and in TauolaStructs.h. It has to remain consistent whenever changes would be introduced.

```
extern "C" struct metyp_ {
    int KEY0[NLT];
    int KEY1[NM1];
    int KEY2[NM2];
    int KEY3[NM3];
    int KEY4[NM4];
    int KEY5[NM5];
    int KEY6[NM6];
};
```

---

[3]Note that it is possible to change channel branching probability dynamically during program execution. It is also possible, however undesireable and error-prone, to change the decay channels definition more than once during program execution.

[4] It can be set with any non-negative numbers. User should ensure correct proportions for channels. For example, if user generates two decay modes only, one of which is 100-times more likely to occur, it makes no difference for the program if respective entries in GAMPRT are set to 1. and 100. or 0.001 and 0.1.

In this structure information on the type of the matrix elements used by each decay channel is stored. As usual, decay channels are grouped by multiplicity. The following types of matrix element calculation are possible:

0 - channel is not initialized,

1 - constant matrix element (flat phase space),

2 - default TAUOLA matrix element and current,

3 - default TAUOLA matrix element, but for case when one, stable particle of spin>0 in final state[5] is present,

4 - default matrix element but with user defined hadronic current,

5 - user defined matrix element.

```
extern "C" struct taudcd_ {
    int  IDFFIN[NMODE][9];
    int  MULPIK[NMODE];
    char NAMES [NMODE][31];
};
```

NAMES[i][31]
   List of names for the decay modes to be used for the output printouts. Up to 31 letters can be used to describe each decay channel[6].

MULPIK[i]
   Multiplicity of stable decay products ($\nu_\tau$ excluded). An overall multiplicity for the channel i is thus MULPIK[i]+1.

IDFFIN[i][j]
   Identifiers of the consecutive stable decay products (except $\nu_\tau$). Entries for j ≥ MULPIK[i]+1 should be set to zero. An exception is for the anomalous decay channels where $\nu_\tau$ is absent. In such cases non-zero IDFFIN[i][j] for j = 3 or j = 4 denotes identifier for the particle to be placed into decay final state instead of $\nu_\tau$. Entry for j = 3 is to be used when identifier is to remain the same for $\tau^-$ and $\tau^+$. Entry for j = 4 should be used when the sign should differ[7].

---
[5]In practice, it is used for $\tau \to \pi^-\pi^0\gamma$ decay channel only.
[6]Note that variable NAMES is defined in FORTRAN as a variable of type CHARACTER*31. As such, it is not a null-terminated C string.

```
extern "C" struct sampl2_ {
    float PROB1[NM2];
    float PROB2[NM2];
    float AM2[NM2];
    float GAM2[NM2];
    float AM3[NM2];
    float GAM3[NM2];
};

extern "C" struct sampl3_ {
    float PROB1[NM3];
    float PROB2[NM3];
    float AMRX[NM3];
    float GAMRX[NM3];
    float AMRA[NM3];
    float GAMRA[NM3];
    float AMRB[NM3];
    float GAMRB[NM3];
};

extern "C" struct sampl4_ {
    float PROB1[NM4];
    float PROB2[NM4];
    float AMRX[NM4];
    float GAMRX[NM4];
    float AMRA[NM4];
    float GAMRA[NM4];
};

extern "C" struct sampl5_ {
    float PROBa2[NM5];
    float PROBOM[NM5];
```

---

[7]Note that this arrangement implies, that anomalous, neutrinoless channels can be initialized only for 3 or 2 particles final states. We think it is sufficient for foreseeable future.

```
    float ama2[NM5];
    float gama2[NM5];
    float AMOM[NM5];
    float GAMOM[NM5];
};
```

The structures collect parameters for the presamplers and channels of multiplicity corresponding respectively from 2 to 5 hadrons in the final state. Those parameters are probabilities for generation sub-channels or coefficients, such as masses and widths of intermediate peaks. Modification of a presampler affects efficiency of generation only. It may be necessary if substantially modified hadronic current is used, for example, if user plans to check the code in intermediate resonances narrow width limit.

## 3.2 Wrappers

Let us document functions that have to be written by the user when an interface to the new software structure is introduced[8]. These wrappers are used to pass parameters from FORTRAN part of the code to the user-provided functions. That is why the format of these functions must be strictly followed. Each function name consist of name of the current (or matrix element) as used in older pure FORTRAN versions of TAUOLA but with _wrap added at the end. For example, wrapper for user routine to replace DAM2PI function is declared as:

```
extern "C" void dam2pi_wrap_
        (int *MNUM, float *PT, float *PN, float *PIM1, float *PIM2,
         float *AMPLIT, float *HV)
```

These wrappers depend on multiplicity of final states but in most cases are prepared accordingly to the same scheme. The exceptions are listed at the end of this Section.

---

[8]An implementation of these functions have been presented in `tauola-c/channel_wrappers.c` file. This implementation demonstrates how parameters of the functions for calculation of matrix element or hadronic currents should be used. If the C++ interface is to be removed, e.g. as the first step of writing another one, possibly to different laguage, use of code of `tauola-no-c/` directory, instead of `tauola-c` may be useful.

Note that such wrappers are executed only if user have registered a custom channel (with matrix element or hadronic current, i.e. KEY2[mnum]=4 or 5) for particular multiplicity and particular mnum (mnum denotes position on the sub-list of decays corresponding to the multiplicity).

The FORTRAN common blocs are not sufficient to describe initialization of TAUOLA. The table, like Tauolapp::taubra_userChannels[channel] of our interface, described in Appendix A is needed. Let us recall essential points only, leaving description of the Interface to the Appendix. The function RegisterChannel of this interface can be used. The position on the list of all channels is calculated from the position of channels with two scalars in final state MNUM as NM4+NM5+NM6+NM3+mnum. Then, a pointer to the user routine is taken from the table:
Tauolapp::taubra_userChannels[channel].
This pointer is set to user routine executed in place of function:
xsec(pt,pn,pim1,pim2,amplit,hv)
If for some reason this pointer is not set, program will exit with an error:
ERROR: dam2pi_wrap: pointer to xsec for channel 3 (mnum 3) not set!"

The remaining part of each function-wrapper is devoted to debugging. It will activate if any of the calculated quantities, matrix element squared or polarimetric vector is of a NaN value.

The following subsubsections list all wrappers used in the project. We use the C/C++ declaration to describe these wrappers as it is less intuitive than the original FORTRAN declaration, which can be deducted much easier.

### 3.2.1 Wrappers for squared Matrix Elements

```
extern "C" void dam1pi_wrap_
    (int *MNUM, float *PNU, float *AMF0, float *PKK, float *AMF1,
    float *GAMM, float *HV)

extern "C" void dam2pi_wrap_
    (int *MNUM, float *PT, float *PN, float *PIM1, float *PIM2,
    float *AMPLIT, float *HV)

extern "C" void dam3pi_wrap_
    (int *MNUM, float *PT, float *PN, float *PIM1, float *PIM2,
```

```
        float *PIM3, float *AMPLIT, float *HV)

  extern "C" void dam4pi_wrap_
      (int *MNUM, float *PT, float *PN, float *PIM1, float *PIM2,
      float *PIM3, float *PIM4, float *AMPLIT, float *HV)

  extern "C" void dam5pi_wrap_
      (int *MNUM, float *PT, float *PN, float *PIM1, float *PIM2,
      float *PIM3, float *PIM4, float *PIM5, float *AMPLIT, float
      *HV)
```

### 3.2.2 Wrappers for Hadronic Currents

```
  extern "C" void curr2_wrap_
      (int *MNUM, float *PIM1, float *PIM2, complex *HADCUR)

  extern "C" void curr3pi_wrap_
      (int *MNUM, float *PIM1, float *PIM2, float *PIM3, complex *HADCUR)

  extern "C" void curr4_wrap_
      (int *MNUM, float *PIM1, float *PIM2, float *PIM3, float *PIM4,
      complex *HADCUR)

  extern "C" void curr5_wrap_
      (int *MNUM, float *PIM1, float *PIM2, float *PIM3, float *PIM4,
      float *PIM5, complex *HADCUR)
```

### 3.2.3 Special cases

```
  extern "C" void dampry_wrap_
      (int *ITDKRC, double *XK0DEC, double *XK, double *XA, double
      *QP, double *XN, double *AMPLIT, double *HV)
```

Matrix element calculation for leptonic decays, including possibly 4-momentum for bremsstrahlung photon and its phase space limit `XK0DEC`.

`extern "C" float sigee_wrap_(float *AMX2, int *MNUM)`
In some decays approximation is used. Only distribution of invariant mass for all hadrons is generated accordingly to function `sigee` as if it was vector state. The final states of hadronic system fragmentation are generated accordingly to flat phase space.

`extern "C" float fconst_wrap_(int *MNUM)`
In this case no matrix element is used at all. Only overall constant is passed to generation. Flat phase space (all mass effects included) distribution is generated.

### 3.3 Generation monitoring variables

All structures presented so far, were for variables initialized at the vey beginning of initialization and left unchanged over the whole generation run. Wrappers are not expected to store information on the whole sample as well. This Subsection presents the variables used to access information which is continuously updated during the program run. Information on generation weights is stored in local variables of routine `DADNEW` for each semileptonic decay separately, and in routines `DADMEL` and `DADMMU` for leptonic decays:

```
      REAL*4 WTMAX(NMODE)
      REAL*8             SWT(NMODE),SSWT(NMODE)
      INTEGER*8 NEVRAW(NMODE),NEVOVR(NMODE),NEVACC(NMODE)
```

The variables denote:

`WTMAX` - maximum weight for the given channel,

`SWT` - sum of all event weights for the given channel,

`SSWT` - sum of all event weights squared for the given channel,

`NEVRAW` - number of raw generated events for the given channel,

`NEVOVR` - number of evens of weights larger than the maximum estimated at the initialization for the given channel,

`NEVACC` - number of accepted events for the given channel.

Further variables regarding the program run are stored in `common block TAUBMC`:

```
COMMON / TAUBMC / GAMPMC(500),GAMPER(500),NEVDEC(500)
```

The variables denote:

`GAMPMC(500)` - width for the given decayc channel calculated from the Monte Carlo generation,

`GAMPER(500)` - error of the `GAMPMC`,

`NEVDEC(500)` - number of generated decays.

Parts of the code and variables of the present Subsection will have to be adjusted at later steps of the program evolution, but as these parts do not affect the use of the presented methods for matrix element replacement we left them in `FORTRAN`.
This setup has to be modified if a multi-threaded parallelization process is to be used. At this moment it does not seem to be necessary, because with the present day matrix elements `TAUOLA` is sufficiently fast. Furthermore, applying parallelization solution as used by `KK` Monte Carlo is straightforward. It would only require to store run information, described in the present Subsection, into histograms.

# A  Example of the interface implementation

The main goal of new `TAUOLA` version was to prepare `FORTRAN` common blocks to facilitate addition of new currents and modification of existing ones[10] in cases when user main program is in `FORTRAN` or in `C++`. New setup allows for initialization or re-initialization of the currents definition at runtime. It also makes it easier to understand in principle modular structure of the algorithms used in the project.

Before describing the implementation of the interface provided with the source code we want to point out that most of the arrangements presented in this Section can be modified or replaced. All of presented methods, have to rewrite information stored in the data structures described in Section 3. We have prepared solutions on the basis of discussion with users, but alternative solutions may be found to be more convenient. The code presented in this Section may thus be replaced rather easily.

The `C++` interface for adding and modifying decay channels in `TAUOLA` is defined in `tauola-c/ChannelForTauolaInterface.c`. Most of the functions provided by this interface work on the instances of the class `ChannelForTauola` (defined in `tauola-c/ChannelForTauola.h`). Functions invoked from the FORTRAN code are defined in `tauola-c/channels_wrappers.c`. These functions are used by FORTRAN code of TAUOLA to access `ChannelForTauola` objects and in this way pass parameters further to user-initialized channels. In particular, to user routines for matrix elements or hadronic current calculation, accessed at the execution time through pointers. This approach allows to program in any language as long as proper wrappers are provided. Scheme of calling functions written in other languages from FORTRAN depends on a particular programming language and a set of compilers used to combine these languages.

---

[10] In principle, there is no need for any interface to modify hadronic currents used in `TAUOLA`. Nonetheless it may be convenient for some users.

Initialization of user defined code is performed in two steps. First, one has to call routine `Tauolapp::setUserRedefinitions(RedefExample);`[11]. Then `iniofc` routine (defined in `tauola-c/channels_wrappers.c`) has to be called to reinitialize TAUOLA with new user-updated currents.

Note that useful information can be found also in comments of the header files located in `tauola-c` subdirectory and in the example programs provided with the distribution. Most notably, an extensive example of use of the interface described in this Appendix is available in `demo-redefine` subdirectory. File `demo-redefine/iniofc.c` contains verbose comments about available functions, their effects and possible uses. All FORTRAN `common blocks` used by the interface are declared as external $\hat{C}$ structures in file `tauola-c/TauolaStructs.h`. Note that in this file parameters `NLT`, `NMODE`, `NM1`, `NM2`, `NM3`, `NM4`, `NM5` and `NM6` are defined. In order for the interface to work correctly, these definitions must coincide with the definitions located in `TAUDCDsize.inc`.

## A.1 Functions defined in `ChannelForTauolaInterface.h`

Let us describe functions presented in `tauola-c/ChannelForTauolaInterface.h` which form an example of such interface. Note that, as pointed before, this solution should serve as a prototype and can be easily replaced if the user wishes to introduce a new one. Especially if some of the functionality of this prototype is not needed. For example, significant part of the code in this implementation is devoted to preventing the coding errors when redefining or adding new channels through strict differentiation between channels of different multiplicity. Strict type checks are applied on such channels and function pointers passed to these channels. Because of that, our solution may be hard to read and it might hinder understanding the main purpose of each function.

---

[11]In example `demo-redefine` it has been shown how this routine can be called from FORTRAN via `subroutine TAUOLAREDEF` defined in `demo-redefine/iniofc.c`

On the other hand, understanding the interface details may not be necessary to perform basic operations and the following explanations should be enough to make a good use of this interface.

`extern ChannelForTauola* taubra_userChannels[500];`
  Holds pointers to user defined `ChannelForTauola` objects. It is a supplement to struct `taubra_` defined in `tauola-c/TauolaStructs.h`.

`extern void* leptonChannelsMEpointers[2];`
  Pointers to matrix elements for leptonic channels.

`extern void (*channelRedefinitionFunction)();`
  Pointer to tauola channel redefinition function.

`void PrintChannelInfo(int channel);`
  Prints current definition of a channel: channel number, branching ratio, multiplicity of hadrons in final state, sub-channel number, matrix element type, and identifiers of final state particles.

`void SetUserRedefinitions(void (*function)());`
  Sets pointer to function reinitializing Tauola channels. This function is the only place where channels can be reinitialized.

`int RegisterChannel(int channel, ChannelForTauola *pointer);`
  Changes information about selected `channel` or adds a new channel. Parameter `channel` defines place on the list of channels where new one is to be added. Setting it to `-1` means that new channel has to be added on the first empty slot. Parameter `pointer` is a pointer to an object defining new or modified channel. Function returns added or modified channel number or 0 in case of failure. If function succeeds, it takes ownership of the object pointed by the `pointer`. Same channel number cannot be registered twice. Use `OverwriteChannel` instead.

`int OverwriteChannel(int channel, ChannelForTauola *pointer);`
  Overwrites information about selected `channel` with new information provided by the `pointer`. Parameter `channel` defines place on the list of channels where new one is to be added. Parameter `pointer` is a pointer to an object defining new or modified channel. If function succeeds, it takes ownership of the object pointed by the `pointer`. Function returns added or modified channel number or 0 in case of failure.

`int ModifyLeptonic(int channel, int me, float br, void *xsec = NULL);`
Changes information about selected leptonic channel. This functionality is by far more delicate than that of function `RegisterChannel`, and user is requested to perform detailed tests of its performance. Parameter `channel` can be chosen only as 1 or 2. Parameter `me` is a matrix element type. See `README` for the meaning. Parameter `br` is a branching ratio of the selected channel. Parameter `current` is a pointer to function calculating the cross-section or current. Pointer `xsec` has to be of type `DAMPRY_POINTER_TYPE` defined in `tauola-c/ChannelForTauola.h`. Function returns modified channel number or 0 in case of failure.

`ChannelForTauola* GetChannel(int channel);`
Get channel information from `FORTRAN` common block. Returns a pointer to `ChannelForTauola` object with channel information filled, ready to register without any changes or to update and register. Such `ChannelForTauola` object can be only registered back on their original place, unless matrix element calculation is reduced to constant (flat phase-space), then it can be registerd on empty slot.

`int SetPresampler2(ChannelForTauola *pointer, float prob1, float prob2, float am2, float gam2, float am3, float gam3);`
Set parameters used to optimize efficiency of 2-scalar mode phase space generator. First argument is a pointer for `ChannelForTauola` for which presampler parameters will be changed. All of the following arguments are new numerical values of presampler parameters. Note: 0<=prob1; 0<=prob2; prob1+prob2<=1.

`int SetPresampler3(ChannelForTauola *pointer, float prob1, float prob2, float amrx, float gamrx, float amra, float gamra, float amrb, float gamrb);`
Set parameters used to optimize efficiency of 3-scalar mode phase space generator. First argument is a pointer for `ChannelForTauola` for which presampler parameters will be changed. All of the following arguments are new numerical values of presampler parameters. Note: 0<=prob1; 0<=prob2; prob1+prob2<=1.

`int SetPresampler4(ChannelForTauola *pointer, float prob1, float prob2, float amrx, float gamrx, float amra, float gamra);`
Set parameters used to optimize efficiency of 4-scalar mode phase space generator. First argument is a pointer for `ChannelForTauola` for which presampler parameters will be changed. All of the following arguments are new numerical values of presampler parameters. Note: 0<=prob1; 0<=prob2; prob1+prob2<=1; only prob1+prob2 is used.

`int SetPresampler5(ChannelForTauola *pointer, float proba2, float probom, float ama2, float gama2, float amom, float gamom);`
Set parameters used to optimize efficiency of 5-scalar mode phase space generator. First argument is a pointer for `ChannelForTauola` for which presampler parameters will be changed. All of the following arguments are new numerical values of presampler parameters. Note 0<=proba2<=1; 0<=probom<=1.

## A.2  Key functionality of the `ChannelForTauola` class

Constructors

Class `ChannelForTauola` provides eleven specialized constructors, one for each type of the user functions that can be passed to TAUOLA[12]. Each of these constructors accept the following parameters:

`float br` - branching ratio
`const vector<int> &ki` - list of the ID of the decay products
`string name` - name of the decay channel
`*_POINTER_TYPE pointer` - pointer to user function

The possible `POINTER_TYPE`s are listed at the top of the `tauola-c/ChannelForTauola.h` file. Initialization of parameters `mulpik` and `me_type` is done accordingly to this pointer type. For channels that are constructed without a pointer to function, a separate constructor has been provided that allows setting `mulpik` and `me_type` manually.

---
[12]Matrix elements for channels of multiplicity from 1 to 5, hadronic currents for channels of multiplicity from 2 to 5, function for a leptonic current, constant function

Accessors
> Get and set methods have been provided for all of the parameters of this class. The most notable are the accessors for `me_type`. The `me_type` can be set to following values:
>
> 0 - channel is not initialized,
> 1 - channel is reduced to constant matrix element, i.e. flat phase space is used,
> 2 - default matrix element is used,
> 3 - default matrix element is used but one stable particle is of spin>0,
> 4 - default matrix element is used with user hadronic current function,
> 5 - user-provided function is used for matrix element calculation.
>
> Note that setter for this parameter can only set it to values 0,1,2 or 3. Values 4 and 5 are reserved for the class constructors and are set based on the function pointer provided to the constructor.

`ChannelForTauola::print()`
> Prints information about this channel.

## B  Installation and examples

The installation procedure of the `tauola-bbb` is similar to the one of original `TAUOLA`. All relevant source code is located in `TAUOLA-FORTRAN/tauola-bbb` directory. Executing `make` in this directory builds all relevant sub-modules and produces the `glib.a` library[13]. Executing `make` in any demo subdirectory builds corresponding example. Running `./go` script in `prod` subdirectory of any demo directory runs the example with the default setup provided with the example.
There are three demo subdirectories prepared:

`demo-babar`

`demo-lfv`

`demo-redefine`

---
[13]Make sure that `HEPEVT` definition located in `TAUOLA-FORTRAN/include/HEPEVT.h` is exactly the same as used in target environment.

These examples are basically copies of each other with minor differences only. We decided to split examples into three folders, for user convenience to grasp properties of new version.
We also provide folders:

`patch-tauolapp`

`patch-KK-face`

`patch-babar-validation`

These Patch-folders include necessary information on how to use new tarball in other projects. The last one is with instruction how to reproduce results of validation for the new initialization.
In the following Subsections we describe briefly the content of the directories.

## B.1 Default BaBar initialization example

Directory `demo-babar` is to demonstrate the default initialization of `tauola-bbb`. This example, produces results validated against BaBar `KKMC` with BaBar `TAUOLA`. It is the simplest example provided with the distribution. It does not use `C++` interface for adding channels. This example is to show how the default `TAUOLA` output looks like.

## B.2 BaBar initialization used for validation against BaBar data

Directory `patch-babar-validation` contains README with a list of modifications that have to be applied to the `tauola-bbb` in order to allow comparison against BaBar data.
File `babar.root` contains an `MC-TESTER` analysis of the 1600 MEvents sample taken from the production files of BaBar experiment. File `tauola-bbb.root` contains an analysis of the sample of the same size, generated using `tauola-bbb` with the modifications described in this directory. Comparison output `babar.vs.tauola-bbb.pdf` is the result of the `MC-TESTER` comparison of these two samples.
The instructions located in this directory, show how to reproduce generation of the `tauola-bbb.root` file. See `MC-TESTER` documentation on how to perform comparison against `babar.root` sample to produce the `pdf` output.

## B.3 Example of adding LFV current to `tauola-bbb`

Directory `demo-lfv` expands the `demo-babar` example with the current described in Section 5, README located in this directory, describes modifications needed to add this new current to the project. In the directory, generation is set for sample of this single decay channel.

## B.4 Example of adding new channels and channel redefinition

Directory `demo-redefine` demonstrates options for the interface implementation described in Appendix A. File `iniofc.c` located in this directory contains extensive description of changes that can be introduced using this interface. This example does not introduce anything valuable in terms of physics content. Its use is to present the technical side of the project.

### B.4.1 Example of $\tau \to \pi\pi\nu_\tau$ current/matrix element replacement

The most useful example for learning how the new interface can be used is to replace one `FORTRAN` channel with it's `C++` version. In directory `demo-redefine` files `pipi0.c, pipi0.h, MEutils.c MEutils.h` contain `C++` version of $\tau \to \pi\pi\nu_\tau$ channel and its current. It is a copy of `FORTRAN` code[14].

User can replace `FORTRAN` hadronic current or the whole matrix element with the `C++` version. We have performed test to check that the `C++` current gives exactly the same results as its `FORTRAN` counterpart. With this, user explores baseline to edit the currents, as well as insight on how hadronic current should look like, if they are written from scratch. We believe, that this example should prepare user for developing his own channels. To turn this example on or off one should inspect `demo-redefine/iniofc.c` file.

## B.5 Using `tauola-bbb` with `Tauola Universal Interface`

Directory `patch-tauolapp` contains README file for the procedure needed to import `tauola-bbb` into the `C++` Interface of Tauola [16]. The steps are fairly simple. They boil down to copying the `tauola-bbb` subdirectory into `TAUOLA/tauola-fortran` subdirectory of the `C++` interface and applying small patches that allow the directory to be used from the `C++` interface.

---

[14] We tried to make it as similar to `FORTRAN` counterpart as possible, so user could compare it step by step. Note that there are some solutions that do not work in `C++` as a straightforward copies, e.g. filling 2-dim tables through simple equation, had to be replaced by two `for` loops.

## B.6 Using `tauola-bbb` with KKMC

Similarily, directory `patch-KK-face` includes `README` file for the procedure needed to use `tauola-bbb` with KKMC [15]. The key element is to correctly pass the branching ratio for the new channels provided by `tauola-bbb`. To achieve this, modifications described in `README` have to be applied to the default KKMC setup.

## B.7 New tar-ball modifications

The distribution tarball discussed in this paper is almost exactly the same as the version announced in [27]. Only minor changes were introduced, mostly affecting documentation (README files and examples of program use) but not the actual code.
In particular:

1. `README-changelog` in main `tauola-bbb` directory has been updated and one incorrect printout in `tauola-c/ChannelForTauola.h` has been fixed.

2. In file `pkorb.f`, lines 47-48 were added to initialize `BRA1` and `BRKS` to zero. It was to avoid undefined variables, which for some compilers may not be set to zero anyway, and resolve potential problem with CLEO parametrization. Note that this parametrization is presently defunct.

3. Instead of single `demo-standalone` of older versions of `TAUOLA`, several new, but similar examples described in previous Subsection have been prepared.

4. We have renamed folders to make clear distinctions between run examples (demo-*), and instructions for patching and validations (patch-*). `README` files were updated and some comments were added to account for the above changes.

5. Prints for outputs have been updated: new version number and release date has been provided.

## A2.5 TauSpinner

Following are excerpts from these four publications highlighting the most important technical aspects of the tool as well as the direction of changes and scope of modifications introduced in later versions of the tool. These excerpts are as follows:

- Appendix of Reference [6] (pages 11-13), which describes the modular approach to `TauSpinner` design. `TauSpinner` was designed as an add-on to `Tauola++` functionality and does not require technical knowledge about `Tauola++`. It does, however, require extensive physics knowledge to properly process and interpret the result of the tool. It also requires data to be correctly prepared before they can be processed.
- Appendix from Reference [92] (fragments of pages 14-15 shown on one page), which describes changes introduced after an example of application of `TauSpinner` to ascertain the spin of new resonances has been provided.
- Section 5 of Reference [91] (pages 11-14), which describes the technical details of how `TauSpinner` can be applied to studies of τ lepton polarization and spin correlations in Z, W and H decays. It also outlines the possible use cases of the package presented in the paper[87].
- Appendix from Reference [90] (pages 15-18), which describes how `TauSpinner` has been expanded to include mechanisms for simulating CP effects in events where Higgs decays to τ pair.

This tool has been described in Section 6.5 of this thesis.

---

[87] The Appendices of this reference contain benchmarks and numerical results that can be recreated using instructions provided in this Section .

## A   Requirements for data files

Data files generated by a MC event generator (or constructed using the tau-embedding method) need to fulfill the following requirements:

1. Four-momenta of the intermediate boson, the taus and the flavor and the four-momenta of the tau decay products need to be available.

2. Flavor of the intermediate boson needs to be available or set by the user.

3. For all types of hard processes, but $Z/\gamma^* \to \tau^-\tau^+$, the four-momenta can be defined in an arbitrary but common frame. For the $Z/\gamma^* \to \tau^-\tau^+$ process, the four-momenta have to be given in the laboratory frame in order to be consistent with the PDFs.

4. The four-momenta of the taus and their decay products need to be known with sufficient precision in order to ensure numerical stability of the algorithm. Six significant digits are recommended.

Note that different MC generators may store the truth information in different ways. It is the responsibility of a user to make sure that all these requirements are fulfilled.

## B   Public version

A generic version of the package can be found in [17]. The main code is written in C++ and relies upon two libraries: Tauola and LHAPDF [18]. A method for reading input information stored using the HepMC [19] format is prepared. Support for any other input format is available upon request.

### B.1   Project organization

The TauSpinner package is organized in the following manner:

- src/tau_reweight_lib.c, src/tau_reweight_lib.h - the core of the algorithm.
- src/Tauola_wrapper.h - wrapper for TAUOLA FORTRAN routines.
- src/SimpleParticle.h - definition of **class** SimpleParticle used as a bridge between the event record (or data file) and the algorithm.
- src/Particle.h - definition of **class** Particle used for boosting and rotation of the particles.
- src/read_particles_from_TAUOLA.c, src/read_particles_from_TAUOLA.h - interface to the HepMC::IO_GenEvent data files used by the example program.
- README - a short manual.

### B.2   The algorithm sequence

The TauSpinner takes the following sequence of steps:

**Initialization of** Tauola**.** It is ensured by invoking:
    Tauola::initialize();

**Initialization of** TauSpinner**.** It is performed by executing:
    void initialize_spinner(**bool** Ipp, **int** Ipol, **double** CMSENE)
    where the argument Ipp passes the information on the type of collision events (Ipp = true sets

*pp* collisions), `Ipol` passes the information on the spin effects included in the input sample (`Ipol=` 0, 1, 2 corresponds to no spin effects, complete spin effects and spin correlations only, respectively) and `CMSENE` sets the collision center of mass energy.

**Reading the data files.** Information on the four-momenta and the flavor of the boson, the final state taus or tau and neutrino pair and the tau decay products is filled and stored in instances of `SimpleParticle` class by the use of the function:
**void** `readParticlesFromTAUOLA_HepMC(HepMC::IO_GenEvent &input_file, SimpleParticle &boson,`
`SimpleParticle &tau, SimpleParticle &tau2, vector<SimpleParticle> &tau_daughters,`
`vector<SimpleParticle> &tau2_daughters)`.
This function should be modified if the input files are not in the `HepMC::IO_GenEvent` format.

**Calculation of the spin weight.** It is performed by the use of the following functions:
**double** `calculateWeightFromParticlesWorHpn(SimpleParticle &boson, SimpleParticle &tau,`
`SimpleParticle &tau2, vector<SimpleParticle> &tau_daughters)` for the $W^\pm \to \tau^\pm \nu$ and $H \to \tau^-\tau^+$ processes
**double** `calculateWeightFromParticlesH(SimpleParticle &boson, SimpleParticle &tau,`
`SimpleParticle &tau2, vector<SimpleParticle> &tau_daughters,`
`vector<SimpleParticle> &tau2_daughters)` for the $Z/\gamma^* \to \tau^-\tau^+$ and $H \to \tau^-\tau^+$ processes.

**Attributing tau helicity states.** For $Z/\gamma^* \to \tau^-\tau^+$ process, the tau helicity states are attributed at the stage of calculation of the spin weight. The information can obtained by calling `getTauSpin()` function.

### B.3 Calculation of the spin weight

For the $\tau^\pm \nu$ final states, the spin weight is calculated in the following steps:

1. The parent boson, the tau, the tau neutrino and the list of tau daughters are identified and boosted to the $\tau^\pm \nu$ rest frame in which the tau is aligned along the *z* axis.

2. The tau daughters are boosted to the tau rest frame. Two angles of spacial orientation of the neutrino from the tau decay, `theta2` and `phi2`, are calculated and stored. The tau daughters are rotated by these angles in order to align the neutrino along the *z* axis.

3. The `Tauola` decay channel is identified.

4. The `Tauola FORTRAN` subroutine is called to perform calculation of the polarimetric vector *h*.

5. The polarimetric vector *h* is rotated back using the `theta2` and `phi2` angles.

6. The spin weight is calculated using eq. 4 and returned to the main program.

For the $\tau^-\tau^+$ final states, the spin weight is calculated in the following steps:

1. The parent boson, the taus and their tau daughters are identified and boosted to the $\tau^-\tau^+$ rest frame in which the taus are aligned along the *z* axis.

2. For each tau:

    Its identified daughters are boosted to its rest frame. Two angles of spacial orientation of the neutrino from the tau decay, `theta2` and `phi2`, are calculated and stored. The tau daughters are rotated by these angles to align the neutrino along the $z$ axis.

    The `Tauola` decay channel is identified.

    The `Tauola FORTRAN` subroutine is called to perform calculation of the polarimetric vector $h$.

    The polarimetric vector $h$ is rotated back using the `theta2` and `phi2` angles

3. In case of the $Z/\gamma^* \to \tau^-\tau^+$ decays:

    The probability $p_\tau^Z$ is calculated using eqs 1-2.

    The spin weight is calculated using eq. 6 and returned to the main program.

4. In case of the $H \to \tau^-\tau^+$ decays:

    The spin weight is calculated using eq. 8 and returned to the main program.

**B.4   The LHAPDF library wrapper**

The evolution of the PDF's is invoked from the wrapper for PDF's:

**double** `f(double x, int ID, double SS, double cmsene)`

where function `f` calls the evolution function `xfx(x, SS, ID)` [18]. The PDF sets need to be available locally. They can be obtained from the LHAPDF project website.

Fragments of pages 14-15 of Ref. [92] **S. Banerjee, J. Kalinowski, W. Kotlarski, T. Przedziński, Z. Wąs.** Ascertaining the spin for new resonances decaying into tau+ tau- at Hadron Colliders.

## A  TauSpinner - changes introduced in version 1.2

Since its first public version, described in [6], two new updates to `TauSpinner` has been introduced. First, `TauSpinner` has been merged into `Tauola++`[17] distribution and now, while working on this paper, it has been extended to add new functionality. In this section we list the changes between version 1.0 and 1.2.

- Merging with `Tauola++`
  `TauSpinner` now comes as an additional library to `Tauola++` distribution. `Tauola++` configuration scripts have been updated to accomodate this setup. As of writing this paper, `Tauola++` v1.1.1, featuring `TauSpinner` v1.2 has been installed in `GENSER`[27, 28] database.

- Two new initialization options - `nonSM2` and `nonSMN`
  The `nonSM2` flag turns on non-Standard Model weight calculation. The `nonSMN` flag, combined with `nonSM2`, allows for calculation of corrections to shape only.

- New functions added.
  An example `examples/tau-reweight-test.cxx` has been updated to show how functions described below can be used in case of spin-2 calculation described in this paper.

    - `set_nonSM_born( double (*fun)(int ID, double S, double cost, int H1, int H2, int key) )`
      Sets function for user-defined born, including new physics. The parameters of this new function are described in `include/TauSpinner/nonSM.h` as well as in the example program.

    - `void setNonSMkey(int key)`
      Sets the value of `nonSM2` flag. Allows turning non-Standard Model calculation on and off for comparison between different models.

    - `double getWtNonSM()`
      Returns non-Standard Model weight `WT3` calculated for the last event processed by `TauSpinner`.

In our example `examples/tau-reweight-test.cxx` fortran function is provided to calculate quark level Born cross section where new physics effects can be switched on and off. Physics model described in the previous sections at the level of quark level annihilation into pair of tau leptons is used. Function:

```
REAL*8 FUNCTION DISTJWK(ID,S,T,H1,H2,KEY)
```

is used in our program with the help of the C++ function:

```
double nonSM_adopt(int ID, double S, double cost, int H1, int H2, int key)
```

Its use is initialized with the method `set_nonSM_born( nonSM_adopt );`
Other user defined function can be used in the same way.

Page 11 of Ref. [91] **A. Kaczmarska, J. Piatlicki, T. Przedziński, E. Richter-Wąs, Z. Wąs.** Application of TauSpinner for Studies on τ-Lepton Polarization and Spin Correlations in Z, W and H Decays at the LHC.

## 5 Technical details

For the purpose of this paper, a directory `TauSpinner/examples/applications`[11] has been added to the previous distributions of `Tauola++`. It contains several tools used to produce the plots for this paper and to obtain necessary results. It was also extended with several tests that help validate `TauSpinner`. If `Tauola++` is configured with all prerequisites needed to compile `TauSpinner` package, as well as `TauSpinner` examples[12], compiling these additional programs should not require any further setup and can be done by executing `make` in `applications` directory.

### 5.1 The `applications` directory

In the following subsection we will briefly describe the sub-directories for this package and their use.

#### 5.1.1 Generating plots

The main program, `applications-plots.cxx`, generates plots which are latter included in the pdf file (like of Appendix A). It uses the same algorithm as the one used in `tau-reweight-test.cxx`; part of the examples for `TauSpinner` included in `Tauola++` tar-ball starting from version of November 2012. In this example code, input file `events.dat` is processed and for each event $WT$ weight is calculated. The set of histograms is filled with weighted (to remove spin effects) and not weighted events, separately for each τ decay mode or τ pair decay mode combination. Histograming and plotting is done using the `ROOT` library [21] (also fits are performed with the help of `RooFit` library).

This program can be used to recreate plots in the Appendices. For this, a datafile with $W$ and $Z$ which decay into τ's is needed. Note that since the template `LaTeX` file is prepared for both $W$ and $Z$ samples, this program can be executed on a single sample file containing both types of events or on two samples with separate $W$ and $Z$ events[13]. Only channels $\tau \to \mu \nu_\mu \nu$, $\tau \to e \nu_e \nu$, $\tau \to \pi \nu$ and $\tau \to \rho \nu$ are analysed. To run the program:

- make sure that `ROOT` configuration is available through `root-config`,
- execute `make` in `TauSpinner/examples/applications` directory,

---

[11]In `Tauola++` v1.1.4, released on 12 Dec 2013, this directory was called `TauSpinner/examples/tauspinner-validation`. All subsequent directories and programs have been renamed following the new convention. In particular, directories: `applications-plots-and-paper`, `applications-rootfiles`, `applications-fits` was respectively called `tauspinner-validation-results`, `tauspinner-validation-plots`, `tauspinner-validation-fit` and programs `applications-plots.cxx`, `applications-comparison.cxx`, `applications-fits.cxx` were called `tauspinner-validation-plots.cxx`, `tauspinner-validation-comparison.cxx`, `tauspinner-validation-fit.cxx`. While the naming of programs and subdirectories changed, the content of the programs remained the same.

[12]Up-to-date instructions can be found on the `Tauola++` website in the documentation to the most-recent version of the package [28].

[13]This program does not produce histograms stored in Appendix B. These plots require change of the `PDGID` of the $Z$ boson so `TauSpinner` can calculate weight as if the intermediate boson is Higgs. This change is omitted from the example provided with the distribution tar-ball for simplicity.

- verify that settings in file `applications-plots.conf` are correct, including the path to input file[14],

- execute `./applications-plots.exe applications-plots.conf`.

A set of plots will be generated in the directory indicated by the configuration file (the default one is `applications-plots--and-paper`) and a breakdown of the τ decay channels found in the sample will be written at the end of running the program. If the input file contains both $W$ and $Z$ decays, two sets of plots will be generated, each accompanied with summary of the $W$ and $Z$ events properties. The program also saves all histograms created during processing time to `out.root` file. This file can be used to archive the results for further analysis or to add fits to the plots.

### 5.1.2 Adding fits

The code for adding fits is provided in the subdirectory `applications-fits`. It is built along with other programs when executing `make` in `applications` directory. This tool adds fits to the histograms generated by `applications-plots.exe` using the formulae (2) and (3), results of the fits are stored in the rootfile. See the `README` file in this directory for details on how input files are processed.

This program uses rootfiles from subdirectory `applications-rootfiles`. They are specified in the default configuration file `applications-fits.conf` as the input files of this program. The resulting plots, with added fit information on polarization, will be stored in `applications-plots-and-paper` directory. Previously generated plots will be overwritten. This can be changed in the configuration file with path to the output directory.

As mentioned in Section 4.3, the fit can be applied not to the whole range but to the interval $(x_1, x_2)$, that is why an option to perform fits only in the limited range of $[x_{min}, x_{max}]$ has been provided in the code and is controlled by the configuration file.

### 5.1.3 Recreating figures 3 and 4

The subdirectory `applications-rootfiles` contains rootfiles of histograms necessary to reproduce all plots shown in our paper. These rootfiles are used by `applications-fits.exe`. Histograms for the plots that are not part of the Appendices are also stored in the rootfiles. Executing `make` will invoke code to generate the plots for Figures 3 and 4. Note that generation of these rootfiles requires different setup and different data samples than for any other plots. While necessary changes are straightforward, including such options would add to the already complex structure of the validation programs, thus they were skipped in the distribution.

### 5.1.4 Additional tests and tools

Two additional subdirectories:

- `test-bornAFB`

- `test-ipol`

were added for further validation of the `TauSpinner` library. These tests are somewhat peripheral to the main topic of the paper, thus they were only documented in the `README` files of the corresponding sub-directories. The result of the first test is briefly discussed in Section 4.5, while the second one has not been presented here. It is, however, included in the package as a validation test of `TauSpinner` options (`Ipol` = 0, 1, 2, 3).

The `applications` directory contains additional programs:

- The `hepmc-tauola-redecay.cxx`, while not being an example for `TauSpinner`, can be used to process existing input file and remove τ decays substituting them with new ones generated by `Tauola++`. This tool can be used to generate unpolarized τ decays needed to verify different `TauSpinner` options (see Section 5.4). Note, that as with `Tauola++`, generation options are limited by the available information stored in the data files.

- The `applications-comparison.cxx`, uses two input files. First one is considered as a reference. For the second one `TauSpinner` weights are used. The same set of histograms is produced for both input files and compared afterwards. This program can be used to validate `TauSpinner` options, as for example in case E described in Section 5.4.

Details of how to use both programs are described in `README` of the directory.

---
[14]Note that example file `examples/events.dat` can be used to verify if the program compiles and runs correctly. However, it contains only a sample of 100 $Z \to \tau^+\tau^- \to \pi^+\pi^- \nu_\tau\nu_\tau$ events.

Page 13 of Ref. [91] **A. Kaczmarska, J. Piatlicki, T. Przedziński, E. Richter-Wąs, Z. Wąs.** Application of TauSpinner for Studies on τ-Lepton Polarization and Spin Correlations in Z, W and H Decays at the LHC.

#### 5.1.5 Generating pdf file

The subdirectory `applications-plots-and-paper` contains the LaTeX files, as well as all other files necessary to prepare Appendices of this paper. Executing `make` in this directory generates the pdf file as of our paper.

Text of Appendices is stored in files: `appendixA.tex` and `appendixB.tex`. The user can thus easily re-attach results of the program run to the documentation of his own project starting from the template `user-analysis.tex`; the `make user-analysis` will include Appendices into short `user-analysis.pdf`.

#### 5.1.6 Final remarks

It is possible to redo, for the sake of documenting results of one own tests, all figures and other numerical results of the Appendix A (that is also of `user-analysis.pdf`). In case the physics assumptions are substantially different than the one used for the present paper, shapes of the obtained distribution may differ as well. In every case the following step have to be followed:

1. generate a sample of $W$ and $Z$ decays to τ; τ decaying to $\mu, e, \pi$ and ρ;

2. run `applications-plots.exe` on this sample;

3. run `applications-fits.exe` on the resulting rootfile and store the output in `applications-plots-and-paper` subdirectory;

4. execute `make` in `applications-plots-and-paper` subdirectory.

Further details on each of these steps, including more technical details on the output and input files, are given in the distribution tar-ball and in the `README` files located in `TauSpinner/examples/applications` directory and all of the subdirectories.

The numerical results of whole paper can also be reconstructed. Scripts for most of the necessary operations are prepared and documented elsewhere in the paper or in `README` files.

### 5.2 Input file formats

Essentially any `HepMC` [17] file (saved in `HepMC::IO_GenEvent` format) can be processed [15] Files with events stored in different format can be either converted to `HepMC` or interfaced using methods described in `TauSpinner` documentation and used in the default example `tau-reweight-test.cxx`.

Note that only the file `applications-plots-and-paper/input-file-info.txt` should be updated by the user with the information on the event sample processed. All other text files will be updated by the appropriate tools described in previous section. The content of these text files is included in the output file of pdf format, as shown in Appendix A.1.

### 5.3 Rounding error recovering algorithm

The τ leptons stored in data files can be ultra-relativistic. This may cause problems for the part of algorithm recalculating matrix elements for τ decays. For our example, there was no problem with errors from rounding numbers, but in general such problems are expected.

The following correcting algorithm is prepared:

1. For each stable τ decay product its energy is recalculated from the mass and momentum.

2. The four-momentum of the τ is recalculated from the sum of four-momenta of its decay products.

3. The algorithm performs check if resulting operation doesn't introduce sizeable modifications, incompatible with rounding error recovery. If it does, a warning message is printed. This may indicate other than rounding error, difficulty with the production file. For example, some decay products not stored (eg. expected as non-observable soft photons).

The algorithm is located in file `applications/CorrectEvent.h`. An example of its use is provided in `applications/hepmc-tauola-redecay.cxx`. By default, this algorithm is turned off.

---

[15] Note however that it is user responsability to verify that HepMC file contains events with correctly structured information for `TauSpinner` to find outgoing τ leptons and their decay products.

Page 14 of Ref. [91] **A. Kaczmarska, J. Piatlicki, T. Przedziński, E. Richter-Wąs, Z. Wąs.** Application of TauSpinner for Studies on τ-Lepton Polarization and Spin Correlations in Z, W and H Decays at the LHC.

## 5.4 Package use cases

This package can be used to validate several `TauSpinner` options representing different applications of `TauSpinner`. Such tests include, but are not limited to:

(A) **Applying longitudinal spin effects:** adding spin effect to an unpolarized sample using weights `WT` calculated by `TauSpinner`. For this purpose, set `Ipol=0` in the configuration file.

(B) **Removing spin effects:** removing spin effects from the polarized sample using weights calculated by `TauSpinner`. This is the default option used for our figures. The weight `1/WT` instead of `WT` should be used.

Note, that regardless of whether `Ipol=0` or `1`, `TauSpinner` works in the same manner. The two options are distinguished at the level of the user program only (use $1/WT$ instead of $WT$ to reweight events), as shown in our demo.

(C) **Working on the input file with spin correlations but without polarization:** initialize `TauSpinner` with `Ipol=2`. In this case `WT` will represent correction necessary for implementation of the full longitudinal spin effects. Analogously, if the sample feature τ polarization, but polarization is missing dependence on the τ leptons directions, `TauSpinner` should be initialized with `Ipol=3` and the missing dependence can be corrected with calculated weight `WT`.

(D) **Replacing spin effects of $Z/\gamma^*$ with the Higgs-like spin-0 state spin correlations**: This could be realized with weights ($\frac{1}{wt_{spin}}$ to remove spin effects of $Z/\gamma^*$ times $wt_{spin}^{\Phi}$ to introduce spin effects of $\Phi$) without modification of event kinematics. Due to large spread in the weights, this method introduces large statistical fluctuations. Alternatively, this can be realised by regenerating τ decays with `Tauola++` configured for the τ pair originating from the scalar state resonance.

(E) **Validation:** test is similar to test [A], we apply spin effect to a sample without polarization. However, for this test we take the polarized sample and replace its τ decays by new, non-polarized ones using `Tauola++`.
This allows to test different `Ipol` options as mentioned in test (C). It requires different setup and use of two input files. The `TauSpinner` should be executed on this new sample. The result should be then compared with the ones from *original* sample. Tools required to perform these steps are described in Section 5.1.4.

The results of the tests B and D are presented in the Appendices of this paper. The details of test C are described in README of `applications/test-ipol` subdirectory. Tools, that can be adapted to perform tests A and E, have been provided as well (see Section 5.1.4).

We have successfully performed tests A-E on samples generated with `Pythia8` + `Photos++` + `Tauola++` (in some cases `Pythia8` alone). Satisfactory results, of similar quality as discussed in our paper, sections 4.3 and 4.5 were always found. Further details for all of the cases listed above are given in the distribution tar-ball.

# A  TauSpinner technical details

## A.1  Initialization

In case of transverse spin effects, the following changes have to be introduced for `TauSpinner` initialization. The rest of its execution is performed as explained in the previous publications [8, 9]. In particular, all weights will still be calculated by calling the function `calculateWeightFromParticlesH`[9].

Let us give a few details concerning different options:

**Case of Higgs bosons**

- At initialization stage:
  `setHiggsParametersTR(-1.0, 1.0, 0.0, 0.0);`  for scalar Higgs or
  `setHiggsParametersTR( 1.0,-1.0, 0.0, 0.0);`  for pseudo-scalar Higgs

- For mixed parity state, use:
  `double theta = 0.2;`
  `setHiggsParametersTR(-cos(2*theta),cos(2*theta),`
  `-sin(2*theta),-sin(2*theta));`

**Case of DY process**[10]

- At initialization stage:
  `setZgamMultipliersTR(1.0,1.0, 1.0, 1.0);`
  the $R_{xx}, R_{yy}, R_{xy}$, and $R_{yx}$ components of the density matrix will be multiplied by these coefficients. The two files named `table1-1.txt, table2-2.txt` have to be present in the directory of the executable main program, exactly as in the case of `Tauola++ universal interface` [20]. If the tables are absent, or if their name is distinct, the transverse components of $R_{ij}$ will be equal to zero.

- At execution stage:
  The method `getZgamParametersTR(RXX, RYY, RXY, RYX);` can be used to port the numerical values of $R_{xx}, R_{yy}, R_{xy}, R_{yx}$ to the user program to monitor the values or to modify them by repeating the following:
  `setZgamMultipliersTR(...);`
  `WT1 = calculateWeightFromParticlesH(...);`
  for each individual event. Motivated by the results of our tests, we have chosen $R_{xy}$ and $R_{yx}$ to be zero in all DY cases.

Further details are given in the following sections and in `README` files of the distribution tar-ball. All the plots of the present paper can be reproduced with the help of new demonstration programs which are included in the `Tauola++ v1.1.5` (or later) distribution tar-ball. We are using the `ROOT`-based `MC-TESTER` package [37] to facilitate histograming and plotting.

---

[9]For the purpose of adding new functionality, `TauSpinner` must access private fields and functions of `TauolaParticlePair` class from `Tauola++ universal interface` library. This has been resolved through a proper `friend` declaration in `TauolaParticlePair.h` header file. No other modifications were needed.

[10]Transverse spin correlations for DY are available with the development version or release version distributions of `Tauola++ v1.1.5` or later.

## A.2 Electroweak corrections

In `Tauola++ universal interface` transverse spin effects are activated together with electroweak corrections of the `SANC` library, see Ref. [25]. Now, this is the case in `TauSpinner` as well. The $R_{xx}$ and $R_{yy}$ components of $\tau^+\tau^-$ pair spin density matrix have been ported to the `TauSpinner` code. Other, non-diagonal components, have been demonstrated to be suppressed to, or below the few percent level and are not taken into account. See the plots of the file `CP-tests/Z-pi/RijS-INTcosthe.root` discussed in the main body of the paper.

## A.3 Installation

As auxiliary material to this paper, the directory `TAUOLA/TauSpinner/examples/CP-tests` has been prepared. The main program `CP-test.cxx` is based on the default example program `tau-reweight-test.cxx` located in the `TAUOLA/TauSpinner/examples` directory. In order to compile it one has to:

- Configure `Tauola++` for compilation with `TauSpinner`, `HepMC` [38] and `MC-TESTER` [37]. See [20] and [8] for instructions on how to define the appropriate paths.

- If using `Tauola++` distribution tarball for LCG[11], available on the project website [40], no additional configuration is required. In case of complete distribution[12], execute `./configure` with appropriate paths in the `TauSpinner/examples` directory.

- Execute `make` in `CP-tests`.

The example located in this directory is prepared to work without modifications. Please see comments in the code for possible options.

## A.4 Executing tests

The directory `CP-tests` contains four tests located in subdirectories `H-rho`, `H-pi`, `Z-rho` and `Z-pi`. Each test can be executed by the command `../CP-test.exe` in one of those subdirectories. Each subdirectory contains a small `HepMC` file `events.dat` with a sample of 100 events to test the installation. See Appendix A.6 for instructions on how to set up generation of larger data samples using `Tauola++` and `Pythia8`. Executing the program with no parameters will process the predefined `events.dat` sample. Any sample can be processed with:

---

[11]With fixed initialization for $\tau$ decay matrix elements and prepared for installation in LCG library [39].
[12]Enabling, in particular, the change of $\tau$ decay matrix elements.

```
../CP-test.exe <data_sample> [<optional_events_limit>]
```

For each subdirectory, an appropriate `MC-TESTER` user analysis script is provided. Appropriate parameters are set with the help of `setHiggsParametersTR` and `setZgamMultipliersTR` in the main program `CP-test.cxx` file. Note that subdirectories `Z-rho` and `Z-pi` contain previously generated tables of electroweak one loop level results for the quark level differential cross section and $R_{ij}$[13], which are needed for these tests (see Appendix A.2).

For tests (and for our figures), `MC-TESTER` data files have been used with the results of processing samples of 1M events for pp@8TeV (or @14TeV) collisions generated with `Pythia8` [41] hard process option `HiggsSM:ffbar2H` ( `WeakSingleBoson:ffbar2gmZ`) turned on. We have checked that variation of the Higgs mass in range 120-125 GeV does not affect results beyond the statistical fluctuations. This applies to the case when ISR, FSR is activated in `Pythia` as well[14]. We have used `LHAPDF` dataset `cteq6ll.LHpdf` [32]. `MC-TESTER` user analysis scripts are adapted to the appropriate Higgs ($Z/\gamma^*$) states.

## A.5  Analyzing results

The result of the test described in previous section is stored in `mc-tester.root` file. It contains all plots defined in user analysis script file[15]. To compare two `MC-TESTER` rootfiles use `compare.sh` located in the `CP-tests` directory:

```
./compare.sh <file1.root> <file2.root>
```

The benchmark distributions will be stored in the section `USER HISTOGRAMS` of the booklet `tester.pdf` produced during comparison. Each directory provides one or more benchmark files that can be used if no changes have been introduced to user analysis script files.

Executing the program in the directory `CP-tests/H-rho` generates Figure 2 if root files `scalar.root`, `thet-0.2.root` are used as inputs. With `scalar-tauframe.root`, `thet-0.2-tauframe.root` used as inputs, the plots for Fig. 3 are generated instead. These plots, to a large degree, coincide with Figure 2 of Ref. [13]. Note the different choice of quantization frames used in these papers. If the directory `CP-tests/H-pi` is used, Fig. 1 is generated, reproducing Fig. 3 from [12] (files `scalar.root`, `pseudoscalar.root` are used). Not only the acollinearity of $\pi^\pm$ directions in the rest frame of the Higgs is histogrammed, but also the acoplanarity angle for the $\pi^-\tau^-$ and $\pi^+\tau^+$ planes of the same frame (it is not included in our paper).

---

[13] See `Z-pi/table1-1.txt`, `Z-pi/table2-2.txt`. Subdirectory `Z-rho` contains symbolic links to these two files. See [20] for instructions on how to generate these tables with modified parametrization.

[14] In this case, a large number of final states can be recognized by `MC-TESTER` – result of FSR activity in Z or H decays. On some platforms, for bigger samples, this may cause buffer overflow problems. Our examples are not expected to work for all possible options of constructing and storing the event records. They are supposed to work for the properly prepared data files only.

[15] User analysis `ROOT` scripts are the `.C` files located in the directory where the test is run. The name of those files ends with `*UserTreeAnalysis.C`). Plot definitions located in these files can be modified as needed, however new plot definitions mean that benchmark files provided with the distribution cannot be used for comparison. Keep in mind that in order for `MC-TESTER` to work correctly `$(MCTESTERLOCATION)` environmental variable must be set.

Numerical results for the DY sample of 1M events are collected in the files: `start60.root` and `transverse60.root` in the `CP-tests/Z-pi` subdirectory, and were used for the plot shown in Fig. 4 for the case with the selection $|\cos\theta_{planes}| > 0.5$, and files `start60-noc.root` and `transverse60-noc.root` without this selection.

## A.6 Generating data files

In this section we describe how data files can be generated for the tests. For this purpose, a subdirectory `CP-tests/generate-datafiles` has been provided with a program specifically configured for generating data files for these tests. Note that using this subdirectory requires a path to `Pythia8` to be provided during the configuration of `Tauola++` in addition to other paths[16]. To compile this program, execute ``make generate'' in the `CP-tests` directory. To run it, execute:

```
./generate.exe <output_file> <pythia8_config> <decay_channel> <events_count>
```

Two `Pythia8` config files have been prepared: `pythia.Z.conf`, `pythia.H.conf`, and they can be used in combination with `Tauola++` decay channel 3 or 4 to provide samples for $\pi$ or $\rho$ respectively[17].

The program is set up to strip event of all particles other than $H$ or $Z$ boson, $\tau$ and $\tau$ decay products. This information is enough for `TauSpinner` to work and output files are relatively small (around 4GB per 1M events).

---

[16]Contrary to `TAUOLA/examples` subdirectory, this program is set up only for `pythia` version 8.180 or later. When using older versions, changes to the use of `pythia8 HepMC` interface may have to be introduced.

[17]Tests described in this paper have been configured for proton-proton collisions at $\sqrt{s} = 14 TeV$, therefore `Pythia8` configuration files use the same setup.

## A2.6 MC-TESTER update

Following is the Appendix of the Reference [15] (pages 21 to 25) showing scope of the changes and describing solution applied during our work on the new version of MC-TESTER.

This tool has been described in Section 6.5 of this thesis.

# A Appendix: `MC-TESTER` setup and input parameters (update for ref. [1])

The values of the parameters used by `MC-TESTER` are controlled using the `SETUP.C` file. Some parameters may also be controlled using `FORTRAN77` interface routines or C++ methods (Section A.2). This provides runtime control over all parameters, yet allowing the user not to have `SETUP.C` at all. One should note that `SETUP.C` always has precedence over the default values set using `F77` or C++ code: it is always looked for in the execution directory.

## A.1 Definition of parameters in the `SETUP.C` file

There are three sets of settings inside `MC-TESTER` to be distinguished: the ones specific to the generation phase, the ones specific to the analysis phase and the ones that are used in both phases[14]. We describe only new features, quoting the scope of their use.

### A.1.1 Setup::UserTreeAnalysis

Type: char*
  Scope: generation
  Default: null
  DESCRIPTION: The name of a function that allows modification of the list of stable particles before histograms for the decay of `MC-TESTER` analyzed object are defined/filled in.
  IMPORTANT: The name that is attributed (eg. "MyUserTree") must be a valid method name existing in a C++ script file located in the working directory. The script must be given the same name as the method, and ended with a ".C" suffix. For example, for the "MyUserTree" method, the script filename would be `MyUserTree.C`. For further information on running and compiling scripts on the fly see `README.UserTreeAnalysis` in the `MC-TESTER/doc/` directory.
  Example of use:
    `Setup::UserTreeAnalysis = "MyUserTree";`
  or for the version compiled and present in `libMCTester` library:
    `Setup::UserTreeAnalysis = "UserTreeAnalysis";`
  In this case, the UserTreeAnalysis.C is not needed, as the built-in UserTreeAnalysis routine will be used.
  Parameters can be passed to the function. For example
    `Setup::UTA_params[0]=0.05;`
    `Setup::UTA_params[1]=0;`
    `Setup::UTA_params[2]=0;`
    `Setup::UTA_params[3]=0;`
    `Setup::UTA_params[4]=22;`
    `Setup::UTA_params[5]=111;`

---

[14]Some parameters from the generation phase (i.e. the description of generators) are stored inside an output data file. However, again for reasons of runtime control, their values may be altered at the analysis time using the `SETUP.C` file in the analysis directory.

```
Setup::UTA_nparams=6;
```
will pass `nparams=6` parameters to the function. For the actual meaning of the parameters if passed into `UserTreeAnalysis` as present in the library, see section 6.

### A.1.2 Setup::mass_power

Type: int
    Scope: generation
    Default: 1
    DESCRIPTION: This option changes the variable passed for histograming, from invariant mass to a power of invariant mass, at the generation step. It also modifies the title displayed on histograms from `Mass(1)` to an appropriate `Mass(value)`, showing that the power of the mass has been changed.
    NOTE: Acceptable values: from 1 to 9. Due to properties of the Lorentz group when this option has value=2 it is particularly suitable for tests of spin polarization, see section 6.1.
    Example of use:
```
  Setup::mass_power=2; //set histograms to invariant mass squared
```

### A.1.3 Setup::mass_scale_on

Type: bool
    Scope: generation
    Default: false
    DESCRIPTION: This option scales invariant masses for all plots of the decay channel to invariant mass constructed from all daughters combined. It scales the X values to the range (0,1).
    NOTE: When using this option consider setting default maximum bin value to 1.1, for nicer graphical representation.
    Example of use:
```
  Setup::mass_scale_on=true; //enables scaling of X axis
```

### A.1.4 Setup::use_log_y

Type: bool
    Scope: analysis
    Default: false
    DESCRIPTION: Enables the use of logarithmic scale in all histograms plotted by MC-TESTER. Turning this option on will draw the histograms in logarithmic scale, and mark a logarithmic scale along the right-hand-side Y axis. This option does not affect SDP calculation or the plot of the ratio of histograms, which remains linear. Its corresponding linear scale is marked on the left-hand Y axis.
    NOTE: This option, combined with previously presented defaults for UserTreeAnalysis can be particularly useful if infrared regulator sensitive particles, such as soft photons are present in the event records. See [20].

Example of use:
```
Setup::use_log_y=true; //enables logarithmic scale on Y axis
```

### A.1.5  Setup::rebin_factor

Type: int
  Scope: analysis
  Default: 1
  DESCRIPTION: One may want to define a large number of bins for the generation scope of MC-TESTER. The number of bins on the actual plots, can be adjusted at the analysis step. The contents of consecutive "rebin_factor" bins are summed together. Calculation of the SDP parameter is appropriately adjusted.
  NOTE: "rebin_factor" must be the natural divider of the number of bins declared during the generation scope of MC-TESTER.
  Example of use:
```
   Setup::rebin_factor=3; //reduces no.  of bins in all histograms by factor of 3
```

## A.2   C++ configuration of MC-TESTER

The configuration of MC-TESTER can be done directly in the main method of the C++ generation program, without the need for a SETUP.C file. This can be accomplished by including the header file Setup.H and setting parameters using the same syntax as described for SETUP.C files (see original documentation [1]). Setup should be done before calling the function MC_Initialize(). Note that if parameters are set in both the generation program and a SETUP.C file, the values present in SETUP.C will be given precedence.

# B   Appendix: updates to versions 1.24.2 and 1.24.3

## B.1   Changes introduced in version 1.24.2

To address the problems that are typically faced when MC-TESTER is installed in a new environment, or a new platform, an automated configuration step has been implemented in version 1.24.2. The configuration files required to set-up/compile/run MC-TESTER may be generated through a dedicated configuration script, which facilitates the GNU autoconf [21].
  To set up MC-TESTER using the new auto-configuration facility, proceed with the following steps:

- Execute ./configure with additional command line options:
  --with-HepMC=<path> provides the path to HepMC installation directory (alternatively HEPMCLOCATION system variable has to be set).
  --with-root=<path> Path to root binaries.
  --with-Pythia8=<path> Path to Pythia version 8.1 or later (this generator is used by

examples only)
`--prefix=<path>` provides the installation path. If this option is used `include/` and `lib/` directories will be copied to this `prefix <path>` when `make install` will be executed. If `--prefix=<path>` is not provided, the default installation directory `/usr/local` will be used

- Execute `make`; this will build `MC-TESTER`.

- To install `MC-TESTER` into the directory specified at step 1) through the `--prefix` parameter, execute `make install`; this will copy the include files and libraries into `include/` and `lib/` sub-directories.

- It is worth to mention that `./configure` scripts only prepare `make.inc` files. These files are rather short and can be easily modified or created by hand: one can also rename `README-NO-CONFIG.txt` to `make.inc` and modify it accordingly to instructions provided inside the file.

Further changes and bug-fixes were implemented too. However they do not require any changes in the way the program is used. Let us nonetheless list them here:

- A bug resulting in faulty functioning of the script `ANALYZE.C` was fixed. Previously, when comparing decay samples which differed by several distinct channels, the program was occasionally crashing.

- A bug resulting in faulty functioning of `UserTreeAnalysis` scripts was fixed. The program was crashing if `MC4Vector` was used inside the script.

- All offending statements resulting in compilation errors if '-ansi -pedantic' flags were activated have been removed now.

## B.2 LCG configuration scripts; available from version 1.24.2

For our project still another configuration/automake system was prepared by Dmitri Konstantinov and Oleg Zenin; members of the LCG project [22].

For the purpose of activation of this set of autotools-based installation scripts enter `platform` directory and execute there `use-LCG-config.sh` script. Then, installation procedure and the names of the configuration script parameters will differ from the one described in our paper. Instruction given in './INSTALL' readme file created by `use-LCG-config.sh` script should be followed. One can also execute `./configure --help`, it will list all options available for the configuration script.

A short information on these scripts can be found in `README` of main directory as well.

## B.3  Merging `MC-TESTER` output files; available from version 1.24.3.

Interest in using the program on distributed systems, such as the grid has been expressed on several occasions. This calls for new functionality: to merge several `mc-tester.root` files into a single one, corresponding to all event samples combined into one.
The `analyze/MERGE.C` script can be used for this purpose:

- Enter the `analyze/` directory.

- Execute `root -b MERGE.C` (or `root -b` and `.L MERGE.C`).

- Type `merge(<output file> , <input directory>/<first file> , [<pattern>])` (the last parameter is optional).

- Copy `<output file>` into `analyze/prod1/mc-tester.root`
  (or `analyze/prod2/mc-tester.root`).

Example:

```
root -b
root [0] .L <path_to_script>/MERGE.C
root [1] merge("out.root","samples/first.root","*.root")
```

The input to the script may consist of just the `<input directory>` path where reside the `.root` files to be merged. Alternatively, the name of the first `MC-TESTER` `.root` file to be merged ( `<input directory>/<first file>` ) can be explicitly given. Generator information will be taken from this first file. The script will search the `<input directory>` to merge all files matching the pattern. If no pattern is provided, the default pattern is `mc-tester_*.root`. The `<output file>` will feature all histograms for decay channels including user defined histograms. Histograms for decay channels of the same name, found in different files will be summed together. The histogram bin count and axis range of the first occurence will be used.

If histograms are found with a distinct axis range or number of bins, compared to other histograms with the same name, then the content of these files is ignored. However, if by mistake, the particular input file contains data for tests of another particle's decays, then all data from this file will be taken. All decay channels for all particles under consideration will be listed in the .pdf file constructed by `MC-TESTER` at the analysis step. Information printed on the front page might then be inconsistent, for example, the overall number of entries or the overall number of channels will represent decays of all particles. If the interest will be expressed in future, an analysis step can be adopted to handle such cases with better front page of the booklet.

If the script is used outside the `MC-TESTER/analyze/` directory, the `MCTESTERLOCATION` system variable needs to be set to the `MC-TESTER` root directory. In addition, user can create and adopt his own copy of `MERGE.C` and use it instead of the default one. Note that our script cannot be used with a version of `MC-TESTER` older than 1.24.3.